\documentclass[a4paper,12pt,reqno,oldfontcommands]{memoir} 

\pdfoutput=1

\usepackage[dvipsnames]{xcolor}
\usepackage[pscoord]{eso-pic}

\newcommand{\placetextbox}[3]{\setbox0=\hbox{#3}\AddToShipoutPictureFG*{\put(\LenToUnit{#1\paperwidth},\LenToUnit{#2\paperheight}){\vtop{{\null}\makebox[0pt][c]{#3}}}}}

\usepackage[]{datetime2}
\usepackage{ifpdf}

\ifdraftdoc 
	\usepackage{draftwatermark}				\SetWatermarkScale{0.5}
	\SetWatermarkText{\mathbf{IGOR}}
\fi

\settrimmedsize{297mm}{210mm}{*}
\settypeblocksize{634pt}{448.13pt}{*} 
\setulmargins{4cm}{*}{*} 
\setlrmargins{*}{*}{1.5} 
\setmarginnotes{17pt}{51pt}{\onelineskip} 
\setheadfoot{\onelineskip}{2\onelineskip} 
\setheaderspaces{*}{2\onelineskip}{*} 
\checkandfixthelayout
\OnehalfSpacing 
\setsecnumdepth{subsection} 
\maxsecnumdepth{subsubsection}

\usepackage[]{graphicx}

\usepackage[brazil, main=english]{babel} \usepackage[T1]{fontenc}
\usepackage[utf8]{inputenc}			

\usepackage{amsfonts}\usepackage{amssymb} 					
\usepackage{amsmath} \usepackage{slashed}
\usepackage{tensor}
\usepackage{utfsym} \usepackage[varvw, varbb]{newtxmath}
\usepackage{newtxtext}

\makeoddfoot{plain}{}{}{\thepage} \makepagestyle{myvf} 
\makeoddfoot{myvf}{}{}{\thepage} 
\makeevenfoot{myvf}{\thepage}{}{} 
\makeheadrule{myvf}{\textwidth}{2\normalrulethickness} 
\makeevenhead{myvf}{\small\textsc{\leftmark}}{}{} 
\makeoddhead{myvf}{}{}{\small\textsc{\rightmark}}
\newcommand{\clearemptydoublepage}{\newpage{\thispagestyle{empty}\cleardoublepage}}
\usepackage{import}
					
\usepackage[square,numbers,sort&compress]{natbib}		

\usepackage{tikz}					\usepackage{float}

\usepackage[font=small]{caption}

\usepackage{memhfixc}					\usepackage{enumerate}					

\usepackage{footnote} 
\usepackage[final]{pdfpages} 
\usepackage[]{booktabs}
\usepackage{fourier-orns}
\usepackage[]{microtype}

\usepackage{bm}
\usepackage[a4paper,top=3cm,left=3cm,right=2cm,bottom=2cm]{geometry} 
\makeatletter
\newif\iffelinenonum
\makechapterstyle{daleif3}{

\setlength\midchapskip{7pt}

}
\makeatother

\setsecheadstyle{\large\bfseries\MakeUppercase} \setsubsecheadstyle{\large\bfseries}

\usepackage{hyperref}              
\usepackage{xurl}						
\ifpdf
\hypersetup{pdfauthor={S\'ergio Martins Filho},pdftitle={Covariant quantization of gauge theories with Lagrange Multipliers},pdfsubject={PhD Thesis},pdfkeywords={quantum equivalence, First-Order formulation, Yang-Mills theory, Gravity, Lagrange multiplier},pdfproducer={LaTeX},pdfcreator={pdfLaTeX @ 2024},
    bookmarksnumbered=true,     
    bookmarksopen=true,         
    bookmarksopenlevel=1,       
    colorlinks=true,            
allcolors=Blue,
breaklinks=true,
}
\fi
\newsubfloat{figure}
\newsubfloat{table}

\makeindex

\begin{document}
\chapterstyle{daleif3}

\frontmatter
\pagenumbering{Roman}
\hypersetup{pageanchor=false}
\begingroup
\let\cleardoublepage\clearpage
\thispagestyle{empty}

\begin{titlingpage}
\begin{SingleSpace}

\begin{center}

{\fontsize{16}{16} \selectfont Universidade de S\~ao Paulo \\}
	\vspace{0.1cm}
	{\fontsize{16}{16} \selectfont Instituto de F\'{i}sica}
    \vspace{3.3cm}

	{\fontsize{22}{22}\selectfont 
        Quantização covariante de teorias de calibre com multiplicadores de Lagrange
\par}
    \vspace{2cm}

    {\fontsize{18}{18}\selectfont S\'ergio Martins Filho
    \par}

    \vspace{2cm}

\end{center}

\leftskip 6cm
\begin{flushright}	
\leftskip 6cm
Orientador:  Prof. Dr. Fernando Tadeu Caldeira Brandt  
\leftskip 6cm
\end{flushright}	

    \vspace{0.8cm}

\par
\leftskip 6cm
\noindent {Tese de doutorado apresentada ao Instituto de F\'{i}sica da Universidade de S\~{a}o Paulo, como requisito parcial para a obten\c{c}\~{a}o do t\'{i}tulo de Doutor em Ci\^{e}ncias.}
\par
\leftskip 0cm
\vskip 2cm

\noindent Banca Examinadora: \\
\noindent Prof. Dr. Fernando Tadeu Caldeira Brandt (IF-USP) \\
Prof. Dr. Oscar José Pinto Éboli  (IF-USP) \\
Prof. Dr. Diogo Rodrigues Boito (IFSC-USP) \\
Prof. Dr. Adriano Antonio Natale (IFT-UNESP) \\
Prof. Dr. George Emanuel Avraam Matsas (IFT-UNESP) \\
\vspace{2.3cm}

\centering
    {S\~ao Paulo \\  2024}
\end{SingleSpace}
\end{titlingpage}
\includepdf[pages=1]{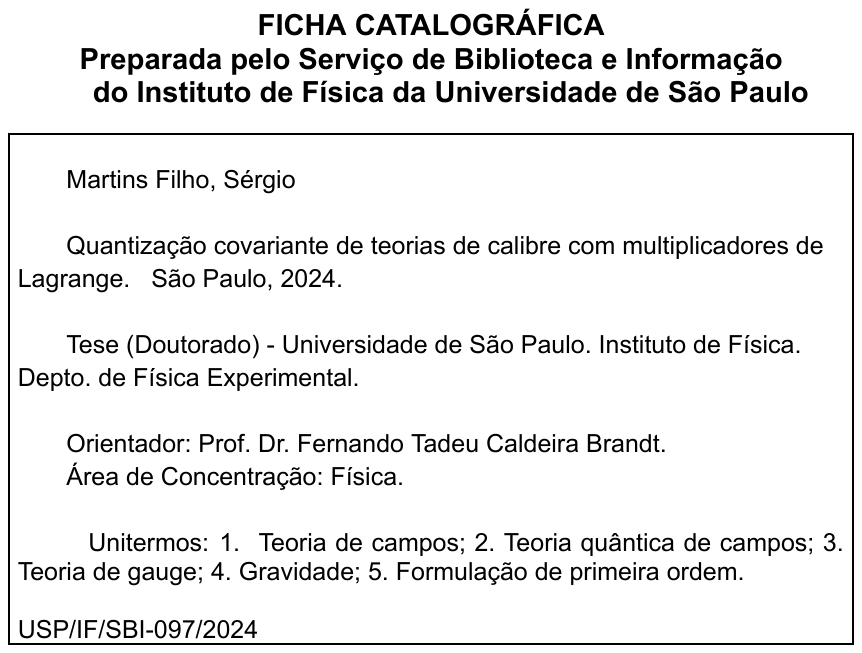}
\endgroup

\begin{titlingpage}
\begin{SingleSpace}

\begin{center}

{\fontsize{16}{16} \selectfont University of S\~ao Paulo \\}
	\vspace{0.1cm}
	{\fontsize{16}{16} \selectfont Physics Institute}
    \vspace{3.3cm}

	{\fontsize{22}{22}\selectfont 
        Covariant quantization of gauge theories with Lagrange multipliers
    \par}
    \vspace{2cm}

    {\fontsize{18}{18}\selectfont S\'ergio Martins Filho \par}

    \vspace{2cm}

\end{center}

\leftskip 6cm
\begin{flushright}	
\leftskip 6cm
Supervisor: Prof. Dr. Fernando Tadeu Caldeira Brandt 
\leftskip 6cm
\end{flushright}	

    \vspace{0.8cm}

\par
\leftskip 6cm
\noindent {Thesis submitted to the Physics Institute of the University of S\~ao Paulo in partial fulfillment of the requirements for the degree of Doctor of Science.}
\par
\leftskip 0cm
\vskip 2cm

\noindent Examining Committee: \\
\noindent Prof. Dr. Fernando Tadeu Caldeira Brandt (IF-USP) \\
Prof. Dr. Oscar José Pinto Éboli  (IF-USP) \\
Prof. Dr. Diogo Rodrigues Boito (IFSC-USP) \\
Prof. Dr. Adriano Antonio Natale (IFT-UNESP) \\
Prof. Dr. George Emanuel Avraam Matsas (IFT-UNESP) \\
\vspace{1.9cm}

\centering
    {S\~ao Paulo \\  2024}
\end{SingleSpace}
\end{titlingpage}
 \cleardoublepage
\hypersetup{pageanchor=true}
\newgeometry{right=2.75cm}
\thispagestyle{empty}
\pdfbookmark[0]{Dedication}{To}
\begin{flushright}
    \par\vspace*{0.925\textheight}{\textit{To Sergei Kuzmin}\par}
{\placetextbox{0.8}{0.253}{\textcolor{black!20}{\resizebox{4cm}{!}{\usym{1F344}}}}}
\end{flushright}
\restoregeometry

\cleardoublepage
\pdfbookmark[0]{Acknowledgments}{Acknowledgments}
\chapter*{Acknowledgments}
\thispagestyle{empty}
        \normalsize
I would like to thank my advisor, Fernando, for his insightful guidance and unwavering support throughout this journey. My deepest appreciation also goes to Josif and Gerry, whose collaboration has been both inspiring and enriching in numerous ways. I am thankful for the opportunity to engage with such amazing and caring individuals, including Pedro, with whom I have shared meaningful discussions on physics over the years. Lastly, I thank all the staff at IFUSP for providing the support and environment that made this work possible.

The present work was conducted with the support of CNPq, 
National Council of Scientific and Technological
Development - Brazil.

\clearemptydoublepage

\thispagestyle{empty}
\vspace*{0.05cm}
\textcolor{black!10}{\resizebox{0.75cm}{!}{\usym{1F34F}}}

\pdfbookmark[0]{Epigraph}{If}
\vspace*{0.125\textheight}
\textcolor{black!25}{\resizebox{0.75cm}{!}{\usym{1F34F}}}

\vspace*{0.45\textheight}
\textcolor{black!55}{\resizebox{0.75cm}{!}{\usym{1F34F}}}

\vspace*{3cm}
\setlength{\epigraphwidth}{6cm}
\epigraph{\textit{If I have seen further, it is by standing on the shoulders of giants.}}{Sir Isaac Newton}

\cleardoublepage

\pdfbookmark[0]{Abstract}{Abstract}
\chapter*{Abstract}
\thispagestyle{empty}
\begin{SingleSpace}
        \normalsize
We revisited the equivalence between the second- and first-order formulations of the Yang-Mills and gravity using the path integral formalism. We demonstrated that structural identities can be derived to relate Green's functions of auxiliary fields, computed in the first-order formulation, to Green's functions of composite fields in the second-order formulation. In Yang-Mills theory, these identities can be verified at the integrand level of the loop integrals. For gravity, the path integral was obtained through the Faddeev-Senjanović procedure. The Senjanović determinant plays a key role in canceling tadpole-like contributions, which vanish in the dimensional regularization scheme but persist at finite temperature. Thus, the equivalence between the two formalisms is maintained at finite temperature. 
We also studied the coupling to matter. In Yang-Mills theory, we investigated both minimal and non-minimal couplings. We derived first-order formulations, equivalent to the respective second-order formulations, by employing a procedure that introduces Lagrange multipliers. This procedure can be easily generalized to gravity. 
We also considered an alternative gravity model, which is both renormalizable and unitary, that uses Lagrange multipliers to restrict the loop expansion to one-loop order. However, this approach leads to a doubling of one-loop contributions due to the additional degrees of freedom associated with Ostrogradsky instabilities. To address this, we proposed a modified formalism that resolves these issues by requiring the path integral to be invariant under field redefinitions. This introduces ghost fields responsible for canceling the extra one-loop contributions arising from the Lagrange multiplier fields, while also removing unphysical degrees of freedom. 
We also demonstrated that the modified formalism and the Faddeev-Popov procedure commute, indicating that the Lagrange multipliers can be viewed as purely quantum fields.
    \end{SingleSpace}
    \par \vspace{0.25cm}
    \noindent\raisebox{.5ex}{\rule{\linewidth}{.8pt}}\par
    \noindent  {\small \textbf{Keywords:} First-Order formulation, Quantum equivalence, Yang-Mills theory, Gravity, Lagrange multiplier}

\clearemptydoublepage

\pdfbookmark[0]{Resumo}{Resumo}
\chapter*{Resumo}

\thispagestyle{empty}
\begin{SingleSpace}
        \normalsize
        Revisitamos a equivalência entre as formulações de segunda- e primeira-ordem das teorias de Yang-Mills e da gravitação usando o formalismo de integrais de trajetória. Demonstramos que identidades estruturais podem ser derivadas para relacionar as funções de Green de campos auxiliares, calculadas na formulação de primeira ordem, com as funções de Green de campos compostos na formulação de segunda ordem. Na teoria de Yang-Mills, essas identidades podem ser verificadas no nível do integrando das integrais de laço. Para a gravitação, a integral de trajetória foi obtida através do procedimento de Faddeev-Senjanovi\'{c}. O determinante de Senjanovi\'{c} desempenha um papel chave no cancelamento das contribuições do tipo \textit{tadpole}, que se anulam no esquema de regularização dimensional, mas persistem à temperatura finita. Portanto, a equivalência entre os dois formalismos é mantida à temperatura finita.
Também estudamos o acoplamento com a matéria. Na teoria de Yang-Mills, investigamos tanto os acoplamentos mínimos quanto os não mínimos. Derivamos formulações de primeira ordem com esses acoplamentos, equivalentes às respectivas formulações de segunda ordem, ao empregar um procedimento que introduz multiplicadores de Lagrange. Esse procedimento pode ser facilmente generalizado para a gravitação.
Consideramos também um modelo alternativo de gravitação, renormalizável e unitário, que usa multiplicadores de Lagrange para restringir a expansão de laço até a ordem de um laço. No entanto, essa abordagem leva à duplicação das contribuições de um laço devido aos graus de liberdade adicionais associados às instabilidades de Ostrogradsky. Para resolver isso, propusemos um formalismo modificado que resolve tais problemas exigindo que a integral de trajetória seja invariante sob redefinições de campo. Isso introduz campos fantasmas responsáveis por cancelar as contribuições extras de um laço provenientes dos campos multiplicadores de Lagrange, enquanto também remove os graus de liberdade não físicos.
Demonstramos também que o formalismo modificado e o procedimento de Faddeev-Popov comutam, indicando que os multiplicadores de Lagrange podem ser vistos como campos puramente quânticos.
    \end{SingleSpace}
\noindent\raisebox{.5ex}{\rule{\linewidth}{.8pt}}\par
    \noindent { \small \textbf{Palavras-chave:} Formulação de primeira ordem, Equivalência quântica, Teoria de Yang-Mills, Gravitação,  Multiplicador de Lagrange }

\clearemptydoublepage

\pdfbookmark[0]{List of publications}{List of publications}
\chapter*{List of publications}
\thispagestyle{empty}
\normalsize
This thesis is based on the following original publications: 
        \begin{itemize}
            \item  F. T. Brandt, J. Frenkel, \textit{S. Martins-Filho}, and D. G. C. McKeon, \href{https://doi.org/10.1103/PhysRevD.101.085013}{Consistency Conditions for the First-Order Formulation of Yang-Mills Theory}, Phys. Rev. D \textbf{101}, 085013 (2020).
            \item  F. T. Brandt, J. Frenkel, \textit{S. Martins-Filho}, and D. G. C. McKeon, \href{https://doi.org/10.1103/PhysRevD.102.045013}{Structural Identities in the First-Order Formulation of Quantum Gravity}, Phys. Rev. D \textbf{102}, 045013 (2020).
            \item  F. T. Brandt, J. Frenkel, \textit{S. Martins-Filho}, and D. G. C. McKeon, \href{https://doi.org/10.1016/j.aop.2021.168426}{On Restricting First Order Form of Gauge Theories to One-Loop Order}, Ann. Phys. \textbf{427}, 168426 (2021).
            \item  D. G. C. McKeon, F. T. Brandt, J. Frenkel, and \textit{S. Martins-Filho}, \href{https://doi.org/10.1016/j.aop.2021.168659}{Restricting Loop Expansions in Gauge Theories Coupled to Matter}, Ann. Phys. \textbf{434}, 168659 (2021).
            \item  F. T. Brandt, J. Frenkel, \textit{S. Martins-Filho}, D. G. C. McKeon, and G. S. S. Sakoda, \href{https://doi.org/10.1139/cjp-2021-0248}{Thermal Gauge Theories with Lagrange Multiplier Fields}, Can. J. Phys. \textbf{100}, 139 (2022).
            \item  F. T. Brandt and \textit{S. Martins-Filho}, \href{https://doi.org/10.1016/j.aop.2023.169323}{Field Redefinition Invariant Lagrange Multiplier Formalism}, Ann. Phys. \textbf{453}, 169323 (2023). 
 \item 
     D. G. C. McKeon, F. T. Brandt, and \textit{S. Martins-Filho}, \href{https://doi.org/10.1140/epjc/s10052-024-12764-z}{Field Redefinition Invariant Lagrange Multiplier Formalism with Gauge Symmetries}, Eur. Phys. J. C \textbf{84}, 399 (2024).
        \end{itemize}

\clearemptydoublepage

\renewcommand{\contentsname}{Contents}
\maxtocdepth{subsection}

\pdfbookmark[0]{List of abbreviations}{List of abbreviations}
\chapter*{List of abbreviations}
\thispagestyle{empty}
\renewcommand{\arraystretch}{1.5}
\begin{tabular}{lp{0.7\textwidth}}
\textbf{1PI}   &One-Particle Irreducible\\
\textbf{BRST}   &Becchi-Rouet-Stora-Tyutin\\
\textbf{DFOGR}   &Diagonal First-Order Gravity\\
\textbf{FOGR}   &First-Order Gravity\\
\textbf{FOYM}   &First-Order Yang-Mills\\
\textbf{FP}   &Faddeev-Popov\\
\textbf{FS}   &Faddeev-Senjanović\\
\textbf{GR}   &General Relativity\\
\textbf{HE}   &Hilbert-Einstein\\
\textbf{HP}   &Hilbert-Palatini\\
\textbf{LM}   &Lagrange Multiplier\\
\textbf{QCD}   &Quantum Chromodynamics\\
\textbf{QED}   &Quantum Electrodynamics\\
\textbf{SOGR}   &Second-Order Gravity\\
\textbf{SOYM}   &Second-Order Yang-Mills\\
\textbf{YM}   &Yang-Mills
\end{tabular}
\renewcommand{\arraystretch}{1}
\clearemptydoublepage

\pdfbookmark[0]{Contents}{Contents}
\pagenumbering{gobble}
\microtypesetup{protrusion=false}
\tableofcontents*
\microtypesetup{protrusion=true}
\clearemptydoublepage

\mainmatter
\setcounter{page}{23}
\chapter{Introduction} \label{section:intro}

Remarkably, all fundamental interactions can be described by gauge theories. The strong and electroweak interactions are well described by \emph{Yang-Mills} (YM) theories. Together with the \emph{Higgs mechanism} (which gives mass to particles), the \emph{Standard Model} is constructed, and it has been an astonishing success \cite{Gaillard:1998ui, ParticleDataGroup:2012pjm,  aiko:2023}. However, the Standard Model is not fully compatible with quantum gravity --- the quantized \emph{General Relativity} (GR). 

On the other hand, gravity can be described as an \emph{effective gauge field theory} \cite{Donoghue:2012zc}, which is consistent with the Standard Model at low energy scales. In this effective theory of gravity, diffeomorphisms are treated as gauge transformations. Hence, the Standard Model and gravity can be treated similarly. 

The standard formulation of these theories is given by \emph{second-order Lagrangians}, which contains at most second-order terms in derivatives. We shall call this formulation the \emph{second-order formulation}. Alternatively, we can describe such theories using \emph{first-order Lagrangians}, which we shall call the \emph{first-order formulation}. 

These formulations are equivalent at the classical level for both pure YM and gravity theories. At the quantum level, it can be shown in several ways: through their canonical Hamiltonian structure \cite{niederle:1983, McKeon:2010nf}, in the Batalin-Vilkovisky formalism \cite{Lavrov:2021pqh} or by directly demonstrating the equivalence of the path integrals of these formulations \cite{Chishtie:2012sq, McKeon:2020lqp, Brandt:2020vre}. Many explicit calculations have shown the equivalence between these formulations, we refer the reader to \cite{McKeon:2020lqp, Brandt:2020vre, martins-filho:2021}.

\section{First-Order formulation of gauge theories}

The first-order formulations offer several advantages \cite{McKeon:1994ds, Brandt:2015nxa}. The Feynman rules are greatly simplified, which significantly reduces the complexity of the quantum correction computations. This approach also allows for a clearer comparison between the Yang-Mills  theory and General Relativity by analyzing their first-order formulations. Moreover, the first-order formulation is more suitable for canonical quantization \cite{niederle:1983}, which can be especially relevant in the context of gravity.

In \cite{McKeon:2020lqp, Brandt:2020vre}, we have established their quantum equivalence through the generating functional $Z$. We also obtained a set of \emph{structural identities} that relates Green's functions computed in the first-order formulation with standard Green's functions of the second-order formulation. 
Here, we provide an investigation of the quantum equivalence through the connected generating functional $W$ and the effective action $ \Gamma $. The \emph{Slavnov-Taylor identities} \cite{Taylor:1971ff, Slavnov:1972fg} are also studied in the first-order formulation of the YM theory.

The structural identities also provide a physical interpretation of the auxiliary fields introduced in the first-order formulation and also reveal that these formulations may be used to obtain Green's functions of composite fields.  However, this was only valid when \emph{dimensional regularization} \cite{Leibbrandt:1975dj} was used in the case of pure gauge theories. 

In \cite{martins-filho:2021}, we have that, in YM theory, these identities hold at the integrand level. Thus, confirming their independence from the regularization scheme. Failing this, an inequivalence could arise at finite temperature, as contributions that are suppressed in dimensional regularization is used could become relevant at finite temperature (see Appendix~\ref{section:tadpoles}). To extend the quantum equivalence to finite temperature, it is necessary to resolve this issue. 

We demonstrate that this can be addressed by introducing an appropriate path integral for the first-order formulation of gravity. Using this generating functional, we verify that the structural identities hold at the integrand level, confirming their independence from the regularization scheme employed. This generating functional is derived by extending a procedure proposed originally for the YM theory. We establish the validity of this procedure for the YM theory from first principles. In order to achieve this, we need to examine the canonical structure of the first-order formulation of YM theory. This shows that quantizing the first-order formulation of the YM theory requires the application of the \emph{Faddeev-Senjanovi\'{c} procedure} (FS) \cite{Senjanovic:1976br}, which generalizes the \emph{Faddeev-Popov procedure} (FP) \cite{Faddeev:1967fc}. 

In \cite{Chishtie:2012sq}, the FS procedure is applied to the first-order formulation of gravity. However, the \emph{Senjanovi\'{c} determinant} is not manifestly covariant. Here, we propose a covariant procedure that leads to a similar determinant that is manifestly covariant. 
By exploring the parallel between YM and gravity, we propose that this determinant is the Senjanovi\'{c} determinant, obtained in \cite{Chishtie:2012sq}, in its manifestly covariant form. 

We can also consider the first-order formulation of the Standard Model. This is related to the coupling of matter fields to the first-order formulation of the YM theory and gravity. In the YM theory, the minimal coupling to matter fields, such as fermions and scalar fields, is realized by the connection $ A_{\mu} $, which is the gauge field. In the first-order formulation, these couplings are independent of the auxiliary field (the curvature tensor). Consequently, the first-order formulation of the YM theory coupled to these matter fields is equivalent to the respective second-order formulation. 

In contrast in gravity, the auxiliary field is the connection $ \Gamma^{\lambda} {}_{\mu \nu} $. When matter fields couple to the connection, it leads to inequivalence between the first- and second-order formulation of gravity, even at the classical level. The most relevant example  is the fermionic field \cite{Lagraa:2010cza}. Recently, \cite{Casadio:2022ozp} showed that the equivalence can be restored by considering a non-minimal coupling of fermions. However, it is only valid in the $(3+1)$-dimensional spacetime. 

Here, we study the non-minimal coupling of fermions to the YM gauge field. We develop a procedure to obtain the appropriate first-order formulation for these curvature-dependent couplings. This procedure can be easily generalized to gravity. Thus, it is important to investigate it in detail. It solves the issue of inequivalence due to connection-dependent couplings in the first-order formulation of gravity in any dimension. Nevertheless, the inequivalence of the \emph{Palatini formalism}, namely the first-order formulation of gravity, has been the focus of several recent studies in cosmology \cite{Jarv:2020qqm, Verner:2020gfa, Karam:2020rpa}. In particular, first-order formulations of $f(R)$ gravity models are considered \cite{Antoniadis:2018ywb, Pinto:2018rfg, Dioguardi:2021fmr}.

Given the potential inequivalence between these formulations, we also investigate an alternative gravity theory. It is obtained from the the so-called \emph{Lagrange Multiplier formalism} proposed in several papers \cite{McKeon:1992rq, Brandt:2018lbe, Brandt:2019ymg, Brandt:2020gms, Brandt:2021qgh, Brandt:2021nev}. 

\section{Lagrange Multiplier formalism}
This formalism consists of introducing Lagrange Multiplier (LM) fields to restrict the path integral to field configurations that satisfy 
the classical equations of motion. The resulting theory is both solvable and unitary, with the  perturbative expansion truncated to one-loop order while keeping the tree-level unaltered. In particular, for gravity, this formalism has GR as classical limit. 

However, the Lagrange multiplier formalism has certain drawbacks, which are related to the doubling of degrees of freedom. This alters the one-loop contributions, which are twice the usual one-loop contributions obtained in the theory without LM fields. These issues are associated with the presence of \emph{Ostrogradsky instabilities} \cite{Ostrogradsky:1850fid, Aoki:2020gfv}. In order to obtain a unitary theory, one has to resort to \emph{indefinite metric quantization} \cite{pauli:1943, Sudarshan:1961vs, Brandt:2021qgh}. In \cite{Brandt:2022kjo}, we proposed extending the standard LM formalism to resolve these issues. 

This \emph{modified LM formalism} is obtained by imposing field redefinition invariance on the path integral of the standard LM formalism. This leads to the introduction of ghost fields, which cancel the additional one-loop contributions arising from the LM fields. We argue that these ghosts are also responsible for removing the unphysical degrees of freedom associated with the LM fields. Thus, the unitarity of this field redefinition invariant LM formalism relies on the cancellation of unphysical states by ghosts. 

We review this formalism, in addition, we shall study certain \emph{Slavnov-Taylor-like} identities that arise from the modified LM formalism due to the ghost sector.  We show that applying this formalism to gauge theories is possible and requires only an extension of the FP procedure. 

\section{Overview}
In Chapter~\ref{section:YMtheory}, we review the second- and first-order formulations of YM theory. We also consider minimal and non-minimal couplings to matter fields. A toy model of the Standard Model is examined. The quantization is shown and we also present the Becchi-Rouet-Stora-Tyutin (BRST) symmetry \cite{Becchi:1974xu, Tyutin:1975qk} of both formulation. 
In Chapter~\ref{chapter:EH}, the quantization of the second- and first-order formulations of gravity are presented. The corresponding BRST symmetries are outlined. We establish the covariant path integral quantization of the first-order gravity theory. We also briefly discuss the diagonalization of the first-order formulation of gravity.  

The quantum equivalence is revisited in Chapter~\ref{section:QE}. We also extend the previous works and examine the quantum equivalence using the effective action $ \Gamma $. Here, we obtain the structural identities. We show that the structural identities can be verified at the integrand level, including in the case of gravity. This result is made possible by the path integral of the first-order formulation of gravity developed in Chapter~\ref{chapter:EH}. 

In Chapter~\ref{section:LMtheory}, we review the standard LM formalism and present the first-order gravity theory in this framework. We also discuss the drawbacks of this formalism in detail. A proposal to resolve these issues is introduced in Chapter~\ref{section:modLMchap}. We provide a diagrammatic analysis to illustrate the major characteristic of the modified LM formalism, namely, the truncation of the perturbative expansion to one-loop order. Moreover, we show that the ghost contributions are responsible for canceling the additional contributions arising from the LM fields. We present the symmetries of this formalism and derive Slavnov-Taylor-like identities. Finally, in Chapter~\ref{chapter:gtmLM}, we extend the modified LM formalism to gauge theories. To illustrate this, we consider the quantization of the first-order formulation of the YM Theory within this framework. We also provide a general proof of the commutation between the FP quantization and the LM formalism. 

In Chapter~\ref{section:discussion}, we present our final considerations and discuss our findings. We also present our future perspective on this research. Finally, in Appendix~\ref{section:DerivationTHETA} we derive the Chern-Simons current associated with the $ \theta $-term in detail. In Appendix~\ref{section:FeynmanRules}, the Feynman rules are given. The first-order formulation of a general scalar theory is presented in Appendix~\ref{section:FOspin0}. In Appendix~\ref{section:tadpoles}, we illustrate with an example that massless \emph{tadpole-like} diagrams (diagrams in which the loop are independent of external momenta), which vanish at zero temperature using dimensional regularization, may become relevant at finite temperature.

Throughout this monograph, we use natural units: $ \hbar = c = k_{B} = 1$ and take the flat metric $ \eta^{\mu \nu}$ with signature $  (+\,- \,-\, -)$. We assume the Einstein sum convention over repeated indices.

 \chapter{Yang-Mills theory} \label{section:YMtheory}

In this chapter, we review the second- and first-order formulations of YM theory. We also study the YM theory coupled to scalar and fermionic matter fields in both formulations, examining both minimal and non-minimal couplings. We develop a procedure to obtain the first-order Lagrangian from the second-order Lagrangian when coupled to matter fields using Lagrangian multipliers. Additionally, we consider the quantization of these theories and present the corresponding BRST symmetries. We also revisit the quantum equivalence using the generating functional $Z$. 

\section{Second-Order Yang-Mills theory} \label{sec:SOYM}

In this section, we review the standard formulation of the YM theory.\footnote{For a classical exposure, we refer to \cite{rubakov:2002}.} 
The YM theory is a general class of gauge theories based on non-Abelian \emph{gauge Lie groups}. The gauge group is the group of gauge transformations of a gauge theory, while a gauge Lie group is a gauge group which also is a Lie group (see Ref. \cite{rubakov:2002} for more detail). A Lie group is also a manifold, which allows us to interpret its structure both algebraically and geometrically. 

The YM theory was originally conceived by Yang and Mills \cite{Yang:1954ek} as a generalization of electrodynamics, which is a gauge theory based on the unitary group $U(1)$ (an Abelian group). Here, we consider a general YM theory in which the gauge group is a non-Abelian compact semisimple Lie group $G(N)$, where $N$ is an integer. In the YM theory, the field strength tensor is expressed as 
\begin{equation}\label{eq:tensor}
    F_{\mu \nu}^{a}(A) = \partial_{\mu} A_{\nu}^{a}  - \partial_{\nu} A_{\mu}^{a} + i g f^{abc} A_{\mu}^{b} A_{\nu}^{c}, 
\end{equation}
where $g$ is the coupling constant and $ f^{abc} $ are the structure constants. 
The structure constants describe the Lie algebra $ \mathfrak{g} $ associated with the gauge group $ G(N)$ and are defined by the commutator of the elements of the Lie algebra as 
\begin{equation}\label{eq:defce}
    [T^{a} , \,  T^{b} ] = if^{abc} T^{c}.
\end{equation}
These elements of the basis of this Lie algebra, $ T^{a} $, are called group generators (elements of the group are ``generated'' by the action of the exponential map over the Lie algebra). When the group $G$ is Abelian, the group generators commute and $ f^{abc} =0$ (which is the case of electrodynamics, equivalently we can set $g=0$ in Eq.~\eqref{eq:tensor}). 

Usually, we consider the special unitary group $SU(N)$\footnote{$SU(N)$ denotes the group of $ N \times N$ unitary matrices  with unit determinant.} since the Standard Model is constructed by gauge theories based on $SU(N)$, quantum chromodynamics (QCD) and the weak interaction theory. The Standard Model itself is based on the semisimple gauge group $G= SU(3) \times SU(2) \times U(1)$.\footnote{Actually, the gauge group is $ G/\bm{\Gamma}$, where $ \bm{\Gamma}$ is a subgroup of $\bm{Z}_6$ \cite{Tong:2017oea}.} When $G=SU(N)$, we have $N^2-1$ group generators and the latin indices $a$, $b$, $\ldots$ run from $1$ to $N^2-1$. In QCD, which is a YM theory with $G=SU(3)$,
these indices are called ``color indices''. Meanwhile, in the weak interaction theory, in which $G=SU(2)$, they are called ``isospin indices''. From now on, we shall refer to them as color indices, although we still consider a general compact semisimple gauge group $G(N)$. 

The dynamics of the pure YM theory can be described by 
\begin{equation}\label{eq:lagYMp1}
        {\mathcal{L}}_{\text{2YM}} = - \frac{1}{4} F^{a}_{\mu \nu} F^{a \, \mu \nu},
\end{equation}
which has the same structure of electrodynamics Lagrangian, although now the field strength tensor $ F^{a}_{\mu \nu} $ is given by Eq.~\eqref{eq:tensor}. This Lagrangian is invariant under the gauge transformations 
\begin{equation}\label{eq:gaugetransformpure}
    A_{\mu}'  = U A_{\mu} U^{-1} - U \partial_{\mu} U^{-1},
\end{equation}
where $U$ is any element of the gauge group $G(N)$. Eq.~\eqref{eq:gaugetransformpure} can be recast as the infinitesimal gauge transformation 
\begin{equation}\label{eq:infgaugetransform}
    \delta_{\zeta} A_{\mu}^{a} = D_{\mu}^{ab} \zeta^{b} 
\end{equation}
in which $ \zeta^{b} $ is a ordinary infinitesimal parameter and $ D_{\mu}^{ab} $ denotes the covariant derivative in the adjoint representation:
\begin{equation}\label{eq:dcovAdj}
    D_{\mu}^{ab} = \partial_{\mu} \delta^{ab} + g f^{apb} A_{\mu}^{p}.
\end{equation}

Geometrically, we can identify the gauge field $ A_{\mu} $ with a connection in the manifold of the gauge group $G(N)$, which is analogous to the connection $ \Gamma^{\lambda} {}_{\mu \nu} $ in GR\@. The connection arises to compare the local geometry of a manifold at two different points, for example, the parallel transport of a tensor along a curve in a manifold. 

Then, we have the notion of curvature of a manifold. 
The \emph{curvature} measures the non-commutativity of the covariant derivative that is equivalent to measuring how curved the manifold is \cite{wald:1984}. 
In the YM theory, we have that 
\begin{equation}\label{eq:curvature}
    [ D_{\mu} , D_{\nu} ] = i F_{\mu \nu},
\end{equation}
which is the field strength tensor.
This leads to Eq.~\eqref{eq:tensor} in the adjoint representation. 
It is enlightening to compare it to the definition of the \emph{Riemann curvature tensor} $ R_{\alpha \lambda \mu \nu} $ in gravity: 
\begin{equation}\label{eq:curvatureGR}
    \tensor{R}{^{\alpha}_{\lambda \mu \nu}} V^{\lambda}
    =[ \mathcal{D}_{\mu} , \, \mathcal{D}_{\nu} ] V^{\alpha},
\end{equation}
where $ \mathcal{D}_{\mu} $ denotes the covariant derivative in gravity and $ V^{\alpha} $ is any contravariant vector.

The (pure) YM theory formulated as in Eq.~\eqref{eq:lagYMp1} is the so-called ``second-order YM'' (SOYM) formalism since second-order derivatives appear in its Lagrangian \eqref{eq:lagYMp1}. We can recast it as a first-order theory, namely the first-order YM (FOYM) theory, by introducing an auxiliary field. This formalism is discussed in detail in Section~\ref{section:FOYM}. In the next section, we shall consider the introduction of matter fields and other couplings in the SOYM theory. 

The classical YM equations can be derived by using the Euler-Lagrange equations which yield
\begin{equation}\label{eq:ELYM2}
    D_{\mu}^{ab} F^{b \, \mu \nu}  =0.
\end{equation}
This is a second-order equation, which is also a consequence of the second-order nature of the SOYM formalism.

\subsection{Minimal coupling}\label{section:YMtheorywithFermions}

In this section, we show how matter fields are minimally coupled to the YM gauge field $ A_{\mu}^{a} $. First, we consider fermionic matter fields to present the minimal coupling prescription. Then, we use the same prescription to couple a scalar field to the YM theory deriving the so-called \emph{scalar YM} theory. This allows us to construct a simple model that resembles the Standard Model. 

In order to obtain the Lagrangian of the YM theory minimally coupled to fermions, we need to introduce the Dirac Lagrangian 
\begin{equation}\label{eq:YM:diracLag}
    {\mathcal{L}}_{\Psi} =\bar{\Psi} \left ( i \gamma^{\mu} \partial_{\mu} - m\right )\Psi,
\end{equation}
where $ \gamma^{\mu} $ are the Dirac matrices and $ \bar{\Psi} =  \Psi^{\dagger} \gamma^{0}$. However, it is not gauge invariant, since 
\begin{equation}\label{eq:noncovariant}
    (\partial_{\mu} \Psi)' = (\partial_{\mu} U) \Psi + U \partial_{\mu}\Psi \neq U \partial_{\mu} \Psi .
\end{equation}
In order to retain the gauge invariance of the YM theory, we have to replace $ \partial_{\mu} $ by the covariant derivative $ D_{\mu} $. 
The covariant derivative satisfies  
\begin{equation}\label{eq:transformcovD}
    D_{\mu}' \Psi ' = U D_{\mu} \Psi,
\end{equation}
transforming covariantly (in the same manner) to $ \Psi $.
Note that matter fields $ \Psi $ are in the fundamental representation of the gauge group: 
\begin{equation}\label{eq:transformMatter}
    \Psi = \begin{pmatrix}
        \psi_{1} \\ \vdots \\ \psi_N
    \end{pmatrix}
    \quad \text{and} \quad \Psi ' = U \Psi,
\end{equation}
transforming as ``vectors'', where $U$ is an element of the gauge group $G(N)$ and the components of $ \Psi $ ($ \psi_{1} $, \ldots, $\psi_N$) are spinor fields. In the fundamental representation, the covariant derivative is given by
\begin{equation}\label{eq:dcov}
    D_{\mu} = \partial_{\mu} - ig A_{\mu}^{p} T^{p}. 
\end{equation}
The prescription 
\begin{equation}\label{eq:YM:minimalcoupling}
    \partial_{\mu} \to D_{\mu}
\end{equation}
is known as \emph{minimal coupling}. 
The resulting Lagrangian is 
\begin{equation}\label{eq:YM:YMandfermionLag}
    {\mathcal{L}}_{\text{mYM}} = - \frac{1}{4} F^{a}_{\mu \nu} F^{a \, \mu \nu} + \bar{\Psi} (i \slashed{D} -m ) \Psi 
\end{equation}
in which we used the Feynman's slash notation $ \slashed{D} \equiv \gamma^{\mu} D_{\mu} $. In QCD, the matter fields $ \Psi $ are the quarks $q$ which are color charged. Note that, the Lagrangian \eqref{eq:YM:YMandfermionLag} is now invariant under the gauge transformations 
\begin{equation} \label{eq:gaugetransfrom}
    A_{\mu}' = U A_{\mu} U^{-1} - U \partial_{\mu} U^{-1} 
    \quad \text{and} \quad \Psi ' = U \Psi,
\end{equation}
where $U$ is any element of the gauge group $G(N)$.

Now, we can consider the case of a scalar matter field $ \Phi $. As the field $ \Psi $, the scalar matter field $ \Phi $ is in the fundamental representation of the gauge group $ G (N)$, then 
\begin{equation}\label{eq:transformMatterScalar}
    \Phi = \begin{pmatrix}
        \phi_1 \\ \vdots \\ \phi_N
    \end{pmatrix}
    \quad \text{and} \quad \Phi ' = U \Phi,
\end{equation}
where the components of $\Phi$ are complex scalar fields $ \phi_{i} $. heplacing the derivatives in the scalar field Lagrangian 
\begin{equation}\label{eq:YM:scalarlag}
    {\mathcal{L}}_{\Phi} = (\partial_{\mu} \Phi)^{\dagger} \partial^{\mu} \Phi - V( \Phi^{\dagger} \Phi )
\end{equation}
by the covariant derivatives Eq.~\eqref{eq:dcov} to retain gauge invariance, and adding the dynamics of the gauge field $ A$, we obtain the scalar YM Lagrangian 
\begin{equation}\label{eq:YM:scalarYMlag}
    {\mathcal{L}}_{\text{sYM}} = - \frac{1}{4} F^{a}_{\mu \nu} F^{a \, \mu \nu} 
    + (D_{\mu} \Phi )^{\dagger} D^{\mu} \Phi - V ( \Phi^{\dagger} \Phi)
\end{equation}
in which a scalar field (in the fundamental representation) is coupled to a YM gauge field. 
If $V ( \Phi^\dagger \Phi ) = m^2 \Phi^{\dagger} \Phi $,  then the Lagrangian \eqref{eq:YM:scalarYMlag} describes a massive Klein-Gordon scalar field. Meanwhile, when $ V ( \Phi^{\dagger} \Phi ) = \mu^{2} \Phi^{\dagger} \Phi - \lambda ( \Phi^{\dagger} \Phi )^{2} $ (with $ \mu^{2}  > 0 $, $ \lambda >0$), the field $ \Phi $ acquires a non-vanishing vacuum expectation value and the gauge field $A$ becomes massive through the Higgs mechanism. The Higgs boson of the Standard Model is described by this model by setting $ G = SU(2)$. 

In the case of a scalar field in the adjoint representation, we have to use the covariant derivative in the same representation given by Eq.~\eqref{eq:dcovAdj}.
The scalar field $ \Phi $ are now matrices that transforms as
\begin{equation}\label{eq:transformadjPhi}
    \Phi ' = U \Phi U^{\dagger}.
\end{equation}
 We are not interested in such matter fields, since, in the Standard Model, the Higgs field is in the fundamental representation of $ SU(2)$. Although in certain grand unifying theories, for example, in $SU(5)$ grand unifying theory \cite{Georgi:1974sy}, the Higgs field lives in the adjoint representation of $ SU(5)$.

Thus, we can construct a model described by the following general Lagrangian 
\begin{equation}\label{eq:YM:lagSM}
    {\mathcal{L}}_{\text{SM}} =
    - \frac{1}{4} F_{\mu \nu}^{a} F^{a \, \mu \nu} + \bar{\Psi} ( i \slashed{D} - m ) \Psi  
    + (D_{\mu} \Phi )^{\dagger} D^{\mu} \Phi - V ( \Phi^{\dagger} \Phi)
\end{equation}
that resembles the Standard Model. In this model, we have a fermionic matter field $ \Psi $ and a scalar field $ \Phi $ (which can play the role of the Higgs boson) minimally coupled to the gauge field $A$. 

The Lagrangian in Eq.~\eqref{eq:YM:lagSM} leads to following equations of motion:  
\begin{subequations}
\label{eq:EoMlagSM}
\begin{align}
\label{eq:EoMlagSMa}
D_{\mu}^{ab} F^{b \, \mu \nu } & = J_{\text{SM}}^{a \, \nu} [ A ],  \quad 
\\
         (i \slashed{D} -m) \Psi = 0 \quad & \text{and} \quad 
\bar{\Psi} (i \overleftarrow{\slashed{D}} +m)  =0,\\
         D^2 \Phi + \frac{ \delta V( \Phi^{\dagger} \Phi )}{ \delta \Phi^{\dagger} } = 0\quad & \text{and} \quad D^2 \Phi^{\dagger} + \frac{ \delta  V( \Phi^{\dagger} \Phi )}{ \delta \Phi }=0.
\end{align}
\end{subequations}
The particular form of $J^{a \, \nu}_{\text{SM}}[A]$ does not play an important role in this work. However, for completeness, it reads
\begin{equation}\label{eq:WAsources}
-ig \bar{\Psi} \gamma^{\nu} T^a \Psi -ig (\partial^{\nu} \Phi^{\dagger}) T^{a}  \Phi   -ig T^{a} \Phi^{\dagger}   \partial^{\nu} \Phi + g^{2} T^{a} \Phi^{\dagger} A^{b \, \nu } T^{b} \Phi + g^{2} A^{b \, \nu } T^{b} \Phi^{\dagger} T^{a} \Phi.
\end{equation}

\subsection{Non-Minimal coupling}\label{section:YMnonMINIMAL}

Here, we briefly discuss two non-minimal couplings: a Pauli-like coupling $ \sigma^{a}_{\mu \nu} F^{a \, \mu \nu} $ and the \emph{$ \theta $-term} 
\begin{equation}\label{eq:YM:thetaterm}
    \frac{\vartheta}{8} \epsilon^{\mu \nu \rho \sigma} F^{a}_{\mu \nu} F^{a}_{\rho \sigma} 
\end{equation}
($ \epsilon^{\mu \nu \rho \sigma} $ is the Levi-Civita tensor, and we defined $ \vartheta \equiv \theta /2\pi^2$ \cite{Forkel:2000sq, Kim:2008hd}). These couplings are not present in the Standard Model. 
The Pauli coupling in quantum electrodynamics (QED) is non-renormalizable, while the $\theta$-term is a $\mathcal{CP}$ (charge conjugation and parity) symmetry-breaking term that should appear in QCD due to a non-trivial vacuum \cite{fried:1993}. However, the parameter $\vartheta$ is very small, on the order of $10^{-11}$, leading to the so-called \emph{strong $\mathcal{CP}$ problem} (see \cite{Kim:2008hd} and references therein). Indeed, there is no evidence of $\mathcal{CP}$ symmetry violation in strong interactions, even though the $\theta$-term is renormalizable and allowed by symmetry.

In gravity, matter fields are coupled to the curvature through the energy-momentum tensor. In contrast, in YM theory, as we have seen, interactions are predominantly curvature-independent (independent of the field strength tensor). These non-minimal couplings are among the few examples of curvature-dependent interactions in YM theory.

In QED, the Pauli interaction term is given by $ \mu_{F}   \bar{\psi} \sigma_{\mu \nu} F^{\mu \nu} \psi$, where $ \sigma_{\mu \nu} = i[ \gamma_{\mu} , \, \gamma_{\nu} ] /2$ and $ \gamma_{\mu} $ denotes gamma matrices. It is associated with an anomalous magnetic moment (intensity measured by $ \mu_F$) of a charged fermion, such as an electron or a muon \cite{das:2006}.  The Pauli interaction is non-renormalizable since its dimension is $ [ \bar{\psi} \sigma_{\mu \nu} {F}^{\mu \nu} \psi ] =5 $. Nevertheless, the field theory described by the Lagrangian 
\begin{equation}\label{eq:YM:lagPI}
    {\mathcal{L}}_{\text{PC}} = 
    - \frac{1}{4} F_{\mu \nu} F^{ \mu \nu} + \bar{\psi} ( i \slashed{D} - m ) \psi + \mu_{F} \bar{\psi} \sigma_{\mu \nu} F^{\mu \nu} \psi 
\end{equation}
can be regarded as an effective field theory. 

On the other hand, the $ \theta $-term is topological, that is, it is a total derivative. The $ \theta $-term defined in Eq.~\eqref{eq:YM:thetaterm} can be rewritten as a total derivative using the identity (see Appendix~\ref{section:DerivationTHETA})
\begin{equation}\label{eq:b1:totalderivative}
     F^{a \, \mu \nu} {\star}{F}^{a}_{\mu \nu} = \partial_{\mu}  \left[\epsilon^{\mu \nu \rho \sigma} \left( A_{\nu}^{a} {F}^{a}_{\rho \sigma} - \frac{2}{3} g f^{a b c} A_{\nu}^{a} A_{\rho}^{b} {A}_{\sigma}^{c} \right)\right],
\end{equation}
where  $ {\star}{F}^{a}_{\mu \nu} $ denotes the dual of $ F^{a}_{\mu \nu} $ given by 
\begin{equation}\label{eq:YM:dualofF}
    {\star}{F}^{a}_{\mu \nu} = \frac{1}{2} \epsilon_{\mu \nu \rho \sigma} F^{a \, \rho \sigma}.
\end{equation}
In a classical field theory, the $ \theta $-term can be disregarded. In QCD, the $ \theta $-term is required, in principle, to account for its non-trivial vacuum. Indeed, we have that 
\begin{equation}\label{eq:integralofTHeta}
    \begin{split}
    \int \mathop{d^{4} x} 
    F^{a \, \mu \nu} {\star}{F}^{a}_{\mu \nu}  
    ={}&\int \mathop{d^{3} x} 
    \epsilon^{\mu \nu \rho \sigma} \left( A_{\nu}^{a} {F}^{a}_{\rho \sigma} - \frac{2}{3} g f^{a b c} A_{\nu}^{a} A_{\rho}^{b} {A}_{\sigma}^{c} \right)
    \\={}&\frac{16\pi^2}{g^2 } \times \text{integer} 
\end{split}
\end{equation}
does not vanish by instantons contributions \cite{Forkel:2000sq}. The integer in Eq.~\eqref{eq:integralofTHeta} is associated with the winding number which appears as a topological charge.

\subsection{Quantization} \label{section:quantizationSOYM}

In this section, we present concisely the path integral quantization of the SOYM theory. It is the standard formulation of the YM theory and its quantization appears in most quantum field theory books \cite{peskin:2018,Ryder:1985wq,das:2006, dasLecturesQuantumField2020}.

The path integral quantization of gauge theory can be realized using the FP procedure. First, we need to fix the gauge. We use the Lorenz gauge condition given by
\begin{equation}\label{eq:gaugeL}
    G^{a} [A] = \partial^{\mu} A^{a}_{\mu }=0.
\end{equation}
This leads to the gauge fixing term  
\begin{equation}\label{eq:laggauge}
    {\mathcal{L}}_{\text{gf}}^{}(A) =-\frac{1}{2 \alpha} ( \partial^{\mu} {A}_{\mu}^{a} )^{2},
\end{equation}
where $ \alpha $ is a real constant. For $ \alpha = 0$, we have the Landau gauge which is classically equivalent to the Lorenz gauge. The Feynman gauge, which simplifies the computation of amplitudes in the perturbative  approach, corresponds to set $ \alpha =1$.

We also have to introduce the ghost term \cite[eq. 7.52]{Ryder:1985wq}
\begin{equation}\label{eq:lagghostL}
    {\mathcal{L}}_{\text{gh}}^{}(A) 
    =-\bar{c}^{a}  \partial^{\mu} D^{ab}_{\mu}  {c}^{b}.
\end{equation}
This term  arises by exponentiating\footnote{We have used that $ \det M = \exp[\mathop{\rm Tr}(\ln{M}) ] $.} the \emph{FP determinant} ($ \zeta $ is the gauge parameter in Eq.~\eqref{eq:infgaugetransform})
\begin{equation}\label{eq:detM}
    \Delta_{\text{FP} } (A)=\det \left ( \frac{\delta {G}^{a} [A^{\zeta } ]}{\delta \zeta}\right ) = \int \mathop{\mathcal{D}c^{a}  }\mathop{\mathcal{D} \bar{c}^{b}  }  \exp\left(- i\int \mathop{d^{} x} \bar{c}^{a}  \partial^{\mu} D^{ab}_{\mu}   {c}^{b}  \right),
\end{equation}
where $ c$ and $ \bar{c} $ (scalar Grassmann fields) are the \emph{FP ghost fields}.

The generating functional (path integral with sources) of SOYM is then given by 
\begin{equation}\label{eq:fgYM2}
    Z_{\text{2YM} } [j] = N \int  \mathcal{D} A^{a}_{\mu} \mathcal{D} c^{a} \mathcal{D} \bar{c}^{a}\exp
    i\int \mathop{d x} \left ( {\mathcal{L}}_{\text{2YM}} + {\mathcal{L}}_{\text{gf}}(A) + {\mathcal{L}}_{\text{gh}}(A)   
        + j^{a \, \mu} {A}_{\mu}^{a} \right ),
\end{equation}
where $N$ is the normalization constant and the ghost sources were omitted. The Lagrangian in Eq.~\eqref{eq:fgYM2} is the so-called effective Lagrangian 
\begin{equation}\label{eq:lageff2}
    \mathcal{L}_{\text{eff}}^{(2)} = \mathcal{L}_{\text{2YM}} + \mathcal{L}_{\text{gf}} (A) + \mathcal{L}_{\text{gh}} (A).
\end{equation}
From Eq.~\eqref{eq:lageff2}, we read off the interactions: $(AAA)$, $(AAAA)$, $( \bar{c} A c )$. The cubic interactions are momentum-dependent and the quartic has no dependency on momentum. The full set of Feynman rules is given in Appendix~\ref{section:FeynmanRules}. 

The introduction of matter fields in Eq.~\eqref{eq:fgYM2} can be done straightforwardly. For example, the generating functional of Standard Model-like Lagrangian~\eqref{eq:YM:lagSM} is  
\begin{equation}\label{eq:fgYMSM}
    Z_{\text{SM} } [j, k,\bar{L} , L] = N \int \mathop{\mathcal{D} \mu_{\text{SM} }}       \exp
    i\int \mathop{d x} \left ( {\mathcal{L}}_{\text{SM}} + {\mathcal{L}}_{\text{gf}}(A) + {\mathcal{L}}_{\text{gh}}(A)   
    + {\mathcal{L}}_{\text{src}} \right ),
\end{equation}
where $ \mathop{\mathcal{D} \mu_{\text{SM}}} =\mathcal{D} A^{a}_{\mu} \mathcal{D} c^{a} \mathcal{D} \bar{c}^{a}
     \mathop{\mathcal{D} \Phi^{\dagger}} 
    \mathop{\mathcal{D} \Phi} \mathop{\mathcal{D} \bar{\Psi}} \mathop{\mathcal{D} \Psi}$  
 and the source term is
\begin{equation}\label{eq:sourceofSM}
    {\mathcal{L}}_{\text{src}} = j^{a \, \mu} A^{a}_{\mu} + k \Phi + \bar{L} \Psi - \bar{\Psi} L
\end{equation}
($j$, $k$ are ordinary sources and $ L$, $ \bar{L} $ are Grassmann sources).

\subsection{BRST symmetry}\label{section:SOYMBRST}

Although the gauge-fixing term in the action in Eq.~\eqref{eq:fgYM2} 
breaks the original gauge invariance Eq.~\eqref{eq:infgaugetransform} of the YM Lagrangian in Eq.~\eqref{eq:lagYMp1}, the action in Eq.~\eqref{eq:fgYM2} is invariant under a global symmetry, namely the well-known BRST symmetry. 

The BRST symmetry of the action in Eq.~\eqref{eq:fgYM2} reads \cite{dasLecturesQuantumField2020}
\begin{subequations}\label{eq:BRSTofSOYM}
    \begin{align}
        & \mathsf{s} A_{\mu}^{a} = D_{\mu}^{ab} c^{b} , \\
        & \mathsf{s} c^{a} =  \frac{g}{2}  f^{abc}c^{b} c^{c} ,\\
        & \mathsf{s} \bar{c}^{a} = - \frac{1}{\alpha} \partial^{\mu} A_{ \mu }^{a} .
    \end{align}
\end{subequations}
The generator of this symmetry is nilpotent $\mathsf{s}^{2} =0$.

The BRST symmetry leads to the Slavnov-Taylor identities, which is the generalization of the Ward-Takahashi identities to non-Abelian gauge theories. To obtain these identities, we extend the source term in Eq.~\eqref{eq:fgYM2} to 
\begin{equation}\label{eq:sourceofSMextended}
    j^{a \, \mu} A_{\mu}^{a} + i( \bar{\eta}^{a} c^{a} - \bar{c}^{a} \eta^{a} ) + K^{a \, \mu} D_{\mu}^{ab} c^{b} + K^{a}  \frac{g}{2} f^{abc} c^{b} c^{c} 
\end{equation}
in order to derive the Zinn-Justin master equation \cite{Zinn-Justin:1974ggz, taylor:1976}
\begin{equation}\label{eq:MEofSTid2sm}
    \int \mathop{d x} \left (\frac{\delta_{R} \Gamma'}{\delta A_{\mu}^{a}} \frac{\delta_{L} \Gamma'}{\delta K^{a \, \mu}} + 
    \frac{\delta_{R} \Gamma ' }{\delta c^{a}} \frac{\delta_{L} \Gamma }{\delta K^{a}} \right) =0,
\end{equation}
where the subscript $L$ ($R$) denotes left (right) differentiation, 
associated with the invariance of the generating functional \eqref{eq:fgYM2} under the BRST symmetry.
We have defined 
\begin{equation}\label{eq:gammatilde}
    \Gamma' = \Gamma - \int \mathop{d x} \mathcal{L}_{\text{gf}} (A),
\end{equation} 
where $ \Gamma $ is the generating functional of the \emph{one-particle irreducible} (1PI) diagrams, namely the effective action. The Slavnov-Taylor identities are derived by taking functional derivatives of the master equation \eqref{eq:MEofSTid2sm}.

\section{First-Order Yang-Mills theory} \label{section:FOYM}

In this section, we review the first-order formulation of YM theories \cite{niederle:1983, McKeon:2020lqp}. In this formulation, the interactions are simpler, and we obtain first-order equations of motion. On the other hand, we must introduce a new field that can be interpreted as the field strength tensor.

The Lagrangian of the FOYM theory is \cite{martins-filho:2021}
\begin{equation}\label{eq:lagYM1}
    {\mathcal{L}}_{\text{1YM}} = \frac{1}{4} \mathcal{F}^{a \, \mu \nu} \mathcal{F}^{a}_{\mu \nu} - \frac{1}{2} \mathcal{F}^{a \, \mu \nu} (\partial_{\mu} A^{a}_{\nu} - \partial_{\nu} A^{a}_{\mu} + g f^{abc} A^{b}_{\mu} A^{c}_{\nu}), 
\end{equation}
where $ \mathcal{\mathcal{F}}^{a}_{\mu \nu} $ is an auxiliary field. This Lagrangian is invariant under the gauge transformations 
\begin{equation}\label{eq:infgaugetransform1YM}
    \delta_{\zeta} A_{\mu}^{a} = D_{\mu}^{ab} \zeta^{b} \quad \text{and} \quad \delta_{\zeta} \mathcal{\mathcal{F}}_{\mu \nu}^{a} = f^{abc} \mathcal{\mathcal{F}}_{\mu \nu}^{b} \zeta^{c}.
\end{equation}
Above expression is similar to Eq.~\eqref{eq:infgaugetransform} supplemented with the transformation of an additional field $ \mathcal{F} $ in the adjoint representation of the gauge group $G(N)$.

The FOYM formulation is related to the SOYM Lagrangian in the following form 
\begin{equation}\label{eq:lagYM21}
    {\mathcal{L}}_{\text{2YM}} = \frac{1}{4} F^{a \, \mu \nu} F^{a}_{\mu \nu} - \frac{1}{2} F^{a \, \mu \nu} (\partial_{\mu} A^{a}_{\nu} - \partial_{\nu} A^{a}_{\mu} + g f^{abc} A^{b}_{\mu} A^{c}_{\nu}) 
\end{equation}
in which we treat the fields $A$ and the field strength tensor $ F$ as independent fields.  It follows from the comparison of Eqs.~\eqref{eq:lagYM1} and \eqref{eq:lagYM21} that the classical equivalence between the FOYM and SOYM holds when 
\begin{equation}\label{eq:eqmF}
    \mathcal{\mathcal{F}}^{a}_{\mu \nu} = F^{a}_{\mu \nu} =\partial_{\mu} A^{a}_{\nu} - \partial_{\nu} A^{a}_{\mu} + g f^{abc} A^{b}_{\mu} A^{c}_{\nu}.
\end{equation}
This is the equation of motion of the auxiliary field described by Eq.~\eqref{eq:lagYM1}. Thus, these formulations are classically equivalent. The equation of motion of the gauge field reads 
\begin{equation}\label{eq:eqmAinYM1}
    D_{\mu}^{ab} \mathcal{\mathcal{F}}^{b \, \mu \nu} =0, 
\end{equation}
which is similar to Eq.~\eqref{eq:ELYM2}. Substituting the equation of motion of the auxiliary field \eqref{eq:eqmF} in Eq.~\eqref{eq:eqmAinYM1} results in the YM equation \eqref{eq:ELYM2}, which verifies the classical equivalence between the field equations in both formalisms. 

\subsection{Canonical analysis}\label{section:CAof1YM}
The equivalence also holds in the Hamiltonian formulation. This can be verified by applying the \emph{Dirac-Bergmann algorithm} \cite{niederle:1983, Kiriushcheva:2005yu}. Introducing the canonical momenta 
\begin{equation}\label{eq:YM:canonicalmomenta}
\pi^{a \, \mu } = \frac{\partial {\mathcal{L}}_{\text{1YM}}}{\partial \partial_{0} A_{\mu}^{a} } \quad \text{and} \quad \Pi^{a \, \mu \nu}  = \frac{\partial {\mathcal{L}}_{\text{1YM}}}{\partial \partial_{0} \mathcal{\mathcal{F}}^{a}_{\mu \nu} }, 
\end{equation}
we identify the primary constraints 
\begin{equation}\label{eq:YM:primaryconstraints}
    \chi_\mu^{a} = \pi^{a}_{\mu } + \mathcal{F}_{0 \mu }^{a}   \approx 0 \quad \text{and} \quad  \quad \Pi^{a}_{\mu \nu } \approx  0 .
\end{equation}
The constraint $ \chi_{0}^{a} $ is the familiar YM first-class constraint that generates the gauge transformations in Eq.~\eqref{eq:infgaugetransform}. 

Before we proceed, let us clarify the notation. The notation $ A \approx B$ denotes a \emph{weak equality}, which means that $A$ is equal to $ B$ when the constraints hold. The indices $i$, $j$, $k$, $\ldots$ run from $1$ to $D-1$, where $D-1$ is the number of spatial dimensions. The greek indices run from $0$ to $D-1$ (as usual), $D$ is the total number of dimensions. The \emph{Poisson bracket} of the phase space functions $A(p,q)$ and $B(p,q)$ is denoted by $ \{A,\,  B\} $. It is defined, in a general way, as 
\begin{equation}\label{eq:defPB}
    \{A,\,  B\} =  \frac{\partial A}{\partial p_{i}} \frac{\partial B}{\partial q_{i}} - \frac{\partial B}{\partial p_{i}} \frac{\partial A}{\partial q_{i}},
\end{equation}
where $ p_{i} $, $ q_{i} $ are the canonical coordinates. We also use the Dirac classification of the constraints in \emph{first- and second-class constraints} (the reader is referred to \cite{henneaux1992quantization}). 

The Dirac-Bergmann algorithm begins with the definition of the canonical Hamiltonian, which is given simply by $ - {\mathcal{L}}_{\text{1YM}} $.  The total Hamiltonian is given by 
\begin{equation}\label{eq:YM:Hamiltonian1}
    \begin{split}
    H_{ \text{YM} } ={}& 
    \lambda_{\mu}^{a} \left(\pi^{\mu \, a} + \mathcal{F}^{0\mu \, a} \right)   + \Lambda^{a}_{\mu \nu} \Pi^{a \,  \mu \nu}   
    - {\mathcal{L}}_{\text{1YM}} \\
    ={}&
    \lambda_{\mu}^{a} \chi^{\mu \, a}  + \Lambda^{a}_{\mu \nu} \Pi^{a \, \mu \nu}   
    + \frac{1}{2} \mathcal{\mathcal{F}}^{a}_{0i} \mathcal{\mathcal{F}}^{a}_{0i}   -  \frac{1}{4} \mathcal{\mathcal{F}}_{ij}^{a} \mathcal{\mathcal{F}}_{ij}^{a} 
    -\mathcal{\mathcal{F}}^{a}_{0i} \partial_{i} A_{0}^{a}-g f^{abc} \mathcal{\mathcal{F}}^{a}_{0i} A_{0}^{b} A_{i}^{c}  + \frac{1}{2} \mathcal{\mathcal{F}}_{ij}^{a} F_{ij}^{a}   .
\end{split}
\end{equation}
Next, we have to check the consistency conditions 
\begin{align}\label{eq:YM:consitencyconditinos1}
    \{\pi_{0}^{a},\,  H_{\text{YM}}\} ={}&0, \\\{\pi^{a}_{i}+\mathcal{\mathcal{F}}^{a}_{0i} ,\,  H_{\text{YM}}\} ={}&0, \\ \{ \Pi_{\mu \nu}^{a},\,  H_{\text{YM}}\} ={}&0
\end{align}
that yields 
\begin{align}\label{eq:YM:secondaryconstraints}
    \varphi^{a}_{0} ={}&  D_{j}^{ab} \mathcal{\mathcal{F}}^{b}_{0j} {\equiv} -D_{j}^{ab} \pi^{j \, b}   \approx 0, \\ \varphi^{a}_{i} ={}& D_{j}  \mathcal{\mathcal{F}}^{a }_{ji}  - g f^{apb} A^{p}_{0} \mathcal{\mathcal{F}}_{0i}^{b} - \Lambda^{a}_{0i} \approx 0, \\ 
    \varPhi_{\mu \nu}^{a} ={}& \mathcal{\mathcal{F}}_{\mu \nu}^{a} - F^{a}_{\mu \nu }  + \lambda^{a}_{\nu} \eta_{\mu 0} - \lambda_{\mu}^{a} \eta_{\nu 0}  \approx 0. 
\end{align}
We have the secondary constraints $ \varphi_{0}^{a} $ and $ \varPhi_{\mu \nu}^{a} $, while $ \varphi_{i}^{a} $ determines the Lagrange multipliers $ \Lambda_{0i}^{a} $. One can check that no more constraints arise as \cite{k.sundermeyer:1982}
\begin{equation}\label{eq:consistencycheck}
    \dot{\varphi}_{0}^{a} \propto \varphi_{0}^{a} \approx 0.
\end{equation}
The consistency condition $ \dot{\varPhi}_{\mu \nu}^{a} \approx 0 $ determines the remaining LM ($ \Lambda_{ij}^{a}$ and $ \lambda_{i} $). The LM $ \lambda_{0}^{a} $ is undetermined since it is associated with the generator of the gauge invariance $ \chi_{0}^{a} $, which is a first-class constraint. The secondary constraint $ \varphi_{0}^{a} $ is also a first-class constraint, the remaining $ \chi_{i}^{a} $, $ \Pi_{\mu \nu}^{a} $, $ \varPhi_{\mu \nu }^{a} $ are second-class constraints. 

The consistency of this result can be verified easily. The second-class constraints $ \varphi_{i}^{a} $ and $ \varPhi_{\mu \nu}^{a}  $ leads to $ {D}-1 + {D}({D}-1)/2 = ({D}-1)({D}+2)/2$ equations between the Lagrange multipliers (multiplied by the dimension of the gauge group $\mathop{\mathrm{dim}} G(N)$), where $D$ is the dimension of the spacetime. We have a total of $ {D} + {D}({D}-1)/2 = {D}({D}+1)/2$ Lagrange multipliers. Thus, these equations leave $ {D}({D}+1)/2 - ({D}-1)({D}+2)/2 = 1$ LM undetermined, which corresponds to $ \lambda_{0}^{a} $. 

Besides that, the number of irreducible second-class constraints should be even. Indeed, we have a total of $ ( {D} -1 ) + ( {D} -1)( {D} -2)/2 + {D} ( {D} -1)/2 = {D} ( {D} -1 )$ irreducible second-class constraints, which is even for any integer $ {D} $. A set of irreducible second-class constraints is given by $ \varPhi_{\mu \nu}^{a} $ and $ \Pi_{\mu \nu}^{a} $, note that the number of constraints in this set is equal to $ 2 ( {D} -1) {D} /2$. It is easy to construct other irreducible sets of second-class constraints using a combination of $ \chi_{i}^{a} $, $ \Pi_{\mu \nu}^{a} $ and $ \varPhi_{\mu \nu}^{a} $. However, $ \varPhi_{\mu \nu}^{a} $ and $ \Pi_{\mu \nu}^{a} $ are already in a manifestly covariant form that shall prove to be very useful later. 

With these results, the degrees of freedom of the FOYM are obtained by using \cite[Eq. (1.60)]{henneaux1992quantization}, which reads
\begin{equation}\label{eq:DoFof1YM}
    \begin{split} 
        N_{\text{DoF}} ={}& \frac{1}{2} N_{\text{total}}  - \frac{1}{2}N_{\text{SC}} - N_{\text{FC}} \\ ={}& \mathop{\mathrm{dim}} G \left[\frac{1}{2}D(D-1) + D - \frac{1}{2}D (D-1) - 2\right] = (D-2)\mathop{\mathrm{dim}} G , 
\end{split} 
\end{equation}
where $N_{\text{DoF}} $, $N_{\text{total}} $, $ N_{\text{SC}} $ and $ N_{\text{FC}} $ denote, respectively, the number of degrees of freedom, the total number of canonical coordinates, the number of second-class and first-class constraints and $ \mathop{\mathrm{dim}} G$ is the dimension of the gauge group $G$. For example, the first-order formulation of the four-dimensional QED would have $2$ degrees of freedom ($D=4$
and the gauge group is one-dimensional) that corresponds to the two polarizations of the photon.
From Eq.~\eqref{eq:DoFof1YM}, we see that the additional degrees of freedom introduced by the auxiliary field $ \mathcal{F} $ match the number of second-class constraints, effectively canceling each other. Therefore, the FOYM and SOYM formulations have the same number of degrees of freedom.

The (reduced) extended Hamiltonian reads 
\begin{equation}\label{eq:YM:Hamiltonian1R}
        \mathcal{H}_{ \text{YM} } = 
    \lambda_{0}^{a} \pi^{a \, 0}   
    -\pi^{a}_{i} D_{i}^{ab} A_{0}^{b}   
    +\frac{1}{2} \pi^{a}_{i} \pi^{a}_{i}    
    + \frac{1}{4} F_{ij}^{a} F_{ij}^{a}.
\end{equation}
This Hamiltonian is not positive-definite due to the gauge freedom associated with $ \lambda_{0} $ and $ A_{0} $. This problem is solved by fixing the gauge, which consists in removing the gauge freedom with new constraints, namely gauge conditions. This turns all first-class constraints into second-class constraints. 

Note that, Eq.~\eqref{eq:YM:Hamiltonian1R} is equal to the Hamiltonian of the SOYM theory \cite{niederle:1983}. 
Thus, we have shown that the canonical structure of the FOYM formalism leads to the classical equivalence to the SOYM formalism, in particular, we have established the equivalence of the Hamiltonian formulations of FOYM and SOYM\@. However, the introduction of the auxiliary field leads to additional second-class constraints. In order to quantize such theories within the framework of path integral quantization, these constraints must be treated appropriately. In Section~\ref{section:quantizationFOYM}, we use the FS procedure to derive the FOYM path integral.

\subsection{Matter fields}\label{section:MFinFOYM}

The models in Section~\ref{section:YMtheorywithFermions} can be written in the FOYM formulation replacing $ {\mathcal{L}}_{\text{2YM}} $ by $ {\mathcal{L}}_{\text{1YM}} $ in Eq.~\eqref{eq:lagYM1}. For instance, we have that  
\begin{equation}\label{eq:YM:lagSM1}
    {\mathcal{L}}_{\text{1SM}} = 
 \frac{1}{4} {\mathcal{F}}_{\mu \nu}^{a} {\mathcal{F}}^{a \, \mu \nu} 
-\frac{1}{2} {\mathcal{F}}_{\mu \nu}^{a} F^{a \, \mu \nu} 
+ \bar{\Psi} ( i \slashed{D} - m ) \Psi  
    + (D_{\mu} \Phi )^{\dagger} D^{\mu} \Phi - V ( \Phi^{\dagger} \Phi).
\end{equation}
The equations of motion are 
\begin{subequations}
\label{eq:EoMlagSM1}
\begin{align}
     D_{\mu}^{ab} \mathcal{F}^{b \, \mu \nu } ={}& {J}^{\nu}_{ \text{SM} } [ A ],  \quad \\
     \mathcal{\mathcal{F}}_{\mu \nu}^{a}  ={}& F_{\mu \nu}^{a} ,  \quad 
\end{align}
\end{subequations}
the remaining equations (for the matter fields) in Eq.~\eqref{eq:EoMlagSM} remain unaltered. It is easy to see that Eqs.~\eqref{eq:EoMlagSM1} are equivalent to Eq.~\eqref{eq:EoMlagSMa}. Hence, the equivalence between the FOYM and SOYM formulations is still valid. This can be verified by the canonical analysis of this model. However, one can convince oneself that the results of Section~\ref{section:CAof1YM} are not altered in the presence of these matter fields.

Hence, we can conclude that a first-order formulation of the Standard Model (the first-order formulation of spin-0 fields is discussed in Appendix~\ref{section:FOspin0}) is classically equivalent to the standard second-order formulation. This is also valid at the quantum level since the results for the pure YM theory in \cite{McKeon:2020lqp, Lavrov:2021pqh, martins-filho:2021} are easily generalized for the Standard Model-like introduced in Section~\ref{section:YMtheorywithFermions}. We shall see this in the following chapter. 

Indeed, any coupling that does not depend on the field strength does not spoil the equivalence between the FOYM and SOYM formulations. 
In Section~\ref{section:YMnonMINIMAL}, we have seen a few examples of such couplings. 
Thus, let us now consider a Pauli-like interaction  $ W^{a}_{\mu \nu} \mathcal{\mathcal{F}}^{a \, \mu \nu} \equiv i f^{abc } \bar{\psi}^{b} \sigma_{ \mu \nu }  \mathcal{\mathcal{F}}^{a \, \mu \nu} \psi^{c}  $, where $ W^{a}_{\mu \nu}  $  transforms in the adjoint of the gauge group $ G(N)$, $ W'_{\mu \nu} \to U W_{\mu \nu} U^{-1}$. 
In the SOYM, we have the Lagrangian  
\begin{equation}\label{eq:lagYM21withPauli}
    \frac{1}{4} F^{a \, \mu \nu} F^{a}_{\mu \nu} - \frac{1}{2} F^{a \, \mu \nu} (\partial_{\mu} A^{a}_{\nu} - \partial_{\nu} A^{a}_{\mu} + g f^{abc} A^{b}_{\mu} A^{c}_{\nu}) + \bar{\psi}^{a} ( i \slashed{D}^{ab} -m \delta^{ab}  ) \psi^{b} + \frac{1}{2} W^{a \, \mu \nu} F_{\mu \nu}^{a},
\end{equation}
which leads to the equations of motion: 
\begin{subequations}
\label{eq:EoMlagPAULI2}
\begin{align}
\label{eq:EoMlagPAULI2a}
 D_{\mu}^{ab} \left(F- W\right)^{b \, \mu \nu } ={}& 0,  \quad 
\\
 \left[i \left(\slashed{D}^{ab} - \frac{1}{2}f^{apb} \sigma_{\mu \nu} F^{p \, \mu \nu} \right) -m \delta^{ab} \right] \psi^{b} ={}&0.
\end{align}
\end{subequations}

However, by following the same approach applied for Eq.~\eqref{eq:lagYM21}, that is, treating the field $ A$ and $ F $ as independent fields leads to   
\begin{subequations}
\label{eq:EoMlagPAULI1}
\begin{align}
\label{eq:EoMlagPAULI1a}
& D_{\mu}^{ab} {\mathcal{F}}^{b \, \mu \nu }  = 0 \quad  \text{and}  \quad  
{\mathcal{F}}^{a}_{\mu \nu} =  (F-W)_{\mu \nu}^{a},
\\
&                                            \left[i \left(\slashed{D}^{ab} - \frac{1}{2} f^{apb} \sigma_{\mu \nu} \mathcal{\mathcal{F}}^{p \, \mu \nu} \right) -m \delta^{ab} \right] \psi^{b} = 0, \end{align}
\end{subequations}
which are not equivalent to Eq.~\eqref{eq:EoMlagPAULI2}. This can be solved by substituting $ W^{a \, \mu \nu} F^{a}_{\mu \nu} $ with $ W^{a \, \mu \nu } ( \partial_{\mu} A^{a}_{\nu} - \partial_{\nu} A^{a}_{\mu} + g f^{ab c} A_{\mu}^{b} A_{\nu}^{c})$, which leads to the first-order formulation of Eq.~\eqref{eq:EoMlagPAULI2}: 
\begin{subequations}
\label{eq:EoMlagPAULI1NEW}
\begin{align}
\label{eq:EoMlagPAULI1aNEW}
& D_{\mu}^{ab} ( \mathcal{F} -W)^{b \, \mu \nu }  = 0 \quad  \text{and}  \quad  
{\mathcal{F}}^{a}_{\mu \nu} =  F_{\mu \nu}^{a},
\\
&                                            \left[i \left(\slashed{D}^{ab} - \frac{1}{2} f^{apb} \sigma_{\mu \nu} F^{p \, \mu \nu} \right) -m \delta^{ab} \right] \psi^{b} = 0\end{align}
\end{subequations}
These equations can be obtained from the first-order Lagrangian 
\begin{equation}\label{eq:lagYM1withPauli}
    \frac{1}{4} {\mathcal{F}}^{a \, \mu \nu} {\mathcal{F}}^{a}_{\mu \nu} - \frac{1}{2} (\mathcal{F}  - W)^{a \, \mu \nu} (\partial_{\mu} A^{a}_{\nu} - \partial_{\nu} A^{a}_{\mu} + g f^{abc} A^{b}_{\mu} A^{c}_{\nu}) +
    \bar{\psi}^{a} ( i \slashed{D}^{ab} -m \delta^{ab}  ) \psi^{b}.
\end{equation}
This shows that the naive approach\footnote{In gravity, this is the so-called Palatini approach.} to obtain first-order equations of motion is not appropriate when the coupling is curvature-dependent. That is, we cannot simply replace (the curvature) $ F $ by an auxiliary field $ \mathcal{F}  $. 

Besides that, we have an ambiguity in the definition of curvature-dependent couplings. This becomes an insurmountable problem when we have to introduce second-order terms, such as the $ \theta$-term \eqref{eq:YM:thetaterm}. In the next section, we propose a systematic approach to obtain the first-order formulation of a second-order Lagrangian. 

\subsection{Lagrange multiplier procedure}\label{section:LMprocedure}

Here, we present a simple approach that can be widely applied to resolve the ambiguity in the derivation of first-order Lagrangians. This procedure employs LM fields to enforce the on-shell equivalence between the first- and second-order formulations of the Lagrangian. 

The procedure is to replace the curvature $ F_{\mu \nu }^{a} $ by the auxiliary field $ \mathcal{F}_{\mu \nu}^{a} $ in the second-order Lagrangian $ {\mathcal{L}}_{2} [F]$ and, at the same time, constrain the curvature to be equal to the auxiliary field. This constraint is responsible to enforce the equivalence between $ {\mathcal{L}}_{2} [F]$ and $ {\mathcal{L}}_{2} [ \mathcal{F} ]$. For this, we introduce an LM field $ \Lambda_{\mu \nu}^{a} $ by adding  
\begin{equation}\label{eq:newconstraint}
    \frac{1}{2 } \Lambda_{\mu \nu}^{a} ( \mathcal{F} -F )_{\mu \nu}^{a} 
\end{equation}
to $ {\mathcal{L}}_{2} [ \mathcal{F} ]$. This leads to the first-order Lagrangian 
\begin{equation}\label{eq:lagFOgeneral}
    {\mathcal{L}}_{1}[F, \mathcal{F} ] = {\mathcal{L}}_{2} [ \mathcal{F} ] +   
    \frac{1}{2 } \Lambda_{\mu \nu}^{a} (  \mathcal{F} -F)_{\mu \nu}^{a}. 
\end{equation}
The LM $ \Lambda_{\mu \nu} $ is a non-dynamical variable and can be eliminated using its equations of motion. However, this yields the second-order formulation. Instead, we can use the equation of motion of the auxiliary field $ \mathcal{F} $. In this case, the particular form of $ {\mathcal{L}} [ \mathcal{F} ]  $ is relevant.

First, let us apply this procedure to the pure SOYM theory. We have that 
\begin{equation}\label{eq:LMPpure}
    {\mathcal{L}}_{1} [ F,\mathcal{F} ] = - \frac{1}{4} \mathcal{F}_{\mu \nu}^{a} \mathcal{F}^{a \, \mu \nu} + \frac{1}{2} \Lambda_{\mu \nu}^{a} ( \mathcal{F} - F )_{\mu \nu}^{a}.
\end{equation}
The equation of motion of the auxiliary field reads 
\begin{equation}\label{eq:LMPpureEoM}
    \mathcal{F}_{\mu \nu}^{a} = \Lambda_{\mu \nu}^{a}.
\end{equation}
Using this equation to remove the LM field, we obtain that 
\begin{equation}\label{eq:LMPpure1}
    {\mathcal{L}}_{1} [ F,\mathcal{F} ] = - \frac{1}{4} \mathcal{F}_{\mu \nu}^{a} \mathcal{F}^{a \, \mu \nu} + \frac{1}{2} \mathcal{F}_{\mu \nu}^{a} ( \mathcal{F} - F )^{a \, \mu \nu}  = {\mathcal{L}}_\text{1YM},
\end{equation}
which shows the consistency of this procedure.

Now, we can try to apply it to the Lagrangian in Eq.~\eqref{eq:lagYM21withPauli}. Replacing the curvature in Eq.~\eqref{eq:lagYM21withPauli} by $ \mathcal{F} $, we get 
\begin{equation}\label{eq:lagwithPauli}
    {\mathcal{L}}_{1} [ F, \mathcal{F} ] = -\frac{1}{4} \mathcal{F}^{a \, \mu \nu} \mathcal{F}^{a}_{\mu \nu}   +
    \bar{\psi}^{a} ( i \slashed{D}^{ab} -m \delta^{ab}  ) \psi^{b} + \frac{1}{2} W^{a \, \mu \nu} \mathcal{F}_{\mu \nu}^{a} + \frac{1}{2} \Lambda_{\mu \nu}^{a} ( \mathcal{F} - F )^{a \, \mu \nu}  .
\end{equation}
Replacing the equation of motion of the auxiliary field 
\begin{equation}\label{eq:EoMforPauli}
    \mathcal{F}_{\mu \nu}^{a} = \Lambda_{\mu \nu}^{a} + W_{\mu \nu}^{a}
\end{equation}
in Eq.~\eqref{eq:lagwithPauli}, we obtain Eq.~\eqref{eq:lagYM1withPauli}. Alternatively,  we can redefine the auxiliary field by the change of variable 
\begin{equation}\label{eq:redofAuxPauli}
    \mathcal{F}_{\mu \nu}^{a} \to \mathcal{F}_{\mu \nu}^{a} + W_{\mu \nu}^{a} 
\end{equation}
in Eq.~\eqref{eq:lagwithPauli}. The equation of motion of the auxiliary field is given by Eq.~\eqref{eq:LMPpureEoM}, and leads to the alternative first-order Lagrangian 
\begin{equation}\label{eq:lagaltpauli}
    {\mathcal{L}}_{1} [ \mathcal{F} ] = {\mathcal{L}}_{\text{1YM}}  +  
    \bar{\psi}^{a} ( i \slashed{D}^{ab} -m \delta^{ab}  ) \psi^{b} + \frac{1}{4} W^{a \, \mu \nu} W^{a}_{\mu \nu}   + \frac{1}{2} W^{a \, \mu \nu} \mathcal{F}_{\mu \nu}^{a}. 
\end{equation}
This alternative formulation is interesting for two reasons. First, it helps us understand why Eq.~\eqref{eq:lagYM21withPauli} fails to lead to the correct equations of motion \eqref{eq:EoMlagPAULI1NEW}. (The term $ W_{\mu \nu}^{a} W^{a \, \mu \nu} /4$ is missing.) Second, it is more suitable to show the equivalence at the quantum level as we shall see in the next chapter. 

For sake of completeness, the first-order formulation of YM theory with the $ \mathcal{CP} $-violating $ \theta $-term is given by 
\begin{equation}\label{eq:lag1thetaterm}
    {\mathcal{L}}_{\text{1CP}} = \frac{1}{4} \mathcal{F}_{\mu \nu}^{a} M^{\mu \nu \alpha \beta} ( \vartheta ) \mathcal{F}_{\alpha \beta}^{a}
- \frac{1}{2} \mathcal{F}_{\mu \nu}^{a} M^{\mu \nu \alpha \beta} ( \vartheta ) {F}_{\alpha \beta}^{a},
\end{equation}
where \begin{equation} \label{eq:defMtheta}
M^{\mu \nu \alpha \beta} ( \vartheta ) 
\equiv I^{\mu \nu \alpha \beta} + \frac{\vartheta}{2}\epsilon^{\mu \nu \alpha \beta} 
\end{equation}  
with the following identity definition\footnote{We have that $ I^{\mu \nu \alpha \beta} = I^{\alpha \beta \mu \nu}$, $ I^{\mu \nu \alpha \beta} = - I^{\nu \mu \alpha \beta} = - I^{\mu \nu \beta \alpha} $.} 
\begin{equation}\label{eq:defI}
    I^{\mu \nu \alpha \beta} \equiv  \frac{\eta^{\mu \alpha} \eta^{\nu \beta} - \eta^{\mu \beta} \eta^{\nu \alpha}}{2}.
\end{equation}

Eq.~\eqref{eq:lag1thetaterm}, obtained through the LM procedure presented here, agrees with the known literature (see Ref.~\cite{Cattaneo:1995xa}). These formulations should be equivalent, although in the first-order formalism, the $\theta$-term is not a total derivative. This discrepancy does not appear to break the equivalence between the pure first- and second-order formulations. However, this scenario could change in the presence of quarks, as the chiral anomaly is crucial for a full understanding of the strong $\mathcal{CP}$ problem.

\subsection{Quantization} \label{section:quantizationFOYM}

We have seen that second-class constraints appear in the FOYM formulation, which leads to the modification of the path integral quantization procedure. We need to use the FS procedure \cite{Senjanovic:1976br}, which is a generalization of the FP procedure. Following \cite{Senjanovic:1976br}, the path integral of a constrained system is given by
\begin{equation}\label{eq:senjanovicPI}
    Z_{\text{FS}} [0] = \int \mathop{\mathcal{D} \mu} \exp i \int \mathop{d x} \left ( p_{i} \dot{q}_{i} - \mathcal{H}(p,q)\right ), 
\end{equation}
where the measure is 
\begin{equation}\label{eq:measureFS}
    \mathop{\mathcal{D} \mu} = \mathop{\mathcal{D} p_{i}} \mathop{\mathcal{D} q_{i}} \underbrace{\delta (F_a) \delta ( \chi_{a} ) |\det \| \{ F_l,\,  \chi_m\}\| |}_{ \text{Faddeev-Popov} } \overbrace{\prod_{b} \delta ( \theta_{b} ) | \det \| \{\theta_{l} , \theta_{m}\} \| |^{1/2}}^{\text{Senjanovi\'{c}}}.
\end{equation}
In the first part of the measure, $ \chi_{a} $ denote the first-class constraints and $ F_{a} $ represent the gauge-fixing constraints. The term $ |\det \| \{F_{l}, \chi_{m}\}\|| $ is the FP determinant. In the second part, $ \theta_{a} $ denotes the second-class constraints, and the term $ |\det \| \{\theta_{l}, \, \theta_{m}\} \| |^{1/2}$ is the Senjanovi\'{c} determinant.

In Eq.~\eqref{eq:senjanovicPI}, $ p_{i} $ and $ q_{i} $ are canonical pairs, and $ \mathcal{H}(p,q) $ is the Hamiltonian of the system. Therefore, the expression $ p_{i} \dot{q}_{i} - \mathcal{H}(p,q) $ should yield the Lagrangian of this system. However, we cannot use the Lagrangian in place of this Legendre transform since the phase space is constrained.

In the canonical analysis of the FOYM, we have seen that the first-class constraints are the same found in the SOYM\@. Therefore, since the first-class sector is the same in both formalisms, the FP procedure is kept unaltered. From Eq.~\eqref{eq:senjanovicPI} and the results in Section~\ref{section:CAof1YM}, we obtain that 
\begin{equation}\label{eq:SenjanovicGF1YMa}
    Z_{\text{1YM}} [0] = \int \mathop{\mathcal{D}_{\text{FP}}A_{\mu}^{a}}   \mathop{\mathcal{D} \mathcal{F}_{\mu \nu}^{a}} \mathop{\mathcal{D} \Pi_{\mu \nu}^{a}} \delta ( \Pi_{\mu \nu}^{a} ) \delta ( \Phi_{ \alpha \beta }^{a} )\Delta_{\text{1YM}}    \exp{i \int \mathop{d x} L},
\end{equation}
where $ \mathop{\mathcal{D}_{\text{FP}} A_{\mu}^{a}} \equiv\mathop{\mathcal{D} A_{\mu}^{a}}  \mathop{\mathcal{D} \bar{c}^{a}} \mathop{\mathcal{D} c^{a}} \Delta_{\text{FP}} (A)$,  
\begin{equation}\label{eq:DeltaSenja1YMY}
    \Delta_{\text{1YM}} \equiv | \det \|\{ \Pi_{\mu \nu}^{a} , \, \Phi_{\alpha \beta }^{b} \} \| |^{1/2} =| \det \{ \Pi_{\mu \nu}^{a} , \, (\mathcal{F}-F)_{\alpha \beta}^{b}/2 \}|= \left|  \det \left(I_{\mu \nu \alpha \beta} \delta^{ab}/2\right)\right|
\end{equation} 
is the Senjanovi\'{c} determinant, and
\begin{equation}\label{eq:defL1YMa}
    L = \Pi_{\mu \nu}^{a} \partial_{0} F^{a \, \mu \nu} + \frac{1}{2} \mathcal{F}_{0i}^{a} \mathcal{F}_{0i}^{a} - \frac{1}{4} \mathcal{F}_{ij}^{a} \mathcal{F}_{ij}^{a}.
\end{equation}
We can integrate $ \Pi_{\mu \nu}^{a} $ yielding 
\begin{equation}\label{eq:SenjanovicGF1YMb}
    Z_{\text{1YM}} [0] = \int \mathop{\mathcal{D}_{\text{FP}}A_{\mu}^{a}}   \mathop{\mathcal{D} \mathcal{F}_{\mu \nu}^{a}}   \Delta_{\text{1YM}} 
\delta  [ (\mathcal{F} - F)_{\mu \nu}^{a}/2    ]
    \exp{i \int \mathop{d x} L'},
\end{equation}
where
\begin{equation}\label{eq:defL1YMb}
    L' =   \frac{1}{2} \mathcal{F}_{0i}^{a} \mathcal{F}_{0i}^{a} - \frac{1}{4} \mathcal{F}_{ij}^{a} \mathcal{F}_{ij}^{a}. 
\end{equation}
Now, if we use the following propriety
\begin{equation}\label{eq:prop1delta}
    \int \mathop{\mathcal{D} \mathcal{F}_{\mu \nu}} \delta  [ (\mathcal{F} - F)_{\mu \nu}^{a}/2    ] \Delta_{\text{1YM}} = 1,
\end{equation}
then $ L ' = {\mathcal{L}}_{\text{2YM}} $, and we obtain that 
\begin{equation}\label{eq:QEofYM}
    Z_{\text{1YM}} [0] = Z_{\text{2YM}} [0].
\end{equation}
This shows that the results obtained from the canonical analysis of the FOYM are valid at the quantum level. The quantum equivalence between SOYM and FOYM formulations is also valid in the presence of sources, the equivalence of these generating functionals is explored in the next chapter. 

Alternatively, we can integrate the $ \delta $-function in Eq.~\eqref{eq:SenjanovicGF1YMb} by using \cite{Senjanovic:1976br}
\begin{equation}\label{eq:prop2delta}
    \delta  [ (\mathcal{F} - F)_{\mu \nu}^{a}/2    ] = \int \mathop{\mathcal{D} \Lambda_{\mu \nu}^{a}} \exp{i \int \mathop{d x} \left ( \Lambda_{\mu \nu}^{a} \frac{\mathcal{F}^{a \, \mu \nu} -F^{a \, \mu \nu}}{2}\right)}.
\end{equation}
This leads to
\begin{equation}\label{eq:newL}
    L' = - \frac{1}{4} \mathcal{F}_{\mu \nu}^{a} \mathcal{F}^{a \, \mu \nu} + \frac{1}{2} \Lambda_{\mu \nu}^{a} ( \mathcal{F} - F )^{a \, \mu \nu},
\end{equation}
which is equal to Eq.~\eqref{eq:LMPpure}. This indicates that we can integrate the LM field $ \Lambda_{\mu \nu}^{a} $ with a field redefinition. Upon the field redefinition
\begin{subequations}
    \begin{align}
        \mathcal{ F}_{\mu \nu}^{a} \to{}&  \mathcal{ F}_{\mu \nu}^{a} +\Lambda_{\mu \nu}^{a}
    \intertext{followed by}
    \Lambda_{\mu \nu}^{a} \to{}&  \Lambda_{\mu \nu}^{a} - \mathcal{F}_{\mu \nu}^{a},
\end{align}
\end{subequations}
Eq.~\eqref{eq:SenjanovicGF1YMb} reads
\begin{equation}\label{eq:SenjanovicGF1YMc}
    Z_{\text{1YM}} [0] = \int \mathop{\mathcal{D}_{\text{FP}}A_{\mu}^{a}}   \mathop{\mathcal{D} \mathcal{F}_{\mu \nu}^{a}}    \mathop{\mathcal{D} {\Lambda_{\mu \nu}^{ a}}} \Delta_{\text{1YM}} 
        \exp{i \int \mathop{d x} \left({\mathcal{L}}_{ \text{1YM} } - \frac{1}{4} \Lambda_{\mu \nu}^{a} \Lambda^{a \, \mu \nu} \right)}.
\end{equation}
We can integrate the LM field $ \Lambda_{\mu \nu}^{a} $, and the final form of FOYM path integral is given by
\begin{equation}\label{eq:GF1YM}
    Z_{\text{1YM}} [0] = N\int 
    \mathop{\mathcal{D} A_{\mu}^{a}}   \mathop{\mathcal{D} \mathcal{F}_{\mu \nu}^{a}} \mathop{\mathcal{D} \bar{c}^{a}} \mathop{\mathcal{D} c^{a}} \Delta_{\text{1YM}}^{1/2}  
    \exp{i \int \mathop{d x} \left({\mathcal{L}}_{ \text{1YM} } + {\mathcal{L}}_{\text{gf}} + {\mathcal{L}}_{\text{gh}} \right)},
\end{equation}
where $N$ is a normalization constant. 

In YM theory, the Senjanovi\'{c} determinant factor can be absorbed into the normalization constant. Thus, the generating functional of the FOYM is given simply by
\begin{equation}\label{eq:fgYM1}
    Z_{\text{1YM}}  [j,J] = N'\int  
    \mathop{\mathcal{D} A^{a}_{\mu} }
    \mathop{\mathcal{D} {\mathcal{F}}_{\mu \nu}^{a}} \mathop{\mathcal{D} \bar{c}^{a}}\mathop{ \mathcal{D} c^{a}}
    \exp
    i\int \mathop{d x} \left ( {\mathcal{L}}_{\text{eff}}^{(1)}  
+ j^{a \, \mu} {A}_{\mu}^{a} + J^{a \, \mu \nu } {\mathcal{F}}_{\mu \nu}^{a} \right ),
\end{equation}
where $ \mathcal{L}_{\text{eff}}^{(1)} $ denote the effective Lagrangian obtained through the FP procedure: 
\begin{equation}\label{eq:lageff1}
    \mathcal{L}_{\text{eff}}^{(1)} = \mathcal{L}_{\text{1YM}} +   {\mathcal{L}}_{\text{gf}}^{} (A) + {\mathcal{L}}_{\text{gh}}^{} (A).
\end{equation}
Note that, our result in Eq.~\eqref{eq:fgYM1} agrees with the literature \cite{niederle:1983, McKeon:1994ds,  martins-filho:2021}. In the context of gravity, the Senjanovi\'{c} determinant cannot be absorbed without further assumptions. Therefore, in order to obtain the appropriate path integral quantization of the first-order theories we need to determine the Senjanovi\'{c} determinant. In fact, the Senjanovi\'{c} determinant in first-order gravity, in a standard covariant form and directly obtained, remains unknown \cite{Chishtie:2012sq}. 

Reading off the interactions from Eq.~\eqref{eq:fgYM1}, we observe some simplifications, the quartic vertex $(AAAA)$ is absent and the cubic vertex $(FAA)$ has no momentum-dependency. The ghost sector remains unaltered. However, we not only have the propagator of the auxiliary field $ \langle 0|T F_{\mu \nu}^{a} F_{\rho \sigma}^{b} | 0 \rangle_{\text{free}} $, but also mixed propagators $ \langle 0|T F_{\mu \nu}^{a} A^{b}_{\rho} | 0 \rangle_{\text{free}} $ and $ \langle 0|T A^{b}_{\rho} F_{\mu \nu}^{a} | 0 \rangle_{\text{free}} $ (the term ``free'' refer to free propagators, that is, full propagators in lowest order). As an alternative, one can redefine the fields in such a way that mixed propagators are eliminated.  This results in a \emph{diagonal first-order} formulation of the YM theory \cite{Brandt:2018avq, martins-filho:2021}. 

The quantization of the FOYM coupled with matter fields (and coupling) such as we have mentioned in Section~\ref{section:MFinFOYM} is similar to the quantization of the pure FOYM\@. However, when the $ \theta $-term is present the Senjanovi\'{c} determinant is modified to
\begin{equation}\label{eq:SDwithTheta}
    \Delta_{\text{1YM}} = \left | \det \left[(I_{\mu \nu \alpha \beta} + \vartheta  \epsilon_{\mu \nu \alpha \beta} /2) \delta^{ab} /2\right]\right |.
\end{equation}

\subsection{Quantization from the SOYM} \label{section:FOYMfromSOYM}

We present an approach to derive a generating functional for the FOYM theory from the generating functional of the SOYM theory, which automatically ensures their quantum equivalence. Although this issue has been addressed in Ref.~\cite{Brandt:2015nxa}, it has not been fully explored, particularly in the context of gravity. In Chapter~\ref{chapter:EH}, we aim to provide a thorough examination of the consequences of applying this approach to gravity and explore its potential application in the presence of matter fields.

Here, we demonstrate that this approach leads to the generating functional obtained by applying the FS procedure, which is the standard generating functional derived from its canonical structure. In addition to providing a physical interpretation, we also consider the inclusion of sources, which further extends this straightforward procedure. Moreover, it is a manifestly covariant approach. Therefore, the covariance of the generating functional is retained. 

Since the canonical structure itself is not manifestly covariant, in general, the path integral obtained from its canonical structure will not be in manifestly covariant form either. This is one of the major advantages of this approach over canonical quantization. Moreover, this approach can be easily generalized. In particular, it can also be applied in the context of gravity as we show in Chapter~\ref{chapter:EH}.  

The procedure starts by introducing the constant factor 
\begin{equation}\label{eq:1YM}
    1 = \int \mathop{\mathcal{D} \mathcal{F}_{\mu \nu}^{a}} \Delta^{1/2}(A,J)  \exp i \int \mathop{d^{}x} \left ( \frac{1}{4} \mathcal{F}_{\mu \nu}^{a} \mathcal{F}^{a \, \mu \nu} 
    + J_{\mu \nu}^{a}  {J}^{a \, \mu \nu }  \right ) 
\end{equation}
into the SOYM generating functional (with an additional source, cf. Eq.~\eqref{eq:func2withoutJ2}) 
\begin{equation}\label{eq:fgYM2withJF}
    Z_{\text{2YM} } [j, J] = N \int  \mathcal{D} A^{a}_{\mu} \mathcal{D} c^{a} \mathcal{D} \bar{c}^{a}\exp
    i\int \mathop{d x} \left ( {\mathcal{L}}_{\text{eff}}^{(2)}          + j^{a \, \mu} {A}_{\mu}^{a}
    + J_{\mu \nu}^{a} F^{a \, \mu \nu} \right ).
\end{equation}
Then, we redefine the auxiliary field $ \mathcal{F} $ 
\begin{equation}\label{eq:}
    \mathcal{F}_{\mu \nu}^{a} \to \mathcal{F}_{\mu \nu}^{a} - F_{\mu \nu}^{a} 
    + 2J_{\mu \nu}^{a} 
\end{equation}
yielding
\begin{equation}\label{eq:fgYM21ab}
    Z_{\text{2YM} } [j,J] = N \int   
    \mathop{\mathcal{D} \mu_{1}}
    \exp
    i\int \mathop{d x}  \left ( {\mathcal{L}}_{\text{eff}}^{(1)} + J_{\mu \nu}^{a} \mathcal{F}^{a \, \mu \nu}  + J_{\mu \nu}^{a} J^{a \, \mu \nu}    + j^{a \, \mu} {A}_{\mu}^{a} \right ),
\end{equation}
where $ \mathop{\mathcal{D} \mu_{1} }  =\mathcal{D} A^{a}_{\mu} \mathop{\mathcal{D} F_{\mu \nu}^{a}} \mathcal{D} c^{a} \mathcal{D} \bar{c}^{a} \Delta^{1/2} (A,J)$. 

We determine the $ \Delta $ factor by integrating the auxiliary field in Eq.~\eqref{eq:1YM}: 
\begin{equation}\label{eq:Deltasolution}
    \Delta^{1/2} (A,J)  |
    \det I_{\mu \nu, \alpha \beta} \delta^{ab} /2 |^{-1/2} \exp i \int \mathop{d x} \left (  J_{\mu \nu}^{a} J^{a \, \mu \nu} \right )  =1,
\end{equation}
consequently, 
\begin{equation}\label{eq:DeltaYM}
    \Delta^{1/2} (A,J) = 
    |\det I_{\mu \nu, \alpha \beta} \delta^{ab} /2 |^{1/2} \exp i \int \mathop{d x} \left ( -J_{\mu \nu}^{a} J^{a \, \mu \nu} \right ).
\end{equation}
When $J=0$, this is similar to the Senjanovi\'{c} determinant in Eq.~\eqref{eq:DeltaSenja1YMY}. 

Replacing Eq.~\eqref{eq:DeltaYM} in Eq.~\eqref{eq:fgYM21ab} reads
\begin{equation}\label{eq:fgYM21a}
    \begin{split}
        Z_{\text{2YM} } [j,J] = N & \int   
    \mathop{\mathcal{D} A_{\mu}^{a}}   \mathop{\mathcal{D} \mathcal{F}_{\mu \nu}^{a}} \mathop{\mathcal{D} \bar{c}^{a}} \mathop{\mathcal{D} c^{a}} \Delta_{\text{1YM}}^{1/2}  
    \\ & \times \exp
    i\int \mathop{d x}  \left ( {\mathcal{L}}_{\text{eff}}^{(1)}    + j^{a \, \mu} {A}_{\mu}^{a} + J_{\mu \nu}^{a} \mathcal{F}^{a \, \mu \nu}   + J^{a}_{\mu \nu} J^{a \, \mu \nu} \right ),
    \end{split}
\end{equation}
which is equal to the generating functional of the FOYM theory in Eq.~\eqref{eq:GF1YM} with sources (with an extra quadratic source term). This shows that this approach leads directly to the resulting functional obtained from the FS procedure and that the covariance is retained.

In the presence of matter fields that are not minimally coupled this approach must be modified. We can consider the Pauli coupling examined in Section~\ref{section:LMprocedure}, for simplicity the sources are set to zero, $j=J=0$. Instead of Eq.~\eqref{eq:1YM}, we now have 
\begin{equation}\label{eq:1YMa}
    1 = \int \mathop{\mathcal{D} \mathcal{F}_{\mu \nu}^{a}} \Delta^{1/2}_{\text{PC} }(A)  \exp i \int \mathop{d^{}x}  \left( \frac{1}{4} \mathcal{F}_{\mu \nu}^{a} \mathcal{F}^{a \, \mu \nu} + \frac{1}{2}\mathcal{F}_{\mu \nu}^{a}  W^{a \, \mu \nu }  \right),
\end{equation}
and consequently 
\begin{equation}\label{eq:DeltaYM1a}
    \Delta^{1/2}_{\text{PC}}  (A) = \Delta^{1/2} (A,0) \exp i \int \mathop{d x} \left ( \frac{1}{4} W_{\mu \nu}^{a} W^{a \, \mu \nu}   \right ),
\end{equation}
are inserted into the SOYM generating functional \eqref{eq:fgYM2}. 
The shift is then given by 
\begin{equation}\label{eq:shiftprocedurea}
    \mathcal{F}_{\mu \nu}^{a} \to \mathcal{F}_{\mu \nu}^{a} - F_{\mu \nu}^{a}. 
\end{equation}
The resulting generating functional is 
\begin{equation}\label{eq:fgYMPAULI}
    \begin{split}
        Z_{\text{PC} } [j,J] ={}& N \int   
    \mathop{\mathcal{D} A_{\mu}^{a}}   \mathop{\mathcal{D} \mathcal{F}_{\mu \nu}^{a}} \mathop{\mathcal{D} \bar{c}^{a}} \mathop{\mathcal{D} c^{a}} \mathop{\mathcal{D} \bar{\psi}^{a} } \mathop{\mathcal{D} \psi^{a}}  \Delta_{\text{PC}}^{1/2}  
    \\ & 
   \times  \exp
    i\int \mathop{d x}  
    \left({\mathcal{L}}_{1} [ \mathcal{F} ] 
        + \mathcal{L}_{\text{gf}} (A) + \mathcal{L}_{\text{gh}} (A) 
        + j^{a \mu} A_{\mu}^{a} + J_{\mu \nu}^{a} \mathcal{F}^{a \, \mu \nu} 
    \right)
,
    \end{split}
\end{equation}
where $ \mathcal{L}_{1} [\mathcal{F} ]$ is the Lagrangian in Eq.~\eqref{eq:lagaltpauli} obtained through the LM procedure proposed in Section~\ref{section:LMprocedure}. This shows the consistency of the procedures proposed in this work. Now is clear why the alternative Lagrangian in Eq.~\eqref{eq:lagaltpauli} is more suitable to demonstrate the equivalence between the SOYM and the FOYM, it results directly from their equivalence.

\subsection{BRST symmetry}\label{section:FOYMBRST}
In this section, we present the BRST symmetry of the FOYM\@. From Eqs.~\eqref{eq:BRSTofSOYM} and \eqref{eq:infgaugetransform1YM}, we have the FOYM BRST symmetry:
\begin{subequations}\label{eq:BRSTofFOYM}
    \begin{align}
        & \mathsf{s} A_{\mu}^{a} = D_{\mu}^{ab} c^{b} , \\
        & \mathsf{s} \mathcal{F}_{\mu \nu }^{a} = g f^{abc} F_{\mu \nu}^{b} c^{c} , \\
        & \mathsf{s} c^{a} =  \frac{g}{2}  f^{abc}c^{b} c^{c} ,\\
        & \mathsf{s} \bar{c}^{a} = -\frac{1}{\alpha} \partial^{\mu} A_{\mu}^{a} .
    \end{align}
\end{subequations}
The nilpotency of the FOYM BRST symmetry can be easily verified using the Jacobi identity: 
\begin{equation}\label{eq:Jacobiidentity}
    f^{acm}f^{lbc} + f^{lcm} f^{bac} + f^{bcm} f^{alc} =0.
\end{equation}
In particular, 
\begin{equation}\label{eq:nilpotencyF}
    \begin{split}
        \mathsf{s}^{2} \mathcal{F}_{\mu \nu}^{a} ={}& g f^{abc} \left ( f^{blm} \mathcal{F}^{l}_{\mu \nu} c^{m} c^{c}  - \frac{1}{2} f^{clm} \mathcal{F}_{\mu \nu}^{b} c^{l} c^{m} \right )\\
={}& g \mathcal{F}_{\mu \nu}^{b} c^{m} c^{l} \left[ f^{acl} f^{cbm} -   \frac{1}{2} ( f^{bcm} f^{alc} + f^{acm} f^{lbc} )  \right]\\ 
        ={}& \frac{g}{2}  \mathcal{F}_{\mu \nu}^{b} c^{m} c^{l} \left( f^{acl} f^{cbm} -     f^{acm} f^{lbc}   \right) =0,
\end{split}
\end{equation}
where we have used the anticommutativity of the ghost field $c$.

This symmetry leads to the Slavnov-Taylor identities. As we have seen in Section~\ref{section:FOYMBRST}, we obtain these identities by extending the source term. In the FOYM, the extended source term is 
\begin{equation}\label{eq:sourceofSMextended1}
    j^{a \, \mu} A_{\mu}^{a} + J^{a \, \mu \nu} \mathcal{F}^{a}_{\mu \nu} +  i( \bar{\eta}^{a} c^{a} - \bar{c}^{a} \eta^{a} ) + K^{a \, \mu} D_{\mu}^{ab} c^{b} + K^{a \, \mu \nu} g f^{abc} \mathcal{F}^{b}_{\mu \nu} c^{c} + K^{a}  \frac{g}{2} f^{abc} c^{b} c^{c}. 
\end{equation}
Consequently, we obtain the master equation \cite{Frenkel:2017xvm} 
\begin{equation}\label{eq:MEofSTid1sm}
    \int \mathop{d x} \left  (
        \frac{\delta_{R} \Gamma'}{\delta A_{\mu}^{a}} \frac{\delta_{L} \Gamma'}{\delta K^{a \, \mu}} + 
        \frac{\delta_{R} \Gamma'}{\delta  \mathcal{F}_{\mu \nu }^{a}} \frac{\delta_{L} \Gamma'}{\delta K^{a \, \mu \nu }} + 
        \frac{\delta_{R} \Gamma' }{\delta c^{a}} \frac{\delta_{L} \Gamma' }{\delta K^{a}} 
    \right) =0
\end{equation}
based on the invariance of the generating functional in Eq.~\eqref{eq:fgYM1} under the FOYM BRST symmetry. Recall that $ \Gamma' $ is defined as the effective action, excluding the gauge fixing terms.

The Slavnov-Taylor identities of the first-order formulation of YM are derived by taking functional derivatives of this master equation.

 \chapter{Gravity}\label{chapter:EH}

In this chapter, we review the quantization of the second-order gravity theory. We establish the covariant path integral quantization of the first-order gravity theory using the procedure developed in Section~\ref{section:LMprocedure}. Additionally, we discuss the diagonalization of the first-order formulation of gravity, which is used for the explicit verification of certain structural identities coming from the quantum equivalence. 

\section{Hilbert-Einstein action}
\label{section:SOHE}

The Hilbert-Einstein (HE) action reads \cite{Brandt:2020vre} 
\begin{equation}\label{eq:11}
    S_{\text{HE}} =  -\frac{1}{ \kappa^{2} } \int \mathop{d^D x} \sqrt{-g} g^{\mu \nu} R_{\mu \nu} (\Gamma),
\end{equation}
where 
\begin{equation}\label{eq:defkappa}
    \kappa^{2} = 16 \pi G_{ \text{N} }
\end{equation}
($G_{\text{N}} $ is the Newton's gravitational constant), $ R_{\mu \nu}( \Gamma ) $ is the Ricci tensor and the dimension $D>2$. 
The \emph{Ricci tensor} is given by \cite{wald:1984}
\begin{equation}\label{eq:13}
    R_{\mu \nu} ( \Gamma ) = \Gamma^{ \rho}{}_{ \mu \rho, \nu} -
    \Gamma^{ \rho}{}_{\mu \nu, \rho } - \Gamma^{\sigma} {}_{ \mu \nu} \Gamma^{ \rho}{}_{
    \sigma \rho} + \Gamma^{\rho}{}_{ \mu \sigma} \Gamma^{\sigma}{}_{ \nu \rho},
\end{equation}
where partial derivatives are represented by a comma $ \partial_{\mu} X = X_{a,\mu} $ and $\Gamma^{\lambda} {}_{\mu \nu}$ is a connection. 

The connection can be used to define the covariant derivative: 
\begin{equation}\label{eq:defCD}
    \nabla_{\alpha}^{\Gamma} V_{\mu} \equiv V_{\mu ; \alpha } = V_{\mu , \alpha } - \tensor{\Gamma}{^{\rho}_{ \mu \alpha }} V_{\rho} \quad \text{and} \quad \nabla_{\alpha}^{\Gamma} V^{\mu} \equiv V^{\mu}{}_{; \alpha} = V^{\mu}{}_{, \alpha} + \tensor{\Gamma}{^{\mu}_{\rho \alpha}} V^{\rho}.
\end{equation}
The definition of the covariant derivative of a tensor of type $(M,N)$ follows immediately 
\begin{equation}\label{eq:defCDgeral}
    \tensor{T}{_{\mu_{1} \cdots \mu_{N} }^{\nu_{1} \cdots \nu_{M} }_{; \alpha} }
    = \tensor{T}{_{\mu_{1} \cdots \mu_{N} }^{\nu_{1} \cdots \nu_{M} }_{, \alpha} } - \tensor{\Gamma}{^{\beta}_{\mu_{1} \alpha}} \tensor{T}{_{ \beta \mu_{2} \cdots \mu_{N} }^{\nu_{1} \cdots \nu_{M} } } - \cdots + \tensor{\Gamma}{^{\nu_{M} }_{ \beta\alpha}} \tensor{T}{_{\mu_{1} \cdots \mu_{N} }^{ \nu_{1} \cdots \nu_{M-1} \beta} }. 
\end{equation}

In GR, the connection is the Levi-Civita connection. This is the only connection that is compatible with the metric (metricity) and is torsion-less \cite{Shapiro:2001rz}. \emph{Metricity} is defined as the vanishing of the covariant derivative of the metric. Using Eq.~\eqref{eq:defCDgeral}, we have that 
\begin{equation}\label{eq:metricity}
 \nabla_{\alpha}^{\Gamma} g_{\mu \nu} = g_{\mu \nu ; \alpha} = g_{\mu \nu , \alpha} - \Gamma^{\rho} {} _{\mu \alpha}g_{\nu \rho} - \Gamma^{\rho} {}_{\nu \alpha} g_{\rho \mu } = 0 \quad \text{(metricity)}.
\end{equation}
Meanwhile, the \emph{torsion} $ \tensor{\mathcal{T}}{^{\rho}_{\mu \nu}}$, defined as 
\begin{equation}\label{q:torsion}
    \tensor{\mathcal{T}}{^{\rho}_{\mu \nu}} = \tensor{\Gamma}{^{\rho}_{\mu \nu}} - \tensor{\Gamma}{^{\rho}_{\nu \mu}},
\end{equation}
also vanishes for the Levi-Civita connection. 
We can obtain the components of the Levi-Civita connection imposing these conditions. This leads to the \emph{Christoffel symbols}
\begin{equation}\label{eq:12}
    \Gamma^{\lambda} {}_{\mu \nu} = \frac{1}{2} g^{\lambda \sigma} \left (
    g_{ \mu \sigma, \nu } + g_{ \nu \sigma , \mu} - g_{ \mu \nu, \sigma} \right
    ).
\end{equation}

The Einstein-Cartan theory is a gravity theory with torsion, $ T^{\rho} {}_{ \mu \nu } \neq 0 $, in which the connection satisfies the metricity condition \cite{Hehl:1976kj, Shapiro:2001rz}. Meanwhile, in metric affine theories both conditions are relaxed \cite{Hehl:1994ue}. Both are classes of theories that are inequivalent to GR\@.\footnote{In contrast, teleparallelism is fully equivalent to the Hilbert-Einstein action theory \cite{Pereira:2013qza}. This is another class of gravity theories in which the gravity is represented by the torsion, not by curvature (which vanishes identically). } On the other hand, the Palatini formulation of GR, a first-order formulation of the HE action, is closely related to the Einstein-Cartan theory. In this work, we shall restrict ourselves to GR described by the HE action (with the Levi-Civita connection).

Using the Euler-Lagrange equations, we obtain from Eq.~\eqref{eq:11} the Einstein equations \cite{wald:1984}:
\begin{equation}\label{eq:fieldEins}
    R_{\mu \nu} - \frac{1}{2} g_{\mu \nu} R = \frac{\kappa^{2}}{2}  T_{\mu \nu},
\end{equation}
where $R= R_{\mu \nu} g^{\mu \nu} $ and  $ T_{\mu \nu} $ is the energy-momentum tensor. 

Instead of the metric $ g_{\mu \nu} $, one can use 
\begin{equation}\label{eq:defh}
h^{\mu \nu} \equiv  \sqrt{-g} g^{\mu \nu}, 
\end{equation}
which is convenient for perturbative expansions of the metric around the Minkowski background. The Hilbert-Einstein action becomes
\begin{equation}\label{eq:aEHh}
    S_{\text{HE}} =  \int \mathop{d^{D}x} {\mathcal{L}}_{\text{HE}},
\end{equation}
where \cite{Capper:1973pv}
\begin{equation}\label{eq:lagEHh}
    {\mathcal{L}}_{\text{HE}} = -\frac{1}{4 \kappa^{2} } \left ( h^{\rho \sigma} h_{\lambda \alpha} h_{\kappa \tau} h^{\alpha \kappa} {}_{, \rho } h^{\lambda \tau} {}_{, \sigma } - \frac{1}{D-2} h^{\rho \sigma} h_{\alpha \kappa} h_{\lambda \tau} h^{\alpha \kappa} {}_{, \rho } h^{\lambda \tau} {}_{, \sigma }- 2 h_{ \alpha \tau} h^{\alpha \kappa} {}_{, \rho } h^{ \rho \tau} {}_{, \kappa }\right ).
\end{equation}
This Lagrangian is invariant under the gauge transformation \cite{Capper:1973pv}
\begin{equation}\label{eq:varh}
    \delta h^{\mu \nu} = \kappa [h^{\mu \rho} \partial_{\rho} \theta^{\nu} + h^{\nu \rho} \partial_{\rho} \theta^{\mu} - \partial_{\rho} ( h^{\mu \nu} \theta^{\rho} )]. 
\end{equation}

Since we are interested in investigating gravity in a perturbative regime, we introduce a background metric. For simplicity, we choose a flat background, then $ h^{\mu \nu} $ in this background reads 
\begin{equation}\label{eq:hwithbg}
    h^{\mu \nu} (x) = \eta^{\mu \nu} + \kappa \phi^{\mu \nu} (x).
\end{equation}
The fluctuations of $ h^{\mu \nu} $ around the background are the graviton field, which is the spin-2 gauge field responsible for mediating gravitational interactions. When expanded in terms of $ \phi^{\mu \nu} $, the quantization of the HE action \eqref{eq:lagEHh} can be carried out using the standard quantization procedure, previously applied in YM theory.

\subsection{Quantization} \label{section:HEquantization}

Here, we proceed with the quantization of the HE action. For more details, the reader is referred to \cite[Appendix B]{Capper:1973pv}. 

The procedure is analogous to the quantization of the YM theory in section~\ref{section:quantizationSOYM}. First, we introduce a gauge-fixing term such as
\begin{equation}\label{eq:laggfEH}
    {\mathcal{L}}_{\text{gf}}^{} (\phi) = -\frac{1}{2 \alpha \kappa^{2} } ( \partial_{\mu} h^{\mu \nu} )^{2} = - \frac{1}{2 \alpha } ( \partial_{\mu} \phi^{\mu \nu} )^{2}
\end{equation}
in which we used that $ \partial_{\mu} h^{\mu \nu } = \partial_{\mu} (\eta^{\mu \nu} + \kappa \phi^{\mu \nu}) = \kappa \partial_{\mu} \phi^{\mu \nu}$. This is the analog of \eqref{eq:laggauge} for gravity. 
For $ \alpha = 1 $, we have the so-called DeDonder gauge, which simplifies considerably any perturbative computation. 

Using Eq.~\eqref{eq:laggfEH}, which corresponds to the gauge condition $ \partial_{\mu} \phi^{\mu \nu} =0$, will lead to the introduction of the corresponding FP ghost fields $d_\nu$, $\bar d_\mu$. The ghost action reads \cite{Brandt:2015nxa} 
\begin{equation}\label{eq:310}
    {\mathcal{L}}_{\textrm{gh}} ( \phi )= 
\bar d_\mu\left\{\partial^2 \eta^{\mu\nu} + \kappa \left[
(\phi^{\rho\sigma}_{,\rho})\partial_\sigma\eta^{\mu\nu}
-(\phi^{\rho\mu}_{,\rho})\partial^{\nu} 
+\phi^{\rho\sigma}\partial_{\rho}\partial_\sigma\eta^{\mu\nu}
-(\partial_\rho\partial^\nu\phi^{\rho\mu})\right]\right\} d_\nu.
\end{equation}

Thus, the effective action of second-order gravity (SOGR) is 
\begin{equation}\label{eq:lageffEH}
    {\mathcal{L}}_{\text{eff}}^{\text{II}} = {\mathcal{L}}_{\text{HE}} + {\mathcal{L}}_{\text{gf}}^{} (\phi ) + {\mathcal{L}}_{\text{gh}}^{} (\phi).
\end{equation}
The path integral quantization follows directly 
\begin{equation}\label{eq:ZEH}
Z_{\text{HE}}[j] =  \int 
 \mathop{\mathcal{D} \phi^{\mu \nu} } 
\mathop{\mathcal{D}\bar{d}_{\mu} } 
\mathop{\mathcal{D} d_{\nu} } 
    \exp{i \left [S_{\text{HE}} + \int \mathop{d^D x} \left ( {\mathcal{L}}_{\text{gf}}^{} (\phi ) + {\mathcal{L}}_{\text{gh}}^{} (\phi)  + {j}_{\mu \nu}^{} \phi^{\mu \nu} \right)\right ]},
\end{equation}
where $S_{\text{HE}} $ is given by Eq.~\eqref{eq:aEHh} with $ h^{\mu \nu} = \eta^{\mu \nu} + \kappa \phi^{\mu \nu}$. 

On the other hand, we also need $ h_{\mu \nu } = g_{\mu \nu} / \sqrt{-g} = (h^{\mu \nu} )^{-1}$. The inverse of Eq.~\eqref{eq:hwithbg} yields the following power series:
\begin{equation}\label{eq:hinv}
    h_{\mu \nu} = \eta_{\mu \nu} - \kappa \phi_{\mu \nu} + \kappa^{2} \phi_{\mu \rho} \phi_{\nu \rho} + \mathcal{O} (\kappa^{3} ).
\end{equation}
This implies that the HE action in Eq.~\eqref{eq:lagEHh} shall lead to infinite momentum-dependent interaction vertices. This is one of the reasons for the non-renormalizability of quantum gravity. Besides that, it considerably complicates the computation of diagrams. The Feynman rules of the SOGR used in this work are presented in appendix~\ref{section:FeynmanRules}. 

\subsection{BRST symmetry}
The BRST symmetry of the action in Eq.~\eqref{eq:ZEH} reads 
\begin{subequations}\label{eq:BRSTofSOGR}
    \begin{align}
        & \mathsf{s} \phi^{\mu \nu}  =  h^{\mu \rho} \partial_{\rho} d^{\nu} +  h^{\nu \rho} \partial_{\rho} d^{\mu}-  \partial_{\rho} (h^{\mu \nu} d^{\rho} ) ,\\
        & \mathsf{s} d^{\mu} = \kappa d^{\rho} d^{\mu}_{, \rho}  ,\\
        & \mathsf{s} \bar{d}^{\mu} = B^{\mu} \quad \text{and} \quad  \mathsf{s} B^{\mu} =0,
    \end{align}
\end{subequations}
where $B^{\mu} $ is a Nakanishi-Lautrup field \cite{lautrup:1967, Nakanishi:1966zz}. The Nakanishi-Lautrup field is necessary to ensure the off-shell nilpotency of the BRST symmetry. 

As we have seen in Section~\ref{section:FOYMBRST}, we obtain the Slavnov-Taylor identities by extending the source term. In the SOGR, the extended source term is 
\begin{equation}\label{eq:sourceof2GRextended}
    j_{\mu \nu} \phi^{\mu \nu} + i( \bar{\eta}_{\mu} d^{\mu} - \bar{d}^{\mu} \eta_{\mu} ) 
    + j_{\mu} B^{\mu}
    +  K_{\mu \nu}\left[ 
 h^{\mu \rho} \partial_{\rho} d^{\nu} +  h^{\nu \rho} \partial_{\rho} d^{\mu}-  \partial_{\rho} (h^{\mu \nu} d^{\rho} ) 
\right]
    +  \kappa K_{\mu} d^{\rho} d^{\mu}_{, \rho}.
\end{equation}
The corresponding Zinn-Justin master equation is given by
\begin{equation}\label{eq:MSofSOGR}
    \int \mathop{d x} \left (\frac{\delta_{R} \Gamma }{\delta \phi^{\mu \nu} } \frac{\delta_{L} \Gamma }{\delta K_{\mu \nu}} + 
    \frac{\delta_{R} \Gamma }{\delta d^{\mu }} \frac{\delta_{L} \Gamma }{\delta K_{\mu }}  + \frac{\delta_{R} \Gamma }{\delta \bar{d}^{\mu }} \frac{\delta_{L} \Gamma }{\delta j_{\mu }}\right) =0,
\end{equation}
which is based on the invariance of the generating functional~\eqref{eq:ZEH} under the SOGR BRST symmetry. The Slavnov-Taylor identities of the second-order formulation of gravity are derived by taking functional derivatives of this master equation.

\section{Hilbert-Palatini action}\label{section:FOEH}

In this section, we examine the first-order formulation of gravity (FOGR) also known as the Hilbert-Palatini (HP) formalism \cite{palatini:1919}. In the standard formalism, the (Levi-Civita) connection $ \Gamma (g^{\mu \nu} )$ is a metric-dependent variable whose components are given by the Christoffel symbol. However, in the HP formalism, metric and the (affine) connection are treated as independent fields.  The affine connection plays a similar role to the auxiliary field $ \mathcal{F} $ in the FOYM formulation.  

As we have done in the SOGR, we replace $ g^{\mu \nu} $ by $ h^{\mu \nu} $ and the connection $ \Gamma $ by \cite{Brandt:2015nxa}
\begin{equation}\label{eq:defG} 
    \mathbb{G}_{\mu \nu}^{\lambda} \equiv \tensor{\Gamma}{^{\lambda}_{\mu \nu}} -     
    \frac{1}{2} \left( \delta_{\mu}^{\lambda} \Gamma^{\alpha}{}_{\nu \alpha } + \delta_{\nu}^{\lambda } \Gamma^{\alpha}{}_{\mu \alpha } \right).
\end{equation}
Replacing $ \mathbb{G}_{\mu \nu}^{\lambda} $ with the auxiliary field $G_{\mu \nu}^{\lambda} $, we arrive at the FOGR Lagrangian
\begin{equation}\label{eq:defagpo}
    {\mathcal{L}}_{\text{HP}}(h) = \frac{h^{\mu \nu}}{\kappa^2} \left (
    G^{\lambda}_{ \mu \nu, \lambda} + \frac{1}{D-1} G^{\lambda}_{ \mu \lambda} G^{\sigma}_{ \nu \sigma} - G^{\lambda}_{ \mu \sigma} G^{\sigma}_{ \nu \lambda} \right ).
\end{equation}
The field $ G^{\lambda}_{\mu \nu} $ is an auxiliary field analogous to $ \mathcal{F}_{\mu \nu}^{a} $, and $ \mathbb{G}_{\mu \nu}^{\lambda}  $ in Eq.~\eqref{eq:defG} is its classical value. The FOGR Lagrangian is invariant under the following gauge transformation \cite{Brandt:2015nxa}
\begin{align}
    \delta h^{\mu \nu} ={}& \kappa \left[h^{\mu \rho} \partial_{\rho} \theta^{\nu} + h^{\nu \rho} \partial_{\rho} \theta^{\mu} - \partial_{\rho} ( h^{\mu \nu} \theta^{\rho} )\right]  \intertext{and} \delta {G}{^{\lambda}_{\mu \nu}} ={}& 
    \kappa \bigg[- \partial_{\mu} \partial_{\nu} \theta^{\lambda} + \frac{1}{2} ( \delta_{\mu}^{\lambda} \partial_{\nu} + \delta_{\nu}^{\lambda} \partial_{\mu} ) \partial_{\rho} \theta^{\rho} - \theta^{\rho} \partial_{\rho}  {G}{^{\lambda}_{\mu \nu}} \nonumber\\  \label{eq:transHP}& \quad + {G}{^{\rho}_{\mu \nu}} \partial_{\rho} \theta^{\lambda} - ( {G}{^{\lambda}_{\mu \rho}} \partial_{\nu} + {G}{^{\lambda}_{\nu \rho}} \partial_{\mu} )\theta^{\rho}\bigg] .
\end{align}

It is important to note that  $ \mathcal{F}_{\mu \nu}^{a} $ is associated with the curvature in the YM theory $ F_{\mu \nu}^{a} $, while $ G^{\lambda}_{\mu \nu} $ is associated with the Levi-Civita connection $ \Gamma^{\lambda}{}_{\mu \nu} $. 
In pure theories, without matter fields, this distinction is not relevant, and the first-order formulation of gravity is quite analogous to the FOYM\@. However, the coupling with matter fields breaks this parallel since the minimal coupling of matter fields is generally related to the connection. Thus, in the first-order formulation of gravity, the coupling is associated with the auxiliary field $G^{\lambda}_{\mu \nu}$. This contrasts with the FOYM, where matter couples to the gauge field (the connection $A_{\mu}^{a}$) rather than the auxiliary field, which is associated with the curvature $F_{\mu \nu}^{a}$.

We can rewrite Eq.~\eqref{eq:defagpo} as
\begin{equation}\label{eq:lagFOEH}
    \mathcal{L}_{ \textrm{HP}} = 
    \frac{1}{2\kappa^{2}}
    G^{\lambda}_{ \mu \nu} \tensor*{M}{*_{\lambda}^{\mu \nu} _{\rho}^{ \pi\tau}}(h) G^{\rho}_{ \pi \tau } - \frac{1}{\kappa^{2} } G^{\lambda}_{\mu \nu} h^{\mu \nu}{}_{, \lambda},
\end{equation}
where
\begin{equation}\label{eq:36BM2016}
    \begin{split}
        \tensor*{M}{*^{\mu\nu}_{\lambda}^{\pi\tau}_{\sigma}}(h)   &= 
        \frac{1}{2}\Big[\frac{1}{D-1}\left( \delta^\nu_\lambda\delta^\tau_\sigma h^{\mu\pi}+
                                                \delta^\mu_\lambda\delta^\tau_\sigma h^{\nu\pi}+
                                                \delta^\nu_\lambda\delta^\pi_\sigma h^{\mu\tau}+
                                                \delta^\mu_\lambda\delta^\pi_\sigma h^{\nu\tau}
\right) 
      \\  & \qquad \quad  -  
\left( 
                                                \delta^\tau_\lambda\delta^\nu_\sigma h^{\mu\pi}+
                                                \delta^\tau_\lambda\delta^\mu_\sigma h^{\nu\pi}+
                                                \delta^\pi_\lambda\delta^\nu_\sigma h^{\mu\tau}+
                                                \delta^\pi_\lambda\delta^\mu_\sigma h^{\nu\tau}
                                        \right) \Big].
\end{split}
\end{equation}
Note that, Eq.~\eqref{eq:lagFOEH} and the FOYM Lagrangian have a similar structure. Expanding the metric $ h^{\mu \nu} $ around a flat background, as in Eq.~\eqref{eq:hwithbg}, Eq.~\eqref{eq:lagFOEH} becomes
\begin{equation}\label{eq:lagFOEHwithg}
    \mathcal{L}_{\text{HP}} = 
    \frac{1}{2 \kappa^{2} } G^{\lambda}_{ \mu \nu} \tensor*{M}{*_{\lambda}^{\mu \nu}_{\rho}^{ \pi\tau}}( \eta + \kappa \phi ) G^{\rho}_{ \pi \tau } - \frac{1}{\kappa } G^{\lambda}_{\mu \nu} \phi^{\mu \nu}{}_{,\lambda}.
\end{equation}
We find that, remarkably, we have a single cubic interaction $(GG \phi )$ which does not depend on momentum, but now we have the mixed propagators $ \langle 0|T G{}^{\lambda}{}_{\mu \nu}  \phi^{\alpha \beta} | 0 \rangle_{\text{free}}  $ and $ \langle 0|T \phi^{\alpha \beta} G{}^{\lambda} {}_{\mu \nu} | 0 \rangle_{\text{free} } $, and the propagator of the auxiliary field $ \langle 0|T G{}^{\lambda}{}_{\mu \nu}G{}^{\rho}{}_{\pi\tau}| 0 \rangle_{\text{free}} $ to compensate it. Nevertheless, the perturbation theory in the FOGR formalism is simplified considerably since infinite vertices are condensed in a finite number of components. 

The Euler-Lagrange equations reads 
\begin{subequations}\label{eq:eomFOGR}
    \allowdisplaybreaks
\begin{equation}\label{eq:eomG}
    G^{\lambda}_{\mu \nu} = \mathbb{G}^{\lambda}_{\mu \nu} =  \tensor*{(M^{-1})}{*_{\mu \nu}^{\lambda}_{ \pi \tau}^{ \rho}} (h) h^{\pi \tau}{}_{, \rho},
\end{equation}
\begin{equation} \label{eq:eomG1}
    \frac{1}{2} {G}{^{\lambda}_{\mu \nu}} {G}{^{\rho}_{\pi \tau}} 
    \tensor*{\mathcal{M}}{*^{\mu \nu}_{\lambda}^{\pi \tau }_{\rho}_{ \alpha \beta }}   
    +  {G}{^{\lambda}_{ \alpha \beta , \lambda} }=0;
\end{equation}
\end{subequations}
where $ \mathbb{G}^{\lambda}_{\mu \nu}$ denotes the classical value of the auxiliary field $G^{\lambda}_{\mu \nu}$, 
\begin{equation}\label{eq:defMcal}
    \tensor*{\mathcal{M}}{*^{\mu \nu}_{\lambda}^{\pi \tau }_{\rho}_{ \alpha \beta }}   
    \equiv \frac{\partial \tensor*{M}{*^{\mu \nu}_{\lambda}^{\pi \rho}_{\sigma}}(h)}{\partial h^{\alpha \beta}}, 
\end{equation}
and 
\begin{equation}\label{eq:35}
    \begin{split}
    \tensor*{(M^{-1})}{*_{\mu \nu}^{\lambda}_{ \pi \tau}^{ \rho}} (h) 
={}&- \frac{1}{2(D-2)} h^{\lambda \rho} h_{\mu \nu} h_{\pi \tau} +
    \frac{1}{4} h^{\lambda \rho} \left ( h_{\pi \mu} h_{\tau \nu} + h_{\pi \nu} h_{\tau \mu}\right ) 
    \\
    & - \frac{1}{4} \left ( h_{\tau \mu} \delta_{\nu}^{\rho} \delta_{\pi}^{\lambda} + h_{\pi \mu} \delta_{\nu}^{\rho} \delta_{\tau}^{\lambda} +  h_{\tau \nu} \delta_{\mu}^{\rho} \delta_{\pi}^{\lambda} +  h_{\pi \nu} \delta_{\mu}^{\rho} \delta_{\tau}^{\lambda}\right ).
\end{split}
\end{equation}
This inverse satisfies $\tensor*{(M^{-1})}{*_{\mu \nu}^{\lambda}_{ \pi \tau}^{ \rho}}    \tensor*{M}{*^{\pi \tau }_{\rho }^{\alpha \beta }_{\gamma}}= ( \delta_{\mu}^{\alpha } \delta_{\nu}^{\beta } + \delta_{\mu}^{\beta } \delta_{\nu}^{\alpha } ) \delta_{\gamma}^{\lambda}/2$.

The Einstein-Hilbert equations are obtained by substituting Eq.~\eqref{eq:eomG} in Eq.~\eqref{eq:eomG1}:
\begin{equation}\label{eq:EHequationGandh}
        \frac{1}{2}   \tensor{h}{^{\mu \nu}_{,\lambda}} 
    \tensor*{(M^{-1})}{*_{\mu \nu}^{\lambda}_{ \pi \tau}^{ \rho}} (h)
    \tensor*{\mathcal{M}}{*_{\rho}^{\pi \tau}_{\gamma}^{\delta \sigma }_{ \alpha \beta }} 
    \tensor*{(M^{-1})}{*_{ \delta \sigma }^{ \gamma }_{ \xi \omega }^{ \chi}} (h) h^{\xi \omega}{}_{, \chi} +  
        \tensor*{(M^{-1})}{*_{\alpha \beta}^{ \lambda }_{\mu \nu}^{ \rho }} (h) h^{\mu \nu} {}_{ , \rho \lambda }=0, 
    \end{equation}
    which is equal to
\begin{equation}\label{eq:EHequationGandh2}
                         \tensor*{(M^{-1})}{*_{ \alpha \beta }^{\lambda}_{ \pi \tau}^{ \rho}} (h)
                         \tensor*{M}{*_{\rho }^{\pi \tau}^{\delta \sigma }_{ \gamma}_{,\lambda  }}(h) 
    \tensor*{(M^{-1})}{*_{ \delta \sigma }^{ \gamma }_{ \xi \omega }^{ \chi}} (h) h^{\xi \omega}{}_{, \chi}=0.
\end{equation}
Since it is pure gravity, the right-hand side (energy-momentum tensor) vanishes.

Moreover, using the classical equation of motion in Eq.~\eqref{eq:eomG} in the FOGR Lagrangian Eq.~\eqref{eq:lagFOEH} to solve the auxiliary field $ G^{\lambda}_{\mu \nu} $ yields 
\begin{equation}\label{eq:eqclass}
    {\mathcal{L}}_{\text{HE}} = 
    - \frac{1}{2 \kappa^{2}} h^{\mu \nu} {}_{, \lambda}
    \tensor*{(M^{-1})}{*_{\mu \nu}^{\lambda}_{ \pi \tau}^{ \rho}} (h) h^{\pi \tau} {}_{, \rho},
\end{equation}
which is equal to the SOGR Lagrangian given in Eq.~\eqref{eq:lagEHh}. This demonstrates the classical equivalence between both formulations of gravity.

\subsection{Quantization} \label{section:QuantizationFOGR}

We can find the appropriate generating functional of the FOGR formalism from the SOGR generating functional. This guarantees the quantum equivalence between these formulations. Besides that, we show that in the first-order formulation, second-class constraints arise. These constraints are treated appropriately within the FS procedure. Meanwhile, the second-order formulation of the gauge theories investigated in this work can be quantized by the standard FP procedure.

In the last chapter, we have shown that the Senjanovi\'{c} determinant in the FOYM is field-independent and can be absorbed by the normalization factor. This is a common characteristic of the Senjanovi\'{c} determinant. However, in certain systems it is not trivial \cite{Chishtie:2011wd}. In the FOGR theory, we find a field-dependent determinant factor, which can be interpreted as the Senjanovi\'{c} determinant.

Even though a trivial Senjanovi\'{c} determinant does not yield any significant contribution at the perturbative level, it may be relevant in certain scenarios. For instance, in quantum field theory at finite temperature, trivial ghosts lead to relevant contributions in the computation of the free energy \cite{Bellac:2011kqa}. We can understand this by the following reasoning. The free energy is proportional to the number of physical degrees of freedom of the theory. Therefore, even the trivial non-interacting ghosts have to be taken into account in order to cancel the unphysical degrees of freedom coming from the gauge field. Thus, we can assume that ghosts may be counted as a negative number of degrees of freedom and need to be accounted for. 

For example, in four-dimensional QED, the gauge field has $4$ components, although only $2$ are physical (transversal components). In the path integral formalism, the FP ghosts $c$ and $ \bar{c} $ are free fields, that is, they are trivial ghosts. Thus, as we have seen, four-dimensional QED should have $ N_{A} - N_{c} - N_{\bar{c}} =4-2=2$ degrees of freedom which is exactly the two possible polarization of the photon, the correct number of physical degrees of freedom of QED. This shows the relevance of finding the appropriate path integral, even when we only have trivial ghosts. 

The standard approach would find all the constraints of the FOGR formalism to compute the Senjanovi\'{c} determinant, but this would lead to a non-covariant path integral \cite{Chishtie:2012sq, Chishtie:2013fna}. A covariant canonical analysis appeared recently in \cite{Romero:2022kjh} (and references within). However, the second-class constraints are solved. Since the second-class constraints are reducible, similar to the second-class in the FOYM theory, it is easier to solve them. A complete analysis, which does not solve the second-class constraints, such as done in \cite{Chishtie:2012sq} has the downside of not being manifestly covariant.

We can avoid these issues by using the procedure proposed in section~\ref{section:FOYMfromSOYM}.
We start the procedure by inserting the factor
\begin{equation}\label{eq:1HE}
    1 =
    \int 
    \mathop{\mathcal{D} G^{\lambda}_{\mu \nu}} 
    \Delta^{1/2}( h,J)  \exp i \int \mathop{d^{D}x} \left ( 
        \frac{1}{2 \kappa^{2} } G^{ \lambda}_{\mu \nu} 
        \tensor*{M}{*^{\mu \nu}_{\lambda}^{\pi \tau}_{\rho}} (h) 
        G^{\rho}_{ \pi \tau} 
    +  \frac{ \kappa^{2}}{2}  J_{\lambda}^{\mu \nu} \tensor*{(M^{-1})}{*_{\mu \nu}^{\lambda}_{\pi \tau}^{\rho}} J_{\rho}^{ \pi \tau}\right ) 
\end{equation}
into the SOGR generating functional
\begin{equation}\label{eq:ZHEwithSourceG}
Z_{\text{HE}}[j] =  N\int 
 \mathop{\mathcal{D} h^{\mu \nu} } 
\mathop{\mathcal{D}\bar{d}_{\mu} } 
\mathop{\mathcal{D} d_{\nu} } 
\exp{i  \int \mathop{d^D x} \left (
{\mathcal{L}}_{\text{eff}}^{\text{II}} (h)  + {j}_{\mu \nu}^{} h^{\mu \nu} +  J^{\lambda }{}_{\mu \nu} \mathbb{G}_{\lambda}^{\mu \nu }(h) \right)}
\end{equation}
with an additional source for the classical value of the auxiliary field $ \mathbb{G}^{\lambda}_{ \mu \nu }(h)$  (see Eq.~\eqref{eq:eomG}). 

Then, we redefine the auxiliary field 
\begin{equation}\label{eq:redfG}
    G^{\lambda}_{\mu \nu} \to G^{\lambda}_{\mu \nu}  
    - \mathbb{G}^{\lambda}_{\mu \nu} + \kappa^{2}  \tensor*{(M^{-1})}{*_{\mu \nu}^{\lambda}_{\pi \tau}^{\rho}} (h)J_{\rho}{}^{\pi \tau}  
\end{equation}
yielding
\begin{equation}\label{eq:fgHE12}
                            N \int   
    \mathop{\mathcal{D} \mu_{\text{I}}}
    \exp
    i\int \mathop{d^D x}  \left ( 
    {\mathcal{L}}_{\text{eff}}^{\text{I}} + J^{\lambda}_{\mu \nu} G_{\lambda}^{\mu \nu}   + \frac{\kappa^{2}}{2} J_{\lambda}^{\mu \nu} \tensor*{(M^{-1})}{*_{\mu \nu}^{\lambda}_{\pi \tau}^{\rho}}(h) J_{\rho}^{ \pi \tau}      + j_{\mu \nu} h^{\mu \nu} \right) ,
\end{equation}
where $ \mathop{\mathcal{D} \mu_{\text{I}} }  =\mathop{\mathcal{D} h^{\mu \nu} } 
    \mathop{\mathcal{D} G^{\lambda}_{\mu \nu} } 
    \mathop{\mathcal{D} \bar{d}_{\mu} } 
    \mathop{\mathcal{D} d_{\nu} }  \Delta ^{1/2}( h, J) $ and 
    \begin{equation}\label{eq:lageffHP}
        \mathcal{L}_{\text{eff}}^{\text{I}} = \mathcal{L}_{\text{HP}}(h)+ \mathcal{L}_{\text{gf} } (h) + \mathcal{L}_{\text{gh}} (h) 
    \end{equation}
    denotes the effective first-order gravity Lagrangian.

    Now, we need to determine $ \Delta^{1/2}  (h,J)$. For this, we integrate the auxiliary field in Eq.~\eqref{eq:1HE} arriving at
\begin{equation}\label{eq:DeltaHE}
    \Delta^{1/2} (h,J) = 
|\det \tensor*{M}{*_{ \lambda}^{\mu \nu}_{\rho}^{\pi \tau}}  (h)|^{1/2} 
\exp i \int \mathop{d^{D} x} \left (  
-\frac{\kappa^{2}}{2} J_{\lambda}^{\mu \nu} \tensor*{(M^{-1})}{*_{\mu \nu}^{\lambda}_{\pi \tau}^{\rho}} (h)J_{\rho}^{ \pi \tau} \right ).
\end{equation}

Using Eq.~\eqref{eq:DeltaHE}, we obtain the resulting generating functional 
\begin{equation}\label{eq:fghHE1}
Z_{\text{HE} } [j,J] = N \int   
\mathop{\mathcal{D} \mu_{\text{I}}'}
    \exp
    i\int \mathop{d^{D} x}  \left ( 
        {\mathcal{L}}_{\text{eff}}^{\text{I}} + J^{\lambda}_{\mu \nu} G_{\lambda}^{\mu \nu} + j_{\mu \nu} h^{\mu \nu}  
+\frac{\kappa^{2}}{2} J_{\lambda}^{\mu \nu} \tensor*{(M^{-1})}{*_{\mu \nu}^{\lambda}_{\pi \tau}^{\rho}} (h)J_{\rho}^{ \pi \tau} 
\right ),
\end{equation}
where
\begin{equation} \label{eq:measureFOGR}
    \mathop{\mathcal{D} \mu_{ \text{I} } '} = \mathop{\mathcal{D} h^{\mu \nu}} \mathop{\mathcal{D} G^{\lambda}_{\mu \nu} }\mathop{\mathcal{D} \bar{d}_{\mu} } \mathop{\mathcal{D} d_{\nu}} |\det \tensor*{M}{*_{ \lambda}^{\mu \nu}_{\rho}^{\pi \tau}}  (h)|^{1/2}.
    \end{equation}
The generating functional~\eqref{eq:fghHE1} is equivalent to that of the second-order gravity (SOGR) formulation. However, the action appearing in Eq.~\eqref{eq:fghHE1} corresponds to the first-order formulation of gravity, namely the HP action. In addition, an extra quadratic term in the source $J_{\mu\nu}^{\lambda}$ arises. 

    This demonstrates the quantum equivalence between these formulations. The determinant factor in the measure \eqref{eq:measureFOGR} may be interpreted as the Senjanovi\'{c} determinant, which arises due to second-class constraints in the FOGR formalism. The second-class constraints, in turn, appear from the presence of additional fields, namely, the auxiliary field $G^{\lambda}_{ \mu \nu} $.

    The manifestly non-covariant Senjanovi\'{c} determinant obtained in \cite{Chishtie:2012sq}, in our notation, reads 
    \begin{equation}\label{eq:SDofFOGR}
        \Delta_{\text{FS}}^{1/2} (H)={} \det \left[h^{00} \left ( \delta_{i}^{l} \delta_{k}^{j} - \frac{1}{D -1} \delta_{i}^{j} \delta_{k}^{l} \right )\right] D_{2}(H/h), 
   \end{equation}
   where
   \begin{equation} \label{eq:D2of2012} 
       \begin{split}
           D_{2} (H/h) = \det \bigg\{\frac{1}{h^{00}}  \bigg[
           \frac{1}{D-1} & \left(   
                \delta^j_i\delta^m_l H^{kn}+
                                                \delta^k_i\delta^m_l H^{jn}+
                                                \delta^j_i\delta^n_l H^{km}+
                                                \delta^k_i\delta^n_l H^{jm}
\right) 
           \\ {}-{}&   
\left( 
                                                \delta^m_i\delta^j_l H^{kn}+
                                                \delta^m_i\delta^k_l H^{jn}+
                                                \delta^n_i\delta^j_l H^{km}+
                                                \delta^n_i\delta^k_l H^{jm}
                                        \right) 
                                \bigg] \bigg\}
\end{split}
    \end{equation}
    with $ H^{ij} \equiv  h^{00} h^{ij} - h^{0i} h^{0j} $. Note that the terms $ h^{00} $ cancel, so that the Senjanovi\'{c} determinant only depends on $H^{ij} $. The first determinant in Eq.~\eqref{eq:SDofFOGR} can be discarded, leaving us with the determinant
    \begin{equation} \label{eq:D2H}
        D_{2} (H) =  \det \tensor*{M}{*_{i}^{jk}_{l}^{mn}} (H),
    \end{equation}
    which resembles the determinant factor appearing in Eq.~\eqref{eq:measureFOGR}. This strongly indicates that the determinant in Eq.~\eqref{eq:measureFOGR} is the manifestly covariant formulation of the Senjanovi\'{c} determinant in Eq.~\eqref{eq:SDofFOGR} found in Ref. \cite{Chishtie:2012sq}. Moreover, we have already shown that this is valid for the first-order formulation of YM theory, which is analogous to first-order gravity. 
    
Introducing a flat background metric $ h^{\mu \nu} = \eta^{\mu \nu} + \kappa \phi^{\mu \nu}$ in~Eq.~\eqref{eq:fghHE1} yields 
\begin{equation}\label{eq:313}
    \begin{split}
        Z_{\text{HP}} [j,J] ={}&N\int   
    \mathop{\mathcal{D} \phi^{\mu \nu} } 
    \mathop{\mathcal{D} G^{\lambda}_{\mu \nu} } 
    \mathop{\mathcal{D} \bar{d}_{\mu} } 
    \mathop{\mathcal{D} d_{\nu} } 
    | \det \tensor*{M}{*^{\mu \nu}_{\lambda}^{\pi \tau}_{\rho}} (h) |^{1/2} 
    \\
                               & \qquad \times \exp{i \left [ \int \mathop{d^D x} \left ( {\mathcal{L}}_{\text{eff}}^{\text{I}} + J_{\lambda}^{\mu \nu} G^{\lambda}_{\mu \nu} + {j}_{\mu \nu}^{} \phi^{\mu \nu} \right)\right ]}, 
\end{split}
\end{equation}
where the effective Lagrangian is 
\begin{equation}\label{eq:leffpog}
    {\mathcal{L}}_{\text{eff}}^{\text{I}} =   \frac{1}{2\kappa^{2}}
    G_{ \mu \nu}^{ \lambda} \tensor*{M}{*_{\lambda}^{\mu \nu}_{\rho}^{ \pi\tau}}(h)
      G_{ \pi \tau }^{ \rho} - \frac{1}{\kappa} G_{\mu \nu}^{\lambda} \phi^{\mu \nu}_{, \lambda} - \frac{1}{2 \alpha} ( \partial_{\mu} \phi^{\mu \nu} )^{2} + {\mathcal{L}}_{\text{gh}}^{} ( \phi ).
\end{equation}
Using the linearity of $ \tensor*{M}{*_{\lambda}^{\mu \nu}_{\rho}^{\pi \tau}} (h)$,
\begin{equation}\label{eq:linM}
   M(h) = M ( \eta ) + \kappa M(\phi ),
\end{equation}
in Eq.~\eqref{eq:leffpog}, we get 
\begin{equation}\label{eq:leffpog1}
    \begin{split}
        {\mathcal{L}}_{\text{eff}}^{\text{I}} &= \frac{1}{2\kappa^{2}}
    G_{ \mu \nu}^{ \lambda} \tensor*{M}{*_{\lambda}^{\mu \nu}_{\rho}^{ \pi\tau}}(\eta )
        G_{ \pi \tau }^{ \rho} - \frac{1}{\kappa } G_{\mu \nu}^{\lambda} \phi^{\mu \nu}_{, \lambda} - \frac{1}{2 \alpha} ( \partial_{\mu} \phi^{\mu \nu} )^{2} + \bar d_\mu \partial^2 \eta^{\mu\nu} d_{\nu} \\ & 
 + \frac{1}{2\kappa}
    G_{ \mu \nu}^{ \lambda} \tensor*{M}{*_{\lambda}^{\mu \nu}_{\rho}^{ \pi \tau}}( \phi )
      G_{ \pi \tau }^{ \rho}
+       
\kappa \bar d_\mu \left[
(\phi^{\rho\sigma}_{,\rho})\partial_\sigma\eta^{\mu\nu}
-(\phi^{\rho\mu}_{,\rho})\partial^{\nu} 
+\phi^{\rho\sigma}\partial_{\rho}\partial_\sigma\eta^{\mu\nu}
-(\partial_\rho\partial^\nu\phi^{\rho\mu})\right] d_\nu.
  \end{split}
\end{equation}
From Eq.~\eqref{eq:leffpog1}, we see that mixed propagators
$  \langle 0|T\phi^{\mu \nu} G_{\alpha \beta}^{\lambda}  |0\rangle_{\text{free}} $ and 
$  \langle 0|T G_{\alpha \beta}^{\lambda} \phi^{\mu \nu}|0   \rangle_{\text{free}} $ arise. 
We also have the momentum-independent vertex $(G \phi \phi )$. 
The complete set of FOGR Feynman rules used in this work is presented in Appendix~\ref{section:FeynmanRules}.

We note that the propagator matrix can be diagonalized by a shift, leading to the \emph{diagonal first-order gravity} (DFOGR) formulation \cite{Brandt:2018avq}, which is reviewed briefly in Section~\ref{sec:DFOEH}. This diagonalization facilitates the explicit verification of the structural identities derived for first-order gravity \cite{Brandt:2020vre}, which is considered in Section~\ref{section:QEGravity}.

\subsection{BRST symmetry}
The BRST symmetry of the action in Eq.~\eqref{eq:313} reads 
\begin{subequations}\label{eq:BRSTofFOGR}
    \begin{align}
        & \mathsf{s}\phi^{\mu \nu}  = h^{\mu \rho} \partial_{\rho} d^{\nu} -h^{\nu \rho} \partial_{\rho} d^{\mu} - \partial_{\rho} ( h^{\mu \nu} d^{\rho} ) ,\\
        & \mathsf{s} {G}{_{\mu \nu }^{\lambda} }=
    \kappa \bigg[-  \partial_{\mu} \partial_{\nu} d^{\lambda} + \frac{1}{2} ( \delta_{\mu}^{\lambda} \partial_{\nu} + \delta_{\nu}^{\lambda} \partial_{\mu} ) \partial_{\rho} d^{\rho} - d^{\rho} \partial_{\rho} {G}{^{\lambda}_{\mu \nu}} \nonumber\\  & \qquad \qquad +   {G}{^{\rho}_{\mu \nu}} \partial_{\rho} d^{\lambda} - ( {G}{^{\lambda}_{\mu \rho}} \partial_{\nu} + {G}{^{\lambda}_{\nu \rho}} \partial_{\mu} )d^{\rho}\bigg],\\
        & \mathsf{s} d^{\mu} = \kappa d^{\rho} d^{\mu}_{, \rho}  ,\\
        & \mathsf{s} \bar{d}^{\mu} = B^{\mu} \quad \text{and} \quad \mathsf{s} B^{\mu} =0,
    \end{align}
\end{subequations}
where $B^{\mu} $ is the Nakanishi-Lautrup field. 

In the FOGR, the extended source term is given by
\begin{equation}\label{eq:sourceof1GRextended}
    \begin{split}
    & j^{\mu \nu} \phi_{\mu \nu} + {J}{_{\lambda}^{\mu \nu}} {G}{^{\lambda}_{\mu \nu}} + i( \bar{\eta}_{\mu} d^{\mu} - \bar{d}^{\mu} \eta_{\mu} ) 
+ j_{\mu} B^{\mu}
 \\ & 
+K_{\mu \nu } \left[
h^{\mu \rho} \partial_{\rho} d^{\nu} -h^{\nu \rho} \partial_{\rho} d^{\mu} - \partial_{\rho} ( h^{\mu \nu} d^{\rho} ) 
\right]
+ \kappa  K_{\mu} d^{\rho} d^{\mu}_{, \rho} 
\\ & 
+ \kappa {K}{_{\lambda}^{\mu \nu}} 
        \bigg[- \partial_{\mu} \partial_{\nu} d^{\lambda} + \frac{1}{2} ( \delta_{\mu}^{\lambda} \partial_{\nu} + \delta_{\nu}^{\lambda} \partial_{\mu} ) \partial_{\rho} d^{\rho} - d^{\rho} {G}{^{\lambda}_{\mu \nu}} 
+ {G}{^{\rho}_{\mu \nu}} \partial_{\rho} d^{\lambda} - ( {G}{^{\lambda}_{\mu \rho}} \partial_{\nu} + {G}{^{\lambda}_{\nu \rho}} \partial_{\mu} )d^{\rho} \bigg].
    \end{split}
\end{equation}
The corresponding Zinn-Justin master equation is given by
\begin{equation}\label{eq:MSofFOGR}
    \int \mathop{d x} \left (
        \frac{\delta_{R} \Gamma }{\delta \phi^{\mu \nu} } \frac{\delta_{L} \Gamma }{\delta K_{\mu \nu}} + 
        \frac{\delta_{R} \Gamma }{\delta {G}{^{\lambda}_{\mu \nu}} } \frac{\delta_{L} \Gamma }{\delta {K}{_{\lambda }^{\mu \nu}}} + 
    \frac{\delta_{R} \Gamma }{\delta d^{\mu }} \frac{\delta_{L} \Gamma }{\delta K_{\mu }}  + \frac{\delta_{R} \Gamma }{\delta \bar{d}^{\mu }} \frac{\delta_{L} \Gamma }{\delta j_{\mu }}\right) =0,
\end{equation}
which is based on the invariance of the generating functional~\eqref{eq:313} under the FOGR BRST symmetry. The Slavnov-Taylor identities of the first-order formulation of gravity are derived by taking functional derivatives of this master equation.

\subsection{Diagonal first-order formulation} \label{sec:DFOEH}
In this section, we briefly examine the DFOGR formalism. In this formulation the mixed propagators are absent. However, we increase the number of interaction vertices. Even so, there is still a significant simplification in this formulation compared to the SOGR theory involves an infinite number of interaction vertices.

The diagonal formulation can be obtained with the field redefinition \cite{martins-filho:2021}
\begin{equation}\label{eq:34}
    G^{\lambda}_{ \mu \nu} = {H}^{\lambda}_{ \mu \nu} + 
    \tensor*{(M^{-1})}{*_{\mu\nu}^{\lambda}_{ \pi \tau}^{ \rho}} (\eta ) h^{\pi \tau}{}_{,\rho} 
\end{equation}
in Eq.~\eqref{eq:lagFOEHwithg}, which leads to the following Lagrangian
\begin{equation}\label{eq:37}
\begin{split}
    {\mathcal{L}}^{ \textrm{Id}}_{ \textrm{HP}} ={}& \frac{1}{2 \kappa^{2} }
    {H}_{ \mu \nu}^{ \lambda} \tensor*{M}{*_{\lambda}^{\mu \nu}_{\rho}^{ \pi\tau}}(\eta)
      {H}_{ \pi \tau }^{ \rho} - \frac{1}{2} {\phi}_{ , \lambda}^{ \mu \nu} 
    \tensor*{(M^{-1})}{*_{\mu \nu}^{\lambda}_{ \pi \tau}^{ \rho}} (\eta ) {\phi}_{ , \rho}^{ \pi \tau} 
                                                \\ &+ \frac{1}{2 \kappa } [ {H}_{ \mu \nu}^{ \lambda} + \kappa {\phi}_{ , \rho}^{ \alpha \beta} \tensor*{(M^{-1})}{*_{\alpha \beta}^{\rho}_{ \mu \nu}^{ \lambda}}(\eta)] \tensor*{M}{*_{\lambda}^{\mu \nu} _{ \sigma }^{ \pi \tau}} (\phi )  [ {H}_{\pi \tau }^{\sigma } + \kappa \tensor*{(M^{-1})}{*_{\pi \tau }^{\sigma }_{ \gamma \delta }^{ \upsilon }}(\eta){\phi}_{ , \upsilon }^{ \gamma \delta}   ].
 \end{split}
\end{equation}
The field redefinition in Eq.~\eqref{eq:34} is simply a shift without any non-local terms. Therefore, the DFOGR is equivalent to the FOGR and SOGR formulation of gravity, even at the quantum level.

In order to quantize the theory, we can use the same measure in Eq.~\eqref{eq:313}. Let us define the effective action of the DFOGR as
\begin{equation}\label{eq:311}
    \mathcal{L}^{\textrm{Id}}_{\textrm{eff}} =
    {\mathcal{L}}_{\textrm{HP}}^{\textrm{Id}} 
    + {\mathcal{L}}_{\textrm{gf}} ( \phi )
+ \mathcal{L}_{\textrm{gh}} ( \phi ).
\end{equation}
Then, the generating functional is given by
\begin{equation}\label{eq:313d}
    \begin{split}
        Z_{\text{HP}}^{\text{Id} }[j,J] ={}&N  \int 
    \mathop{\mathcal{D} \phi^{\mu \nu} } 
    \mathop{\mathcal{D} H_{\mu \nu}^{\lambda} }
    \mathop{\mathcal{D}\bar{d}_{\mu} } 
    \mathop{\mathcal{D} d_{\nu} } 
    | \det \tensor*{M}{*^{\mu \nu}_{\lambda}^{\pi \tau}_{\rho}} (h) |^{1/2} 
 \\ & \qquad   
    \times \exp{i  \int \mathop{d^D x} \left ({\mathcal{L}}_{\text{eff}}^{\text{Id}} + J_{\lambda}^{\mu \nu} {H}_{\mu \nu}^{\lambda} + {j}_{\mu \nu}^{} \phi^{\mu \nu} \right)}.
    \end{split}
\end{equation}
From Eq.~\eqref{eq:37}, we observe that the auxiliary field and the metric are no longer coupled. Consequently, there are no mixed propagators. On the other hand, we have two additional vertices $(H \phi \phi )$ and $( \phi \phi \phi )$ that are momentum-dependent.

In the next chapter, we explore the consequences of the quantum equivalence in greater detail. We derive identities that relate Green's function in the first-order formulation to those of composite fields computed in the second-order formulation. These structural identities can be regarded as identities of the first-order gravity itself, suggesting a novel physical interpretation of the auxiliary field.

 \chapter{Quantum equivalence} \label{section:QE}

In this chapter, we review the quantum equivalence between the first- and second-order formulations of YM and gravity theories, following the results of Ref.~\cite{martins-filho:2021}. We also explore this quantum equivalence through the effective action, \( \Gamma \), which generates the 1PI Green's functions. Finally, we demonstrate that the contributions from the Senjanovi\'{c} determinant play a crucial role in verifying structural identities between the Green's functions of the first- and second-order formulations of YM and gravity theories, as originally derived in Ref.~\cite{Brandt:2020vre}, at the integrand level.

\section{Yang-Mills theory}\label{section:QuantumEquivalenceYM}

By shifting the auxiliary field in the generating functional of the FOYM in Eq.~\eqref{eq:fgYM1} 
\begin{equation}\label{eq:shiftEsf}
    \mathcal{F}^{a}_{\mu \nu} \to \mathring{\mathcal{F}}^{a}_{\mu \nu}   +( \partial_{\mu} A^{a}_{\nu} - \partial_{\nu} A^{a}_{\mu} + g f^{abc} A^{b}_{\mu} A^{c}_{\nu}),
\end{equation}
which must leave it unchanged, we find 
\begin{equation}\label{eq:fgYM1sfshifted}
        Z_{\text{1YM} }' [0] = 
        N \int \mathcal{D} \mathring{\mathcal{F}}_{\mu \nu}^{a} \exp i\int \mathop{dx} \left ( \frac{1}{4}  \mathring{\mathcal{F}}^{a}_{\mu \nu} \mathring{\mathcal{F}}^{ a \, \mu \nu }\right) 
        \int  \mathop{\mathcal{D} A^{a}_{\mu}}  \mathcal{D} c^{a} \mathcal{D} \bar{c}^{a} \exp i\int \mathop{dx}  {\mathcal{L}}_{\text{eff}}^{(2)} 
                                 .
\end{equation}
We observe that the effective action turns out to be the  SOYM effective action \eqref{eq:lageff2}. 
Absorbing the first integral in the normalization constant defining 
\begin{equation}\label{eq:YM:normalizationconstantredefinition}
    N' = 
        N \int \mathcal{D} \mathring{\mathcal{F}}_{\mu \nu}^{a} \exp i\int \mathop{dx} \left ( \frac{1}{4}  \mathring{\mathcal{F}}^{a}_{\mu \nu} \mathring{\mathcal{F}}^{{ a \, \mu} \nu }\right),
\end{equation}
we get 
\begin{equation}\label{eq:equivsf12}
    Z^{\prime}_{\text{1YM}}[0] = Z_{\text{1YM}} [0] = Z_{\text{2YM}} [0].
\end{equation}
This demonstrate the quantum equivalence between the first and second-order formulation of Yang-Mills theory, which is consistent with our discussion in Section~\ref{section:quantizationFOYM}. 

With the inclusion of sources, we generalize the shift~\eqref{eq:shiftEsf}, as shown in Ref. \cite{McKeon:2020lqp}, to
\begin{equation}\label{eq:shiftws}
    {\mathcal{F}}^{a}_{\mu \nu} \to \mathring{\mathcal{F}}^{a}_{\mu \nu} + {F}^{a}_{\mu \nu} -2 J^{a}_{\mu \nu},
\end{equation}
the generating functional \eqref{eq:fgYM1} is then given by
\begin{equation}\label{eq:fgYM1wsp2}
Z^{\prime}_{\text{1YM} } [j,J] = 
     N' \int  \mathop{\mathcal{D} A^{a}_{\mu} } \mathcal{D} c^{a} \mathcal{D} \bar{c}^{a}  \exp i\int \mathop{dx}    \big[{\mathcal{L}}_{\text{eff}}^{(2)}   + j^{a \, \mu} {A}_{ \mu}^{a} + {J}_{}^{{ a \, \mu \nu}} {F}_{\mu \nu}^{a} - J^{a \, \mu \nu} {J}_{\mu \nu}^{a}  \big].
\end{equation}
When $J=0$, 
\begin{equation}\label{eq:equivwJ}
    Z_{\text{1YM}} [j,0]= Z'_{\text{1YM}} [j,0] = Z_{\text{2YM}} [j],
\end{equation}
which establishes that the quantum equivalence holds when the sources are present.

Functional derivatives on both sides of Eq.~\eqref{eq:equivwJ} with respect to $j$  generates Green's function of the field $ A$, respectively, in FOYM and SOYM theory:
\begin{equation}\label{eq:eqvA}
    \begin{split}
                                                    \left. (-i)^n \frac{\delta^{n} Z_{\text{1YM}}[j,J] }{\delta j^{a_1}_{\mu_1} (x_1) \cdots \delta {j}_{\mu_{n} }^{a_{n}}(x_n) } \right  |_{J=j=0}
        ={}& \left. (-i)^n \frac{\delta^{n} Z_{\text{2YM}} [j]}{\delta j^{a_1}_{\mu_1} (x_1) \cdots \delta {j}_{\mu_{n} }^{a_{n}}(x_n) } \right |_{j=0} \\ 
            \equiv{}& \langle 0|T A^{a_1 \, \mu_1}(x_1) \cdots A^{{ a_n \, \mu_n}} (x_n)| 0 \rangle .
    \end{split}
\end{equation}
Therefore, the $n$-point Green's functions of the gauge field $A$ are the same in both formulations. This result has been explicitly verified for $n=4$ at the tree-level \cite{martins-filho:2021}, $n=2$ (up to the one-loop order) in Ref. \cite{niederle:1983} (by using the background field method in  \cite{McKeon:1994ds}) and at finite temperature \cite{vasconcelos:2020}.

\subsection{Structural identities}\label{section:SIofYM}

Now, for $ J \neq 0 $, 
\begin{equation}\label{eq:equivws12}
    Z_{\text{1YM}} [j,J] = Z^{\prime}_{\text{1YM}}[j,J] \equiv  Z_{\text{2YM}} [j,J].
\end{equation}
We can obtain structural identities \cite{McKeon:2020lqp} that relate Green's functions of the auxiliary field $F$ with Green's function in the SOYM formalism of the field strength tensor $ {F} $. As we have shown, the field strength tensor $ {F}_{\mu \nu}^{a} (x) = \partial_{\mu} A^{a}_{\nu}(x) - \partial_{\nu} A^{a}_{\mu}(x) + g f^{abc} A^{b}_{\mu} (x)A^{c}_{\nu}(x)$ 
is a composite operator, which is equal to the auxiliary field $ \mathcal{F}_{\mu \nu}^{a} (x)$ at the classical level. These structural identities are the realization of this equality at the quantum level.

Some of these structural identities are (see \cite{McKeon:2020lqp,martins-filho:2021} for a detailed derivation)
\begin{subequations}\label{eq:YM:SI}
    \begin{align} \label{eq:YM:SI1}
      \langle 0| T \mathcal{F}^{a \, \mu \nu} (x)  |0  \rangle ={}& \langle 0|T {F}^{a \, \mu \nu} (x)| 0 \rangle,
        \\\label{eq:YM:SI2}
          \langle 0|T \mathcal{F}^{a \, \mu \nu} (x) A^{b \, \rho} (y)| 0 \rangle
        ={}&\langle 0|T {F}^{a \, \mu \nu} (x) A^{b \, \rho} (y)| 0 \rangle,
        \\\label{eq:YM:SI3}
            \langle 0|T \mathcal{F}^{a \, \mu \nu} (x) \mathcal{F}^{b \, \rho \sigma } (y)  | 0 \rangle ={}&  \langle 0| T {F}^{a \, \mu \nu} (x) {F}^{b \, \rho \sigma} (y)  |0  \rangle +2i\delta^{ab} I^{\mu \nu \rho \sigma } \delta (x-y).
    \end{align}
\end{subequations}
In principle, the left-hand side of Eq.~\eqref{eq:YM:SI} is computed in the FOYM formalism while the right-hand side is in the SOYM\@. However, from Eq.~\eqref{eq:eqvA}, we see that any Green's function in the SOYM can also be computed equivalently in the FOYM\@. Consequently,  we can also state that these structural identities are identities between FOYM Green's functions. For instance, we have that
\begin{equation}\label{eq:YM:fid1}
    \left \langle 0| T ( \mathcal{F} - {F} )^{a \, \mu \nu} (x)  |0 \right \rangle = 0
\end{equation}
in which Green's function of the composite operator is also computed in the FOYM formalism. Therefore, the composite field Green's functions can be easily evaluated in the FOYM formalism. It also relates the renormalization of the auxiliary field $ \mathcal{F} $ with the renormalization of the composite field. 

We can easily generalize the structural identities Eq.~\eqref{eq:YM:SI}, or even obtain more involved ones as 
\begin{equation}\label{eq:is3p2}
    \begin{split}
         \langle 0|T \mathcal{F}^{a \, \mu \nu} (x) \mathcal{F}^{b \, \rho \sigma } (y) \mathcal{F}^{c \, \alpha \beta } (z) A^{d \, \gamma} (w)| 0 \rangle ={}&
         \langle 0|T  {F}^{a \, \mu \nu}(x) {F}^{b \, \rho \sigma }(y) {F}^{c \, \alpha \beta } (z) A^{d \, \gamma} (w)| 0 \rangle 
                                                                                                                                 \\ &  + 2 i \delta^{ab} I^{\mu \nu \rho \sigma} \delta (x-y) \langle 0|T {F}^{c \, \alpha \beta } (z) A^{d \, \gamma} (w) | 0 \rangle 
                                                                                                                                 \\ &  + 2 i \delta^{bc} I^{\rho \sigma \alpha \beta } \delta (y-z) \langle 0|T {F}^{a \, \mu \nu  } (x) A^{d \, \gamma} (w) | 0 \rangle 
                                                                                                                                 \\ &  + 2 i \delta^{ac} I^{\mu \nu \alpha \beta } \delta (x-z) \langle 0|T {F}^{b \, \rho \sigma } (y) A^{d \, \gamma} (w) | 0 \rangle .
 \end{split}
\end{equation}
Note the presence of the contact term in Eq.~\eqref{eq:YM:SI3} revealing the quantum nature of the auxiliary $ F$. These terms also appear in the Ward-Takahashi identities \cite{Takahashi:1957xn} and in the Dyson-Schwinger equations \cite{Dyson:1949ha, Schwinger:1951ex}. 

Although these structural identities are not the usual Slanov-Taylor identities (derived in Chapter~\ref{section:YMtheory}) they can be obtained alternatively by a similar procedure using the symmetry \cite{Lavrov:2021pqh} 
\begin{equation}\label{eq:YM:trivialsymmetry}
    \delta \mathcal{F}^{a}_{\mu \nu} = f^{abc} ( \mathcal{F}^{b}_{\mu \nu} - {F}^{b}_{\mu \nu} ) \zeta^{c} , \quad \delta A^{a}_{\mu} =0
\end{equation}
of the FOYM\@.\footnote{We thank Prof. Josif Frenkel for this clarification.} We can recast Eq.~\eqref{eq:YM:trivialsymmetry} as
\begin{equation}\label{eq:YM:proving}
    \delta \mathcal{F}_{\mu \nu}^{a} = 2 \eta^{ab} \frac{\delta S_{\text{1YM}}}{\delta \mathcal{F}^{b \, \mu \nu}},
\end{equation}
where $ \eta^{ab} = f^{abc} \zeta^{c} = - \eta^{ba} $. 
Symmetries of this form are referred to as \emph{trivial symmetries} \cite{henneaux1992quantization}, since the variation of the action vanishes trivially:
\begin{equation}\label{eq:variationofL}
    \delta {S_{\text{1YM}}} = \frac{\delta S_{\text{1YM}}}{\delta \mathcal{F}^{b \, \mu \nu} } \delta \mathcal{F}^{b \, \mu \nu} =2 \eta^{bc}  
\frac{\delta S_{\text{1YM}}}{\delta \mathcal{F}^{b \, \mu \nu} } 
\frac{\delta S_{\text{1YM}}}{\delta \mathcal{F}^{c \, \mu \nu} } =0,
\end{equation}
a condition that is independent of structure of the action.

The invariance of Eq.~\eqref{eq:fgYM1} under the symmetry in Eq.~\eqref{eq:YM:trivialsymmetry} means that 
\begin{equation}\label{eq:Wardp1}
    \begin{split}
        \delta Z_{\text{1YM}} [j,J;K] ={}&
\int \mathop{\mathcal{D} \mathcal{F}_{\mu \nu}^{a}} \mathop{\mathcal{D} A_{\mu}^{a}}   J^{a}_{\mu \nu} \delta \mathcal{F}^{a \, \mu \nu}  \exp{i (S_{\text{1YM}} + S_{\text{src}}[j,J;K]} )
\\
        ={}& \int \mathop{d x} f^{abc} J_{\mu \nu}^{a}\left (  \frac{\delta Z_{1YM} }{\delta J_{\mu \nu}^{b} }  -  \frac{\delta Z_{\text{1YM}} }{\delta K_{\mu \nu}^{b}} \right ) \zeta^{c} =0 
,
\end{split}
\end{equation}
where $ S_{\text{1YM}} $ is the FOYM action. We extended the source term to 
\begin{equation}\label{eq:Wardp2}
    S_{\text{src}} [j,J;K]=\int \mathop{d x} (j_{\mu}^{a} A^{a \, \mu} + J_{\mu \nu}^{a} \mathcal{F}^{a \, \mu \nu} + K_{\mu \nu}^{a} F^{a \, \mu \nu}  ),
\end{equation}
introducing the source $ K_{\mu \nu}^{a} $ for $ F^{a \, \mu \nu} $.

Taking functional derivatives of Eq.~\eqref{eq:Wardp1} with respect to the sources $j$ and $J$  leads to structural identities, such as Eqs.~\eqref{eq:YM:SI1} and \eqref{eq:YM:SI2}. However, the fundamental identity \eqref{eq:eqvA} does not follow from Eq.~\eqref{eq:Wardp1}. Hence, these Slanov-Taylor-like identities arising from the trivial symmetry in Eq.~\eqref{eq:YM:trivialsymmetry} are insufficient to ensure the quantum equivalence.

\subsection{Structural identities at one-loop order} 

Here, we briefly review the verification of the structural identities in Eq.~\eqref{eq:YM:SI}. This is done in great detail in \cite{martins-filho:2021}, and in the appendix of Ref. \cite{McKeon:2020lqp}. 

The explicit verification of the structural identities at the tree-level is straightforward. For instance, the left-hand side of Eq.~\eqref{eq:YM:SI2} reads 
\begin{equation}\label{eq:LHSSI2}
    \langle 0|T F^{a \, \mu \nu} A^{b \, \rho}| 0 \rangle = \partial^{\mu} \langle 0|T A^{a \, \nu} A^{b \, \rho}| 0 \rangle - \partial^{\nu} \langle 0|T A^{a \, \mu} A^{b \, \rho}| 0 \rangle.
\end{equation}
By using that $  \partial^{\mu} = ip^{\mu} $, and the standard form of the $ \langle 0|T A^{a \mu}  (-p) A^{b \nu} (p)|0  \rangle_{\text{free}} $, which is given by 
\begin{equation}\label{eq:propA}
    - \frac{i}{p^{2}} \left(\eta^{\mu \nu}  - (1- \alpha ) \frac{p^{\mu} p^{\nu}}{p^{2}} \right)\delta^{ab}
\end{equation}
in the Lorenz gauge, we find that 
\begin{equation}\label{eq:LHSSI2a}
    \langle 0|T F^{a \, \mu \nu} A^{b \, \rho}| 0 \rangle = \frac{ p^{\mu} \eta^{\nu \rho } - p^{ \nu } \eta^{\mu \rho }}{p^{2}} \delta^{ab} = \langle 0|T \mathcal{F}^{a \, \mu \nu} A^{b \, \rho} | 0 \rangle
\end{equation}
(cf. Eq.~\eqref{fig:FRpgFA}).

\begin{figure}[ht] 
\centering
\includegraphics[scale=0.7]{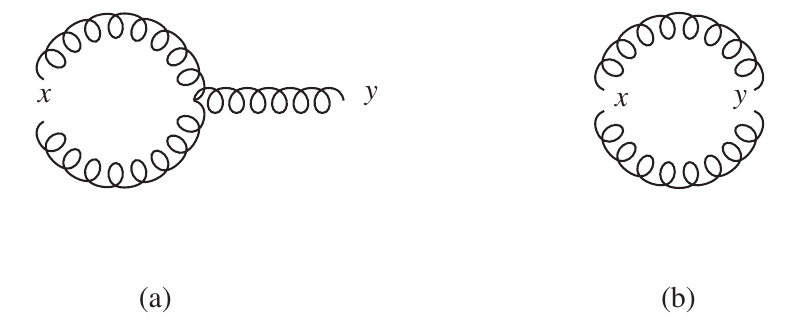}
\caption{Pinched diagrams. Diagrams (a) and (b) respectively represent the contributions to the left-hand side of the structural identities~\eqref{eq:YM:SI1} and~\eqref{eq:YM:SI2} at one-loop order.} \label{fig:HH1p}
\end{figure}

\begin{figure} [ht]
    \centering
\includegraphics[scale=0.75]{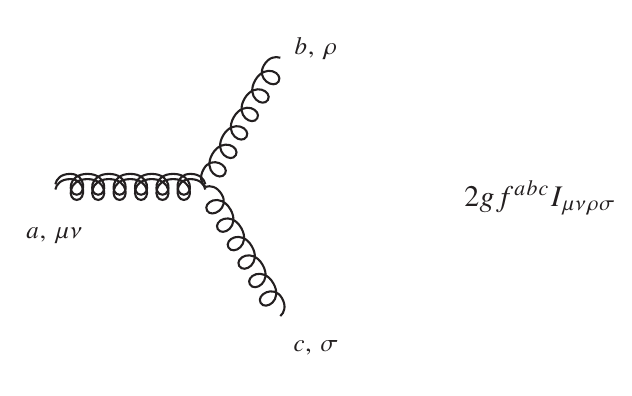}
\caption{Extended Feynman rule for the interaction vertex with a composite field $ \mathbb{H}_{\mu \nu}^{a} $, which is represented by a double spring line.}
\label{fig:FRhAA}
\end{figure}

On the other hand, at one-loop order we have to compute \emph{pinched diagrams} (see Fig.~\ref{fig:HH1p}). Composite fields, such as the curvature $ F_{\mu \nu}^{a} (x)$, in which two or more fields are taken in the same space-time point, lead to ultraviolet singularities. 
The standard procedure to compute these diagrams is using a limit process, for instance, the diagram in Fig.~\ref{fig:HH1p}(a) is given by ``pinching'' their external legs:
\begin{equation}\label{eq:limitpinched}
    \displaystyle \lim_{z \to x} \langle 0|T A(x) A (z) A(y)| 0 \rangle = \langle 0|T A(x)^2 A (z)| 0 \rangle,
\end{equation}
where $ A(x)^2 $ is a composite field.
In \cite{martins-filho:2021}, the Feynman rules were extended by introducing the vertex $(\mathbb{H}AA)$ shown in Fig.~\ref{fig:FRhAA}, where \begin{equation}\label{eq:def:auxH} \mathbb{H}_{\mu \nu}^{a} (x)= g f^{abc} A_{\mu}^{b} (x)A_{\nu}^{c}(x) \end{equation}  is the composite part of $ F_{\mu \nu}^{a} (x)$. 

\begin{figure}[ht] 
\centering
\includegraphics[scale=0.70]{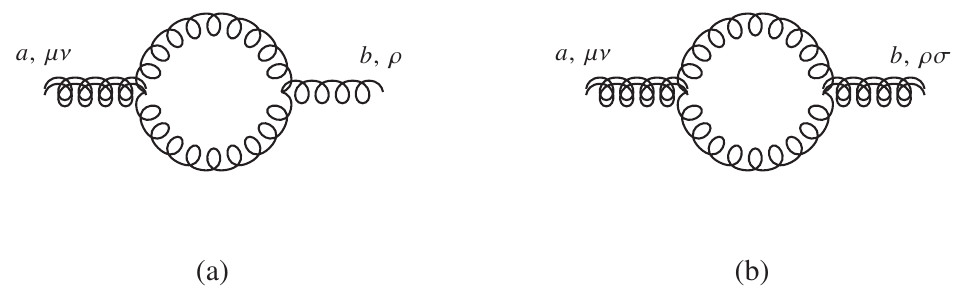}
\caption[]{Pinched diagrams of Fig.~\ref{fig:HH1p} represented by extended Feynman rules. The symmetry factors are the same of those of the original pinched diagrams.}
\label{fig:pinchedYM}
\end{figure}

This vertex allows us to construct diagrams that contribute to Green's functions of the composite field $\mathbb{H}_{\mu \nu}^{a} $. For instance, the pinched diagrams in Fig.~\ref{fig:HH1p} are represented by the standard diagrams in Fig.~\ref{fig:pinchedYM} using the extended Feynman rule in Fig.~\ref{fig:FRhAA}. The computation of the pinched diagrams can be done using standard methods (this also includes the symmetry factor). 

Contributions to the right-hand side of the structural identities \eqref{eq:YM:SI} are given by pinched diagrams, while the left-hand side is the usual Green's function computed in the FOYM\@. 
The extended Feynman rule $(hAA)$ can be written as \cite{martins-filho:2021} 
\begin{equation}\label{eq:hAAintermsofHAA}
    (\mathbb{H}_{\mu \nu}^{a} A_{\rho}^{b} A_{\sigma}^{c} ) 
    \equiv 
    \langle 0|T H_{\mu \nu}^{a} H_{\alpha \beta}^{d} |0\rangle_{\text{free}}   ( \mathcal{F}_{\alpha \beta}^{d} A_{\rho}^{b} A_{\sigma}^{c} ) = 2 i ( \mathcal{F}_{ \mu \nu }^{a} A_{\rho}^{b} A_{\sigma}^{c} ),
\end{equation}
where $ ( \mathcal{F}_{\mu \nu}^{a} A_{\rho}^{b} A_{\sigma}^{c}) $ is the $(FAA)$ vertex defined in Eq.~\eqref{eq:vertYM1} and $\langle 0|T H_{\mu \nu}^{a} H_{\alpha \beta}^{d} |0\rangle_{\text{free}} = 2i I_{\mu \nu \alpha \beta} $ is the free propagator of the auxiliary field $ H_{\mu \nu}^{a} $ obtained by the diagonalization of the propagator matrix of the FOYM theory \cite{Brandt:2018avq}.
Using this rule, one can derive equivalent diagrams that contribute to Green's functions of the FOYM \cite{martins-filho:2021}.

\begin{figure}[ht]
    \centering
    \includegraphics[scale=0.85]{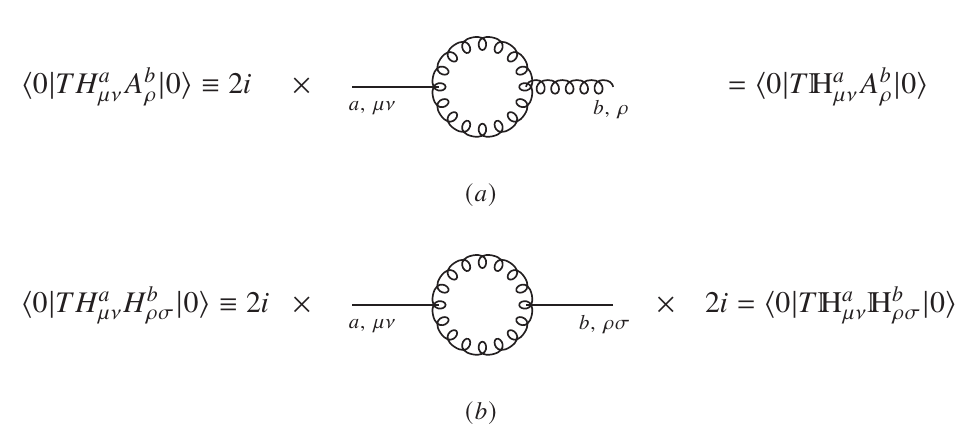}
    \caption{Green's functions (at one-loop order) of the FOYM formalism equivalent to the pinched diagrams in Fig.~\ref{fig:pinchedYM}.}
    \label{fig:HAandHH}
\end{figure}

For instance, it is shown in Fig.~\ref{fig:HAandHH} the equivalence of the pinched diagrams in Fig.~\ref{fig:pinchedYM} to the contributions of the Green's functions of the diagonal FOYM theory: (a) the mixed propagator $ \langle 0|T H_{\mu \nu}^{a}  A_{\rho}^{b} | 0 \rangle $ and (b) the propagator of the auxiliary field $ \langle 0|T H_{\mu \nu}^{a} H_{\rho \sigma}^{b} | 0 \rangle $.
In the non-diagonal FOYM theory, the pinched contributions of the left-hand side of Eq.~\eqref{eq:YM:SI} combine these diagrams. 

Hence, the structural identities hold at the integrand level. No explicit diagram computations are needed for their verification. This is interesting since they arise directly from the quantum equivalence between the first- and second-order formulations of the YM theory (and gravity). The generality of the quantum equivalence itself is reflected in the generality of the structural identities. 

Moreover, verifying these identities at the integrand level confirms their validity at any dimensions and gauge. Furthermore, they are independent of the regularization scheme \cite{martins-filho:2021}. 
In Section~\ref{section:QEGravity}, we show that this is also valid, under certain conditions, in the context of gravity. This shall extend even further the work in \cite{martins-filho:2021}, illuminating the quantum equivalence of the first- and second-order formulation of gravity.

\subsection{Effective action}\label{section:QEusingWandGamma}

The quantum equivalence derived using the generating functional $ Z [j] $ can also be appreciated through other functionals, such as the connected generating functional $ W[j] $ and the effective action $ \Gamma [ {\phi}_c ] $.

By the definition of the connected generating functional 
\begin{equation}\label{eq:gf:Wj}
    W [ j ] = -i \ln Z[j]
\end{equation}
and from Eq.~\eqref{eq:equivws12}, we get 
\begin{equation}\label{eq:qe:W}
    W_{\text{1YM}} [j,J] = W_{\text{1YM}}'[j,J] =W_{\text{2YM}} [j,J].
\end{equation}
A functional derivative in both sides of Eq.~\eqref{eq:qe:W} with respect to $j$ leads to the connected Green's function $ \left \langle \phi\right\rangle_{\mathsf{C}} $. Meanwhile, a functional derivative in the left-hand side of Eq.~\eqref{eq:qe:W} with respect to $J$  gives $ \left \langle \mathcal{F}\right\rangle_{\mathsf{C}} $ and, in the right-hand side, yields the connected Green's function of the composite field  $ F$, which is equal to the classical value of $ \mathcal{F} $. That is, we obtain the same structural identities derived using $Z $, however, now only connected diagrams contribute: 
\begin{equation}\label{eq:gf:connectedstructuralid}
    \left \langle \phi_{1} \cdots \phi_{n} \mathcal{F}_{1} \cdots \mathcal{F}_{m}\right\rangle_{\mathsf{C}} 
    =
    \left \langle \phi_{1} \cdots \phi_{n} {F}_{1} \cdots {F}_{m}\right\rangle_{\mathsf{C}} + 
    \text{contact terms,} 
\end{equation}
the subscript $ \mathsf{C} $ denotes connected Green's functions.

Let us introduce the generating functional 
\begin{equation} \label{eq:func2withoutJ2}
        Z^{(2)}_{\text{YM}} [j,J] = N \int  \mathcal{D} A^{a}_{\mu}  \mathcal{D} c^{a} \mathcal{D} \bar{c}^{a}  \exp i\int \mathop{dx}    \left( {\mathcal{L}}_{\text{eff}}^{(2)} +  j^{a \, \mu} {A}_{ \mu}^{a} + {J}_{}^{{ a \, \mu \nu}} {F}_{\mu \nu}^{a}   \right).
\end{equation}
We have that
\begin{equation} \label{eq:Z1and2relation}
    Z^{(2)}_{\text{YM}} =
    Z_{\text{1YM}} '[j,J] \exp \left(i \int \mathop{d x} J_{\mu \nu}^{a} J^{a \, \mu \nu} \right),
\end{equation}
taking the logarithm of the expression above leads to
\begin{equation}\label{eq:W2andW1}
    W^{(2)}_{\text{YM}} [j,J]  = W_{\text{1YM}} [j,J] + \int \mathop{d x} J_{\mu \nu}^{a} J^{a \, \mu \nu},
\end{equation}
where we have used Eq.~\eqref{eq:qe:W}. 
The left-hand side of this relation leads to Green's functions involving the field $A$ and the composite field $F$ in the SOYM, while the right-side leads to Green's functions with external fields $A$ and the auxiliary field $ \mathcal{F} $. The unusual quadratic dependence on the source is now factored, and Green's functions are derived in a standard way. 

Thus, Eq.~\eqref{eq:W2andW1} can be used to derive structural identities straightforwardly.
For example, the functional derivative $ \delta^{2} / \delta J_{\alpha \beta}^{a} \delta J_{\mu \nu}^{b} $ applied in Eq.~\eqref{eq:W2andW1} leads directly to a particular case of Eq.~\eqref{eq:gf:connectedstructuralid} analogous to Eq.~\eqref{eq:YM:SI3} (see Eq.~\eqref{eq:SIgamma111A}). 

Now we can discuss the quantum equivalence using effective actions.
The effective action $ \Gamma [ \phi_{\mathsf{c}} ]$ can be constructed using a Legendre transform of the functional $W[J]$. The conjugated variable to the source $J$ is the classical field $ \phi_{\mathsf{c}} $, the functional derivative of $ W[J]$ with respect to $ J $. 
Hence, let us define the classical fields 
\begin{subequations}\label{eq:definitionclassical}
    \allowdisplaybreaks
    \begin{align}
        \label{eq:v1}
        \frac{\delta W_{\text{YM} }^{(2)}[j,J] }{\delta j_{\mu}^{a} } ={}&
A^{a \, \mu}_{\mathsf{c}},
    \\ \label{eq:u1}
    \frac{\delta W_{\text{YM} }^{(2)} [j,J] }{\delta J_{\mu \nu }^{a} } ={}&
    F_{\mathsf{c}}^{a \, \mu \nu} + \mathfrak{F}^{a \, \mu \nu},    \\ \label{eq:u2}
    \frac{\delta W_{\text{1YM} }{[j,J]} }{\delta j_{\mu }^{a} } ={}& 
    A^{a \, \mu}_{\mathsf{c}},\\
    \frac{\delta W_{\text{1YM}} {[j,J]} }{\delta J_{\mu \nu }^{a} } ={}& 
    \mathcal{F}^{a \, \mu \nu}_{\mathsf{c}},
    \\
    \label{eq:v2}
        \frac{\delta W_{\text{2YM}} [j] }{\delta j_{\mu}^{a}} ={}&
A^{a \, \mu}_{\mathsf{c}},
    \end{align}
\end{subequations}
where $ F_{\mathsf{c}}^{a \, \mu \nu} \equiv  F^{a \, \mu \nu} ( A_{\mathsf{c}} )$. The subscript $\mathsf{c}$ will be omitted from now on.

This allows us to construct the following effective actions: 
\begin{align}
    \allowdisplaybreaks
    \label{eq:effectiveaction1YM}
    \Gamma_{\text{1YM}} [A, \mathcal{F} ] ={}& W_{\text{1YM}} [j,J] - j_{\mu}^{a} {A}^{a \, \mu} - J_{\mu \nu}^{a} \mathcal{F}^{a \, \mu \nu}, 
    \\
    \label{eq:effectiveaction12YM}
    \Gamma^{(2)}_{\text{YM}}   [A, \mathfrak{F} ] ={}& W_{\text{YM}}^{(2)} [j,J] - j_{\mu}^{a} A^{a \, \mu}  - J_{\mu \nu}^{a} F^{a \, \mu \nu} - J_{\mu \nu}^{a} \mathfrak{F}^{a \, \mu \nu} ,
    \\
    \label{eq:effectiveaction2YM}
    \Gamma_{\text{2YM}} [A] ={}& W_{\text{2YM}} [j] - j_{\mu}^{a} A^{a \, \mu}.
\end{align}
The effective action $ \Gamma^{(2)}_{\text{YM}} [A, \mathfrak{F} ]$ is the effective action of the composite field $ F^{a \, \mu \nu} $, that is, the curvature.  

The quantum equations of motion are
\begin{subequations}\label{eq:EoMquantumall}
\allowdisplaybreaks
\begin{align}\label{eq:EoMquantum}
    \frac{\delta \Gamma^{(2)}_{\text{YM}} }{\delta A^{a \, \mu}(x)} ={}& - j_{\mu}^{a} (x) - 2D^{ab \, \nu} J^{b}_{\nu \mu} (x), &  \frac{\delta \Gamma^{(2)}_{\text{YM }} }{\delta \mathfrak{F}^{a \, \mu \nu} (x)} = -J_{\mu \nu}^{a};\\
    \frac{\delta \Gamma_{\text{1YM}} }{\delta A^{a \, \mu}(x)} ={}& - j_{\mu}^{a} (x) , & \frac{\delta \Gamma_{\text{1YM }} }{\delta \mathcal {F}^{a \, \mu \nu} (x)} = -J_{\mu \nu}^{a}.
\end{align}
\end{subequations}

Differentiating Eq.~\eqref{eq:W2andW1} with respect to $ J^{a \, \mu \nu} $  and recalling the definition of the classical fields in Eq.~\eqref{eq:definitionclassical}, we obtain the fundamental identity 
\begin{equation}\label{eq:Fandf}
    F_{\mu \nu}^{a} + \mathfrak{F}_{\mu \nu}^{a} = \mathcal{F}_{\mu \nu}^{a} + 2 J_{\mu \nu}^{a},
\end{equation}
which is consistent with Eq.~\eqref{eq:shiftws} when $ \mathfrak{F} =0$.

From Eqs.~\eqref{eq:effectiveaction1YM} and \eqref{eq:effectiveaction12YM}, we find that 
\begin{equation}\label{eq:G1andG12}
    \Gamma_{\text{1YM}} [A , \mathcal{F} ]= \Gamma^{(2)}_{\text{YM}} [A , \mathfrak{F} ] + J_{\mu \nu}^{a} F^{a \, \mu \nu} + J_{\mu \nu}^{a} \mathfrak{F}^{a \, \mu \nu}  - J_{\mu \nu}^{a} J^{a \, \mu \nu}  - J_{\mu \nu}^{a} \mathcal{F}^{a \, \mu \nu}.
\end{equation}
Finally, by replacing Eq.~\eqref{eq:Fandf} in Eq.~\eqref{eq:G1andG12}, we obtain that
\begin{equation}\label{eq:qe:Gamma}
    \Gamma_{\text{1YM}} [A , \mathcal{F} ]    =\Gamma_{\text{YM}}^{(2)}  [A, \mathcal{F} -F + 2 J ]  + J_{\mu \nu}^{a} J^{a \, \mu \nu},
\end{equation}
or alternatively,
\begin{equation}\label{eq:qe:Gammaalt}
    \Gamma_{\text{1YM}} [A , \mathcal{F} ]       =\Gamma_{\text{YM}}^{(2)}  \left[A, \mathfrak{F}  \right] + \frac{1}{4} (\mathfrak{F}_{\mu \nu}^{a} + F_{\mu \nu }^{a} - \mathcal{F}_{\mu \nu}^{a} ) ^{2} .
\end{equation}
This shows the quantum equivalence between the FOYM and SOYM analogously to Eqs.~\eqref{eq:equivws12} and \eqref{eq:qe:W}. 
Setting $j=J=0$, we have that \cite{Cornwall:1974vz}
\begin{equation}\label{eq:equiv2and12a}
    \Gamma_{\text{1YM}}  \left[A, \bar{F}(A) \right]   = \Gamma^{(2)}_{\text{YM}} [A, \bar{F}(A)-F]=\Gamma_{\text{2YM}}  [A], 
\end{equation}
where $ \bar{F}(A)$ is the value of $ \mathcal{F} $ in which
\begin{equation} \label{eq:minimum}
    \frac{\delta \Gamma_{\text{1YM}}[A, \mathcal{F} ]}{\delta \mathcal{F}}=0.
\end{equation}
Note that, Eq.~\eqref{eq:equiv2and12a} is the analog of Eq.~\eqref{eq:equivwJ}. It is the quantum realization of the classical equivalence between the action of the SOYM and the FOYM\@.

In order to obtain structural identities, we can use the relation 
\begin{equation}\label{eq:Gammaequiv}
    \Gamma^{(2)}_{\text{YM}} [A, J]=\Gamma_{\text{1YM}} [ A , \mathcal{F} ] + J_{\mu \nu}^{a} \mathcal{F}^{a \, \mu \nu} + J_{\mu \nu}^{a} J^{a \, \mu \nu}.
\end{equation}
The effective action
\begin{equation} \label{eq:HALFeffectiveaction}
    \Gamma^{(2)}_{\text{YM}} [A, J] = W_{\text{YM}}^{(2)} [j,J] - j_{\mu}^{a} A^{a \, \mu} 
\end{equation}
is the standard effective action of the SOYM in which we introduced a source $ J$ to the composite field $F$. 
The Legendre transform of the effective action $ \Gamma^{(2)}_{\text{YM}} [A , J ]$ with respect to $J$ (with the conjugate variables \eqref{eq:definitionclassical}) results in $ \Gamma^{(2)}_{\text{YM}}[ A , \mathfrak{F} ]$. 
For instance, taking $ \delta^{2} / \delta J_{\mu \nu}^{a}(x) \delta J_{\alpha \beta}^{b}(y)$ and setting $j=J=0$ on both sides of Eq.~\eqref{eq:Gammaequiv} yields 
\begin{equation}\label{eq:SIgamma111A}
    - \langle 0|T F^{a \, \mu \nu}(x) F^{b \, \mu \nu}(y)| 0 \rangle_{\mathsf{C}}  
    = 2 I^{\mu \nu \alpha \beta} \delta (x-y) 
    - \langle 0|T \mathcal{F}^{a \, \mu \nu}(x) \mathcal{F}^{b \, \mu \nu}(y)| 0 \rangle_{\mathsf{C}}.
\end{equation}
Recall that these fields are not the classical fields defined in Eq.~\eqref{eq:definitionclassical}. The relations are given by  $ \langle 0|T F^{a \, \mu \nu} | 0 \rangle_{\mathsf{C}, \, J} = F_{\mathsf{c}}^{a \, \mu \nu} + \mathfrak{F}^{a \, \mu \nu}  -2J^{a \, \mu \nu} $ and $ \langle 0|T \mathcal{F}^{a \, \mu \nu}| 0 \rangle_{\mathsf{C}, \, J} = \mathcal{F}^{a \, \mu \nu}_{\mathsf{c}} $.

The equivalence between the FOYM and SOYM effective actions indicates that the auxiliary field $ \mathcal{F} $ contributes to the 1PI Green's functions of the gauge field $A$. We can understand the physical meaning of Eq.~\eqref{eq:equiv2and12a} by solving Eq.~\eqref{eq:minimum}. At the lowest order in the loop expansion, $ \Gamma_{\text{1YM}} [A , \mathcal{F}] = S_{\text{1YM}} $. Therefore, the solution to Eq.~\eqref{eq:minimum} is given by the classical solution of the equation of motion~\eqref{eq:eqmF}, leading to 
\begin{equation}\label{eq:solutionminimum}
      \bar{F}_{\mu \nu}^{a} (A)= F_{\mu \nu}^{a}.
\end{equation}
Indeed, we already know that substituting the auxiliary field $ \mathcal{F} $ with the curvature $ F$ in the FOYM action yields the SOYM action. Next, we will use Eq.~\eqref{eq:solutionminimum} to compute the effective action $ \Gamma_{\text{1YM}}  [A, F]$ at one-loop order. This result can be compared with the effective action of the SOYM\@.

\subsection{Self-Energy}\label{section:SEYM}

At one-loop order, the effective action $ \Gamma_{\text{1YM}} [A, \bar{F} = F] $ is given by
\begin{equation}\label{eq:effectiveaction1loop}
\int \mathop{d x} \left(
    \frac{1}{2}A 
     \vcenter{\hbox{\includegraphics[scale=0.45]{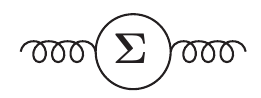}}} 
A
    + A 
     \vcenter{\hbox{\includegraphics[scale=0.45]{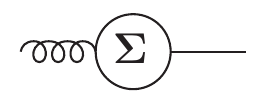}}} 
    F
    + F 
     \vcenter{\hbox{\includegraphics[scale=0.45]{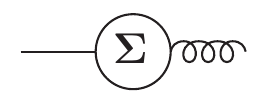}}} 
    A
    + \frac{1}{2}F 
     \vcenter{\hbox{\includegraphics[scale=0.45]{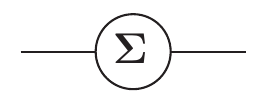}}} 
    F + \cdots 
    \right)
    ,
\end{equation}
where $ \Sigma $ denotes the one-loop self-energy. By the quantum equivalence \eqref{eq:equiv2and12a}, this is equal to the effective action of the SOYM $ \Gamma_{\text{2YM}} [A]$, which is given by
\begin{equation}\label{eq:effective2ymoneloop}
  \int \mathop{d x} \left(
    \frac{1}{2}A 
     \vcenter{\hbox{\includegraphics[scale=0.45]{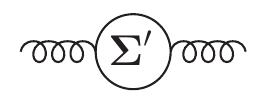}}} 
 A + \cdots \right)
\end{equation}
at one-loop order. 
The dots denote terms of higher powers in the fields $A$ and $F$. These terms are not relevant to the computation of the self-energy. Recall that the \emph{self-energy} is obtained by taking the second-order functional derivative of the effective action and setting the quantum equations of motion. Thus, the resulting contribution coming from terms of order higher than two vanish, since the quantum equations of motion lead to $ \langle 0|T A| 0 \rangle = \langle 0|T F| 0 \rangle = \langle 0|T \mathcal{F}| 0 \rangle =0$ (sources vanishing).

Taking the functional derivative $ \delta^{2} / \delta A_{\mu}^{a} \delta A_{\nu}^{b}  $ of Eq.~\eqref{eq:effectiveaction1loop} (and setting the quantum equations of motion) yields
\begin{equation}\label{eq:1loopSE1YM}
    \Sigma_{\mu \nu}^{ab}  +     \Sigma_{ \mu }^{ac\, \alpha \beta} Q_{\alpha \beta \, \nu}^{cb} + Q_{\mu \, \alpha \beta }^{ac}   \Sigma^{bc \, \alpha \beta} {}_{ \nu}   + Q^{ac}_{\mu\, \alpha \beta }  \Sigma^{cd \, \alpha \beta \rho \sigma } Q^{db}_{\rho \sigma \, \nu }\,   
\end{equation}
where we defined 
\begin{equation}\label{eq:defQ}
    Q_{ \nu\, \alpha \beta }^{cb} \equiv   - \frac{1}{2}\frac{\delta F_{\alpha \beta}^{c}}{ \delta A^{b \, \nu}} =  -\partial^{\sigma} I_{\sigma \nu \, \alpha \beta} \delta^{cb} \quad \text{and} \quad    Q_{\alpha \beta \, \nu}^{cb} =-Q_{\nu \, \alpha \beta}^{cb}.
\end{equation}
We also defined the identity $ I $ in Eq.~\eqref{eq:defI} and $ \bm{\Sigma}_{\bullet \bullet} $ are the components of the matrix: 
\begin{equation}\label{eq:selfenergy}
     \begin{pmatrix}
        \bm{\Sigma}_{AA} & \bm{\Sigma}_{A\mathcal{F}} \\
        \bm{\Sigma}_{\mathcal{F}A} & \bm{\Sigma}_{\mathcal{F}\mathcal{F}} 
    \end{pmatrix}_{\text{components}} 
    =
     \begin{pmatrix}
    \Sigma_{\mu \nu}^{ab}     
         & 
             \Sigma_{ \nu \, \alpha \beta}^{cb} 
      \\ 
         \Sigma_{\rho \sigma  \, \mu}^{ad} &
        \Sigma^{cd}_{\alpha \beta \rho \sigma} 
    \end{pmatrix}.
\end{equation}
Computing Eq.~\eqref{eq:1loopSE1YM} leads to the SOYM self-energy $ \Sigma'$, which verifies Eq.~\eqref{eq:qe:Gamma} at one-loop order.

On the other hand, it is well-known that the self-energy of the SOYM $ \Sigma' $ differs from the $AA$ self-energy in the FOYM, denoted by $ \Sigma_{AA} $ \cite{Brandt:2018avq}. The discussion in this section clarifies it, the self-energy of the SOYM can be obtained from the FOYM\@. Let us show another approach, we shall show that the effective $AA$ self-energy \cite{Frank:2006yh} is given by Eq.~\eqref{eq:1loopSE1YM}. This is not a coincidence, the self-energy is related to the pole of the propagator with is associated with the particle's mass. The naive self-energy $ \Sigma_{AA} $ is not the field $A$ propagator pole, it is the effective self-energy.

We can compute the effective self-energies using the relation between the full propagator $G$, the (free) propagator $G_0$, and the self-energy $ \Sigma $, $ G^{-1}= G_{0}^{-1} + \Sigma $ \cite{das:2006}.
In the FOYM formulation,  we have that 
\begin{equation}\label{eq:FOYMGG0S}
    \mathbf{G}^{-1} = \begin{pmatrix}
        \mathbf{A} & \mathbf{Q}_{L} \\
        \mathbf{Q}_{R} & \mathbf{D}
    \end{pmatrix}
    + 
    \begin{pmatrix}
        \bm{\Sigma}_{AA} & \bm{\Sigma}_{A\mathcal{F}} \\
        \bm{\Sigma}_{\mathcal{F}A} & \bm{\Sigma}_{\mathcal{F}\mathcal{F}} 
    \end{pmatrix},
\end{equation}
where the components of the last matrix $ \mathbf{\Sigma }$  are the \emph{naive self-energies}, which are obtained summing all 1PI diagrams.
The first matrix in the above expression is the Hessian of the FOYM action 
\begin{equation}\label{eq:hessionFOYM}
    \begin{pmatrix}
         \mathbf{A} & \mathbf{Q}_{L} \\
        \mathbf{Q}_{R} & \mathbf{D}
    \end{pmatrix}
    \equiv \begin{pmatrix}
        \dfrac{\delta^{2} S_{\text{1YM}} }{ \delta A_{\mu}^{a} \delta A_{\nu}^{b} }&
        \dfrac{\delta^{2} S_{\text{1YM}} }{ \delta A_{\nu}^{b} \delta \mathcal{F}_{\alpha \beta}^{c}  } \\ 
        \dfrac{\delta^{2} S_{\text{1YM}} }{ \delta \mathcal{F}_{\rho \sigma}^{d}  \delta A_{\mu}^{a}}  &  
        \dfrac{\delta^{2} S_{\text{1YM}} }{ \delta \mathcal{F}_{\rho \sigma}^{d}  \delta \mathcal{F}_{\alpha \beta}^{c}  }
    \end{pmatrix}.
\end{equation}
The bold symbols are tensors in which the indices have been omitted for clarity. 

The effective one-loop self-energy is defined by 
\begin{equation}\label{eq:oneloopEFFSE}
    \bm{\Sigma}_{\text{eff}} = (\mathbf{G_0})^{-1}  -\mathbf{\mathcal{G}_{0}^{-1}}\left[\begin{pmatrix}
        \mathbf{A} & \mathbf{Q}_{L} \\
        \mathbf{Q}_{R} & \mathbf{D}
    \end{pmatrix}
    + 
    \begin{pmatrix}
        \bm{\Sigma}_{AA} & \bm{\Sigma}_{A\mathcal{F}} \\
\bm{\Sigma}_{A\mathcal{F}} & \bm{\Sigma}_{\mathcal{F}\mathcal{F}} \end{pmatrix}\right]^{-1} \mathbf{\mathcal{G}_{0}^{-1}}- \mathcal{O} ( \bm{\Sigma}^{2} ),
\end{equation}
where $ \mathcal{O} ( \bm{\Sigma}^{2} )$ denotes all contributions of order greater than $ \bm{\Sigma} $ (one-loop).
The \emph{effective self-energy} can be defined, at any order, as the sum of all 1PI diagrams that contributes to the amputated propagator $ \mathbf{\mathcal{G}_{0}^{1-} {G} \mathcal{G}_{0}^{-1}}$. Note that, there is a very subtle difference between the free propagator terms that appear in Eq.~\eqref{eq:oneloopEFFSE}. We have $ \mathbf{\mathcal{G}_{0}^{-1}} $ that denotes the matrix whose components are the inverse of the free propagators, while $ (\mathbf{G_{0}} )^{-1}$ is the inverse of the matrix of free propagators (the components are given in Eq.~\eqref{eq:hessionFOYM}). 

This should clarify the difference between the naive self-energy defined by Eq.~\eqref{eq:FOYMGG0S}, and the \emph{proper self-energy}\footnote{It can also be interpreted as an effective self-energy \cite{Frank:2006yh}.} defined by
\begin{equation} \label{eq:defaltEFF}
    \bm{\Sigma}_{\text{eff}} \equiv \mathbf{G^{-1}} - \mathbf{\mathcal{G}_{0}^{-1}},
\end{equation}
which is the appropriate definition for the self-energy in theories such as the FOYM in which the fields are mixed (the Hessian is not diagonal). Indeed, the proper self-energy can also be obtained directly from the effective action. The two-point 1PI Green's functions are 
\begin{equation}\label{eq:matrix2p1PI}
    \frac{\delta^{2} \Gamma_{\text{1YM}} }{\delta {}\bm{\Phi} \delta {}\bm{\Phi}} = - \frac{\delta {}\mathbf{J} }{\delta {}\bm{\Phi}}, \quad {}\bm{\Phi} = \begin{pmatrix}
        A_{\mathsf{c}} \\ 
        \mathcal{F}_{\mathsf{c} }
    \end{pmatrix}
    \quad \text{and} \quad {}\mathbf{J} = \begin{pmatrix}
        j \\ J
    \end{pmatrix}.
\end{equation}
One can see that by the definition of the classical fields, we have that 
\begin{equation}\label{eq:invofJmatrix}
    - \frac{\delta {}\mathbf{J} }{\delta {}\bm{\Phi}} = - \left ( \frac{\delta^{2} W_{\text{1YM}} }{\delta {}\mathbf{J} \delta {}\mathbf{J}}\right )^{-1}= \mathbf{G}^{-1}.
\end{equation}
This allows us rewrite the relation Eq.~\eqref{eq:matrix2p1PI} and the definition of the self-energy (see \cite{das:2006}) as 
\begin{equation}\label{eq:properSE}
    \bm{\Sigma} = \frac{\delta^{2} \Gamma_{\text{1YM}} }{\delta {}\bm{\Phi} \delta {}\bm{\Phi}} - \mathbf{\mathcal{G}_{0}^{-1}} = \mathbf{G}^{-1}- \mathbf{\mathcal{G}_{0}^{-1}},
\end{equation}
which is the same definition of the effective or proper self-energy in Eq.~\eqref{eq:defaltEFF}. Hence, the components of the naive self-energy in Eq.~\eqref{eq:FOYMGG0S} can be seen as ``partial self-energies''. Now, we show that the proper $AA$ self-energy $ ({}\bm{\Sigma}_{\text{eff}} )_{AA} $ is obtained from these partial self-energies.

We can proceed by obtaining the full propagator by the inverse of the block matrix 
\begin{equation}\label{eq:invblock}
    \mathbf{G} =\begin{pmatrix}
        \mathbf{A} + \bm{\Sigma}_{AA} & \mathbf{Q}_{L} + \bm{\Sigma}_{ A\mathcal{F} } \\
        \mathbf{Q}_{R} + \bm{\Sigma}_{ \mathcal{F} A} & \mathbf{D} + \bm{\Sigma}_{\mathcal{F} \mathcal{F} }  
    \end{pmatrix}^{-1}.
\end{equation}
For this we can use the blockwise matrix inversion, for example, 
\begin{equation}\label{eq:bmatrixinversion}
    \begin{pmatrix}
        \mathbf{A} & \mathbf{Q}_{L}  \\
        \mathbf{Q}_{R} & \mathbf{D} 
    \end{pmatrix}^{-1} = 
    \begin{pmatrix}
        \mathbf{X}^{-1} & - \mathbf{X}^{-1} \mathbf{Q}_{L}  \mathbf{D}^{-1}\\
        - \mathbf{D}^{-1} \mathbf{Q}_{R}  \mathbf{X}^{-1} & 
        \mathbf{D}^{-1} + \mathbf{D}^{-1} \mathbf{Q}_{R}  \mathbf{X}^{-1} \mathbf{Q}_{L}  \mathbf{D}^{-1} 
    \end{pmatrix},
\end{equation}
where $ \mathbf{X} $ is the Schur complement: 
\begin{equation}\label{eq:schurcomplement}
    \mathbf{X} = \mathbf{A} - \mathbf{Q}_{L} \mathbf{D}^{-1} \mathbf{Q}_{R}. 
\end{equation}
Since we are only interested in finding the proper (effective) self-energy of the gauge field $A$ at one-loop order, the computation of the inverse of $ \mathbf{G} $ can be skipped. 

In Eq.~\eqref{eq:bmatrixinversion} we have the matrix of the propagators of the FOYM $ \mathbf{G_{0}} $, which is the inverse of the Hessian. From Eq.~\eqref{eq:oneloopEFFSE}, we have that 
\begin{equation}\label{eq:effAA1loop}
    [\bm{\Sigma}_{\text{eff}}]_{AA} = \mathbf{X} -\mathbf{X} \bar{\mathbf{X}}^{-1} \mathbf{X}  \Rightarrow \bar{\mathbf{X}} = 
     \left( \mathbf{I} - [\bm{\Sigma}_{\text{eff}}]_{AA} \mathbf{X}^{-1} \right)^{-1}\mathbf{X}.
\end{equation}
Restricting the above relation to one-loop order leads to 
\begin{equation}\label{eq:effeasy}
    \bar{\mathbf{X}} = \mathbf{X} +   [ \bm{\Sigma}_{\text{eff}} ]_{AA} . 
\end{equation}
Computing the Schur complement $\bar{\mathbf{X}} $ of Eq.~\eqref{eq:invblock} (similar to Eq.~\eqref{eq:schurcomplement}) yields 
\begin{equation}\label{eq:schurSE}
    \bar{\mathbf{X}} = \mathbf{A} + \bm{\Sigma}_{AA} - (\mathbf{Q}_{L} + \bm{\Sigma}_{A \mathcal{F}} ) ( \mathbf{D} + \bm{\Sigma}_{\mathcal{F} \mathcal{F}} )^{-1} ( \mathbf{Q}_{R} + \bm{\Sigma}_{\mathcal{F} A} ).
\end{equation}
Using that $ ( \mathbf{D} + \bm{\Sigma}_{\mathcal{F} \mathcal{F}} )^{-1} = \mathbf{D} - \bm{\Sigma}_{\mathcal{F} \mathcal{F}} + \mathcal{O} ( \bm{\Sigma}^{2}_{\mathcal{F} \mathcal{F}} )$, we have 
\begin{equation}
\label{eq:schurSE1loop}
\bar{\mathbf{X}} = \mathbf{X} + 
\bm{\Sigma}_{AA} +  \mathbf{Q}_{L} \mathbf{D}^{-1}\bm{\Sigma}_{\mathcal{F} \mathcal{F}} \mathbf{D}^{-1}\mathbf{Q}_{R}  - \bm{\Sigma}_{A \mathcal{F}} \mathbf{D}^{-1}\mathbf{Q}_{R} - \mathbf{Q}_{L} \mathbf{D}^{-1}\bm{\Sigma}_{\mathcal{F} A},
\end{equation}
therefore, from Eq.~\eqref{eq:effeasy}, we obtain 
\begin{equation}\label{eq:effbyQplus}
    [ \bm{\Sigma}_{ \text{eff} } ]_{AA} = 
\bm{\Sigma}_{AA} - \bm{\Sigma}_{A \mathcal{F}} \mathbf{D}^{-1}\mathbf{Q}_{R} - \mathbf{Q}_{L} \mathbf{D}^{-1}\bm{\Sigma}_{\mathcal{F} A}
+  \mathbf{Q}_{L} \mathbf{D}^{-1}\bm{\Sigma}_{\mathcal{F} \mathcal{F}} \mathbf{D}^{-1}\mathbf{Q}_{R}.
\end{equation}

Eq.~\eqref{eq:effbyQplus} has the same structure of the self-energy \eqref{eq:1loopSE1YM} derived from Eq.~\eqref{eq:effectiveaction1loop}. Indeed, it is easy to verify that the proper $AA$ self-energy \eqref{eq:effbyQplus} is equal to Eq.~\eqref{eq:1loopSE1YM}. From Eq.~\eqref{eq:lagYM1}, one can show that $ (\mathbf{Q}_{L}  \mathbf{D}^{-1})_{\mu \, \alpha \beta}^{ab}  = -Q_{\mu \,  \alpha \beta} \delta^{ab} $ and $ (\mathbf{D}^{-1} \mathbf{Q}_{R})_{\alpha \beta \,  \mu}^{ab}  = -Q_{\alpha \beta \,  \mu} \delta^{ab} $. Now, using this in Eq.~\eqref{eq:effbyQplus} leads to Eq.~\eqref{eq:1loopSE1YM}.
Consequently, it also coincides with the self-energy of the gauge field $A$ in the SOYM theory. 

Thus, we have shown that we can obtain the proper self-energy in the FOYM theory using two distinct approaches. Moreover, the proper self-energy of the field $A$ in the FOYM formulation, obtained with these approaches, coincides with the respective self-energy computed in the SOYM formulation. 

\section{Gravity}\label{section:QEGravity}

In this section, we concisely review the quantum equivalence between the first- and second-order formulations of gravity. Here, we consider the case of pure gravity. 
The structural identities for gravity are presented here, and their explicit verification at one-loop order is examined. We aim to extend the findings in \cite{martins-filho:2021} to demonstrate that, with the appropriate quantization of the FOGR theory, the first- and second-order formalism are fully equivalent at the quantum level. This equivalence is reflected in the verification of the structural identities at the integrand level, indicating that dimensional regularization is not a necessary condition for establishing equivalence. For more details, we refer the reader to Refs. \cite{Brandt:2020vre, martins-filho:2021}.

\subsection{Hilbert-Palatini action}

Shifting the auxiliary field in Eq.~\eqref{eq:313} by its classical value
\begin{equation}\label{eq:shiftwsEH12}
     G_{\mu \nu}^{\lambda} = \mathring{{G}}_{\mu \nu}^{\lambda} + \tensor*{(M^{-1})}{*_{\mu \nu}^{\lambda}_{ \pi \tau}^{ \rho}}  (h) (h^{\pi \tau}_{, \rho} - \kappa^{2}  J^{\pi \tau}_{\rho} ) 
\end{equation}
we arrive at the equivalent generating functional
\begin{equation}\label{eq:fgEH12ws}
    \begin{split}
        Z_{\text{HP}} [j,J] ={}& N \int 
        \mathop{\mathcal{D} \phi^{\mu \nu}} 
        \mathop{\mathcal{D} d_{\nu} } \mathop{\mathcal{D}
            \bar{d}_{\mu} } 
            \mathop{\mathcal{D} \mathring{{G}}_{\mu \nu}^{\lambda } } 
            | \det \tensor*{M}{*_{\lambda}^{\mu \nu}_{\rho}^{ \pi\tau}}(h) |^{1/2}\\ 
                                           & \quad \quad \quad \times \exp i  \int \mathop{d^D x}  \left( {\mathcal{L}}_{\text{eff}}^{\text{II}} + \frac{1}{2 \kappa^{2} }{ \mathring{{G}} }_{ \mu \nu}^{ \lambda} \tensor*{M}{*_{\lambda}^{\mu \nu}_{\rho}^{ \pi \tau}}(h){ \mathring{{G}} }_{ \pi \tau }^{ \rho} + 
                                           \mathcal{L}_{\text{HP-src}} \right),
\end{split}
\end{equation}
where 
\begin{equation}\label{eq:sourceHP}
    \mathcal{L}_{\text{HP-src}} =                                         {J}_{\lambda}^{\mu \nu} \tensor*{(M^{-1})}{*_{\mu \nu}^{\lambda}_{ \pi \tau}^{ \rho}}  (h) h^{\pi \tau}_{, \rho}  - \frac{\kappa^{2} }{2} J^{\mu \nu}_{\lambda} \tensor*{( M^{-1})}{*_{\mu \nu}^{\lambda}_{ \pi \tau}^{ \rho}} (h) J^{\pi \tau}_{\rho} +  j_{\mu \nu} \phi^{\mu \nu}  . 
\end{equation}
Integrating the auxiliary field $ \mathring{{G}}_{\mu \nu}^{\lambda} $ results in the determinant $ | \det \tensor*{M}{*^{\mu \nu}_{\lambda}^{\pi \tau}_{\rho}} (h) |^{-1/2} $. In previous works \cite{martins-filho:2021, Brandt:2020vre, Brandt:2020gms} (and references within), it is argued that this determinant is trivial when using dimensional regularization. 
However, as we discussed previously, these determinants can play a relevant role in the path integral formalism. Moreover, this argument is not valid at finite temperature, since tadpole-like contributions (loop integrals which are independent of external momenta), which is the case of this determinant, do not vanish. In Appendix~\ref{section:tadpoles}, we illustrate this with a simple example using the $ \phi^4$ theory.

Nevertheless, we have introduced the Senjanovi\'{c} determinant $ |\allowbreak \det  \tensor*{M}{*^{\mu \nu}_{\lambda}^{\pi \tau}_{\rho}} (h)|^{+1/2} $ in the measure (see Eq.~\eqref{eq:fgEH12ws}) that neatly cancels this determinant. Consequently, we demonstrate the equivalence between the generating functionals without requiring any additional condition. This reinforces our interpretation of the determinant in Eq.~\eqref{eq:fgEH12ws} as the Senjanovi\'{c} determinant, which arises due to the auxiliary field $ G^{\lambda} {}_{\mu \nu} $. The resulting generating functional is 
\begin{equation}\label{eq:fgEH12wswG}
    \begin{split}
        Z_{\text{HP}} [j,J] ={}&  N\int 
            \mathop{\mathcal{D} \phi^{\mu \nu} } 
        \mathop{\mathcal{D} d_{\nu} } \mathop{\mathcal{D}
            \bar{d}_{\mu} }  
            \exp i   \int \mathop{d^D x}  \bigg( {\mathcal{L}}_{\text{eff}}^{\text{II}}   + j_{\mu \nu} \phi^{\mu \nu} 
                             + {J}_{\lambda}^{\mu \nu} \tensor*{(M^{-1})}{*_{\mu \nu}^{\lambda}_{ \pi \tau}^{ \rho}}  (h) h^{\pi \tau}_{, \rho} 
                            \\ & - \frac{\kappa^{2}}{2} J^{\mu \nu}_{\lambda} \tensor*{( M^{-1})}{*_{\mu \nu}^{\lambda}_{ \pi \tau}^{ \rho}} (h) J^{\pi \tau}_{\rho}  \bigg) .
\end{split}
\end{equation}
Note that the action in the above expression is the SOGR action in Eq.~\eqref{eq:lageffEH}. Besides that, the source term is rather involved and unusual as we see a quadratic dependence on $J$. 

When $J=0$,  
\begin{equation}\label{eq:equivwjEH}
    Z_{\text{HP}} [j,0] = Z_{\text{HE}} [j],
\end{equation}
which shows transparently the equivalence between these formalisms.
Consequently,
\begin{equation}\label{eq:eqvphi}
    \begin{split}
        \langle 0|T \phi^{\mu_1 \nu_1}(x_1) \cdots \phi^{{ \mu_{n} \nu_{n} }} (x_n)| 0 \rangle & \equiv  
        \left . (-i)^n \frac{\delta^{n} Z_{\text{HE}} [j]}{\delta j_{\mu_1 \nu_{1} } (x_1) \cdots \delta {j}_{\mu_{n} \nu_{n} }(x_n) } \right |_{j=0} \\ 
                                                                                               &= \left .(-i)^n \frac{\delta^{n} Z_{\text{HP}}[j,J] }{\delta j_{\mu_1 \nu_{1} } (x_1) \cdots \delta {j}_{\mu_{n} \nu_{n} }(x_n)  }\right  |_{J=j=0} .
    \end{split}
\end{equation}

When $J \neq 0$, the functional derivatives with respect to $J^{\mu \nu}_{\lambda} $  of the right-hand side of Eq.~\eqref{eq:fgEH12wswG} leads to Green's functions of the composite field $ \mathbb{G}^{\lambda}_{\mu \nu} (x)$ computed in the SOGR formulation, which is the classical value of the auxiliary field $ G_{\mu \nu}^{\lambda} (x)$. The same functional derivative in the left-hand side of the same expression would lead to Green's functions of the auxiliary field $ G_{\mu \nu}^{\lambda} $. One can easily verify that the following structural identities can be derived: 
\begin{subequations}\label{eq:SIEH}
\begin{align}\label{eq:id1EH1}
    & \langle 0|T G_{\mu \nu}^{\lambda} (x) \phi^{\alpha \beta } (y)| 0 \rangle = \kappa \langle 0|T \tensor*{(M^{-1})}{*_{\mu \nu}^{\lambda}_{ \pi \tau}^{ \rho}}  [ \eta + \kappa \phi (x)] \phi^{\pi \tau}_{, \rho} (x) \phi^{\alpha \beta }  (y) | 0 \rangle,\\
    \label{eq:id2EH1}
        & \langle 0|T G_{\mu \nu }^{ \lambda } (x) G_{ \alpha \beta }^{\gamma } (y)| 0 \rangle = 
        i \kappa^{2}  \langle 0|T \tensor*{(M^{-1})}{*_{\mu \nu}^{\lambda}_{ \alpha  \beta }^{ \gamma }}  [\eta + \kappa \phi (x)] | 0 \rangle  \delta (x-y) \\ \nonumber & + \kappa^{2} \langle 0|T  
        \tensor*{(M^{-1})}{*_{\mu \nu}^{\lambda}_{ \pi_2 \tau_2}^{ \rho_2}}  [ \eta + \kappa \phi (x)] \phi^{\pi_2 \tau_2}_{, \rho_2} (x) 
        \tensor*{(M^{-1})}{*_{\alpha \beta}^{\gamma}_{ \pi_1\tau_1}^{ \rho_1}}  [ \eta + \kappa \phi (y)] \phi^{\pi_1\tau_1}_{, \rho_1} (y) 
        | 0 \rangle. 
    \end{align}
\end{subequations}
These identities are analog of the Eqs.~\eqref{eq:YM:SI2} and \eqref{eq:YM:SI3}, respectively.

\subsection{Diagonal first-order formulation}

The quantum equivalence of the DFOGR formulation is immediate.
Consider the shift
\begin{equation}\label{eq:314}
    {H}_{ \mu \nu}^{ \lambda} \rightarrow \mathring{H}_{ \mu \nu}^{ \lambda} 
    + \mathbb{H}_{\mu \nu}^{\lambda}  -
      \kappa^{2} \tensor*{(M^{-1})}{*_{\mu\nu}^{\lambda}_{ \pi \tau}^{ \rho}}  (\eta + \kappa
    \phi ) {J}_{\rho}^{\pi \tau}, 
\end{equation}
where 
\begin{equation}\label{eq:defD}
\mathbb{H}_{\mu \nu}^{\lambda} (x) \equiv \tensor*{[M^{-1}(h) - M^{-1}(\eta )]}{*_{\mu \nu}^{\lambda}_{ \pi \tau}^{ \rho}}  h^{\pi \tau}_{, \rho} (x)
\end{equation}
is the classical value of the field $ H_{\mu \nu}^{\lambda} $. Upon this shift, the generating functional of the DFOGR~\eqref{eq:313d} can be rewritten as 
\begin{equation}\label{eq:315a}
    \begin{split} 
        Z^{\text{Id}}_{\text{HP}} ={}&
        N\int 
        \mathop{\mathcal{D} \phi^{\mu \nu}} \mathop{\mathcal{D} d_{\nu} } \mathop{\mathcal{D} \bar{d}_{\mu} }
        \exp i \int \mathop{d^D x}   \Big(
         \mathcal{L}_{\textrm{eff}}^{\textrm{II}} +    \kappa {J}_{\lambda}^{\mu \nu}
         \tensor*{[M^{-1}(\eta + \kappa \phi ) - M^{-1}(\eta ) ]}{*_{\mu \nu}^{\lambda}_{ \pi \tau}^{ \rho}} 
    \phi^{\pi\tau}_{,\rho}  \\ & - \frac{\kappa^{2} }{2} {J}_{\lambda}^{\mu \nu} \tensor*{(M^{-1})}{*_{\mu\nu}^{\lambda}_{ \pi \tau}^{ \rho}}  (\eta + \kappa \phi ) {J}_{\rho}^{\pi \tau}+ {j}_{\mu \nu}^{} \phi^{\mu \nu} \Big).
    \end{split}
\end{equation}

Likewise the HP formalism, we have that
\begin{equation}\label{eq:eqv1d}
    Z_{\text{HE}}^{\text{Id}} [j,0] = Z_{\text{HE}}^{\text{II}} [j].
\end{equation}
When $J \neq 0$, we can obtain structural identities by applying functional derivatives on both sides of Eq.~\eqref{eq:315a}.
For instance, the functional derivative with respect to $ J^{\lambda}_{\mu\nu}$ yields
\begin{equation}\label{eq:fidEHex}
 \langle 0 | T H_{\mu \nu}^{\lambda} (x) | 0  \rangle=
 \langle 0 | T \mathbb{H}_{\mu \nu}^{\lambda} (x) |0  \rangle.
\end{equation}

Note that the right-hand side can be computed in both the FOGR and SOGR formulation. Thus, this identity can be seen as an identity between Green's functions of the FOGR theory. Consequently, it is the quantum realization of the classical equation of motion of $ H_{\mu \nu}^{\lambda} $~\eqref{eq:defD}. 

We also have \cite{Brandt:2020vre}
\begin{equation}\label{eq:is1EH}
        \langle 0|T H_{\mu \nu}^{\lambda} (x) \phi^{\pi \tau} (y)| 0 \rangle
    =\langle 0|T  \mathbb{H}_{\mu \nu}^{\lambda} (x)\phi ^{\pi \tau } (y)| 0 \rangle
\end{equation}
and
\begin{equation}\label{eq:is2EH}
        \langle 0|T H_{\mu \nu}^{\lambda} (x) H_{\pi \tau }^{\rho} (y)  | 0 \rangle 
=   \langle 0 | T \mathbb{H}_{\mu \nu}^{\lambda} (x)  \mathbb{H}_{\pi \tau}^{\rho} (y) |0  \rangle
+i\kappa^{2} \delta (x-y) \langle 0| T \tensor*{(M^{-1})}{*_{\mu\nu}^{\lambda}_{ \pi \tau}^{ \rho}}  [\eta + \kappa \phi(x)] |0 \rangle.
\end{equation}
The contact term in the above expression arises due to the quadratic dependence on $J$ in Eq.~\eqref{eq:315a}. These identities relate the FOGR Green's function with Green's function of the composite field \eqref{eq:defD} computed in the SOGR\@.

One can easily generalize these identities, for example:
\begin{equation}\label{eq:is1EHgeral}
    \langle 0|T H_{\mu \nu}^{\lambda} (x) \phi^{\pi_1\tau_1} (y_1) \cdots \phi^{\pi_n \tau_{n}} (y_n)| 0 \rangle
    =\langle 0|T  \mathbb{H}_{\mu \nu}^{\lambda} (x)\phi^{\pi_1\tau_1} (y_1) \cdots \phi^{\pi_n \tau_{n}} (y_n)| 0 \rangle.
\end{equation}
Moreover, we can obtain identities between the FOGR and the DFOGR formulations, such as
\begin{equation}\label{eq:1drl1e1dEH}
    \begin{split}
     \langle 0|T G_{\mu \nu}^{\lambda} (x) \phi^{\alpha \beta } (y)| 0 \rangle -
    \langle 0|T H_{\mu \nu}^{\lambda} (x) \phi^{\alpha \beta } (y)| 0 \rangle 
={}& \langle 0|T [\mathbb{G} - \mathbb{H} ]\tensor*{}{*_{\mu \nu}^{\lambda}_{ \pi \tau}^{ \rho}}  (x)  \phi^{\alpha \beta }  (y) | 0 \rangle 
    \\
    ={}&
    \langle 0|T  \tensor*{(M^{-1})}{*_{\mu \nu}^{\lambda}_{ \pi \tau}^{ \rho}}  (\eta) h^{\pi \tau}_{, \rho} (x) \phi^{\alpha \beta }  (y) | 0 \rangle,
\end{split}
\end{equation}
which follows from the definition of the classical values (see Eqs.~\eqref{eq:eomG} and \eqref{eq:defD}).

\subsection{Structural identities at one-loop order} \label{section:SIoneloopGR}

The identities \eqref{eq:is1EH} and \eqref{eq:is2EH} are rewritten as
\begin{subequations}
\begin{align}\label{eq:41}
    \langle 0|T {H}_{\mu \nu}^{\lambda} ( x) {\phi}^{\pi \tau} (
    y) | 0 \rangle ={}&
     \langle 0|T
    \mathbb{H}_{\mu \nu}^{\lambda} (x){\phi}^{\pi \tau} ( y)| 0 \rangle,
    \\ \label{eq:51}
    \langle 0|T {H}_{\mu \nu}^{\lambda} ( x) {H}_{\pi \tau}^{\rho } (
    y)  | 0 \rangle ={}&i \kappa^{2} \langle 0|T \tensor*{( M^{-1})}{*_{\mu \nu}^{\lambda}_{\pi \tau}^{\rho}} [h(x)] | 0 \rangle\, 
\delta (x - y) 
+ \langle 0|T \mathbb{H}_{\mu \nu}^{ \lambda } (x) \mathbb{H}_{\pi \tau}^{\rho} (y) | 0 \rangle,
\end{align}
\end{subequations}
where $ \mathbb{H}_{\mu \nu}^{\lambda} $ is the composite field defined in Eq.~\eqref{eq:defD}.
Using the expansion  
\begin{equation}\label{eq:43}
M^{-1}(\eta+\kappa\phi)-m^{-1}=
-\kappa m^{-1} M(\phi) m^{-1}  
+\kappa^2 m^{-1} M(\phi) m^{-1} M(\phi)  m^{-1}  
+ \cdots,
\end{equation}
where 
\begin{equation}\label{eq:defmsmall}
    m \equiv M ( \eta ),
\end{equation}
one easily verify Eqs.~\eqref{eq:41} and \eqref{eq:51} at the tree-level.
For example, the right-hand side of the identity \eqref{eq:51} yields $ i \kappa^{2} m^{-1} $, which is equal to the propagator of the auxiliary field $ H_{\mu \nu }^{\lambda} $ (see Eq.~\eqref{eq:propHEH}). The only contribution from the left-hand side comes from the contact term, $ i \kappa^{2} m^{-1} $, as expected. At one-loop order pinched diagrams must be considered which leads to an involved algebraic computation. Here we show that we can verify these identities even at the integrand level extending the approach used for YM theory to gravity.

\begin{figure}[ht]
\centering
\includegraphics[width=0.8\textwidth]{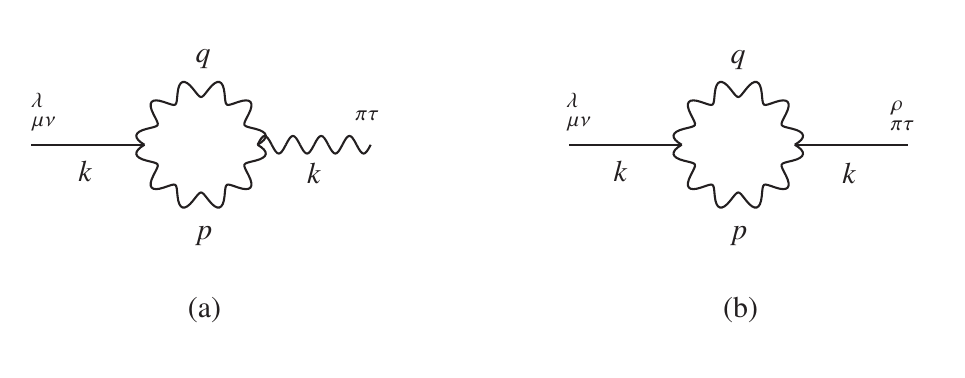} 
\caption[]{In diagram (a), we have the one-loop contribution to the mixed propagator $ \langle 0| TH^\lambda_{\mu\nu} \phi^{\pi\tau}|0\rangle $. In diagram (b), we have the one-loop contribution to the propagator $ \left \langle 0|TH_{\mu \nu}^{\lambda} H_{\pi \tau}^{\rho}|0\right\rangle $. The momenta satisfy the condition $q=p+k$.}
\label{fig1}
\end{figure}

The one-loop contributions to the left-hand side of the structural identities Eqs.~\eqref{eq:41} and \eqref{eq:51} are shown in Fig.~\ref{fig1}.

\begin{figure}[ht]
\centering
\includegraphics[width=0.8\textwidth]{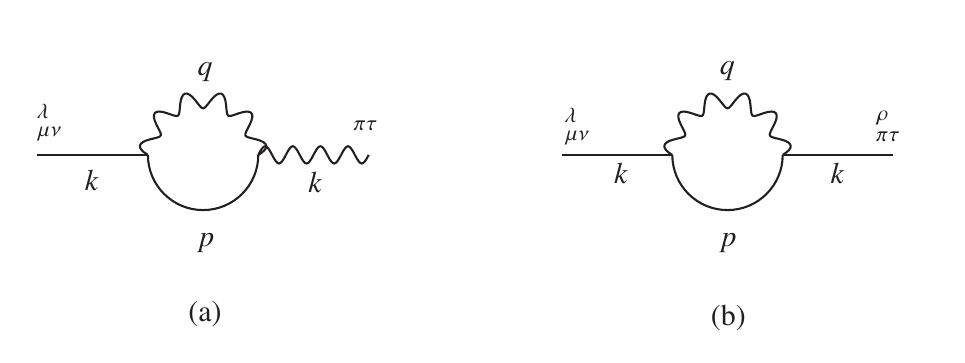} 
\caption{Tadpole-like contributions coming from the naive FOGR path integral (without the Senjanović determinant).}
\label{fig3b}
\end{figure}

In \cite{Brandt:2020vre, martins-filho:2021}, we give a thorough examination of these Green's functions. We have obtained the one-loop contributions for any dimension $D$ and any gauge parameter $ \alpha $. Before we proceed, note that the tadpole-like contribution shown in Fig.~\ref{fig3b} is canceled by the Senjanovi\'{c} determinant that we proposed in Eq.~\eqref{eq:fgEH12ws}. Thus, we do not have to appeal to any particular regularization scheme to show that these contributions must vanish. This statement also holds when we show the quantum equivalence of the FOGR and the SOGR generating functionals.

Now, we have to consider the pinched diagrams arising from the right-hand side of these structural identities.
The right-hand side of Eq.~\eqref{eq:41}, at one-loop order ($ \kappa^{2} $), becomes
\begin{equation}\label{eq:ladodireito1s1}
    - \kappa^{2}  \langle 0|T  \tensor*{[m^{-1} M(\phi) m^{-1} ]}{*_{\mu \nu}^{\lambda}_{\alpha \beta }^{\gamma}}(x) \phi^{\alpha \beta}_{, \gamma }(x)   \phi^{\pi \tau} (y) | 0 \rangle,
\end{equation}
where the expression \eqref{eq:43} is used. 

\begin{figure}[ht]
\centering
\includegraphics[scale=0.75]{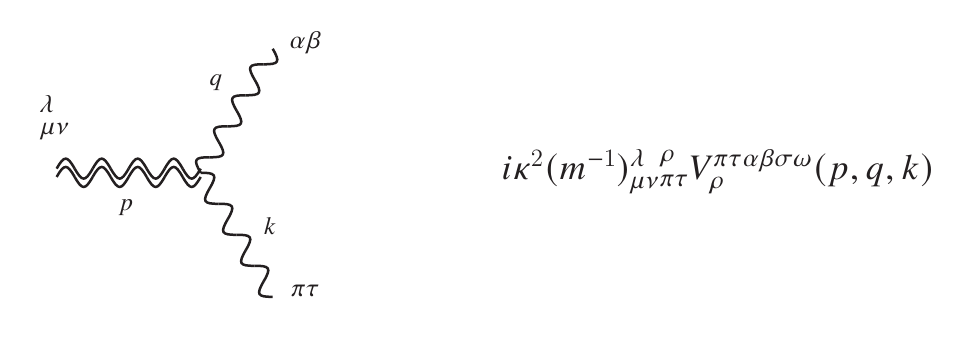} 
\caption{Extended Feynman rule for the interaction vertex with a composite field $ \mathbb{H}_{\mu \nu}^{\lambda} $, which is represented by a double wavy line. The definition of $ V_{\rho}^{\pi \tau} {}^{\alpha \beta \sigma \omega} (p,q,k)$ is given in Eq.~\eqref{eq:FRHphiphieq}.}
\label{fig:FRDeltahh}
\end{figure}

Accordingly to \cite{Brandt:2020vre}, we define 
\begin{equation}\label{eq:defMEH}
    \kappa \mathfrak{M}{}{}_{\mu \nu}^{ \lambda} {}_{\alpha \beta}^{\gamma } {}_{ \rho \sigma } \phi^{\rho \sigma} 
    \equiv - \kappa \tensor*{[m^{-1} M(\phi) m^{-1} ]}{*_{\mu \nu}^{\lambda}_{\alpha \beta }^{\gamma}}
\end{equation}
and rewrite Eq.~\eqref{eq:ladodireito1s1} as
\begin{equation}\label{eq:ladodireito1s11}
    \kappa^{2} \mathfrak{M}_{\mu \nu}^{\lambda} {}_{\alpha \beta }^{\gamma}{}_{\rho \sigma }
\langle 0|T   \phi^{\rho \sigma } (x)\phi^{\alpha \beta}_{, \gamma}(x)   \phi^{\pi \tau} (y) | 0 \rangle.
\end{equation}
This can be computed using the same limiting procedure as for the pinched diagrams in YM theory. On the other hand, amputating each graviton propagator from the following structural identity
\begin{equation}\label{eq:DeltaandSI0}
    \langle 0|T \mathbb{H}_{\mu \nu}^{\lambda} \phi^{\alpha \beta} \phi^{\sigma \omega} | 0 \rangle = \langle 0|T H_{\mu \nu }^{\lambda} \phi^{\alpha \beta} \phi^{\sigma \omega} | 0 \rangle 
\end{equation}
results in
\begin{equation}\label{eq:DeltaandSI}
    ( \mathbb{H}_{\mu \nu}^{\lambda} \phi^{\alpha \beta } \phi^{\sigma \omega} )  = {i \kappa^{2} \tensor*{(m^{-1})}{*_{\mu \nu}^{\lambda}_{\pi \tau}^{\rho}}}( H^{\pi \tau }_{\rho} \phi^{\alpha \beta } \phi^{\sigma \omega  } ),
\end{equation}
where $i \kappa^{2} \tensor*{(m^{-1})}{*_{\mu \nu}^{\lambda}_{\pi \tau}^{\rho}}$ is the free propagator $\langle 0|TH_{\mu \nu}^{\lambda} H_{\pi \tau}^{\rho} |0\rangle_{\text{free}}$. At tree-level, Eq.~\eqref{eq:DeltaandSI} determines the extended Feynman rule for the composite field $ \mathbb{H}_{\mu \nu}^{\lambda} $, which is shown in Fig.~\ref{fig:FRDeltahh}. 

\begin{figure}[!ht]
\centering
\includegraphics[scale=0.7]{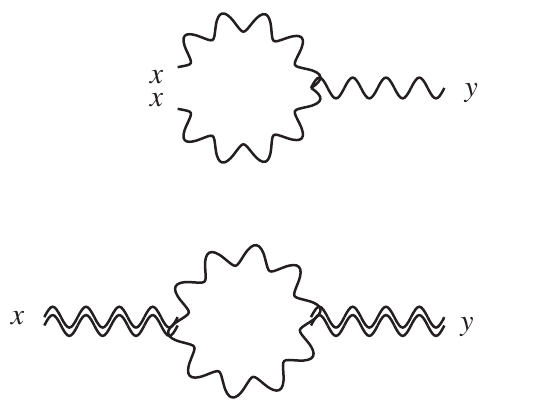} 
\caption{Pinched diagram that contributes to the right-hand side of Eq.~\eqref{eq:41} at one-loop order. We also present the same diagram using the extended Feynman rule in Fig.~\ref{fig:FRDeltahh}.}
\label{fig2}
\end{figure}

\begin{figure}[!ht]
\centering
\includegraphics[width=0.85\textwidth]{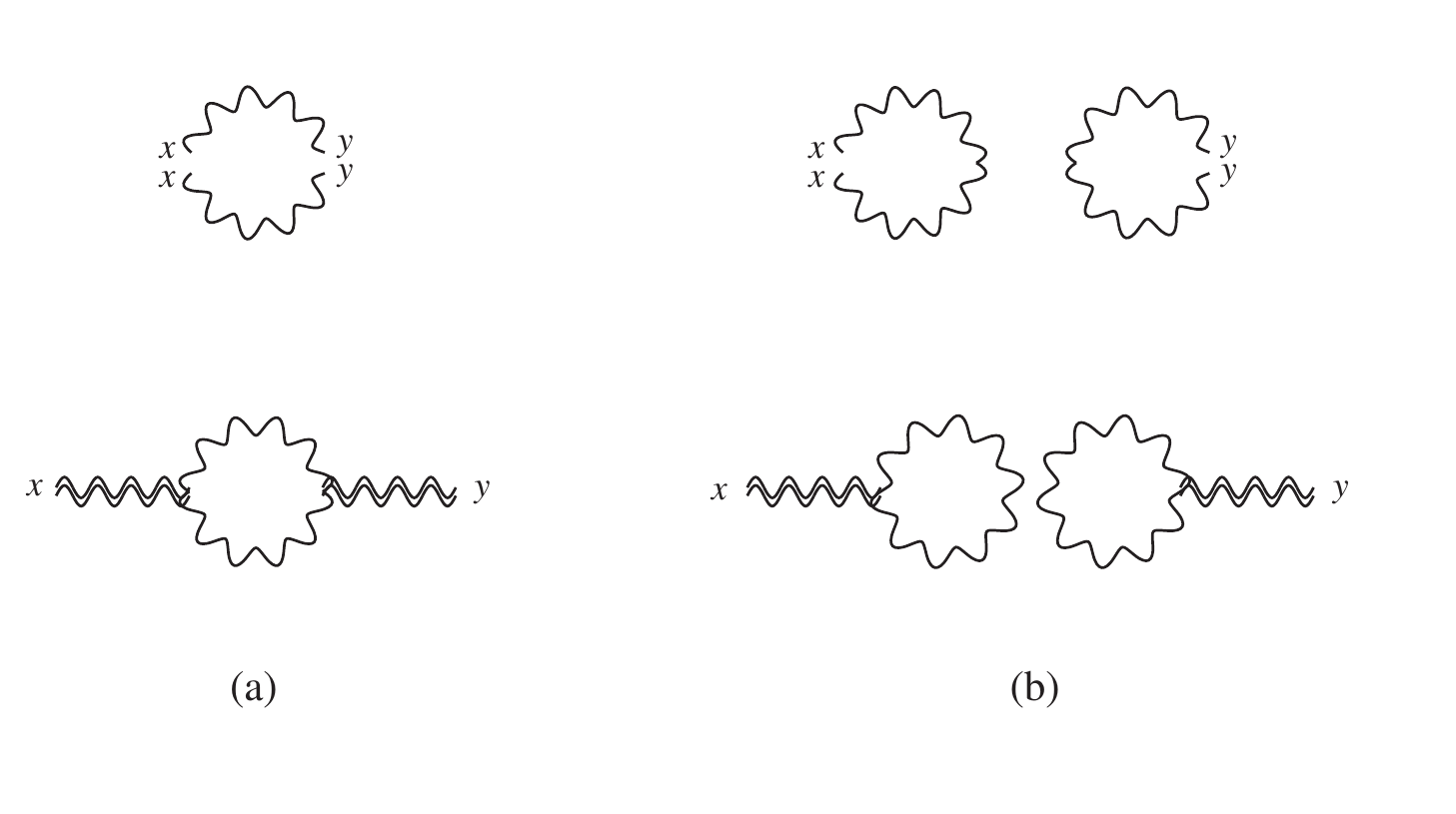} 
\caption{Pinched diagram, at one-loop order, which contributes to the right-hand side of Eq.~\eqref{eq:51}. We also present the same diagrams using the extended Feynman rule in Fig.~\ref{fig:FRDeltahh}.}
\label{fig5}
\end{figure}

\begin{figure}[ht]
\centering
\subbottom[Structural identity~\eqref{eq:41}. ]{\includegraphics[width=0.85\textwidth]{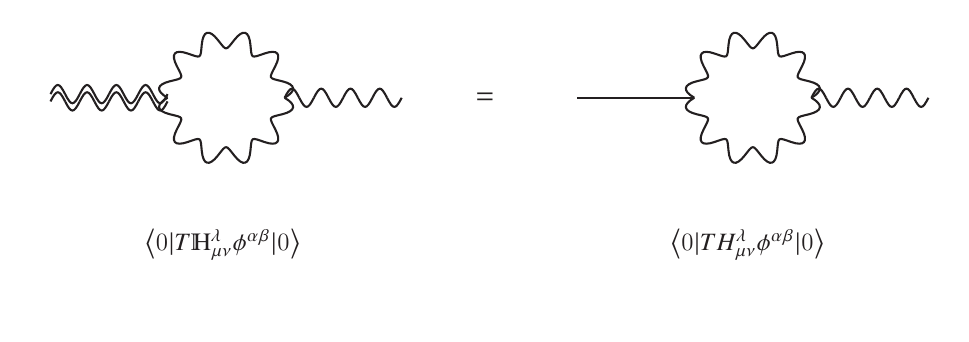}}
\subbottom[Structural identity~\eqref{eq:51}.]{\includegraphics[width=0.85\textwidth]{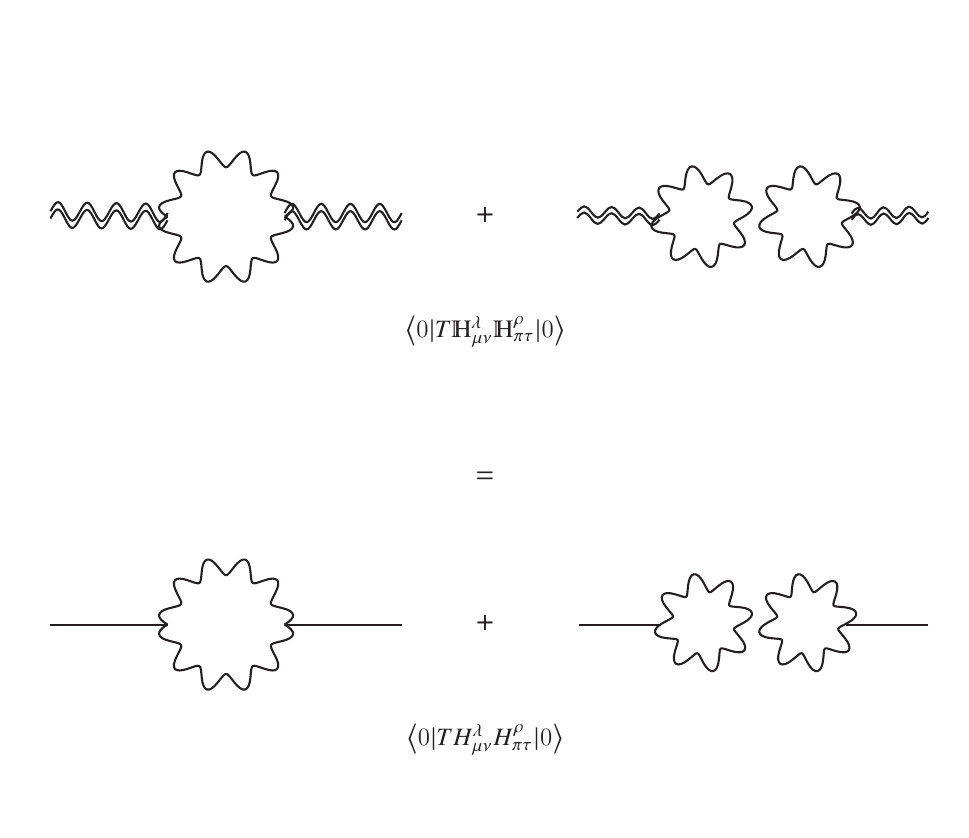}}
\caption{ Diagrammatic representation of the structural identities at one-loop order. Pinched diagrams are fully equivalent to FOGR diagrams.}
\label{figfinal}
\end{figure}

Using this extended rule, we can easily compute the pinched diagrams in gravity. However, this is not necessary. From Eq.~\eqref{eq:DeltaandSI} can be inferred that the pinched diagrams in Figs~\ref{fig2} and~\ref{fig5} should be equivalent to diagrams in the FOGR formalism (see Fig.~\ref{figfinal}).  In Fig.~\ref{figfinal}, we observe that each diagram on the left-hand side is equivalent to the respective diagram on the right-hand side. 
Thus, we verify the structural identities \eqref{eq:41} and \eqref{eq:51} at one-loop order at the integrand level. 

We shall remark that this is only possible when we consider the proper path integral for the first-order formulation of gravity. The determinant in Eq.~\eqref{eq:fgEH12ws} is responsible for the cancellation of spurious contributions, such as the tadpole-like contributions in Fig.~\ref{fig3b}. 
In \cite{martins-filho:2021, Brandt:2020vre} these diagrams were not relevant since dimensional regularization was employed to compute them. 
By using different schemes, or at finite temperature, these diagrams could possibly give a non-vanishing contribution (see Appendix~\ref{section:tadpoles}). This possibility could reveal the inequivalence between the FOGR and the SOGR formalisms. 

In this work, we addressed this issue by accounting for the second-class constraints in the FOGR leading to the proposed Senjanovi\'{c} determinant in Eq.~\eqref{eq:measureFOGR}. 
This determinant is not only consistent with the results presented in this section but also essential for verifying the structural identities at the integrand level. 
Moreover, it is necessary for establishing the quantum equivalence, as we have demonstrated in the previous section and confirmed here through a diagrammatic analysis.

 \chapter{Lagrange multiplier formalism} \label{section:LMtheory}
In this chapter, we present the \emph{standard Lagrange multiplier formalism} \cite{McKeon:1992rq, Chishtie:2012sq, Brandt:2018lbe, Brandt:2019ymg, Brandt:2020gms, Brandt:2021qgh, Brandt:2021nev} in which LM fields are introduced to restrict the path integral to fields configurations that satisfy the classical equations of motion. In this formalism, the radiative corrections of orders higher than one-loop are suppressed. The resulting path integral describes a solvable theory, which, in general, is renormalizable. Moreover, it is consistent with the unitary condition \cite{Brandt:2021qgh}. On the other hand, the degrees of freedom are doubled with the presence of the LM fields. These LM fields contribute at one-loop order doubling the usual one-loop contribution. In the next chapter, we consider a extension of this formalism in which this doubling is absent \cite{Brandt:2022kjo}. 

Clearly, this formalism is not suited to the Standard Model since higher-loop effects are required to describe nature well. For example, the computation of the electron anomalous magnetic moment at to tenth-order \cite{Aoyama:2017uqe} or the lambda shift at third-order \cite{Eides:2021wuv} shows the relevance of higher-order loops to obtain more accurate theoretical values which reflects the results of the experiments \cite{Hanneke:2008tm, Crivelli:2018vfe}. On the other side, it is perfectly suited as a quantum gravity alternative. The formulation of gravity in this formalism has GR as its classical limit and is renormalizable. 

It is interesting to remark that this formalism, proposed by McKeon and Sherry \cite{McKeon:1992rq}, is a generalization of the structure of lower dimensional gravity theories, in which fields appear linearly, to $(3+1)$-gravity. These fields play the role of LM fields imposing the equations of motion. Let us take, for example, $(2+1)$-dimensional gravity. The HE action in three dimensions is \cite{Witten:1988hc} 
\begin{equation}\label{eq:3DEH}
    S_{3\text{HE}} = \frac{1}{2} \int_{M} \epsilon^{ijk} \epsilon_{ABC} \tensor{e}{_{i}^{A}}  \left( \partial_{j} \tensor{\omega}{_{k}^{BC}}- \partial_{k} \tensor{\omega}{_{j}^{BC}} +  [ \tensor{\omega}{_{j}}\, , \tensor{\omega}{_{k}} ]^{BC} \right), 
\end{equation}
where $M$ is a space-time manifold of dimension three, $ \tensor{e}{_{i}^{A}}$ is the vierbein and $ \tensor{\omega}{_{k}^{BC}}$ is the spin connection. In the above action, the vierbein and the spin connection are treated as independent gauge fields \cite{Witten:1988hc}. However, in four-dimensional second-order gravity,
the Levi-Civita spin connection can be written as \cite[Eq. (12.1.5)]{Green:2008qa} 
\begin{equation}\label{eq:spinconnection}
    \omega_{ j}^{AB} = 
    \frac{1}{2} e^{i A} \left ( \tensor{e}{_{i}^{B}_{, j }} - \tensor{e}{_{j}^{B}_{, i}}\right ) 
    - 
    \frac{1}{2} e^{i B} \left ( \tensor{e}{_{i}^{A}_{, j }} - \tensor{e}{_{j}^{A}_{, i}}\right ) 
    - \frac{1}{2} e^{k A} e^{l B} \left ( \tensor{e}{_{l C, k }}- \tensor{e}{_{k C , l}}\right ) \tensor{e}{_{j}^{C}}.
\end{equation}

In Eq.~\eqref{eq:3DEH}, it is the vierbein $ \tensor{e}{_{i}^{A}} $ that appears linearly. It imposes the Einstein equations in the vacuum $ \tensor{R}{_{jk}^{BC}}=0$, with no cosmological constant $ \Lambda = 0$. In $1+1$ dimensional gravity, it is the dilaton field $ \phi $ that appears linearly. This is a scalar field that imposes the two-dimensional Einstein equations $ R + \Lambda =0$ \cite{Chamseddine:1989yz}, where $ R = g^{\mu \nu} R_{\mu \nu} $ is the Ricci scalar.

In $(3+1)$-dimensional gravity, or simply GR, this structure does not emerge naturally from its Lagrangian formulation. Thus, the standard LM formalism consists of imposing the equations of motion by introducing a new field $ \lambda_{\mu \nu} $. The HE action is supplemented with the LM action 
\begin{equation}\label{eq:LMactionHE}
    S_{ \lambda } = 
    \int \mathop{d^{4} x} \sqrt{-g}  \lambda_{\mu \nu} \left(R^{\mu \nu} - \frac{1}{2}R g^{\mu \nu} \right) 
\end{equation}
so that $ \lambda_{\mu \nu} $ can act similarly to the dilation in two-dimensional gravity (or the vierbein in Eq.~\eqref{eq:3DEH}).

\section{General formalism}

In this section, we describe the LM formalism in general terms. For this, we consider a scalar theory with $N$ commuting fields $ \phi_{i} $. This theory is described by a non-singular action 
\begin{equation}\label{eq:lm.actionS}
    S [ \phi ] = \int \mathop{d x} \mathcal{L} ( \phi ).
\end{equation}
The indices will be omitted for simplicity. The equations of motion are derived through Hamilton's principle which reads 
\begin{equation}\label{eq:lm.eomS}
    \frac{\delta S [\phi ] }{\delta \phi} =0.
\end{equation}

Since we assumed that the action $ S [\phi ]$ is non-singular, that is, the system is not constrained, the quantization within the path integral formalism is straightforward. The generating functional of the general action in Eq.~\eqref{eq:lm.actionS} is 
\begin{equation}\label{eq:lm.genfuncS}
    Z [j] = 
    \int \mathop{\mathcal{D} \phi} \exp{i \int \mathop{d x}  \left( \mathcal{L} (\phi ) + j \phi \right) }.
\end{equation}

In the framework of the standard LM formalism, we supplement the action Eq.~\eqref{eq:lm.actionS} with 
\begin{equation}\label{eq:lm.actionLM}
    S_{\lambda} = \int \mathop{d x} \lambda \frac{\delta S [\phi ] }{\delta \phi}. 
\end{equation}
At the classical level, the field $ \lambda $ plays the role of an LM which imposes the  equations of motion as a constraint. In section~\ref{section:Instabilities}, we investigate the classical behavior of this system in more detail. Now, we shall proceed by presenting the standard quantization of the LM theory obtained in the LM formalism.

The standard quantization procedure leads to the generating functional 
\begin{equation}\label{eq:lm.genfuncSLM}
    \mathsf{Z}_{\text{LM}} [j,k] = 
    \int \mathop{\mathcal{D} \phi} \mathop{\mathcal{D} \lambda} \exp{i \int \mathop{d x}  \left[ \mathcal{L} (\phi ) + \lambda \left(\frac{\delta S }{\delta \phi} + j\right)+ j \phi + k \lambda \right] }.
\end{equation}
Note that the LM field $ \lambda $ couples to both sources $j$ and $k$. This is required as the sources should be introduced in the equations of motion. Or equivalently, the action in Eq.~\eqref{eq:lm.actionLM} must include the sources.

In section~\ref{section:LMfermion}, we show that the standard LM formalism can be extended to fermionic systems.

\subsection{Perturbative properties}\label{section:GPofsLM}

One can integrate the LM field $ \lambda $ in Eq.~\eqref{eq:lm.genfuncSLM} which leads to 
\begin{equation}\label{eq:lm.genfuncSLMdelta}
    \mathsf{Z}_{\text{LM}} [j,k] = 
    \int \mathop{\mathcal{D} \phi}  \mathop{\delta}\left ( \frac{\delta S }{\delta \phi} + j + k\right )\exp{i \int \mathop{d x}    \left[ \mathcal{L} (\phi ) + j \phi  \right] }.
\end{equation}
Now, it is transparent that the LM fields $ \lambda $ constraints the path integral to field configurations $ \tilde{\phi}(j,k) $ that satisfy 
\begin{equation}\label{eq:lm.constraintJandK}
    \left. \frac{\delta S }{\delta \phi} \right|_{\phi = \tilde{\phi}}   + j + k  =0.
\end{equation}

We can try to integrate the field $ \phi $ in Eq.~\eqref{eq:lm.genfuncSLMdelta} using the functional analog of the following property of the $ \delta $-function 
\begin{equation}\label{eq:deltafuncp1}
    \delta \left ( f(x)\right ) = \sum_{i}^{} | f' (\tilde{x}_{i}  ) |^{-1} \delta \left ( x - \tilde{x}_{i} \right ),
\end{equation}
where $ \tilde{x}_{i} $ satisfy $ f( \tilde{x}_{i} )=0$. In our case, it becomes 
\begin{equation}\label{eq:lm.deltafuncp1}
    \mathop{\delta} \left ( \frac{\delta S }{\delta \phi} + j + k\right ) = \sum_{\tilde{\phi}}^{} \left| \det \frac{\delta^{2} S [ \tilde{\phi} ]}{\delta \phi \delta \phi} \right|^{-1}\mathop{\delta} \left ( \phi - \tilde{\phi}\right ).
\end{equation}
We assume that the Hessian of the action $ S [ \phi ]$ is always positive. Hence, the absolute value can be discarded. 

Replacing Eq.~\eqref{eq:lm.deltafuncp1} in Eq.~\eqref{eq:lm.genfuncSLMdelta} yields 
\begin{equation}\label{eq:lm.genfuncExact}
    \mathsf{Z}_{\text{LM }} [j,k] = \sum_{\tilde{\phi}}^{} 
    \det \left(\frac{\delta^{2} S [ \tilde{\phi} ]}{ \delta \phi \delta \phi}\right)^{-1} \exp i \left ( S [ \tilde{\phi} ] + \int \mathop{d x} j \tilde{\phi} (j,k)\right ). 
\end{equation}
Thus, we obtained an exact form for the generating functional in the LM formalism. This shows that the resulting theory is solvable. The dependence on the source $k$ is entirely on $ \tilde{\phi} (j,k)$. 

The exponential in Eq.~\eqref{eq:lm.genfuncExact} leads to tree-level diagrams, while the determinant is the square of the usual one-loop determinant $ \det [\mathcal{L}^{\prime \prime} ( \phi )]^{-1/2} $. Thus, the tree-level is kept unaltered. However, the one-loop contributions comes out twice the usual obtained from Eq.~\eqref{eq:lm.genfuncS}. The higher-order loop diagrams are absent.

This result also can be seen in a diagrammatic way \cite{Brandt:2019ymg, Brandt:2022kjo}. Expanding the Lagrangian $ \mathcal{L} (\phi )$ as 
\begin{equation}\label{eq:lm.expandedLag}
    \mathcal{L} (\phi) = \frac{1}{2} a_{ij} \phi_{i} \phi_{j}  + \frac{1}{3!} a_{ijk} \phi_{i} \phi_{j} \phi_{k} + \frac{1}{4!} a_{ijkl} \phi_{i} \phi_{j} \phi_{k} \phi_{l} +  \cdots,
\end{equation}
the Lagrangian in the framework of the standard LM theory is 
\begin{equation}\label{eq:lm.expandedLagLM}
    \begin{split}
        \mathcal{L}_{\text{LM}} (\phi) ={}& 
        \frac{1}{2} a_{ij} \phi_{i} \phi_{j}  + \frac{1}{3!} a_{ijk} \phi_{i} \phi_{j} \phi_{k} + \frac{1}{4!} a_{ijkl} \phi_{i} \phi_{j} \phi_{k} \phi_{l} +  \cdots \\
                                          &     + a_{ij} \lambda_{i} \phi_{j} + \frac{1}{2}a_{ijk}  \lambda_{i} \phi_{j} \phi_{k} + \frac{1}{3!} a_{ijkl} \lambda_{i} \phi_{j} \phi_{k} \phi_{l} + \cdots. 
\end{split}
\end{equation}
We can read off the Feynman rules from the above expression straightforwardly. The propagators are given by the inverse of the matrix 
\begin{equation}\label{eq:lm.FRprop}
    \begin{pmatrix}
        a_{ij} & a_{ij} \\
        a_{ij} & 0
    \end{pmatrix}^{-1}
    = \begin{pmatrix}
        0 & a_{ij}^{-1}\\
        a_{ij}^{-1}& - a^{-1}_{ij} 
    \end{pmatrix}. 
\end{equation}
Thus, we have the mixed propagators $  \langle \lambda_{i} \phi_{j}\rangle =  \langle \phi_{i} \lambda_{j}\rangle = a_{ij}^{-1} $ and the propagator of the LM field $  \langle \lambda_{i} \lambda_{j}\rangle =- a_{ij}^{-1} $. Note that, the physical field $ \phi $ does not propagate anymore. 

Besides that, the negative sign in the propagator of the LM fields suggests the presence of instabilities (ghosts) in the theory \cite{McKeon:1992rq, Brandt:2022kjo}.
The origins of the instabilities in the LM theory have not yet been fully established. We interpret them as Ostrogradsky instabilities \cite{Ostrogradsky:1850fid}.\footnote{Ostrogradsky instabilities are related to the presence of higher derivative terms in the Lagrangian. In the LM formalism, the equations of motion, usually of second-order, are introduced in the Lagrangian using an LM field.} For clarity, we illustrate this argument by using an example in section~\ref{section:Instabilities}, which can be generalized to any field theory in a straightforward manner.

We already know that: 
\begin{enumerate}[(i)]
    \item There is no $ \phi $ propagator.
    \item The interactions have a single LM field $ \lambda $ in the external legs.
\end{enumerate}
The tree-level remains unchanged. At one-loop order,  only mixed propagators appear in the internal lines, with $ \phi $ being the only field present in the external legs. Other configurations are forbidden, thus higher-order loop diagrams do not arise. The doubling of one-loop diagrams is attributed to the combinatorial factor \cite{Brandt:2019ymg}. 

\subsection{Fermionic systems}\label{section:LMfermion}
Let us consider now a fermionic system described by the complex spinor fields $ \psi $ and $ \bar{\psi} $. The action of this system is given by 
\begin{equation}\label{eq:lm.actionSpsi}
    S [ \psi , \bar{\psi} ] = \int \mathop{d^{}x} \mathcal{L} [ \psi , \bar{\psi} ].
\end{equation}
The Euler-Lagrange equations are 
\begin{equation}\label{eq:lm.eomSpsi}
    \frac{\delta S }{\delta \psi} = 0 \quad \text{and} \quad \frac{\delta S }{\delta \bar{\psi}} =0.
\end{equation}
If the generating functional of this system is given by 
\begin{equation}\label{eq:lm.genfuncSpsi}
    Z [ \eta , \bar{\eta} ] = \int \mathop{\mathcal{D} \bar{\psi}} \mathop{\mathcal{D} \psi} \exp{i \int \mathop{d x} \left(\mathcal{L} [ \psi , \bar{\psi} ] + \eta \bar{\psi} + \bar{\eta} \psi \right)},
\end{equation}
then, in the framework of the LM theory, it would become
\begin{equation}\label{eq:lm.genfuncSpsiExact}
    \begin{split}
        \mathsf{Z}_{\text{LM}} [ \eta , \bar{\eta}, \xi , \bar{\xi}  ] ={}& \int  \mathop{\mathcal{D} \bar{\psi}} \mathop{\mathcal{D} \psi} \mathop{\mathcal{D} \bar{\lambda}} \mathop{\mathcal{D} \lambda}  \exp{i} \int \mathop{d x}   \bigg(\mathcal{L} [ \psi , \bar{\psi} ] +  \lambda \frac{\delta S }{\delta \psi} + \bar{\lambda} \frac{\delta S }{\delta \bar{\psi}} \\ & + \bar{\psi} \eta+ \bar{\eta} \psi+ \bar{\eta} \lambda + \bar{\lambda} \eta + \bar{\lambda} \xi + \bar{\xi} \lambda ),
    \end{split}
\end{equation}
where all the sources and the LM fields $ \lambda $, $ \bar{\lambda} $ are Grassmann fields, as the fields $ \psi $ and $ \bar{\psi} $. 

We can perform a similar analysis for this generating functional as we have done for the bosonic fields. The only difference is that the determinants appear with an opposite sign. The exact form of the generating functional is 
\begin{equation} \label{eq:lm.genfuncExactpsi}
    \mathsf{Z}_{\text{LM}} [ \eta , \bar{\eta}, \xi , \bar{\xi}  ] =\sum_{\tilde{\psi}}^{}  \sum_{\tilde{\bar{\psi}}}^{} \det \left( \frac{\delta_{L} \delta_{R}  S [ \tilde{\psi} , \tilde{\bar{\psi}} ] }{\delta \psi \delta \bar{\psi} } \right)^{2}       
    \exp{i \int \mathop{d x} \left(\mathcal{L} [ \tilde{\psi} , \tilde{\bar{\psi}} ] + \tilde{\bar{\psi}} \eta+ \bar{\eta} \tilde{\psi} \right)}.
\end{equation}
The determinant in Eq.~\eqref{eq:lm.genfuncExactpsi} comes from 
\begin{equation}\label{eq:lm.detpsibarpiso}
    \begin{vmatrix}
        \dfrac{\updelta^{2} S [ \psi , \bar{\psi} ] }{ \delta \psi \delta \psi }&
        \dfrac{\updelta^{2} S [ \psi , \bar{\psi} ] }{ \delta \psi \delta \bar{\psi}} \\
    \dfrac{\updelta^{2} S [ \psi , \bar{\psi} ] }{ \delta \bar{\psi} \delta \psi }&
    \dfrac{\updelta^{2} S [ \psi , \bar{\psi} ] }{ \delta \bar{\psi}  \delta \bar{\psi} }
    \end{vmatrix} 
    =
\begin{vmatrix}
         0&
         \dfrac{\updelta^{2} S [ \psi , \bar{\psi} ] }{ \delta \psi \delta \bar{\psi}} \\
         \dfrac{\updelta^{2} S [ \psi , \bar{\psi} ] }{ \delta \bar{\psi} \delta \psi }&
      0
    \end{vmatrix},
\end{equation}
where $ \updelta^{2} \equiv \delta_{L} \delta_{R} $. The determinant in Eq.~\eqref{eq:lm.detpsibarpiso} is the square of the usual fermionic one-loop contribution.

\section{Gauge theories}\label{section:gaugeSLM}

The LM formalism for gauge theories requires more attention, since, in this context, the LM field turns out to be a gauge field itself. Hence, the FP procedure must be altered to accommodate the new gauge invariances. 

Let us assume that $ S [ \phi ] $ is invariant under the gauge transformations 
\begin{equation}\label{eq:lm.gaugetrans}
    \delta \phi_{i} = H_{ij} \zeta_{j}, 
\end{equation}
then the LM action is invariant under 
\begin{equation}\label{eq:lm.gaugetransLM}
    \delta \phi_{i} = H_{ij} \zeta_{j} \quad \text{and} \quad \delta \lambda_{i} = H_{ij, l} \lambda_{l} \zeta_{j}.
\end{equation}
We remind the reader that the comma represents the partial derivative.
In addition, we have a novel gauge invariance due to the LM field: 
\begin{equation}\label{eq:lm.gaugetransLambda}
    \delta \lambda_{i} = H_{ij} \zeta_{j} .
\end{equation}

The variation of the action $ S [ \phi ] $, under Eq.~\eqref{eq:lm.gaugetrans}, is
\begin{equation}\label{eq:lm.variationS}
    \delta S= \int \mathop{d x} \frac{\delta S }{\delta \phi} \delta \phi = 0. 
\end{equation}
When supplemented by LM fields, it becomes 
\begin{equation}\label{eq:lm.variationSLM}
    \delta S_{\text{LM}} = \int \mathop{d x} \left(\frac{\delta S }{\delta \phi} \delta \phi + \delta \lambda \frac{\delta S }{\delta \phi} + \frac{\delta^{2} S }{\delta \phi \delta \phi} \lambda \delta \phi\right). 
\end{equation}
Clearly, Eq.~\eqref{eq:lm.gaugetransLambda} is valid, when $ \delta \phi = 0$. Now, by differentiating Eq.~\eqref{eq:lm.variationS} we obtain 
\begin{equation}\label{eq:lm.diffvaiationallS}
    \int \mathop{d x} \left(\frac{\delta^{2} S }{\delta \phi \delta \phi} \delta \phi + \frac{\delta S }{\delta \phi} \frac{\delta \delta \phi }{\delta \phi}\right) =0. 
\end{equation}
Thus, as long $ \delta \lambda_{i} = \lambda_{l} \delta ( \delta \phi_{i} )/ \delta \phi_{l} = \lambda_{l} H_{ij, l} \zeta_{l} $, Eq.~\eqref{eq:lm.variationSLM} is invariant under Eq.~\eqref{eq:lm.gaugetrans}. This shows the consistency of Eq.~\eqref{eq:lm.gaugetransLM}.  

The standard method is to account for these gauge invariances and build up the appropriate FP procedure, we refer the reader to Refs. \cite{McKeon:1992rq, Brandt:2019ymg, Brandt:2020gms}. 

\subsection{Alternative method}\label{sect:gravity521}
Alternatively, one can introduce the LM fields after quantizing the physical field \( \phi \). We show that this method yields the same generating functional by applying it directly to the HP generating functional \eqref{eq:313}. In this method, we also need to introduce LM fields for the FP ghost fields, which are anticommuting. 
This shows that the standard LM formalism and the FP procedure commute. In Chapter~\ref{chapter:gtmLM}, we provide a  general demonstration. 

We also show that the LM formalism is consistent with the quantum equivalence between the FOGR and the SOGR formulation previously examined. We shall obtain the SOGR formulation through the FOGR formulation. 

Let us supplement the generating functional in Eq.~\eqref{eq:313} with the LM fields $ \lambda_{\mu \nu} $ and $ \Lambda_{\mu \nu}^{\lambda} $. The bosonic LM action is 
\begin{equation}\label{eq:lm.hp.actionLMb}
    \begin{split}
        S_{b\text{LM}} ={}& \frac{1}{\kappa^{2}} \int \mathop{d x} \left [
            \tensor{\Lambda}{^{\rho}_{\pi \tau }} \left( \tensor*{M}{*^{\pi \tau}_{\rho}^{\mu \nu}_{\lambda}} (h) G^{\lambda}_{\mu \nu} -  \kappa \phi^{\pi \tau}{}_{, \rho}\right) 
    \right]\\
                         & + \int \mathop{d x}   
                         \bigg[ \frac{1}{\kappa} \lambda^{\alpha \beta} \left(\frac{1}{2} \tensor{G}{^{\lambda}_{\mu \nu}} \tensor{G}{^{\rho}_{\pi \tau}} \tensor*{\mathcal{M}}{*^{\mu \nu}_{\lambda}^{\pi \tau }_{\rho}_{ \alpha \beta }}   +  \tensor{G}{^{\lambda}_{ \alpha \beta , \lambda} }\right)
        - \frac{1}{\alpha} ( \partial_{\mu} \phi^{\mu \nu} ) \eta_{\nu \rho} ( \partial_{\sigma} \lambda^{\sigma \rho } ) \\ 
                         & 
                         +       
\kappa \bar d_\mu \left[
(\lambda^{\rho\sigma}_{,\rho})\partial_\sigma\eta^{\mu\nu}
-(\lambda^{\rho\mu}_{,\rho})\partial^{\nu} 
+\lambda^{\rho\sigma}\partial_{\rho}\partial_\sigma\eta^{\mu\nu}
-(\partial_\rho\partial^\nu\lambda^{\rho\mu})\right] d_\nu 
\bigg].
\end{split}
\end{equation}
The LM fields $ c_{\mu} $ and $ \bar{c}_{\mu} $ associated with the FP ghosts $ d_{\mu} $ and $ \bar{d}_{\mu} $ leads to the fermionic LM action:
\begin{equation}\label{eq:lm.hp.actionLMf}
    \begin{split}
        S_{f\text{LM}} ={}& \int \mathop{d x} \kappa  \big\{\!\left[
\bar{c}_\mu(\phi^{\rho\sigma}_{,\rho})\partial_\sigma\eta^{\mu\nu}
-(\phi^{\rho\mu}_{,\rho})\partial^{\nu} 
+\phi^{\rho\sigma}\partial_{\rho}\partial_\sigma\eta^{\mu\nu}
-(\partial_\rho\partial^\nu\phi^{\rho\mu})\right] d_\nu 
\\
                          & 
                          +  \bar{d}_\mu \left[
(\phi^{\rho\sigma}_{,\rho})\partial_\sigma\eta^{\mu\nu}
-(\phi^{\rho\mu}_{,\rho})\partial^{\nu} 
+\phi^{\rho\sigma}\partial_{\rho}\partial_\sigma\eta^{\mu\nu}
-(\partial_\rho\partial^\nu\phi^{\rho\mu})\right] c_\nu
\big\}. 
\end{split}
\end{equation}
The generating functional of the FOGR theory in the framework of the LM formalism is then given by 
\begin{equation}\label{eq:lm.genfuncFOGR}
    \begin{split}
        \mathsf{Z}_{\text{HP}} [j,J,k,K] = \int &  
    \mathop{\mathcal{D} \phi^{\mu \nu} } 
    \mathop{\mathcal{D} G^{\lambda}_{\mu \nu} } 
    \mathop{\mathcal{D} \lambda^{\mu \nu} } 
    \mathop{\mathcal{D} \Lambda^{\lambda}_{\mu \nu} } 
    \mathop{\mathcal{D} \bar{d}_{\mu} } 
    \mathop{\mathcal{D} d_{\nu} } 
    \mathop{\mathcal{D} \bar{e}_{\mu} } 
    \mathop{\mathcal{D} e_{\nu} } 
    | \det \tensor*{M}{*^{\mu \nu}_{\lambda}^{\pi \tau}_{\rho}} (h) | 
    \\
                                        & \quad \times \exp{i \left(   S_{\text{eff}}^{\text{I}}   + S_{b \text{LM}} + S_{f \text{LM}} + S_{\text{sc}}[j,J,k,K]\right)}, 
\end{split}
\end{equation}
where $S_{\text{eff}}^{\text{I}} $ is the effective action of the FOGR theory and $ S_{\text{sc}}[j,J,k,K] $ denotes the source action: 
\begin{equation}\label{eq:lm.scaction}
    S_{\text{sc}}[j,J,k,K] = j_{\mu \nu } \phi^{\mu \nu } 
    + j_{\mu \nu} \lambda^{ \mu \nu} 
    + J_{\lambda}^{\mu \nu} G^{\lambda}_{\mu \nu}  
+ J_{\lambda}^{\mu \nu} \Lambda^{\lambda}_{\mu \nu}  
    + k_{\mu \nu} \lambda^{ \mu \nu} 
+ K_{\lambda}^{\mu \nu} \Lambda^{\lambda}_{\mu \nu}.
\end{equation}
The Senjanovi\'{c} determinant in the LM formalism is the square of the usual Senjanovi\'{c} determinant that arises when the LM fields are absent. The LM fields duplicates the constraints of a constrained system, which leads to the square of the FP and the FS determinant.

\begin{figure}[ht]
    \centering
    \includegraphics[scale=0.85]{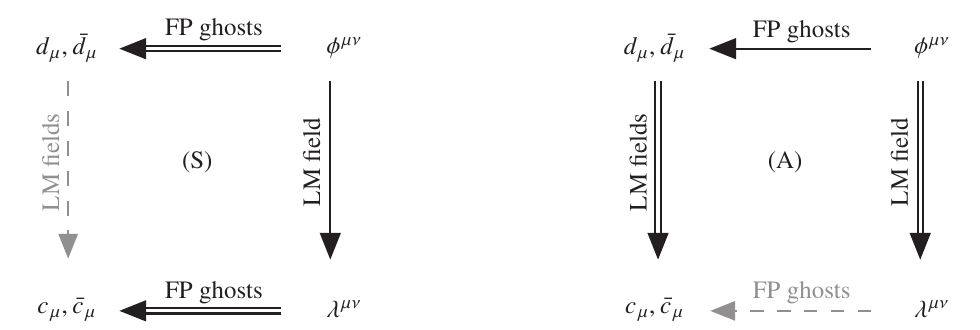}
    \caption{In the standard method (S), the introduction of the LM field $ \lambda $ is followed by the FP procedure. In the alternative method (A), we start with the FP procedure and then we introduce the LM fields $ \lambda^{\mu \nu} $ and $ c_{\mu} $, $ \bar{c}_{\mu} $. The order is represented by the number of solid lines, the dashed lines indicate the physical interpretation.}\label{fig:duality}
\end{figure}

The generating functional in Eq.~\eqref{eq:lm.genfuncFOGR} is the same obtained in \cite{Brandt:2020gms}. Thus, the alternative method may be used to obtain the generating functional of gauge theories without requiring changing the FP procedure. Moreover, it confirms that the anticommuting fields $ c_{\mu} $ and $ \bar{c}_{\mu} $ can be interpreted as LM fields of the FP ghosts. In \cite{Brandt:2020gms}, these fields are the FP ghosts associated with the LM field $ \lambda^{\mu \nu} $. Therefore, we see a duality: the fields $ c_{\mu} $, $ \bar{c}_{\mu} $ are the FP ghosts associated with the LM field $ \lambda^{\mu \nu} $, or alternatively, the LM fields of the FP ghosts $ d_{\mu} $, $ \bar{d}_{\mu} $ are the fields $ c_{\mu} $, $ \bar{c}_{\mu}$. This is shown in Fig.~\ref{fig:duality}. 

Now, we can proceed to obtain the SOGR theory from Eq.~\eqref{eq:lm.genfuncFOGR}. We integrate the LM field $ \Lambda^{\lambda}_{\mu \nu} $ associated with the auxiliary field $ G^{\lambda}_{\mu \nu} $ yielding 
\begin{equation}\label{eq:lm.deltafuncFOGR}
    \mathop{\delta} \left ( \tensor*{M}{*^{\pi \tau}_{\rho}^{\mu \nu}_{\lambda}} (h) G^{\lambda}_{\mu \nu} -  \kappa \phi^{\pi \tau}{}_{, \rho}\right) = \left(\det \tensor*{M}{*_{\rho}^{\pi \tau}^{\mu \nu}_{\lambda}} \right)^{-1} \mathop{\delta}  ( G^{\lambda}_{\mu \nu} - \mathbb{G}^{\lambda}_{\mu \nu}  ),
\end{equation}
where $ \mathbb{G}^{\lambda}_{\mu \nu} $ is the classical value of the auxiliary field defined in Eq.~\eqref{eq:eomG}). The determinant in Eq.~\eqref{eq:lm.deltafuncFOGR} cancels the square of the Senjanovi\'{c} determinant in Eq.~\eqref{eq:lm.genfuncFOGR}.  Integrating the auxiliary field using the $ \delta $-function allows us to replace the auxiliary field with its classical value. This results in 
\begin{equation}\label{eq:72sm}
    \begin{split}
        & \mathsf{Z}_{\text{HE}} [j,J,k,K]   \\ ={}&
        \int \mathcal{D} {\phi}^{}_{\mu \nu}\mathcal{D} {\lambda}^{}_{\mu \nu} \exp  \bigg[ i \int \mathop{dx} \bigg( -\frac{1}{2 \kappa^{2}}\mathbb{G}^{\lambda}_{\mu \nu}[h, K] \tensor*{M}{*^{\mu \nu}_{\lambda}^{\pi \tau}_{\sigma}}( \chi ) (  \mathbb{G} [h, K]- 2 \mathbb{G}  [\chi , J] ){}^{\sigma}_{\pi \tau }
                \\ & - \frac{1}{2 \alpha} \left [ ( \partial_{\mu} \phi^{\mu \nu} )^{2} + 2 ( \partial_{\mu} \phi^{\mu \nu} ) \eta_{\nu \rho} (\partial_{\sigma} \lambda^{\sigma \rho} )\right ] 
            +  j_{\mu \nu} \phi^{\mu \nu} + k_{\mu \nu} \lambda^{\mu \nu}  \bigg) \bigg] {\Xi}_{\text{FP}}^{} (\phi, \lambda ),
    \end{split} 
\end{equation}
where $ \chi^{\mu \nu} = \eta^{\mu \nu} + \kappa (\phi^{\mu \nu} + \lambda^{\mu \nu})$. The FP sector is kept unaltered and is denoted by $ \Xi_{\text{FP}} ( \phi , \lambda )$. Setting $J=K=0$ in Eq.~\eqref{eq:72sm} leads to the SOGR generating functional in the framework of the LM formalism \cite{Brandt:2020gms}.

Both Eq.~\eqref{eq:lm.genfuncFOGR} and Eq.~\eqref{eq:72sm} leads to solvable, renormalizable gravity theories \cite{Brandt:2020gms}. The perturbative expansion is truncated at one-loop order. The tree diagrams are kept unaltered, thus GR is obtained in the classical limit. The one-loop diagrams result in twice the usual contribution, which arises in the theory without LM fields.

\section{Instabilities in the standard Lagrange multiplier theory}\label{section:Instabilities}

In general, the classical LM theory is given by a constrained system described by a Lagrangian $ L( x , \dot{x} ) $ (and action $S[x]$) with the Euler-Lagrange equation
\begin{equation}\label{eq:ELeq}
    \alpha_{0} \equiv  \frac{\delta S[x]}{\delta x } =\frac{d}{d t}\left( \frac{\partial L }{\partial \dot{x} } \right) - \frac{\partial L }{\partial x} =0
\end{equation}
as the constraint. Let us consider a general, standard, one-dimensional Lagrangian
\begin{equation}\label{eq:exLag}
    L (x, \dot{x}) = \frac{\dot{x}^{2}}{2} - V(x),
\end{equation}
where $ V(x)$ denotes the potential, and dots denote times derivatives. The constraint is  
\begin{equation}\label{eq:EoMofLag}
    \alpha_0 = \ddot x + \partial_{x} V(x)=0,
\end{equation}
which is the equation of motion obtained by substituting Eq.~\eqref{eq:exLag} in Eq.~\eqref{eq:ELeq}. The constraint $ \alpha_{0} $ is highly non-trivial since it is acceleration-dependent. It can be recast as the set of non-holonomic constraints: 
\begin{equation}\label{eq:setofNHC}
    \dot{Q} + \partial_{x} V(x) =0 \quad \text{and} \quad Q - \dot{x} =0,
\end{equation}
where we introduced a new variable $Q$. Thus, the price that we have to pay is the doubling of the degrees of freedom. It clarifies the origin of the doubling of degrees of freedom in the LM theory, which is directly related to the presence of instabilities \cite{Brandt:2022kjo}.

The LM theory is then given by the non-holonomic system described by the Lagrangian $L(x,Q)$ with the non-holonomic constraints \eqref{eq:setofNHC}. There are two methods to approach non-holonomic systems: the standard non-holonomic mechanics based on the D'Alembert-Lagrange principle (see the reviews \cite{borisov:2016, deleon:2012}) and vakonomic mechanics\footnote{It stands for mechanics of variational axiomatic kind.} \cite{kozlov:2010}. Vakonomic mechanics is based on the generalization of Hamilton's principle of least action for non-holonomic systems. By following the vakonomic approach, we obtain the standard LM theory. In this approach, the extended Lagrangian 
\begin{equation}\label{eq:EXLMLAG}
    L_{\text{LM}} (x, \dot{x} ;y)= L(x, \dot{x} ) + y \left( \ddot{x} + \partial_{x} V(x)\right)
\end{equation}
describes the dynamics of the LM theory. 
Despite the apparent higher derivative nature of the Lagrangian, the Ostrogradsky theorem \cite{Ostrogradsky:1850fid, Pagani:1987ue} cannot be applied due to its degeneracy. However, the resulting Hamiltonian still exhibits Ostrogradsky instabilities, which can be remedied by removing the higher derivative terms using a total derivative resulting in\footnote{In this approach, the LM field $ \lambda $ becomes dynamical.}
\begin{equation}\label{eq:EXsLMLAG}
    L_{\text{LM}} (x, \dot{x} ;y)= L(x, \dot{x} ) - \dot{y} \dot{x} +y \partial_{x} V(x).
\end{equation}
This shows that the LM theory is not a higher derivative theory in this approach.
From Eq.~\eqref{eq:EXsLMLAG}, we can obtain the Hamiltonian 
\begin{equation}\label{eq:ExsLMH}
    H = - p_{x} p_{y} - \frac{p_{y}^{2}}{2} + V(x) - y \partial_{x} V(x), 
\end{equation}
where $ p_{x} $ and $ p_y$ are the canonical momenta of $x$ and $y$. The Hamiltonian depends linearly on the momenta revealing the presence of Ostrogradsky-like instabilities in the standard LM theory. The Ostrogradsky instability leads to an unbound Hamiltonian as we can see in Eq.~\eqref{eq:ExsLMH}, and therefore the quantum standard LM theory must be plagued by Ostrogradsky-like ghosts. 

One may try to reduce the phase space to remove the unphysical degrees of freedom and obtain a bounded Hamiltonian, which in our case could be done using a physical sector such as $(x, p_y)$. Thus, the quantization of the Hamiltonian \eqref{eq:ExsLMH} would lead to any quantum correction since $x$ and $ p_y$ commutes. This approach removes the Ostrogradsky-like instabilities, but the resulting theory is trivial. It would be simply the classical theory stated within the framework of quantum theory (known as the KvNS formalism \cite{koopman:1931, v.neumann:1932, Sudarshan:1961vs}). 

Therefore, it seems that the quantum LM theory would not be consistent. However, it is important to remark that vakonomic mechanics is not equivalent to non-holonomic mechanics. In fact, the vakonomic mechanics generally does not lead to the same equations of motion obtained with Newtonian mechanics \cite{ray:1966, ray:1966a, flannery:2005, lemos:2022}. Moreover, experimental results support the non-holonomic mechanics \cite{lewis:1995}. 

This suggests that the standard path integral quantization of the LM theory must be modified, as the standard approach may be inappropriate for describing an LM non-holonomic system, even at the classical level. Furthermore, the quantization of non-holonomic systems within non-holonomic mechanics cannot be done systematically \cite{bloch:2008}. In principle, these systems are neither Hamiltonian nor variational. Although, in some cases, a Hamiltonian can be obtained and standard quantization procedures can be used \cite{bloch:2008, abudfilho:1983}. Thus, to conclude, the consistency of the quantization of the standard LM theory must be examined carefully. This is considered in the next chapter.

In the next chapter, we propose a modification of the LM formalism in which the doubling of one-loop contributions is absent. In addition, the spurious degrees of freedom due to LM fields are also removed.

\chapter{Modified Lagrange multiplier formalism}\label{section:modLMchap}

In this chapter, we present a modification of the standard LM formalism, which was proposed in Ref. \cite{Brandt:2022kjo}. 
We demonstrate that we can remove the doubling of degrees of freedom and the consequent additional one-loop contributions, by restoring the field redefinition invariance of the LM theory's path integral. This proposal resolves the known drawbacks of the standard LM formalism while maintaining the renormalizability and unitarity of the resulting theory.  

\section{Field redefinitions}\label{section:FieldRef}

An expected propriety of any quantum path integral is the invariance under field redefinitions. Quantum fields are not observable quantities and can be redefined without affecting the physical quantities such as the $S$-matrix. In particular, for the $S$-matrix, this is known as the equivalence theorem \cite{Kamefuchi:1961sb}. 

Field redefinitions in the quantum path integral are analogous to a change of variables of an ordinary integral. We will show this by using the general generating functional in Eq.~\eqref{eq:lm.genfuncS}. Under the field redefinition 
\begin{equation}\label{eq:fieldRed}
    \phi \to \phi ' = F [ \phi ]
\end{equation}
the generating functional~\eqref{eq:lm.genfuncS} transforms as 
\begin{equation}\label{eq:transgenfunc}
    Z [j] \to Z'[j]= 
    \int \mathop{\mathcal{D} \phi'} \det {}\mathbb{J}_{\phi}  \exp{i \int \mathop{d x}  \left( \mathcal{L} (F^{-1}[\phi']) + j F^{-1}[\phi'] \right) },
\end{equation}
where
\begin{equation}\label{eq:def.J}
    \det {}\mathbb{J}_{\phi} \equiv \frac{\delta \phi }{\delta \phi'}
\end{equation}
denotes the Jacobian determinant associated with the field redefinition in Eq.~\eqref{eq:fieldRed}. Note that, the source term is modified. In principle, it is equal to $ j F^{-1} [ \phi'] $. However, it is necessary so that the transformed generating functional in Eq.~\eqref{eq:transgenfunc} yields the same Green functions, that is, 
\begin{equation}\label{eq:invariance}
    i^{n} \langle 0|T \phi (x_{1} )\cdots \phi ( x_{n} )| 0 \rangle =  \frac{\delta^{n} Z[j] }{\delta j(x_1) \cdots \delta j(x_n)} = \frac{\delta^{n} Z'[j] }{\delta j(x_1) \cdots \delta j(x_n) }.
\end{equation}

In the LM formalism, any field redefinition of the physical field \( \phi \) must be accompanied by a corresponding field redefinition of its associated LM field \( \lambda \) \cite{Brandt:2022kjo}. This can be understood as follows: the LM field \( \lambda \) enforces the equations of motion, which change under the field redefinition of \( \phi \). Therefore, the invariance of the form of the LM action\footnote{We define $ S[F^{-1}[\phi']] = S'[\phi'] $.}
\begin{equation}\label{eq:structureLMaction}
    S [ \phi ] + \lambda \frac{\delta S [ \phi ]}{\delta \phi} \to S'[\phi']+ \lambda ' \frac{\delta S'[\phi'] }{\delta \phi '} 
\end{equation}
is maintained as long 
\begin{equation}\label{eq:FieldRedlamb}
    \lambda ' = \lambda \frac{\delta \phi ' }{\delta \phi}.
\end{equation}
Consequently, the generating functional \eqref{eq:lm.genfuncS} in the framework of the LM formalism (in Eq.~\eqref{eq:lm.genfuncSLM}) transforms as 
\begin{equation}\label{eq:lm.transformZ}
        \mathsf{Z}_{\text{LM}} [j,k]  \to{}
        \mathsf{Z}_{\text{LM}} '[j,k],
    \end{equation}
    where
\begin{equation}\label{eq:lm.transformedZ}
        \mathsf{Z}'_{\text{LM}} [j,k]  
        = \int \mathop{\mathcal{D} \phi'} \mathop{\mathcal{D} \lambda '} \det {}\mathbb{J}_{\phi}^{2}   \exp{i \int \mathop{d x}  \left( \mathcal{L}' (\phi') + \lambda' \frac{\delta S'[\phi']}{\delta \phi '} + j F^{-1}[\phi'] + j \lambda' + k \lambda'   \right) }.
\end{equation}
Therefore, we see that the behavior of the generating functional of the standard LM theory under field redefinitions differs from the standard (see Eq.~\eqref{eq:transgenfunc}). 

Note that there is no doubling of the Jacobian determinant factor for field redefinitions of \( \lambda \). This reveals an inconsistency arising from the lack of field redefinition invariance in the path integral within the framework of the standard LM formalism. In \cite{Brandt:2022kjo}, we have shown that this issue can be resolved by introducing ghost fields. Furthermore, these ghost fields are responsible for canceling the extra one-loop contributions due to the LM field \( \lambda \). In the next section, we present the \emph{modified LM formalism} in more detail.

\section{Field redefinition invariant Lagrange multiplier formalism}\label{section:modLM}

In this section, we present the field redefinition invariant LM formalism \cite{Brandt:2022kjo}. It is a modification of the standard LM formalism studied in chapter~\ref{section:modLMchap}. In this section, we restrict our presentation to non-singular theories. 

We proposed in \cite{Brandt:2022kjo} to introduce 
\begin{equation}\label{eq:mlm.determinant}
    \mathit{\Delta} [\phi ]= \det \left ( \frac{\delta^{2}S [ \phi ] }{\delta \phi \delta \phi}\right )^{1/2} 
\end{equation}
in the measure of Eq.~\eqref{eq:lm.genfuncSLM}. Now the measure of integration of the path integral in the LM formalism transforms as 
\begin{equation}\label{eq:mlm.transmeasure}
    \mathop{\mathcal{D} \phi} \mathop{\mathcal{D} \lambda} \mathit{\Delta} [ \phi ] \to 
    \mathop{\mathcal{D} \phi '} \mathop{\mathcal{D} \lambda '} \mathit{\Delta} '[ \phi '] \det {}\mathbb{J}_{\phi}^{2} 
    = 
    \mathop{\mathcal{D} \phi '} \mathop{\mathcal{D} \lambda '} \mathit{\Delta} [ \phi '] \det {}\mathbb{J}_{\phi},
\end{equation}
which is the expected behavior of the path integral under field redefinitions (cf. Eq.~\eqref{eq:transgenfunc}). Thus, the field redefinition invariance of the path integral in the framework of the LM formalism is restored.

The determinant $ \mathit{\Delta} $ is the pfaffian of the Hessian of the action. We assume that the Hessian is always positive, since, in principle, an absolute value should appear in Eq.~\eqref{eq:mlm.determinant}. Interestingly, this determinant only satisfies the transformation law 
\begin{equation}\label{eq:mlm.translaw}
    \mathit{\Delta} [ \phi ] \to \mathit{\Delta} '[ \phi ']= \det {}\mathbb{J}_{\phi} \mathit{\Delta} [ \phi '] 
\end{equation}
in the measure of integration of the path integral of the LM formalism. The determinant in Eq.~\eqref{eq:mlm.determinant} transform as  \cite{Brandt:2022kjo}
\begin{equation}\label{eq:mlm:translaw}
    \mathit{\Delta} [ \phi ] \to \mathit{\Delta} '[ \phi ']= \det {}\mathbb{J}_{\phi} \Delta [ \phi '] \det K [\phi ]
\end{equation}
under the field redefinition~\eqref{eq:fieldRed}, where 
\begin{equation}\label{eq:mlm:detK}
    K [ \phi ] = 
    1 - \left(\frac{\delta^{2} S'[\phi']}{\delta \phi ' \delta \phi '}\right)^{-1}  \frac{\delta {}\mathbb{J}_{\phi}}{\delta \phi ' } {}\mathbb{J}_{\phi}^{-1}\frac{\delta S'[\phi']}{\delta \phi'}.
\end{equation}
Note that, the second term on the right-hand side of the above expression is proportional to the equations of motion. In the LM formalism, they are imposed to vanish. Hence, $ K [ \phi ] = 1$, and can be discarded.

Thus, the field redefinition invariant path integral of the LM formalism is 
\begin{equation}\label{eq:GenFuncLM_Modified}
    \mathcal{Z}_{\text{LM}}[0] = \int \mathop{\mathcal{D} \phi} \mathop{\mathcal{D} \lambda} \det \left ( \frac{\delta^{2} S [\phi]}{\delta \phi \delta \phi}\right )^{+1/2} \exp i  S_{\text{LM}} [\phi ],
\end{equation}
where 
\begin{equation}\label{eq:def:S_LM}
    S_{ \text{LM} } [\phi ]= 
    \int \mathop{dx}\left( \mathcal{L} ( \phi ) + \lambda  \frac{\delta S [\phi]   }{\delta \phi} \right)
\end{equation}
is the action of the standard LM theory.

The determinant in Eq.~\eqref{eq:GenFuncLM_Modified} can be cast in local form through the introduction of ghost fields. First, we rewrite it as
\begin{equation}\label{eq:DeltaInTermsOfFieldsBefore2}
    \begin{split}
        \det \left ( \frac{\delta^{2} S [\phi] }{\delta \phi \delta \phi}\right )^{+1/2} =
                     \det \left(\frac{\delta^{2} S_{} [ \phi ] }{\delta \phi \delta \phi}\right)\det \left(\frac{\delta^{2} S [ \phi ] }{\delta \phi \delta \phi}\right)^{-1/2}.
    \end{split} 
\end{equation}
Next, we proceed by exponentiating the determinants in Eq.~\eqref{eq:DeltaInTermsOfFieldsBefore2}, which results in 
\begin{equation}\label{eq:DeltaInTermsOfFields}
    \mathit{\Delta} [ \phi ] 
                     =  \int \mathop{\mathcal{D}{\bar{\theta}}} \mathop{\mathcal{D }\theta } \mathop{\mathcal{D} \chi} \exp {i} \int \mathop{d^{}x} \left ( \bar{\theta} \frac{\delta^{2 }S [\phi ] }{\delta \phi \delta \phi} \theta + \frac{1}{2} \chi \frac{\delta^{2} S [\phi] }{\delta \phi \delta \phi} \chi\right ),
\end{equation}
where $ \bar{\theta} , \theta $ and $\chi$ are scalar fermionic and bosonic fields, respectively.

It is worth mentioning the similarity with the Lee-Yang ghosts \cite{Bastianelli:1991be, Bastianelli:1998jb}, which are necessary for preserving general covariance of the path integral in the worldline formalism. In this formalism, the determinant $ \sqrt{\det g^{\mu \nu}}  $ introduced in the measure of integration of the path integral leads to the Lee-Yang ghosts. It is analogous to the determinant considered in Eq.~\eqref{eq:mlm.determinant}.

From Eq.~\eqref{eq:DeltaInTermsOfFields}, we obtain the ghost action 
\begin{equation}\label{eq:def:Sgh}
    S_{\text{gh}}[ \phi ] = 
 \int \mathop{dx}\left(  \bar{\theta} \frac{\delta^{2 }S [\phi ] }{\delta \phi \delta \phi} \theta + \frac{1}{2} \chi \frac{\delta^{2} S [\phi] }{\delta \phi \delta \phi} \chi\right).
\end{equation}
Then, we can proceed and define
\begin{equation}\label{eq:def:Seff}
    \begin{split}
        S_{\text{eff}} [\phi ] &=   S_{\text{LM}}[ \phi ]+ S_{\text{gh}}[ \phi ] \\   
                                                                                              &= \int \mathop{dx}\left( \mathcal{L} ( \phi ) + \lambda  \frac{\delta S
    [\phi ]   }{\delta \phi}  + \bar{\theta} \frac{\delta^{2 }S [\phi ] }{\delta \phi \delta \phi} \theta + \frac{1}{2} \chi \frac{\delta^{2} S [\phi] }{\delta \phi \delta \phi} \chi\right),
\end{split}
\end{equation}
which can be interpreted as an effective action of the modified LM formalism.
Replacing Eqs.~\eqref{eq:DeltaInTermsOfFieldsBefore2}---\eqref{eq:def:Seff} into Eq.~\eqref{eq:GenFuncLM_Modified} yields 
\begin{equation}\label{eq:GenFuncLM_ModifiedLocal}
    \mathcal{Z}_{\text{LM}}[0] = \int \mathop{\mathcal{D} \phi} \mathop{\mathcal{D} \lambda} \mathop{\mathcal{D}\bar{\theta}}  \mathop{\mathcal{D}\theta } \mathop{\mathcal{D} \chi  } 
    \exp {i} S_{\text{eff} } [\phi ],
\end{equation}
which is the proper path integral for the LM theory \cite{Brandt:2022kjo}. 

We have that 
\begin{equation}\label{eq:transGenFuncmLM_local}
    \mathcal{Z}_{\text{LM}}[0] 
    \to
    \mathcal{Z'}_{\text{LM}}[0] 
    = \int \mathop{\mathcal{D} \phi'} \mathop{\mathcal{D} \lambda'} \mathop{\mathcal{D}\bar{\theta}'}  \mathop{\mathcal{D}\theta'} \mathop{\mathcal{D} \chi'} \det {}\mathbb{J}_{\phi}  
    \exp {i} S_{\text{eff} }^{\prime}[\phi'],
\end{equation}
which agrees with the expected behavior shown in Eqs.~\eqref{eq:transgenfunc} and \eqref{eq:mlm.transmeasure}. 
In Eq.~\eqref{eq:transGenFuncmLM_local}, we see that all the fields associated with the physical field $ \phi $ change accordingly. Thus, under the field redefinition \eqref{eq:fieldRed}, these fields transforms as
\begin{subequations}\label{eq:transformationsofmLM}
    \allowdisplaybreaks
    \begin{align}
        \lambda' ={}& \lambda \frac{\delta F[\phi]}{\delta \phi}, \\
        \chi' ={}& \chi \frac{\delta F[\phi]}{\delta \phi} , \\
        \theta ' ={}& \theta \frac{\delta F[\phi]}{\delta \phi} , \\  \bar{\theta} ' ={}& \bar{\theta} \frac{\delta F[\phi]}{\delta \phi}; 
    \end{align}
\end{subequations}
while other independent fields remain unaltered. This is the analog of Eq.~\eqref{eq:FieldRedlamb} in the modified LM formalism. The transformation of the bosonic fields leads to a Jacobian determinant factor of $ \det {}\mathbb{J}_{\phi} $, and fermionic fields leads to the inverse $ \det {}\mathbb{J}_{\phi}^{-1}$. 

In \cite{Brandt:2022kjo}, it is noted that the effective action in Eq.~\eqref{eq:def:Seff} has a ghost-number symmetry. This symmetry also appears in gauge theories. We can use this similarity to count, naively, the degrees of freedom of the modified LM theory. We assume that the ghost fields $ \theta$ and $ \bar{\theta} $, as the FP ghosts, cancel degrees of freedom.\footnote{Indeed, the FP ghosts cancel degrees of freedom, the unphysical degrees of freedom associated with longitudinal modes.} Let us denote the degrees of freedom of a field $ I$ by $ N_I$ and the total degrees of freedom of the fermionic ghost fields by $ N_{\text{gh}} =N_{\theta} + N_{\bar{\theta}}$. The number of degrees of freedom associated with the modified LM sector is
\begin{equation}\label{eq:DOF_GHOSTS}
     N_{\lambda} - N_{ \text{gh}} + N_{\chi} = 0,
\end{equation}
since $ N_{\lambda} = N_{\chi} = N_{ \text{gh}}/2 =  N_{\phi} $.
Consequently, the modified LM theory has 
\begin{equation}\label{eq:DOF_M_LM}
        N_{\phi} +N_{\lambda} - N_{ \text{gh}} + N_{\chi}  =  N_{\phi} 
\end{equation}
degrees of freedom.\footnote{We should remark that this result is also consistent with the free energy density at finite temperature \cite{Bellac:2011kqa}. A one-loop analysis shows transparently the relation between the free energy with the number of degrees of freedom.}

Thus, the degrees of freedom in the modified LM formalism are not doubled. It is kept unaltered. Now, let us show that the doubling of the one-loop contributions is also absent. 

First, we conveniently rewrite Eq.~\eqref{eq:GenFuncLM_ModifiedLocal} 
as 
\begin{equation}\label{eq:def:PathIntLM}
        \mathcal{Z}_{\text{LM}}[0] =  \int \mathop{\mathcal{D} \phi} \det \left[ \mathcal{L}^{\prime\prime}( \phi )\right]\det \left[ \mathcal{L}^{\prime\prime}( \phi )\right]^{-1/2} 
        \mathop{\delta} \left [ \mathcal{L}^{\prime} ( \phi )\right ]
    \exp {i} \int \mathop{d x} \mathcal{L} ( \phi )
\end{equation}
in which the LM field $ \lambda $ and the ghost fields are integrated.
Using the functional analog of the Eq.~\eqref{eq:deltafuncp1}:
\begin{equation}\label{eq:mlm.deltaanalog}
    \delta \left ( \mathcal{L} ' (\phi )\right ) = \sum_{ \tilde{\phi}}^{} \det [\mathcal{L}^{\prime\prime} ( \tilde{\phi} )]^{-1} \mathop{\delta}  ( \phi - \tilde{\phi} ) 
\end{equation}
in Eq.~\eqref{eq:def:PathIntLM}, we obtain
\begin{equation}\label{A.17c}\mathcal{Z}_{\text{LM}} [0]= \sum_{\tilde{\phi}(x)} \det[ \mathcal{L}^{\prime\prime}(\tilde{\phi})]^{-1/2} \exp {i}  
    \int \mathop{d x}   \mathcal{L}(\tilde{\phi}).
\end{equation}
This is the exact generating functional of the modified LM theory, as shown in Eq.~\eqref{eq:lm.genfuncExact}. The exponential factor is similar, but the determinant is no longer squared compared with the usual one-loop determinant.

The field redefinition invariant formulation of the LM formalism leads to the same tree-level and one-loop contributions obtained with the generating functional in Eq.~\eqref{eq:lm.genfuncS}. However, the higher-loop order contributions vanish. Remarkably, the modification proposed in \cite{Brandt:2022kjo} establishes a formalism in which the radiative effects are restricted to one-loop order, while the doubling of degrees of freedom and one-loop contributions are absent.

\subsection{Diagrammatic analysis}

Here, we present a diagrammatic analysis of the modified LM formalism. For an analysis in more detail, we refer to \cite{Brandt:2022kjo}.

Substituting Eq.~\eqref{eq:lm.expandedLag} in Eq.~\eqref{eq:def:Sgh} yields 
\begin{equation}\label{eq:Lgh}
    \begin{split}
        \mathcal{L}_{\text{gh}} ( \phi ) ={}&    \bar{\theta}_{i} 
    \left ( a^{(2)}_{ij} + a^{(3)}_{ijk} \phi_{k} + \frac{1}{2!} a^{(4)}_{ijkl} \phi_{k} \phi_{l} + \cdots \right ) 
    \theta_{j} \\ 
                                         & + 
                                         \frac{1}{2} \chi_{i} \left ( a^{(2)}_{ij} + a^{(3)}_{ijk} \phi_{k} + \frac{1}{2!} a^{(4)}_{ijkl} \phi_{k} \phi_{l} + \cdots \right ) \chi_{j} . 
\end{split}
\end{equation}
This is the Lagrangian of the ghost fields. We see that the propagators of the ghosts are equal to $ a_{ij} $, which is equal to the mixed propagators $ \left \langle \phi_{i} \lambda_{j}\right\rangle $. The ghost sector is independent of the LM field $ \lambda $. Now we can examine how the ghost fields will affect perturbatively the LM formalism. 

The perturbation theory at the lowest order, tree-level, is kept unaltered since ghosts can only appear in closed loops. This is a characteristic shared by most known ghost fields, such as the FP, Senjanovi\'{c}, and Lee-Yang ghosts. The ghost fields do not interact with the LM field $ \lambda $. Thus, their contributions only affect the Green's function with $ \phi $ in the external legs. Moreover, since the propagator $ \phi $ vanishes, we cannot have any ghost diagram of order higher than one-loop. Therefore, the ghost fields contribute only at one-loop order. 

\begin{figure*}[ht]
    \centering
    \includegraphics[scale=0.83]{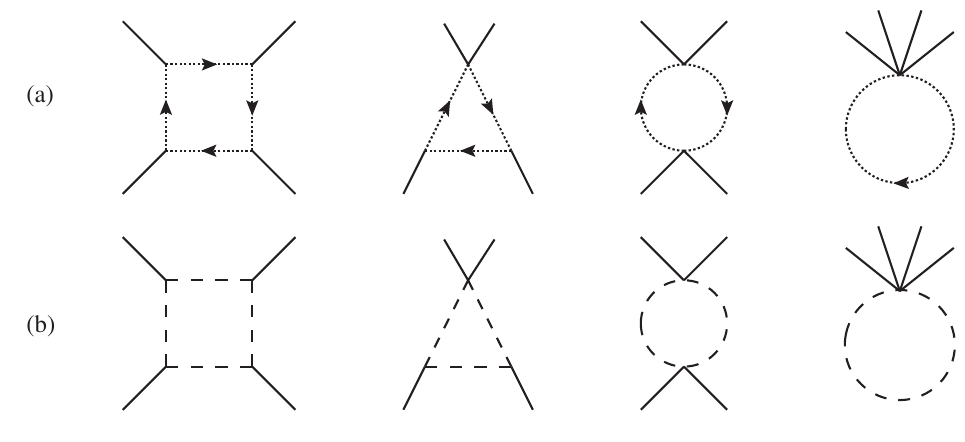}
    \caption{Ghost contributions to 4-point amplitude $ ( \phi_{i} \phi_{j} \phi_{k} \phi_{l} )$ in the modified LM theory. Pointed and dashed lines represent respectively the ghost fields $ \bar{\theta}$, $ \theta $,  and $\chi $.}\label{fig:modLM}
\end{figure*}

\begin{figure*}[ht!]
    \centering
    \includegraphics[scale=0.75]{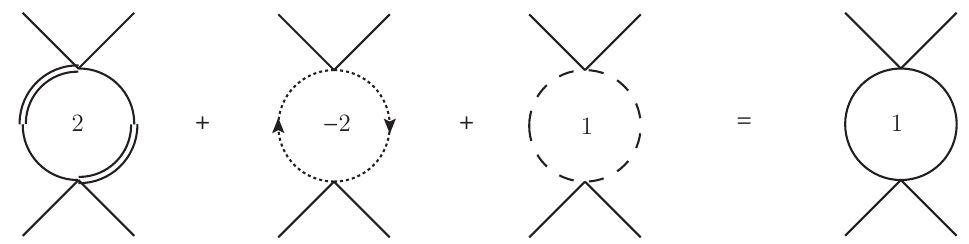}
    \caption{One-loop contribution in the modified LM theory to 4-point amplitude $( \phi_{i} \phi_{j} \phi_{k} \phi_{l} )$. Ghost contributions are responsible for canceling the extra contributions due to the LM field.  The relative factors are indicated.}\label{fig:NEW}
\end{figure*}

In fig.~\ref{fig:modLM}, we present some of the ghost contributions coming from Eq.~\eqref{eq:Lgh}.
The fermionic loops in Fig.~\ref{fig:modLM} (a) lead to a negative sign compared with the bosonic loops in Fig.~\ref{fig:modLM} (b). In addition, the combinatorial factor is also twice. The resulting contribution is exactly the required to remove the additional one-loop contributions coming from the LM field $ \lambda$ (see Fig.~\ref{fig:NEW}). 

The results in this section are more transparent in the diagonalized form of the modified LM action (which is realized by the shift $ \phi \to \varphi - \lambda $). We have shown in \cite{Brandt:2020gms} that diagrams of order higher than one-loop can be constructed in the diagonal LM theory, but the resulting contribution vanishes. 

\begin{figure}[ht]
    \centering
    \includegraphics[width=0.90\textwidth]{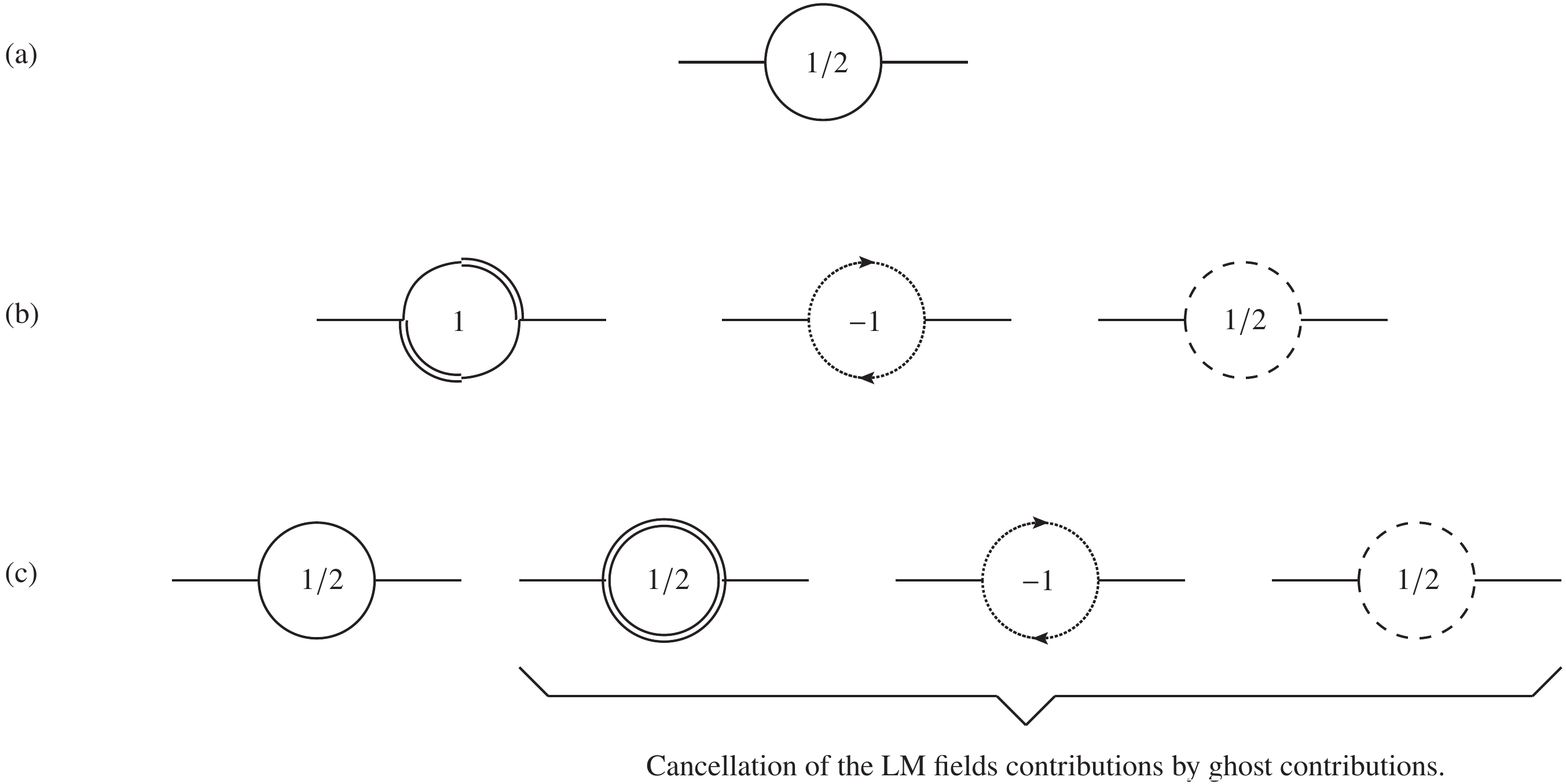}
    \caption{One-loop contributions to the two-point amplitudes $( \phi_{i} \phi_{j} )$. Diagram (a) from the original theory, diagrams (b) of the modified LM theory, and diagrams (c) of the diagonalized formulation. The relative factor between the diagrams is indicated. In the diagonalized modified LM theory the cancellation between the LM field and ghost contributions is evident.}\label{fig:diag2}
\end{figure}

In Fig.~\ref{fig:diag2} we compare the one-loop diagrams from the original theory described by the action $S[ \phi ]$, modified LM theory, and the diagonalized LM theory. Inside each loop, we have the respective overall factor that must be compared with the one-loop diagram from the original theory. In each row, the sum of the overall factors yields $1/2$, including the diagonalized modified LM theory agreeing with the previous diagrammatic analysis presented in section~\ref{section:modLM}. In Fig.~\ref{fig:diag2} (c) (diagonalized modified LM theory) we see that the ghost contributions are responsible for canceling the extra contribution coming from the LM field $ \lambda $.

\begin{figure}[ht]
    \centering
\includegraphics[width=0.90\textwidth]{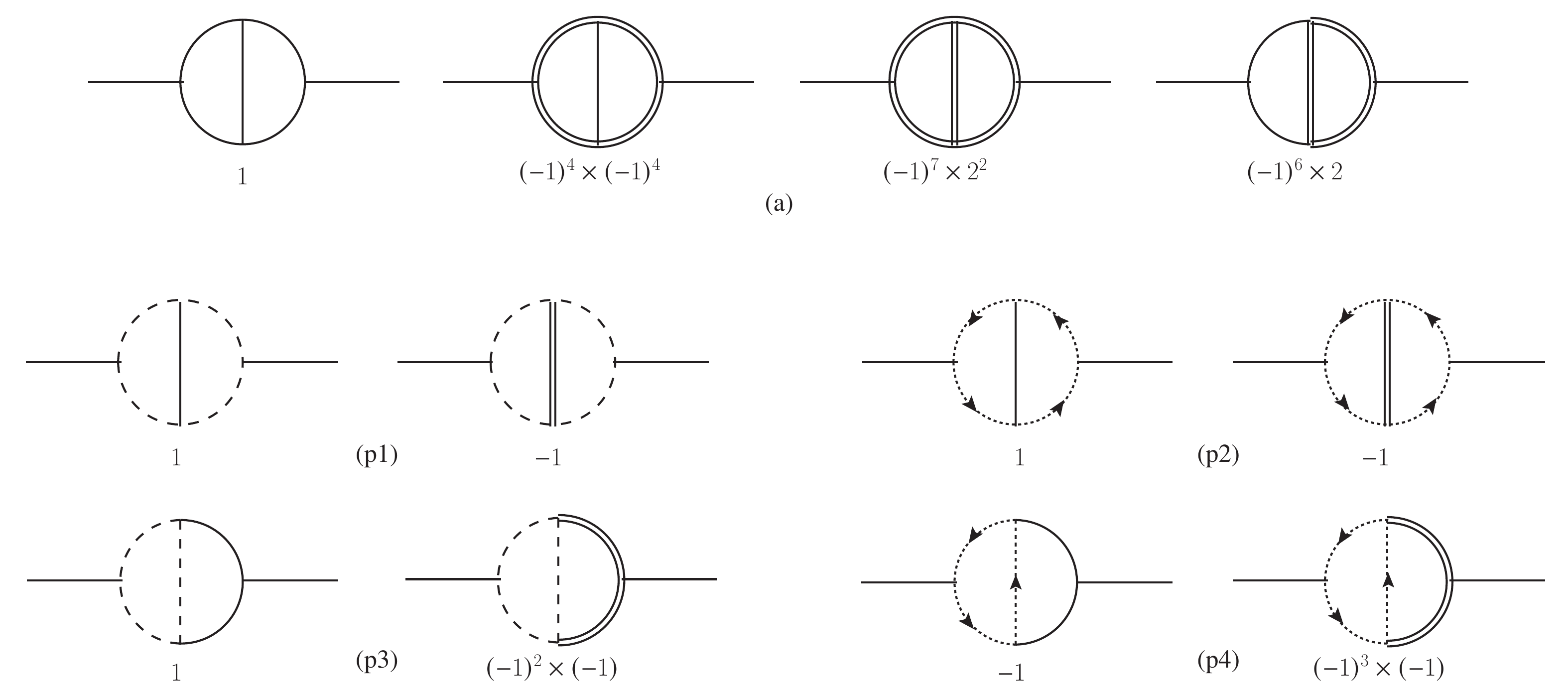}
\caption{The 1PI two-loop diagrams (a, p) with ghost-loops in the modified diagonal LM theory. The sum of the diagrams (a) from the standard diagonal LM theory vanishes, while ghost diagrams from the modified formalism cancel within each pair (p). The pairs (p1, p2) cancel by the oppositive sign of internal propagators of $ \phi $ and $ \lambda $, while the pairs (p3, p4) canceling due to the extra minus sign of the $ (\lambda \lambda \phi) $ vertex \cite{Brandt:2020gms}. We indicate the relative factor between the diagram.}\label{fig:diagGH}
\end{figure}

\begin{figure}[ht]
    \centering
    \includegraphics[width=0.9\textwidth]{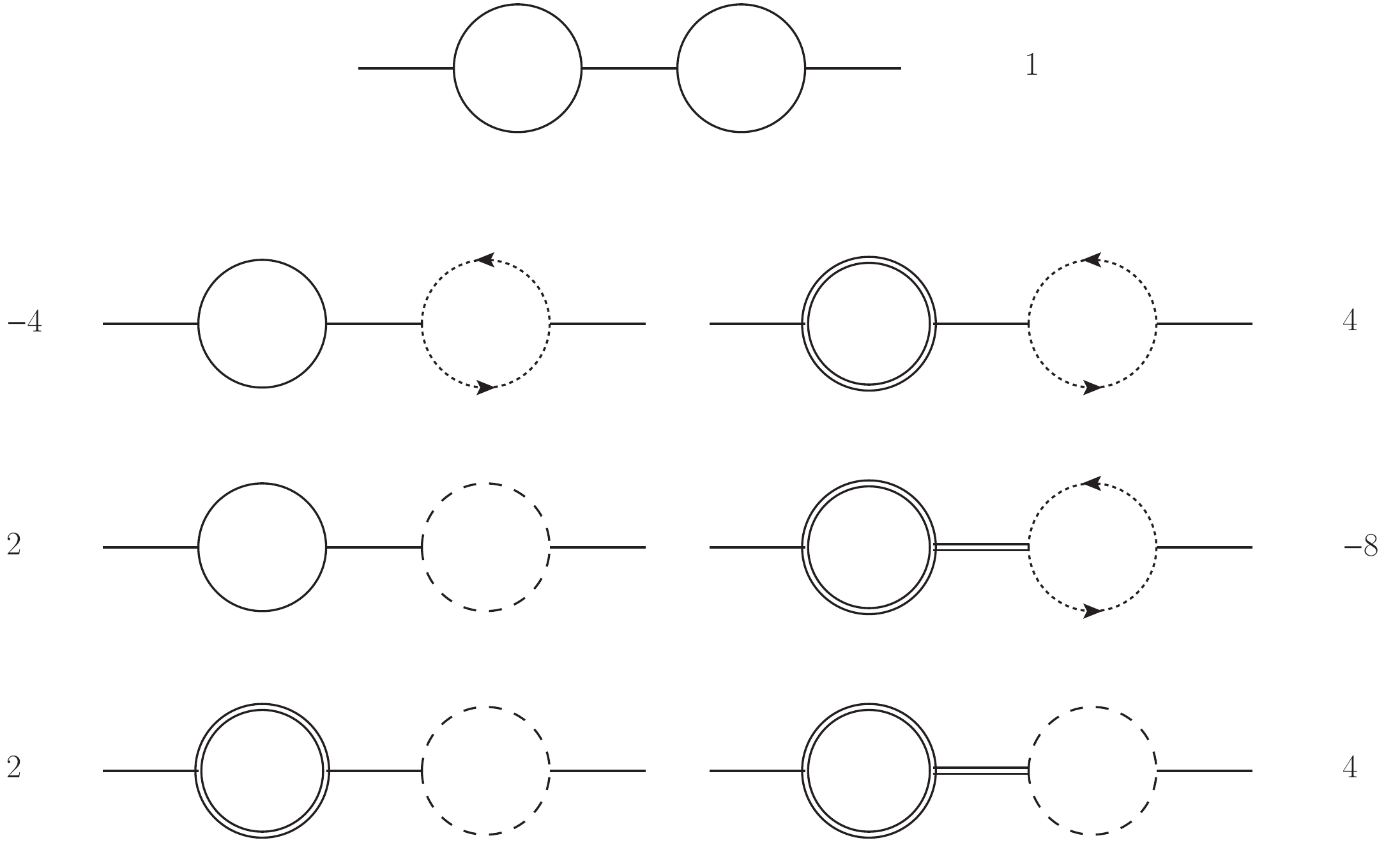}
    \caption{Reducible two-loop order diagrams with ghost-loops in the modified diagonal LM theory. The sum of the diagrams vanishes in each column. Relative factors are indicated.}
    \label{fig:diagGHReducible}
\end{figure}

In the modified LM theory the quantum corrections are restricted to the one-loop order. While in the diagonalized formulation loop diagrams of higher order appear which, for consistency, must add to zero. 
In \cite{Brandt:2020gms} we checked the consistency of the diagonalized standard LM theory using a scalar $ \phi^{3} $ model showing that all two-loop order contributions vanish. The same analysis is applied to the ghost contributions of the diagonalized modified LM theory. Together with the analysis found in \cite{Brandt:2020gms} the  consistency check for the diagonalized formulation of the modified LM theory is completed.
In Figs.~\ref{fig:diagGH} and \ref{fig:diagGHReducible}, we analyze two specific topologies of two-loop order diagrams from the scalar $ \phi^{3} $ model. It is shown that the total contribution vanishes, in particular the contribution coming from ghost diagrams. In the diagonalized modified LM theory ghosts also interact with the LM field $ \lambda $ which is essential for the consistency of the diagonalized modified LM theory.

\section{Additional symmetries}\label{section:symmetries}
In this short section, we present the symmetries of the modified LM theory. We also comment on the Slavnov-Taylor-like identities obtained from the BRST-like symmetry. 

The effective action of the modified LM theory in Eq.~\eqref{eq:def:Seff} is invariant under 
\begin{equation}\label{eq:transform:ghost}
    \delta \bar{\theta} = - \sigma \bar{\theta} , \quad \delta \theta = \sigma \theta ,  \quad  \delta \phi = \delta \lambda = \delta \chi =0;
\end{equation}
where $ \sigma $ is a commuting parameter. 
This symmetry is related to the conservation of Lee-Yang-like \emph{ghost charge} $Q_{\text{gh}}$, which is directly related to the \emph{ghost-number} \cite{Brandt:2022kjo}. It can be shown that $ Q_{\text{gh}} \theta = \sigma $, $ Q_{\text{gh}} \bar{\theta}=- \sigma $ while $ Q_{\text{gh}} \phi = Q_{\text{gh}} \lambda = Q_{\text{gh}} \chi =0$.

In addition, the modified LM theory is also invariant under the \emph{BRST-like} symmetry
\begin{equation}\label{eq:transform:ghostS}
    \delta \theta =  \chi \eta, \quad 
    \delta \chi = \eta \bar{\theta},
    \quad \delta \phi = \delta \lambda = \delta \bar{\theta} =0,
\end{equation}
and the anti-BRST symmetry
\begin{equation}\label{eq:transform:ghostS2}
    \bar{\delta} \bar{\theta} =   \chi \bar{\eta}, \quad \bar{\delta} \chi = \theta \bar{\eta},
     \quad \bar{\delta} \phi = -\bar{\delta} \lambda = \bar{\delta} \theta =0,
\end{equation}
where ($ \bar{\eta} $) $ \eta $ is an anticommuting parameter of the (anti)BRST symmetry.

\subsection{Slavnov-Taylor-Like identities}

The master equations derived from the (anti)BRST-like symmetry are simply given by 
\begin{subequations}\label{eq:mlm.master}
    \begin{align}
        \label{eq:mlm.master1}
     \int \mathop{d x} \left ( \chi \frac{\delta \Gamma_{\text{mLM}} }{\delta \theta}  + \bar{\theta} \frac{\delta \Gamma_{\text{mLM}} }{\delta \chi} \right ) ={}&0,
    \\
    \label{eq:mlm.master2}
     \int \mathop{d x} \left ( \chi \frac{\delta \Gamma_{\text{mLM}} }{\delta \bar{ \theta}}  - {\theta} \frac{\delta \Gamma_{\text{mLM}} }{\delta \chi} \right ) ={}&0,
 \end{align}
\end{subequations}
where $ \Gamma_{\text{mLM}} $ is the effective action of the modified LM formalism.

We can now obtain the Slavnov-Taylor-like identities involving the 1PI Green functions. For instance, taking the functional derivative $ \delta^{2} / \delta \chi \delta \bar{\theta} $ of  Eq.~\eqref{eq:mlm.master1} gives 
\begin{equation}\label{eq:ST1aLM}
    \frac{\delta^{2} \Gamma_{\text{mLM}} }{\delta \bar{\theta} \delta \theta} + \frac{\delta^{2} \Gamma_{\text{mLM}} }{\delta \chi \delta \chi} =0.
\end{equation}
Eq.~\eqref{eq:ST1aLM} is clearly satisfied in all orders. 
We also find other identities, such as:
\begin{align}\label{eq:STnaLM}
    \frac{\delta^{2} \Gamma_{\text{mLM}}  }{\delta \phi \delta \bar{\theta} } 
    ={}& 
    \frac{\delta^{2} \Gamma_{\text{mLM}}  }{\delta \phi  \delta {\theta} } = 0, \\
    \frac{\delta^{2} \Gamma_{\text{mLM}}  }{\delta \chi \delta \chi \delta \chi  } ={}&
    \frac{\delta^{2} \Gamma_{\text{mLM}}  }{\delta \chi \cdots \delta \chi \delta \chi  }=0.
\end{align}

The Slavnov-Taylor-Like identities implies that 
\begin{equation}\label{eq:mLMaction}
    \Gamma_{\text{mLM}} = \Gamma_{\text{sLM}}  + \left ( \frac{1}{2} \chi^{2} + \bar{\theta} \theta \right ) \frac{\delta^{2} S_{0} }{\delta \phi \delta \phi},
\end{equation}
where 
$ \Gamma_{\text{sLM} } $ is the effective action of the standard LM formalism.

\chapter{Modified Lagrange multiplier formalism with gauge symmetries}\label{chapter:gtmLM}

In this chapter, we consider the case of a gauge theory in the framework of the modified LM formalism. We show that the Lee-Yang-like ghosts are also gauge fields, requiring careful consideration in the quantization of these theories. First, we review our results in \cite{McKeon:2024psy}, where we extended the FP procedure to accommodate the additional invariances introduced by the LM field $ \lambda $ and the associated ghosts $ \chi $, $ \bar{\theta} $ and $ \theta $. We also present the corresponding BRST transformations.

\section{Gauge invariances}\label{section:giMLM}
The classical action 
\begin{equation}\label{eq:clS0}
    S_{0} = \int \mathop{d x} \mathcal{L}_{0} ( \phi ),
\end{equation}
which is invariant under the gauge transformation 
\begin{equation}\label{eq:gtransS0}
    \delta \phi_{i} = \phi '_{i} - \phi_{i}  = H_{ij} ( \phi ) \xi_{j},
\end{equation}
in the framework of the modified LM theory is supplemented by an LM field $ \lambda_{i} $ and the Lee-Yang ghosts $ \bar{\theta}_{i} $, $ \theta_{i} $, $ \chi_{i} $: 
\begin{equation}\label{eq:mLMS}
    S_{\text{mLM}} = S_{0} + \frac{\delta S }{\delta \phi_{i}} \lambda_{i} + \frac{\delta^{2} S }{\delta \phi_{i} \delta \phi_{j}} \left( \bar{\theta}_{i} \theta_{j} + \frac{1}{2} \chi_{i} \chi_{j} \right).
\end{equation}

The invariance in Eq.~\eqref{eq:gtransS0} implies that 
\begin{equation}\label{eq:invaS0}
    \delta S_{0} = \int \mathop{d x} \mathcal{L}_{0, i} H_{ij} \xi_{j} =0,
\end{equation}
where commas denote derivatives with respect to $ \phi $: $ H_{I,k} \equiv \frac{\delta H_{I} }{\delta \phi_{k}} $, 
which indicates that the action \eqref{eq:mLMS} is invariant under the gauge transformation: 
\begin{equation}\label{eq:oneinvaS0}
    \delta \lambda_{i}  =  H_{ij} ( \phi_{i} ) \zeta_{j} .
\end{equation}

Moreover, from Eq.~\eqref{eq:gtransS0}, we obtain that 
\begin{equation}\label{eq:13a}
    \frac{\partial \mathcal{L}_{0}}{\partial \phi_{i}} = \frac{\partial \phi_{j} '}{\partial \phi_{i}} \frac{\partial \mathcal{L}_{0}}{\partial \phi_{j} '} = \left ( \delta_{ji} + \frac{\partial H_{jk}}{\partial \phi_{i}} \xi_{k}\right ) \frac{\partial \mathcal{L}_{0}}{\partial \phi_{j} '} ,
\end{equation}
and 
\begin{equation}\label{eq:14}
    \frac{\partial^{2} \mathcal{L}_{0}}{\partial \phi_{i} \partial \phi_{j}} = \frac{\partial \phi_{m} '}{\partial \phi_{i}} \frac{\partial}{\partial \phi_{m} '} \left [ \frac{\partial \mathcal{L}_{0}}{\partial \phi_{j} '} + \frac{\partial H_{nk}}{\partial \phi_{j}} \xi_{k} \frac{\partial \mathcal{L}_{0}}{\partial \phi_{n} '}\right ]
\end{equation}
that can be expanded to 
\begin{equation}\label{eq:15}
    \frac{\partial^{2} \mathcal{L}_{0}}{\partial \phi_{i} \partial \phi_{j}} =
    \frac{\partial^{2} \mathcal{L}_{0}}{\partial \phi_{i} ' \partial \phi_{j} '} +  \frac{\partial^{2} H_{lk}}{\partial \phi_{i} \partial \phi_{j}}  \frac{\partial \mathcal{L}_{0}}{\partial \phi_{l} '} \xi_{k}+ 
    \left ( \frac{\partial H_{lk}}{\partial \phi_{i}} \frac{\partial^{2} \mathcal{L}_{0}}{\partial \phi_{j} ' \partial \phi_{l} '}  + \frac{\partial H_{lk}}{\partial \phi_{j}} \frac{\partial^{2} \mathcal{L}_{0}}{\partial \phi_{i} ' \partial \phi_{l} '}\right ) \xi_{k} + O ( \xi^{2} ).
\end{equation}

By comparing the structure in Eq.~\eqref{eq:15} to the structure of the action \eqref{eq:mLMS}, we see that Eq.~\eqref{eq:gtransS0} is accompanied with 
 \begin{subequations}\label{eq:16}
 \begin{align}
 \delta \lambda_{i}  ={}&  \left [ \frac{\partial H_{ik}}{\partial \phi_{j}} \lambda_{j} + \frac{\partial^{2} H_{ik}}{\partial \phi_{m} \partial \phi_{n}}\left ( \bar{\theta}_{m} \theta_{n} + \frac{1}{2} \chi_{m} \chi_{n}\right )\right ] \xi_{k} , \\
 \delta \bar{\theta}_{i} ={}&\frac{\partial H_{ik}}{\partial \phi_{j}} \bar{\theta}_{j} \xi_{k}, \\
 \delta \theta_{i}  ={}& \frac{\partial H_{ik}}{\partial \phi_{j}} \theta_{j} \xi_{k},\\
 \delta \chi_{i} ={}&\frac{\partial H_{ik}}{\partial \phi_{j}} \chi_{j} \xi_{k}.
     \end{align}
 \end{subequations} 

From Eq.~\eqref{eq:gtransS0}, we get 
 \begin{equation}\label{eq:18}
     ( \mathcal{L}_{0 , i} H_{ik} )_{,j} = \mathcal{L}_{0 , ij} H_{ik} + \mathcal{L}_{0 , i} H_{ik,j} =0,
 \end{equation}
which implies that there are also the gauge transformations
 \begin{subequations}\label{eq:17}
    \begin{align}
 \label{eq:17a}
 \delta \lambda_{i}  ={}& H_{ik, j} \left ( \chi_{j} \sigma_{k} + \bar{\theta}_{j} \tau_{k} - \theta_{j} \pi_{k} \right ), \\ 
\label{eq:17b}
        \delta \bar{\theta}_{i} ={}& H_{ij} \pi_{j}, \\
 \label{eq:17c}
       \delta  \theta_{i} ={}& H_{ij} \tau_{j} ,\\
 \label{eq:17d}
        \delta \chi_{i} ={}& H_{ij} \sigma_{j} ;
     \end{align}
 \end{subequations}
where $ \pi_{i} $ and $ \tau_{i} $ are Grassmann gauge parameters. 

Thus, we find out that there are five independent gauge transformations. Each invariance is associated with a field: $ \phi_{i} $, $ \lambda_{i} $, $ \chi_{i} $, $ \bar{\theta}_{i} $ and $ \theta_{i} $. Therefore, we see that both the LM field and the Lee-Yang ghost fields are now gauge fields. 

To quantize the action \eqref{eq:mLMS}, we extend the FP to accommodate these five gauge transformations. In order to use the FP, we impose the closure of the gauge algebra, that is, 
\begin{equation}\label{eq:closure}
    [ \delta_{\xi_1} , \delta_{\xi_2} ] \phi_{i} = \delta_{\xi_{3}} \phi_{i} \equiv  f_{mn|j} H_{ij} \xi_{1}^{m} \xi_{2}^{n}, 
\end{equation}
and that the condition 
\begin{equation}\label{eq:conditions74}
    H_{ij,pq} = f_{mn|i, j} = 0
\end{equation}
is satisfied by $ H_{ij} ( \phi )$ and the structure constants $ f_{mn|i} $,
which implies that the original theory is a gauge theory of YM type.\footnote{In \cite{McKeon:2024psy}, it is shown in Appendix C, that the closure of the algebra and the condition Eq.~\eqref{eq:conditions74} is retained in the framework of the modified LM theory.}
The gauge gravity theories based on the Hilbert-Einstein action or Hilbert-Palatini action that are treated here are also theories of this type. Another example of this type of theory is the gauge theory of the Poincaré group based on Einstein-Cartan action \cite{Brandt:2024rsy}. This theory describes more general spacetimes in which torsion is present. 

\section{Faddeev-Popov quantization}\label{section:quantizationgaugemLM}

The path integral 
\begin{equation}\label{eq:21paper}
    \mathcal{Z}[0] = \int \mathop{\mathcal{D} \phi_{i}} \mathop{\mathcal{D} \lambda_{i}} \mathop{\mathcal{D} \chi_{i}} \mathop{\mathcal{D} \bar{\theta}_{i} \mathop{\mathcal{D} \theta_{i}}} \exp{i \int \mathop{d^{}x} \left [ \mathcal{L}_{0} ( \phi ) + \mathcal{L}_{0 , i} \lambda_{i} + \mathcal{L}_{0 , ij} ( \phi ) \left( \bar{\theta}_{ i} \theta_{j} + \frac{1}{2} \chi_{i} \chi_{j}\right)\right ]} 
\end{equation}
is not well-defined, since the action in Eq.~\eqref{eq:21paper} is singular. We need to fix the gauge and introduce the FP ghost contribution.  Thus, we insert the unity: 
 \begin{equation}\label{eq:22paper}
     \begin{split}
         1 ={}&\int \mathop{\mathcal{D} \xi_{i}} \mathop{\mathcal{D} \zeta_{i}} \mathop{\mathcal{D} \sigma_{i}} \mathop{\mathcal{D} \pi_{i}} \mathop{\mathcal{D} \tau_{i}}   
              \delta \left \{ F_{ij} \left [ \begin{pmatrix}
                 \phi_{j} \\ 
                 \lambda_{j} \\
                 \chi_{j} \\
                 \bar{\theta}_{j} \\
                 \theta_{j} 
             \end{pmatrix}
             + 
             M_{jp} 
\begin{pmatrix}
    \zeta_{p} \\ 
    \xi_{p} \\
    \sigma_{p} \\
    \pi_{p} \\
    \tau_{p} 
\end{pmatrix}
\right ] - \begin{pmatrix}
    p_{i} \\
    q_{i} \\
    r_{i} \\
    s_{i} \\
    t_{i}
\end{pmatrix}
\right \} \mathop{\mathrm{Sdet} } {M}_{jp}, 
     \end{split}
 \end{equation}
where 
\begin{equation}\label{eq:FPmat}
    M_{jp}= 
\begin{pmatrix}
     0 & H_{jp} & 0 & 0 & 0 \\ 
     H_{jp} & H_{jp, k} \lambda_{k}   & H_{jp,k} \chi_{k}  & -H_{jp,k} \theta_{k} &H_{jp,k} {\bar{\theta}}_{k}  \\
     0 & H_{jp,k} \chi_{k} & H_{jp} & 0 & 0 \\
     0 & H_{jp,k} \bar{\theta}_{k} & 0 & H_{jp} & 0 \\
     0 & H_{jp,k} \theta_{k} & 0 & 0 & H_{jp} \\ 
     \end{pmatrix},
\end{equation}
$ \mathop{\mathrm {Sdet}} {M}_{jp}$ is the FP superdeterminant and $ F_{ij} $ is the gauge fixing condition.

Since the FP determinant is gauge invariant, we perform the gauge transformations~\eqref{eq:gtransS0}, \eqref{eq:15}, \eqref{eq:oneinvaS0}, \eqref{eq:17} with $ (- \xi_{i} , - \zeta_{i} , - \sigma_{i} , - \pi_{i} , - \tau_{i} )$ and then insert the constant 
 \begin{equation}\label{eq:24}
     \int \mathop{\mathcal{D} p_{i}} \mathop{\mathcal{D} q_{i}} \mathop{\mathcal{D} r_{i}} \mathop{\mathcal{D} s_{i}} \mathop{\mathcal{D} t_{i}} \exp i \int \mathop{d^{}x} \left ( - \frac{1}{2 \alpha} p_{i}^{2} - \frac{1}{\alpha} p_{i} q_{i} - \frac{1}{2 \alpha} r_{i}^{2} - \frac{1}{\alpha} s_{i} t_{i}\right )
 \end{equation}
 which extends the gauge condition $ F_{ij} $ in a manner analogous to the $R_{\xi} $ gauges.

The path integral is now independent of $ ( \xi_{i} , \zeta_{i}, \sigma_{i} , \pi_{i} , \tau_{i} ) $, the $ \delta $-functions in Eq.~\eqref{eq:22paper} becomes 
\begin{equation}\label{eq:22alt}
              \delta \left [ F_{ij}  \begin{pmatrix}
                 \phi_{j} \\ 
                 \lambda_{j} \\
                 \chi_{j} \\
                 \bar{\theta}_{j} \\
                 \theta_{j} 
             \end{pmatrix}
 - \begin{pmatrix}
    p_{i} \\
    q_{i} \\
    r_{i} \\
    s_{i} \\
    t_{i}
\end{pmatrix}
\right ]
\end{equation}
which allows us to integrate over $ ( p_{i} , q_{i} , r_{i} , s_{i} , t_{i} )$. This leads to 
\begin{equation}\label{eq:23paper}
\mathop{\mathrm{Sdet} } {M}_{jp}
\exp i \int \mathop{d^{}x} \left ( - \frac{1}{2 \alpha} (F_{ij} \phi_{j} )^{2} - \frac{1}{\alpha} (F_{ij} \phi_{j} ) ( F_{ik} \lambda_{k} ) - \frac{1}{2 \alpha} (F_{ij} \chi_{j} )^{2} - \frac{1}{\alpha} (F_{ij} \bar{\theta}_{j} )  (F_{ik} \theta_{k} )\right ),
 \end{equation}

 The path integral \eqref{eq:21paper} is now given by  
 \begin{equation}\label{eq:25}
         \mathcal{Z} [0] = \int \mathop{\mathcal{D} \phi_{i}} \mathop{\mathcal{D} \lambda_{i}} \mathop{\mathcal{D} \chi_{i}} \mathop{\mathcal{D} \bar{\theta}_{i}} \mathop{\mathcal{D} \theta_{i}} 
         \exp \left [ i \int \mathop{d x} \left ( \mathcal{L}_{\text{EFF}} + \mathcal{L}_{\text{GF}}\right )\right ] \mathop{\mathrm{Sdet}} {M}_{jp},
 \end{equation}
 where 
 \begin{equation}\label{eq:lageffmLM}
     \mathcal{L}_{\text{EFF}} =
      \mathcal{L}_{0} ( \phi_{i} ) + \mathcal{L}_{0 , i} \lambda_{i} + \mathcal{L}_{0 , ij} ( \phi_{i} ) \left( \bar{\theta}_{ i} \theta_{j} + \frac{1}{2} \chi_{i} \chi_{j}\right),
 \end{equation}
     $\mathcal{L}_{\text{GF}}$ is the gauge fixing Lagrangian
 \begin{equation}\label{eq:gflagmLM}
      - \frac{1}{2 \alpha} (F_{ij} \phi_{j} )^{2} - \frac{1}{\alpha} (F_{ij} \phi_{j} ) ( F_{ik} \lambda_{k} ) - \frac{1}{2 \alpha} (F_{ij} \chi_{j} )^{2} - \frac{1}{\alpha} (F_{ij} \bar{\theta}_{j} )  (F_{ik} \theta_{k} ),
 \end{equation}
 and $ \mathop{\mathrm{Sdet}} {M}_{jp}$ is the FP superdeterminant of Eq.~\eqref{eq:22paper}. 

The FP superdeterminant can be exponentiated using Grassmann ghost fields $ ( \bar{c}_{i} , c_{i} )$, $ ( \bar{d}_{i} , d_{i} )$, $ ( \bar{e}_{i} , e_{i} )$ and complex ghost fields $ ( \tilde{\gamma }_{i} , \gamma_{i} )$, $ ( \tilde{ \varepsilon  }_{i} , \varepsilon_{i} )$ \cite{berezin:1987}. Note that, we denote the complex conjugate of $A_{i}  $ by $ \tilde{A}_{i} $. We find that 
 \begin{equation}\label{eq:26}
         \mathop{\mathrm{ Sdet}} {M}_{jp} ={} \int 
     \mathop{\mathcal{D} \bar{c}_{i}} \mathop{\mathcal{D} c_{i}}  
     \mathop{\mathcal{D} \bar{d}_{i}} \mathop{\mathcal{D} d_{i}}  
     \mathop{\mathcal{D} \bar{e}_{i}} \mathop{\mathcal{D} e_{i}}  
     \mathop{\mathcal{D} \tilde{\gamma}_{i}} \mathop{\mathcal{D} \gamma_{i}}  
         \mathop{\mathcal{D} \tilde{\varepsilon}_{i}} \mathop{\mathcal{D} \varepsilon_{i}}   
     \exp i \int \mathop{d^{}x} 
     \mathcal{L}_{\text{GH}},
 \end{equation}
 where
 \begin{equation}\label{eq:26a}
     \mathcal{L}_{\text{GH}}=
     \begin{pmatrix}
         \bar{d}_{i} \\ \bar{c}_{i} \\ \bar{e}_{i} \\ \tilde{\gamma}_{i} \\ \tilde{\varepsilon}_{i}
     \end{pmatrix}^{\mathrm{T}} 
     F_{ij}  
      \begin{pmatrix}
     0 & H_{jp} & 0 & 0 & 0 \\ 
     H_{jp} & H_{jp} + H_{jp,k} \lambda_{k}   & H_{jp,k} \chi_{k}  & -H_{jp,k} \theta_{k} &H_{jp,k} {\bar{\theta}}_{k}  \\
     0 & H_{jp,k} \chi_{k} & H_{jp} & 0 & 0 \\
     0 & H_{jp,k} \bar{\theta}_{k} & 0 & H_{jp} & 0 \\
     0 & H_{jp,k} \theta_{k} & 0 & 0 & H_{jp} \\ 
     \end{pmatrix}
\begin{pmatrix}
    d_{p} \\ 
    c_{p} \\
    e_{p} \\
    \gamma_{p} \\
    \varepsilon_{p} 
\end{pmatrix},
 \end{equation}

Note that, we have used 
\begin{equation}\label{eq:GF:mLM:supermatrixMprop}
    \mathop{\mathrm{Sdet}}  
\begin{pmatrix}
    0 & A & 0 & 0 & 0 \\ 
    A & B & C  & -D & E  \\
    0 & C & A & 0 & 0 \\
    0 & E  & 0 & A & 0 \\
    0 & D & 0 & 0 & A \\ 
    \end{pmatrix}
=   \mathop{\mathrm{Sdet}}  
\begin{pmatrix}
    0 & A & 0 & 0 & 0 \\ 
    A & A+B & C  & -D & E  \\
    0 & C & A & 0 & 0 \\
    0 & E  & 0 & A & 0 \\
    0 & D & 0 & 0 & A \\ 
    \end{pmatrix},
\end{equation}
which results in the following ghost action: 
\begin{equation}\label{eq:27alt}
    \begin{split}
        \mathcal{L}_{\text{GH}} ={}& \bar{d}_{i} F_{ij} H_{jp} c_{p} + \bar{c}_{i} F_{ij} H_{jp} d_{p} + \bar{c}_{i} F_{ij}  H_{jp}  c_{p} + \bar{e}_{i} F_{ij} H_{jp} e_{p} + \tilde{\gamma}_{i} F_{ij} H_{jp} \gamma_{p} + \tilde{\varepsilon}_{i} F_{ij} H_{jp} \varepsilon_{p} 
                                \\ & 
                                + \bar{c}_{i} F_{ij} H_{jp,k} \lambda_{k}  c_{p}
                                + \bar{c}_{i} F_{ij} H_{jp, k} \chi_{k} e_{p} - \bar{c}_{i} F_{ij} H_{jp,k} \theta_{k} \gamma_{p}  + \bar{c}_{i} F_{ij} H_{jp,k} \bar{\theta}_{k } \varepsilon_{p} 
                                \\ & + \bar{e}_{i} F_{ij} H_{jp,k} \chi_{k} c_{p} 
                                + \tilde{\gamma}_{i} F_{ij} H_{jp, k} \bar{\theta_{k}} c_{p} + \tilde{\varepsilon}_{i} F_{ij} H_{jp,k} \theta_{k} c_{p}.
\end{split}
\end{equation}
This is required to obtain a total effective Lagrangian
\begin{equation}\label{eq:27}
    \mathcal{L}_{\text{T}} = \mathcal{L}_{\text{EFF}} + \mathcal{L}_{\text{GF}} + \mathcal{L}_{\text{GH}}
\end{equation}
that is invariant under a standard ``BRST'' transformation \cite{Becchi:1974xu, Tyutin:1975qk}. 

\section{BRST invariance}\label{section:BRSTZinnJustin}

Using the standard procedure, one can find that the following BRST transformation 
\begin{subequations}\label{eq:28}
     \begin{align}
    \allowdisplaybreaks
         \label{eq:28a}
         \mathsf{s}   \phi_{i} ={}& H_{ij} c_{j}  , \\
         \label{eq:28b}
         \mathsf{s} \lambda_{i} ={}& 
         H_{ij} d_{j}  +   H_{ij,k}\lambda_{k}     c_j  + H_{ij,k}( \chi_{k} e_{j}   +   \bar{\theta}_{k} \varepsilon_{j}  -  \theta_{k}{\gamma}_{j})  
          \\ 
         \label{eq:28c}
         \mathsf{s}  \chi_{i} ={}&H_{ij} e_{j}  +  H_{ij,k}\chi_{k} c_{j}   , \\
         \label{eq:28d}
         \mathsf{s}  \bar{\theta}_{i} ={}&H_{ij} \gamma_{j}  +  H_{ij,k} \bar{\theta}_{k} c_{j}  ,\\
         \label{eq:28e}
         \mathsf{s}  \theta_{i} ={}& H_{ij} \varepsilon_{j}  +  H_{ij,k} {\theta}_{k} c_{j} , 
         \\
         \mathsf{s} c_{j} ={}& - \frac{1}{2} f_{mn | j} c_{m} c_{n} ,\\
       \mathsf{s} \gamma_{j} ={}&- f_{mn |j} c_{m} \gamma_{n}  , \\
        \mathsf{s} \varepsilon_{j} ={}&- f_{mn |j} c_{m} \varepsilon_{n} ,
        \\ 
        \mathsf{s} \bar{c}_{i} ={}& - \frac{1}{\alpha}F_{ij} ( \phi_{j} + \lambda_{j} ) , \ \mathsf{s} \bar{d}_{i} = - \frac{1}{\alpha}F_{ij} \lambda_{j}   , \ \mathsf{s} \bar{e}_{i} = - \frac{1}{\alpha}F_{ij} \chi_{j}   , \ \mathsf{s} \tilde{\gamma}_{i} = - \frac{1}{\alpha}F_{ij} {\theta}_{j}   , \ \mathsf{s} \tilde{\varepsilon}_{i} =  \frac{1}{\alpha}F_{ij} \bar{\theta}_{j}  ;
     \end{align}
\end{subequations}
leaves the total Lagrangian \eqref{eq:27} invariant \cite{McKeon:2024psy}. By construction, the BRST transformation \eqref{eq:28} is nilpotent $ \mathsf{s}^{2} =0$.

\section{First-order Yang-Mills theory}\label{section:YMinthemodLM}

In the Ref.~\cite{McKeon:2024psy}, the SOYM theory is quantized in the modified LM formalism.  
Here, we consider the quantization of the FOYM theory in the framework of the modified LM formalism. 
Since the first-class constraints remain unaltered with the introduction of the auxiliary fields, the contributions from the FP procedure are consistent across both formulations. 
This remains valid even in the presence of the LM fields from the LM theory, which shows that the modified LM formalism is consistent with these different formulations of the YM theory.

The classical FOYM Lagrangian reads
\begin{equation}\label{eq:61}
    \mathfrak{L}_{0}= - \frac{1}{4} \mathcal{F}_{\mu \nu}^{a} \mathcal{F}^{a \, \mu \nu} - \frac{1}{2} \mathcal{F}_{\mu \nu}^{a} F^{a \, \mu \nu} [A_{\mu}^{a} ] ,
\end{equation}
where we defined 
\begin{equation}\label{eq:62}
    F_{\mu \nu}^{a}[B_{\mu}^{a} ]= \partial_{\mu} B_{\nu}^{a} - \partial_{\nu} B_{\mu}^{a} + g f^{abc} B_{\mu}^{b} B_{\nu}^{c},
\end{equation}
and $ \mathcal{F}_{\mu \nu }^{a} $ is the auxiliary field.
It is invariant under the gauge transformation 
\begin{equation}\label{eq:63}
    \delta A_{\mu}^{a}  =  D_{\mu}^{ab} (A) \xi^{b}, \quad  \delta F_{\mu \nu}^{a} = g f^{abc} F_{\mu \nu}^{b} \xi^{c} \quad \text{and} \quad \delta \mathcal{F}_{\mu \nu}^{a} = g f^{abc} \mathcal{F}_{\mu \nu}^{b} \xi^{c}.
\end{equation}
We can identify $ H_{ij} ( \phi_{i})$ with the covariant derivative, which is given by 
\begin{equation}\label{eq:defDgeral} 
    D_{\mu}^{ab} (B) = \partial_{\mu} \delta^{ab} + g f^{apb} B_{\mu}^{b} . 
\end{equation}
We also have $ f_{mn |i} \mapsto f^{abc} $ (the structural constants~\eqref{eq:defce}).

The effective action of the FOYM in the framework of the modified LM theory is given by
\begin{equation}\label{eq:64}
    \begin{split}
        \mathfrak{L}_{\text{EFF}} ={}&
        \mathfrak{L}_{\text{0}}
        + \lambda_{\mu}^{a} D^{ab }_{\nu}(A)  \mathcal{F}^{ b \nu \mu}  
        + \frac{1}{2} \Lambda_{\mu \nu}^{a} ( \mathcal{F}_{\mu \nu}^{a} - F_{\mu \nu}^{a} )
        - \frac{1}{4} \chi_{\mu \nu}^{a} \chi^{a \, \mu \nu} +\frac{1}{2} \chi_{\mu \nu}^{a} F^{a \, \mu \nu} [ \chi_{\mu}^{a} ]
                                  \\ & 
                                  - \frac{1}{2} \bar{\theta}_{\mu \nu}^{a} \theta^{a \, \mu \nu} + \frac{1}{2} \bar{\theta}_{\mu \nu} ^{a} F^{a \, \mu \nu} [ \theta_{\mu}^{a} ] 
                                  + \frac{1}{2} F^{a \, \mu \nu} [ \bar{\theta}_{\mu}^{a} ] \theta_{\mu \nu}^{a}.
    \end{split}
\end{equation}

We will employ the Lorenz gauge fixing:
 \begin{equation}\label{eq:68}
     F_{ij} A_{j}  \mapsto
     ( \partial^{\mu} \delta^{ab} ) A_{\mu}^{b} = \partial^{\mu} A_{\mu}^{a} = \partial^{\mu} \lambda_{\mu}^{a} = \partial^{\mu} \chi_{\mu}^{a} = \partial^{\mu} \bar{\theta}_{\mu}^{a} = \partial^{\mu} \theta_{\mu}^{a} =0
 \end{equation}
 which leaves us with the following gauge fixing Lagrangian
\begin{equation}\label{eq:69}
    \mathfrak{L}_{\text{GF}} = - \frac{1}{2 \alpha} (\partial \cdot A^{a} )^{2} - \frac{1}{\alpha} (\partial \cdot \lambda^{a} ) (\partial \cdot A^{a} ) - \frac{1}{2 \alpha } (\partial \cdot \chi^{a} )^{2} - \frac{1}{\alpha} (\partial \cdot \bar{\theta}^{a} )(\partial \cdot \theta^{a} ),
\end{equation}
where $ \alpha $ is a gauge parameter and $ \partial \cdot X_{I} \equiv \partial_{\mu} X^{\mu}_{I} $ ($I$ are internal indices).  
The corresponding ghost Lagrangian is  
\begin{equation}\label{eq:610}
    \begin{split}
        \mathfrak{L}_{\text{GH}} ={}&\bar{c}^{a} \partial \cdot D^{ab} (A + \lambda ) c^{b} + \bar{d}^{a} \partial \cdot D^{ab} (A) c^{b} + \bar{c}^{a} \partial \cdot D^{ab} (A) d^{b} 
    \\ &  + \bar{e}^{a} \partial \cdot D^{ab} (A)e^{b} + \tilde{\gamma}^{a} \partial \cdot D^{ab} (A)\gamma^{b} + \tilde{\varepsilon}^{a} \partial \cdot D^{ab} (A)\varepsilon^{b} \\ &  
   +       \bar{c}^{a} g f^{apb} \partial \cdot \chi^p e^{b} + \bar{e}^{a} g f^{apb} \partial \cdot \chi^p c^{b}    
        -\bar{c}^{a}  g f^{apb} \partial \cdot \theta^p \gamma^{b}\\ &  + \tilde{\gamma}^{a} g f^{apb} \partial \cdot {\bar{\theta}}^p c^{b}  
        +\bar{c}^{a} g f^{apb} \partial \cdot {\bar{\theta}}^p \varepsilon^{b} + \tilde{\varepsilon}^{a} g f^{apb} \partial \cdot \theta^p c^{b},
\end{split}
\end{equation}
which is the same of the SOYM within the framework of the modified LM formalism \cite{McKeon:2024psy}.

The generating functional of the FOYM theory, in the modified LM theory, is given by
\begin{equation}\label{eq:611}
    \begin{split}
        \mathcal{Z} [ {}\mathbf{J}, {}\bm{\mathcal{J}}  , {}\bm{\eta} , \bar{{}\bm{\eta}} ]   ={}& \int \mathop{\mathcal{D} {}\mathbf{A}_{\mu}^{a}} \mathop{\mathcal{D} {}\bm{\mathcal{F}}_{\mu \nu}^{a}} \mathop{\mathcal{D} {}\mathbf{c}^{a} } \mathop{\mathcal{D} \bar{{}\mathbf{c}}^{a} } \\
                                                                                             & \quad  \exp i \int \mathop{d x} \left ( \mathfrak{L}_{ \text{EFF} } + \mathfrak{L}_{\text{GF} } + \mathfrak{L}_{\text{GH}} + \bar{{}\mathbf{J}}^{a \, \mu} {}\mathbf{A}_{\mu}^{a} + {}\bm{\mathcal{J} }^{a \, \mu \nu} {}\bm{\mathcal{F}}_{\mu \nu}^{a } +    \bar{{}\bm{\eta}}^{a} {}\mathbf{c}^{a} +  \bar{{}\mathbf{c}}^{a} {}\bm{\eta}^{a}  \right ), 
\end{split} 
\end{equation}
where we have employed the compact notation: 
\begin{align*}\label{eq:612}
    \allowdisplaybreaks
 & {}\mathbf{A}_{\mu}^{a} = ( A_{\mu}^{a} , \lambda_{\mu}^{a} , \chi_{\mu}^{a} , \bar{\theta}_{\mu}^{a} , \theta_{\mu}^{a} ), & {}\bm{ \mathcal{F} }_{\mu \nu}^{a} = (\mathcal{F}_{\mu \nu}^{a}, \Lambda_{\mu \nu}^{a} , \chi_{\mu \nu}^{a} , \bar{\theta}_{\mu \nu}^{a} , \theta_{\mu \nu}^{a} )
 \\ & 
    {}\mathbf{c}^{a} = ( c^{a} , d^{a} , e^{a} , \gamma^{a} , \varepsilon^{a} ), &  
 {}\bm{\bar{c}}^{a} = ( \bar{c}^{a} , \bar{d}^{a} , \bar{e}^{a} , \tilde{\gamma}^{a} , \tilde{\varepsilon}^{a} ) 
 \intertext{and}
                                                                                      & \mathop{\mathcal{D} {}\mathbf{A}_{\mu}^{a} } \equiv\mathop{\mathcal{D} A_{\mu}^{a} } \mathop{\mathcal{D} \lambda_{\mu}^{a}} \mathop{\mathcal{D} \chi_{\mu}^{a}} \mathop{\mathcal{D} \bar{\theta}_{\mu}^{a}} \mathop{\mathcal{D} \theta_{\mu}^{a}}, 
    & 
    \mathop{\mathcal{D} \bm{\mathcal{F}}_{\mu \nu}^{a}} 
    \equiv \mathop{\mathcal{D} \mathcal{F}_{\mu \nu}^{a}} 
     \mathop{\mathcal{D} {\Lambda}_{\mu \nu}^{a}} 
     \mathop{\mathcal{D} \chi_{\mu \nu}^{a}} 
     \mathop{\mathcal{D} \bar{\theta}_{\mu \nu}^{a}} 
     \mathop{\mathcal{D} \theta_{\mu \nu}^{a}},
    \\
    & \mathop{\mathcal{D} {}\mathbf{c}^{a} } \equiv 
    \mathop{\mathcal{D} c^{a}}     
    \mathop{\mathcal{D} d^{a}}          
    \mathop{\mathcal{D} e^{a}}            
    \mathop{\mathcal{D} \gamma^{a}}   
    \mathop{\mathcal{D} \varepsilon^{a}},  
    & 
    \mathop{\mathcal{D} \bar{{}\mathbf{c}}^{a} } \equiv 
    \mathop{\mathcal{D} \bar{c}^{a}} 
    \mathop{\mathcal{D} \bar{d}^{a}} 
    \mathop{\mathcal{D} \bar{e}^{a}} 
    \mathop{\mathcal{D} \tilde{\gamma}^{a}} 
    \mathop{\mathcal{D} \tilde{\varepsilon}^{a}}.
\end{align*}
The sources terms are defined as
\begin{subequations}\label{eq:613}
    \allowdisplaybreaks
    \begin{align}
        \bar{{}\mathbf{J}}^{a \, \mu} {}\mathbf{A}_{\mu}^{a} \equiv{}& J^{a \, \mu} (A_{\mu}^{a} + \lambda_{\mu}^{a} )+ K^{a \, \mu} \lambda_{\mu}^{a} + L^{a \, \mu} \chi_{\mu}^{a} + \bar{\eta}^{a \, \mu} \bar{\theta}_{\mu}^{a} + \bar{\kappa}^{a \, \mu} \theta_{\mu}^{a}, \\
    {}\bm{\mathcal{\bar{J}}}^{a \, \mu \nu } {}\bm{\mathcal{F} }_{\mu \nu}^{a} \equiv{}& \mathcal{J} ^{a \, \mu} (\mathcal{F}_{\mu \nu }^{a} + \Lambda_{\mu\nu }^{a} )+ \mathcal{K}^{a \, \mu \nu } \Lambda_{\mu \nu }^{a} + \mathcal{L}^{a \, \mu\nu } \chi_{\mu \nu }^{a} +  \bar{\theta}_{\mu \nu }^{a} \Omega^{a \, \mu \nu}  + \bar{\Omega}^{a \, \mu \nu } \theta_{\mu \nu }^{a}, \\
        \bar{{}\bm{\eta}}^{a} {}\mathbf{c}^{a} \equiv{}& \bar{\eta}^{a} (c^{a} + d^{a} )+ \bar{\kappa}^{a} d^{a}  + \bar{\upsilon}^{a} e^{a}   + \tilde{J}^{a} \gamma^{a} + \tilde{K}^{a} \varepsilon^{a}, \\
        \bar{{}\mathbf{c}}^{a} {}\bm{\eta}^{a} \equiv{}& ( \bar{c}^{a} + \bar{d}^{a} ) \eta^{a} + \bar{d}^{a} \kappa^{a} + \bar{e}^{a} \upsilon^{a} + \tilde{\gamma}^{a} J^{a} + \tilde{\varepsilon}^{a} K^{a},
\end{align}
\end{subequations}
where ($ J^{a \, \mu} $, $ K^{a \, \mu} $, $ L^{a \, \mu} $, $ \mathcal{J}^{a \, \mu \nu} $, $ \mathcal{K}^{a \, \mu \nu} $, $ \mathcal{L}^{a \, \mu \nu} $) are real ordinary sources, ($ {J}^{a} $, $ {K}^{a} $) are complex ordinary sources, and ($ \eta^{a \, \mu } $, $ \kappa^{a \, \mu} $, $ \Omega^{a \mu \nu} $, $ \bar{\Omega}^{a \, \mu \nu}$ , $ \eta^{a} $, $ \kappa^{a} $, $ \upsilon^{a} $) are Grassmann sources. The Feynman rules derived from Eq.~\eqref{eq:611} are presented in appendix~\ref{section:FeynmanRulesmLMYM}.

The BRST transformation that leaves the action in Eq.~\eqref{eq:611} invariant reads 
{\allowdisplaybreaks
\begin{subequations}\label{eq:BRSTYM}
     \begin{align}
         \mathsf{s}   A_{\mu}^{a}  ={}& D_{\mu}^{ab}(A) c^{b}    , \\
         \mathsf{s} \lambda_{\mu}^{a} ={}& 
         D_{\mu}^{ab} (A)d^{b}   
         +   g f^{abc} \lambda^{b}_{\mu}  c^{c}   
       + g f^{abc} ( \chi^b_{\mu} e^c   +   \bar{\theta}^b_{\mu}  \varepsilon^c  -  \theta^b_{\mu} {\gamma}^c)  
         , \\ 
         \mathsf{s}  \chi_{\mu}^a ={}& D_{\mu}^{ab} (A)e^b  +  g f^{abc}\chi^b c^c   , \\
         \mathsf{s}  \bar{\theta}_{\mu}^a ={}& D_{\mu}^{ab}(A) \gamma^b  +  g f^{abc} \bar{\theta}^b c^c  ,\\
         \mathsf{s}  \theta_{\mu}^a ={}&D_{\mu}^{ab} (A)\varepsilon^b  +  g f^{abc} {\theta}^b c^c ,
         \\
         \mathsf{s} \mathcal{F}_{\mu \nu}^{a} ={}& g f^{abc} F_{\mu \nu}^{b} c^{c} , \\
         \mathsf{s} \Lambda_{\mu \nu}^{a} ={}& g f^{abc} \Lambda_{\mu \nu}^{b} d^{c}  + g f^{abc} F_{\mu \nu}^{b} c^{c} , \\
         \mathsf{s} \chi_{\mu \nu}^{a} ={}&  g f^{abc} \chi_{\mu \nu}^{b} c^{c} , \\
         \mathsf{s} \bar{\theta}_{\mu \nu}^{a} ={}&  g f^{abc} \bar{\theta}_{\mu \nu}^{b} c^{c} , \\
         \mathsf{s} \theta_{\mu \nu}^{a} ={}&  g f^{abc} \theta_{\mu \nu}^{b} c^{c} , \\
         \mathsf{s} c_{\mu}^a ={}&  -\frac{g}{2} f^{abc}  c^b c^c \\        
        \mathsf{s} e_{\mu}^a ={}& - g f^{abc}  c^b e^c  \\
        \mathsf{s} \gamma_{\mu}^a ={}&- g f^{abc} c^b \gamma^c  , \\
        \mathsf{s} \varepsilon_{\mu}^a ={}&- gf^{abc} c^b \varepsilon^c  \\
        \mathsf{s} d_{\mu}^a ={}&- gf^{abc} \left ( d^b c^c + \frac{1}{2} e^b e^c + \gamma^b \varepsilon^c\right ) ,
     \end{align}
\end{subequations}
and
\begin{subequations}
    \begin{equation}
\label{eq:614}
\begin{split} 
    &\mathsf{s} \bar{c}^a ={}- \frac{1}{\alpha} \partial \cdot {(A + \lambda )}^{a}  , 
                        \quad \mathsf{s}\bar{d}^a ={}  -\frac{1}{\alpha}\partial \cdot A^{a}  , 
                        \quad \mathsf{s} \bar{e}^a ={} -\frac{1}{\alpha}\partial \cdot \chi^{a}  , \\
                    & \mathsf{s} \tilde{\gamma}^a ={}  -\frac{1}{\alpha} \partial \cdot \theta^{a}   , 
                    \quad \mathsf{s} \tilde{\varepsilon}^a ={}  \frac{1}{\alpha} \partial \cdot \bar{\theta}^a .
\end{split}
\end{equation}
\end{subequations}
}

\section{Alternative method for a general gauge theory}\label{section:aainMLM}

In this section, we revisit the alternative method (see Section~\ref{sect:gravity521}) in the framework of the modified LM theory.  In this method, the LM fields are introduced after the action is quantized through the FP procedure. We demonstrate that the LM formalism and the FP procedure commute, which means that the LM fields may be introduced either before or after the quantization of the original theory. Consequently, LM fields can be viewed as either pure quantum fields or classical fields.

\subsection{Second-order Yang-Mills theory}
In this section, we consider the YM theory in the second-order formulation. 
Using the FP procedure, we obtain a nonsingular action to the SOYM theory \cite{McKeon:2024psy}
\begin{equation}\label{eq:altYM}
    \mathfrak{S}_{\text{FP}} = \int \mathop{d x} \left (
        - \frac{1}{4} (\partial_{\mu} A_{\nu}^{a}  - \partial_{\nu} A_{\mu}^{a} +  g f^{abc} A_{\mu}^{b} A_{\nu}^{c})^{2} 
        - \frac{1}{2 \alpha} ( \partial_{\mu} A^{a \, \mu} )^{2} + \bar{c}^{a} \partial^{\mu} D_{\mu}^{ab} c^{b}
\right ), 
\end{equation}
where the Lorenz gauge was employed. Following the alternative method presented in Sec.~\ref{sect:gravity521}, we now introduce LM fields to each field in the action \eqref{eq:altYM}. This results in the following terms 
\begin{equation}\label{eq:mLM:LagYM}
    \begin{split}
        gf^{apb} \bar{c}^{a} \partial^{\mu} \lambda^{p}_{\mu} c^{b} 
        + \lambda_{\nu}^{a} D^{ab }_{\mu} (A)F^{b \, \mu \nu }  - \frac{\partial^{\mu} \lambda_{\mu}^{a} \partial^{\nu} A^{a}_{\nu} }{ \alpha} + \bar{d}^{a} \partial^{\mu} D_{\mu}^{ab}(A) c^{b} + \bar{c}^{a} \partial^{\mu} D_{\mu}^{ab} (A) d^{b},
    \end{split}
\end{equation}
where $ F^a_{\mu \nu}$ is the curvature~\eqref{eq:tensor} and $ \lambda_{\mu}^{a} $, $ \bar{d}^{a} $, $ d^{a} $ are respectively the LM fields corresponding to the fields $ A_{\mu}^{a} $, $ \bar{c}^{a} $ and $ c^{a} $. 

In the modified LM formalism, the introduction of the LM fields is compensated by inserting the superdeterminant factor\footnote{Recall that $ \updelta^{2} \equiv \delta_{L} \delta_{R} $.}
\begin{equation}\label{eq:mLM:YM:Delta}
    \left( \mathop{\mathrm{Sdet}} \frac{\updelta^{2}\mathfrak{S}_{ \text{FP} } }{\delta {\Phi}_{\text{YM}} \delta \Phi_{\text{YM}}}\right)^{+1/2} \equiv \mathop{\mathrm{Sdet}} \mathbf{H}_{\text{YM}}^{+1/2} ,
\end{equation}
into the path integral, where the superfield $\Phi_{\text{YM}} \equiv  (A, c , \bar{c} )$ contains all of the fields of the nonsingular action~\eqref{eq:altYM}. One can show that the Hessian in Eq.~\eqref{eq:mLM:YM:Delta} is given by 
\begin{equation}\label{eq:HessianYM}
    \mathbf{H}_{\text{YM}}  = \begin{pmatrix}
        B^{uv}_{\mu \nu} & -g f^{aul} ({\partial}^{\mu}\bar{c}^{a}) & - g f^{muc}  c^{c} \overleftarrow{\partial^{\mu}}\\
        g f^{avp}  ({\partial^{\nu}} \bar{c}^{a}) & 0 & - {D}{_{\rho}^{mp}(A)  }\overleftarrow{\partial^{\rho}} 
        \\
        g f^{qvc}  \overrightarrow{\partial^{\nu}}c^{c} &
        \overrightarrow{\partial^{\rho}} {D}_{\rho}^{ql} (A)
                                                        & 0
    \end{pmatrix}^{1/2} ,
\end{equation}
with \begin{equation}\label{eq:YM:segundaderivadaAA}
    B_{\mu \nu}^{uv} \equiv  \frac{\updelta^{2} \mathfrak{S}_{\text{FP}} }{\delta A^{u \, \mu} \delta A^{v \, \nu}}.
\end{equation} 
Alternatively, one may use
\begin{equation}\label{eq:mLM:YM:Delta3}
    \left( \mathop{\mathrm{Sdet}} \frac{\updelta^{2}\mathfrak{S}_{ \text{FP} } }{\delta \bar{\Psi}_{\text{YM}} \delta \Psi_{\text{YM}}}\right)^{+1/2} 
  = 
\begin{Vmatrix}
    \frac{1}{2}B^{uv}_{\mu \nu} & -g f^{aul} ({\partial}^{\mu}\bar{c}^{a})
                        \\
    g f^{qvc}   \overrightarrow{\partial^{\nu}} c^{c} & 
    \overrightarrow{\partial^{\rho} } D_{\rho}^{ql} (A)
    \end{Vmatrix}^{1/2}
\end{equation}
with $ \Psi_{ \text{YM} } = ( A, c)$ and $ \bar{\Psi}_{\text{YM}} = (A, \bar{c} )$.

Exponentiating the superdeterminant factor, we find 
\begin{equation}\label{eq:altlocaldet}
    \begin{split}
\int 
   \mathop{\mathcal{D} \mu_{\text{mYM}}}
      \exp i\int \mathop{d x}  
      \left[
    \begin{pmatrix}
        \chi_{\mu}^{u} &  \bar{e}^{q} 
    \end{pmatrix}
    \mathbf{R}_{\text{YM}}     \begin{pmatrix}
        \chi^{v}_{\nu} \\
        e^{l} 
    \end{pmatrix}   
    + 
    \begin{pmatrix}
        \theta_{\mu}^{u} & \varepsilon^{p} & \tilde{\gamma}^{q}  
    \end{pmatrix}
    \mathbf{H}_{\text{YM}}     \begin{pmatrix}
        -\bar{\theta}^{v}_{\nu} \\
        \gamma^{l} \\
        -\tilde{\varepsilon}^{m} 
    \end{pmatrix}
   \right ]
\end{split},
\end{equation}
where $ \mathop{\mathcal{D} \mu_{\text{mYM}}} \equiv \mathop{\mathcal{D} \chi_{\mu}^{a}} \mathop{\mathcal{D} \bar{e}^{a}} \mathop{\mathcal{D} e^{a}} \mathop{\mathcal{D} \bar{\theta}_{\mu}^{a}} \mathop{\mathcal{D} \theta_{\mu}^{a}}  \mathop{\mathcal{D} \tilde{\gamma}^{a}} \mathop{\mathcal{D} \gamma^{a}} \mathop{\mathcal{D} \tilde{\varepsilon}^{a}} \mathop{\mathcal{D} \varepsilon^{a}}$, $ \chi_{\mu}^{a} $ is a real field; $ \bar{e}^{a} $, $ e^{a} $, $\bar{\theta}_{\mu}^{a} $, $ \theta_{\mu}^{a} $ are Grassmann fields; $ \bar{\gamma}^{a} $, $ \gamma^{a} $, $ \bar{\varepsilon}^{a} $, $ \varepsilon^{a} $ are complex fields. The matrix $ \mathbf{H}_{\text{YM}}  $ is defined in Eq.~\eqref{eq:HessianYM} and $\mathbf{R}_{\text{YM}} $  is the Hessian in Eq.~\eqref{eq:mLM:YM:Delta3}. 

The total contribution from the Lee-Yang ghost fields is
\begin{equation}\label{eq:altLeeYAngghost}
    \int 
\mathop{\mathcal{D} \chi_{\mu}^{a}} \mathop{\mathcal{D} \bar{e}^{a}} \mathop{\mathcal{D} e^{a}} 
    \mathop{\mathcal{D} \bar{\theta}_{\mu}^{a}} \mathop{\mathcal{D} \theta_{\mu}^{a}}  \mathop{\mathcal{D} \tilde{\gamma}^{a}} \mathop{\mathcal{D} \gamma^{a}} \mathop{\mathcal{D} \tilde{\varepsilon}^{a}} \mathop{\mathcal{D} \varepsilon^{a}}
    \exp i \int \mathop{d x} \mathcal{L}_{\text{LY}},
\end{equation}
where 
\begin{equation}\label{eq:altLYghaction}
    \begin{split}
        \mathcal{L}_{\text{LY}} ={}&
    \frac{1}{2} \chi_{\mu}^{u} B_{\mu \nu}^{uv} \chi_{\nu}^{v} + \bar{\theta}_{\mu}^{u} B_{\mu \nu}^{uv} \theta_{\nu}^{v} + \bar{e}^{u}  
    \partial^{\mu} D_{\mu}^{uv} (A)
    e^{v} + 
    \tilde{\gamma}^{u}  
    \partial^{\mu} D_{\mu}^{uv} (A)
+ \tilde{\varepsilon}^{u}  
    \partial^{\mu} D_{\mu}^{uv} (A)
    \varepsilon^{v}
                        \\ & 
                        -g f^{aul} \chi_{\mu}^{u}  (\partial^{\mu} \bar{c}^{a}) e^{l} 
                        + g f^{qvc}   \bar{e}^{q} \partial^{\nu} {c}^{c} \chi_{\nu}^{v} 
        - g f^{aul} \theta_{\mu}^{u}   (\partial^{\mu} \bar{c}^{a} ) \gamma ^{l} 
        +g f^{muc} \partial^{\nu} (\theta_{\mu}^{u}c^{c} ) \tilde{\varepsilon}^{m} 
                        \\ & 
                        - g f^{avp} \varepsilon^{p} (\partial^{\nu} \bar{c}^{a} ) \theta_{\nu}^{v}  
                        - g f^{qvc} \tilde{\gamma}^{q} \partial^{\nu} c^{c} \theta_{\nu}^{v}. 
    \end{split} 
\end{equation}
Thus, one can show that the total effective Lagrangian can be expressed as 
{\allowdisplaybreaks
\begin{align}\label{eq:altYMcomplete}\nonumber
        & - \frac{1}{4} \left[ (\partial_{\mu} A_{\nu}^{a}  
        - \partial_{\nu} A_{\mu}^{a} 
        +g f^{abc} A_{\mu}^{a} A_{\nu}^{b})^{2} 
 +   ( \partial_{\mu} \chi_{\nu}^{a}  - \partial_{\nu} \chi_{\mu}^{a} )^{2} 
+  2( \partial_{\mu} \bar{\theta}_{\nu}^{a}  - \partial_{\nu} \bar{\theta}_{\mu}^{a} ) ( \partial^{\mu} \theta^{ a \, \nu }  - \partial^{\nu} \theta^{a \, \mu } )
\right]
     \\ \nonumber & 
 + \lambda_{\mu}^{a} D^{ab }_{\nu}  F^{ b \nu \mu}  
       - \frac{1}{2 \alpha} \left[(\partial \cdot A^{a} )^{2} + 2 (\partial \cdot \lambda^{a} ) (\partial \cdot A^{a} )  +(\partial \cdot \chi^{a} )^{2} 
    + 2 (\partial \cdot \bar{\theta}^{a} )(\partial \cdot \theta^{a} )  \right]\\ \nonumber & 
- g f^{abc} ( \partial_{\mu} \chi_{\nu}^{a}  - \partial_{\nu} \chi_{\mu}^{a} ) A^{b \, \mu} \chi^{c \, \nu} 
     - \frac{g}{2} f^{abc}  (\partial_{\mu} A_{\nu}^{a} - \partial_{\nu} A_{\mu}^{a} )    \chi^{b \, \mu} \chi^{c \, \nu} 
\\ \nonumber &  - \frac{g^{2}}{2}  \left[ f^{abc} f^{ade} A_{\mu}^{b} \chi_{\nu}^{c} A^{d \, \mu} \chi^{e \, \nu}
                                  + f^{abc} f^{ade} A_{\mu}^{b} \chi_{\nu}^{c} \chi^{d \, \mu} A^{e \, \nu}
                                  + f^{abc} f^{ade} A_{\mu}^{b} A_{\nu}^{c} \chi^{d \, \mu} \chi^{e \, \nu} \right]
     \\ \nonumber &    
- g \left[f^{abc} ( \partial_{\mu} \bar{\theta}_{\nu}^{a}  - \partial_{\nu} \bar{\theta}_{\mu}^{a} ) A^{b \, \mu} \theta^{c \, \nu} 
+ f^{abc}  A^{b \, \mu} \bar{\theta}^{c \, \nu}  ( \partial_{\mu} \theta_{\nu}^{a}  - \partial_{\nu} \theta_{\mu}^{a} )
+  f^{abc}  (\partial_{\mu} A_{\nu}^{a} - \partial_{\nu} A_{\mu}^{a} )    \bar{\theta}^{b \, \mu} \theta^{c \, \nu} \right]
     \\ \nonumber & 
                                  - g^2   \left[f^{abc} f^{ade} A_{\mu}^{b} \bar{\theta}_{\nu}^{c} A^{d \, \mu} \theta^{e \, \nu}
                                  +  f^{abc} f^{ade} A_{\mu}^{b} \bar{\theta}_{\nu}^{c} \theta^{d \, \mu} A^{e \, \nu}
                                  +f^{abc} f^{ade} A_{\mu}^{b} A_{\nu}^{c} \bar{\theta}^{d \, \mu} \theta^{e \, \nu}\right] 
     \\ \nonumber & 
+ \bar{c}^{a} \partial \cdot D^{ab} (A + \lambda ) c^{b} + \bar{d}^{a} \partial \cdot D^{ab} (A) c^{b} + \bar{c}^{a} \partial \cdot D^{ab} (A) d^{b} 
    + \bar{e}^{a} \partial \cdot D^{ab} (A)e^{b} \\ \nonumber & + \tilde{\gamma}^{a} \partial \cdot D^{ab} (A)\gamma^{b} 
        + \tilde{\varepsilon}^{a} \partial \cdot D^{ab} (A)\varepsilon^{b}  
 + \bar{c}^{a} g f^{apb} \partial \cdot \chi^p e^{b} + \bar{e}^{a} g f^{apb} \partial \cdot \chi^p c^{b}    
\\ \nonumber&  -\bar{c}^{a}  g f^{apb} \partial \cdot \theta^p \gamma^{b} + \tilde{\gamma}^{a} g f^{apb} \partial \cdot {\bar{\theta}}^p c^{b}  
                                +\bar{c}^{a} g f^{apb} \partial \cdot {\bar{\theta}}^p \varepsilon^{b} + \tilde{\varepsilon}^{a} g f^{apb} \partial \cdot \theta^p c^{b}
,\\ & 
\end{align}
}
which is the same obtained in \cite{McKeon:2024psy}.

Thus, we explicitly verified that the alternative method introduced in Section~\ref{sect:gravity521} also holds in the framework of the modified LM formalism. 

\subsection{General gauge theory}
Now, we consider a more general proof for the commutation of the LM formalism and the FP quantization. We use a general gauge theory of YM type. 

We will start with the quantization. Let us consider a gauge theory with the classical action given by
\begin{equation}\label{eq:clS0alt}
    S_{0} = \int \mathop{d x} \mathcal{L}_{0} ( \phi ).
\end{equation}
It is invariant under the gauge transformation 
\begin{equation}\label{eq:gtransS0alt}
    \delta \phi_{i}   = H_{iJ} ( \phi ) \xi_{J}.
\end{equation}
We assume that the gauge algebra is closed and that the condition~\eqref{eq:conditions74} is satisfied by $ H_{iJ} ( \phi )$ and the structure constants $ f_{mn|i} $.

To quantize this theory, we have to fix the gauge using some gauge condition $ F_{Ij} \phi_{j} = 0$.\footnote{We assume that the gauge fixing is linear, that is, $ F_{Ij,k} =0$.} Employing the FP quantization, we arrive at the FP action:
\begin{equation}\label{eq:alteff}
    S_{\text{FP}} = S_{0} + \int \mathop{d x} \left[ \mathcal{L}_{\text{gf}}( \phi )+ \mathcal{L}_{\text{gh}}( \phi ) \right],
\end{equation}
where 
\begin{equation}\label{eq:lagaltGF}
    \mathcal{L}_{\text{gf}} ( \phi )= - \frac{1}{2 \alpha} (F_{Ij} \phi_{j} )^{2}
\end{equation}
and 
\begin{equation}\label{eq:lagaltGH}
    \mathcal{L}_{\text{gh}} ( \phi )= \bar{c}_{I} F_{Ij} H_{jK} c_{K}.
\end{equation}

Now, we proceed with the LM formalism and supplement the effective action in Eq.~\eqref{eq:alteff} with the LM field $ \lambda_{i} $ (commuting field) and $ d_{i} $, $ \bar{d}_{i} $ (anticommuting fields). This yields 
\begin{equation}\label{eq:altSLMaction}
    S_{\text{sLM}} = \text{S}_{\text{FP}} + S_{\lambda} + S_{d, \bar{d}} ,
\end{equation}
where 
\begin{equation}\label{eq:altSlamb}
    S_{\lambda} = \lambda_{i} \frac{\delta S_{\text{FP} } }{\delta \phi_{i}} = 
    \lambda_{i} \frac{\delta S_{0} }{\delta \phi_{i}} - \frac{1}{\alpha} F_{Ij} \phi_{j} F_{Ik} \lambda_{k} + \bar{c}_{I} F_{Ij} H_{jK, l} \lambda_{l} c_{K}  
\end{equation}
and 
\begin{equation}\label{eq:altSdbard}
    S_{d , \bar{d}} = \bar{d}_{I} F_{Ij} H_{jK} c_{K} + \bar{c}_{I} F_{Ij} H_{jK} d_{K}.
\end{equation}
The action~\eqref{eq:altSLMaction} is identical to the action of the standard LM formalism obtained in Ref.~\cite[Eq. 5.21]{Brandt:2020gms}. 

In~\cite{Brandt:2020gms}, the standard method is used.  First, the classical action is supplemented with LM fields and then quantized using an extension of the FP procedure. Here, we have interchanged this order. This demonstrates that the FP quantization and the standard LM formalism commute, as we have explicitly verified for gravity in Section~\ref{sect:gravity521}.

Finally, we introduce the superdeterminant factor 
\begin{equation} \label{eq:altSdet1over2}
    \mathop{\mathrm{Sdet}} \mathbf{H}^{+1/2}\equiv   \mathop{\mathrm{Sdet}} \left( \frac{\updelta^{2} S_{ \text{FP} } }{\delta {\Phi}\delta \Phi }\right)^{+1/2}   \qquad \big[\Phi \equiv  (\phi, c , \bar{c} )\big]
\end{equation}
to obtain the generating functional of the action~\eqref{eq:clS0alt} in the framework of the modified LM theory: 
\begin{equation} \label{eq:altgenZ}
\mathcal{Z}[0]= \int \mathop{\mathcal{D} \phi_i} \mathop{\mathcal{D} c_I} \mathop{\mathcal{D} \bar{c}_I} \mathop{\mathcal{D} \lambda_i}
\mathop{\mathcal{D} d_I} \mathop{\mathcal{D} \bar{d}_I} 
\mathop{\mathrm{Sdet}} \mathbf{H}^{+1/2} \exp i S_{\text{sLM}}. 
\end{equation}
We rewrite the superdeterminant factor as
\begin{equation} \label{eq:altSdet}
\mathop{\mathrm{Sdet}} \left( \frac{\updelta^{2}S_{ \text{FP} } }{\delta {\Phi}\delta \Phi }\right)^{+1/2}=
 \mathop{\mathrm{Sdet}} \mathbf{H}^{-1/2} 
\mathop{\mathrm{Sdet}} \mathbf{H}^{+1},
\end{equation}
where 
\begin{equation}\label{eq:Hessianalt}
    \mathbf{H}  
   = \begin{pmatrix}
        B_{ij}&   \bar{c}_{I} \overleftrightarrow{O}_{IL | i}  &  -c_{K} \overleftarrow{O}\!_{MK | i}                               \\
        -\bar{c}_{I} \overleftrightarrow{O}_{IP | j}  & 0 & -  F_{Ma} H_{aP}   \\
        \overrightarrow{O}\!_{QK | j}c_{K}    &  F_{Qa} H_{aL}  & 0
    \end{pmatrix}^{1/2} , \qquad \left(
    B_{ij}\equiv  \frac{\updelta^{2} S_{\text{FP}} }{\delta \phi_i \delta \phi_j}
    \right)
\end{equation}
and $ O_{IL | i} \equiv F_{Ik} H_{kL, i} $. The arrows indicate in which fields the operator $O $ acts. The operator $ \overleftrightarrow{O}$ does not act on $ \bar{c}_{I} $.

The superdeterminant factors in Eq.~\eqref{eq:altSdet} can be exponentiated using ghost fields yielding
\begin{equation}\label{eq:localdetalt1}
    \begin{split}
\int 
   \mathop{\mathcal{D} \mu_{\text{LY}}}
      \exp i\int \mathop{d x}  
      \left[
          \frac{1}{2} 
    \begin{pmatrix}
        \chi_{i} &  {e}_{P} &  \bar{e}_{Q}  
    \end{pmatrix}
    \mathbf{H}    \begin{pmatrix}
        \chi_{j} \\
        e_{L} \\ 
        \bar{e}_{M}
    \end{pmatrix}   
    - 
\begin{pmatrix}
    \theta_{i} &  \varepsilon_{P} &  \tilde{\gamma}_{Q}  
    \end{pmatrix}
        \mathbf{H}    
\begin{pmatrix}
        \bar{\theta}_j \\
       - \gamma_{L} \\ 
        \tilde{\varepsilon}_{M} 
    \end{pmatrix}
   \right ]
\end{split},
\end{equation}
where $ \mathop{\mathcal{D} \mu_{\text{LY}}} \equiv \mathop{\mathcal{D} \chi_{i}} \mathop{\mathcal{D} \bar{e}_{I}} \mathop{\mathcal{D} e_{I}} \mathop{\mathcal{D} \bar{\theta}_{i}} \mathop{\mathcal{D} \theta_{i}}  \mathop{\mathcal{D} \tilde{\gamma}_{I}} \mathop{\mathcal{D} \gamma_{I}} \mathop{\mathcal{D} \tilde{\varepsilon}_{I}} \mathop{\mathcal{D} \varepsilon_{I}}$, $ \chi_{i} $ is a real field; $ \bar{e}_{I} $, $ e_{I} $, $\bar{\theta}_{i} $, $ \theta_{i} $ are Grassmann fields; $ \bar{\gamma}_{I} $, $ \gamma_{I} $, $ \bar{\varepsilon}_{I} $, $ \varepsilon_{I} $ are complex fields. The Lee-Yang ghost Lagrangian~\eqref{eq:localdetalt1} can be rewritten as 
\begin{equation}\label{eq:altYL}
    \begin{split}
        \mathcal{L}_{\text{LY}} ( \phi )={}&
    \frac{1}{2} \chi_{i} B_{ij} \chi_{j} + \bar{\theta}_{i} B_{ij} \theta_{j} + \bar{e}_{I} F_{Il} H_{lJ} e_{J} +  
\tilde{\gamma}_{I} F_{Il} H_{lJ} \gamma_{J}+\tilde{\varepsilon}_{I} F_{Il} H_{lJ} \varepsilon_{J} \\ 
    & +\bar{c}_{I} F_{Ik} H_{kL, i} \chi_{i} e_{L}+ \bar{e}_{M}  F_{M l} H_{lK, i} \chi_{i} c_{K} - \bar{c}_{I} F_{Ik} H_{kL, i} \theta_{i} \gamma_{L} \\ & + \tilde{\varepsilon}_{M} F_{Ml} H_{lK,i} \theta_{i} c_{K}   +\bar{c}_{I} F_{Ik} H_{kP, j}  \bar{\theta}_{j} \varepsilon_{P} + \tilde{\gamma}_{Q} F_{Ql} H_{lK,j} \bar{\theta}_{j} c_{K}.
    \end{split}
\end{equation}

Adding the Lee-Yang ghost Lagrangian~\eqref{eq:altYM} to the standard LM Lagrangian (see Eq.~\eqref{eq:altSLMaction}) results in the effective Lagrangian of the modified LM formalism:
\begin{equation}\label{eq:altTotal}
    \begin{split} 
    & \mathcal{L}_{\text{T}} = \mathcal{L}_{0} (\phi ) + \mathcal{L}_{\text{gf}} ( \phi )+ \mathcal{L}_{\text{gh}} ( \phi )
    + 
        \lambda_{i} \frac{\delta S_{0} }{\delta \phi_{i}} - \frac{1}{\alpha} F_{Ij} \phi_{j} F_{Ik} \lambda_{k} \\ & + \bar{c}_{I} F_{Ij} H_{jK, l} \lambda_{l} c_{K} +  
    \bar{d}_{I} F_{Ij} H_{jK} c_{K} + \bar{c}_{I} F_{Ij} H_{jK} d_{K}
    + \mathcal{L}_{\text{LY}} (\phi ).
    \end{split}
\end{equation}
This Lagrangian~\eqref{eq:altTotal} is equal to the total effective action obtained with the standard method in Eq.~\eqref{eq:27}, which confirms that the LM formalism commutes with the FP quantization formalism.

\section{Additional symmetries}\label{section:symmetriesalt}
Here, we extend the symmetries shown in Section~\ref{section:symmetries}. The Lagrangian of the modified LM theory in Eq.~\eqref{eq:altTotal} is invariant under 
\begin{subequations}\label{eq:Ghostalt}
    \allowdisplaybreaks
\begin{align}\label{eq:transform:ghost:alt}
    & \delta \bar{\theta}_{i}  = - \sigma_{\text{LM}}\bar{\theta}_{i} , &&   \delta \theta_{i} = \sigma_{\text{LM}}\theta_{i}\\
    & \delta \bar{c}_{I}  = - \sigma_{\text{FP}} \bar{c}_{I} , && \delta c_{I} = -\sigma_{\text{FP}} c_{I },\\  
    & \delta \bar{d}_{I}  = - \sigma_{\text{FP}} \bar{d}_{I} , && \delta d_{I} = -\sigma_{\text{FP}} d_{I }, \\
    & \delta \bar{e}_{I}  = - \sigma_{\text{FP}} \bar{e}_{I} , && \delta e_{I} = \sigma_{\text{FP}} e_{I} ,\\ 
    & \delta \tilde{\gamma}_{I}  = - (\sigma_{\text{FP}} - \sigma_{\text{LM}})\tilde{\gamma}_{I} , && \delta \gamma_{I} =  (\sigma_{\text{FP}} - \sigma_{\text{LM}})\gamma_{I} ,\\ 
    & \delta \tilde{\varepsilon}_{I}  = - (\sigma_{\text{FP}} + \sigma_{\text{LM}})  \tilde{\varepsilon}_{I} , && \delta \varepsilon_{I} = (\sigma_{\text{FP}} + \sigma_{\text{LM}})\varepsilon_{I} ,  \\ 
    & \delta \phi_{i} = \delta \lambda_{i} = \delta \chi_{i} =0;
\end{align}
\end{subequations}
where $ \sigma_{\text{LM}} $ and $ \sigma_{\text{FP}} $ are commuting parameters. We find that the ghost number symmetries of both the modified LM and the FP ghost fields are compatible. This extended symmetry allows us to classify each field by its charge.  We observe that the commuting complex fields $ ( \tilde{\gamma}_{I}, \gamma_{i} ) $ and $ ( \tilde{\varepsilon}_{I} , \epsilon_{I} )$ carry both the LM and FP ghost charges, respectively, $ \sigma_{\text{LM}} $ and  $ \sigma_{\text{FP}} $. 

We also have the extended (anti)BRST-like symmetry:
\begin{align}\label{eq:transform:ghostSalt}
    & \delta \theta_{i}  =  -\rho_{+} \chi_{i}  \eta , &&\delta \bar{\theta}_{i} =  \rho_{-} \chi_{i} \eta , &&
    \delta \chi_{i} = ( \rho_{-} \theta_{i} +\bar{\theta}_{i}  \rho_{+} ) \eta , \\
    &\delta {{\varepsilon}}_I =  \rho_{+} {e}_I \eta ,  &&\delta \gamma_I =  \rho_{-} e_{I}  \eta , && \delta e_I = -( {{\varepsilon}}_I \rho_{-}+ \rho_{+} {\gamma}_I ) \eta , \\
    & \delta \tilde{\gamma}_I =  \rho_{+} \bar{e}_{I}  \eta , && 
   \delta \tilde{{\varepsilon}}_I =  \rho_{-} \bar{e}_I \eta , && \delta \bar{e}_I = ( \tilde{{\varepsilon}}_I \rho_{+} + \rho_{-} \tilde{\gamma}_I ) \eta , \\
    & \delta \phi_{i} = \delta \lambda_{i} =0, && \delta c_{I} = \delta d_{I} =   0, &&
    \delta \bar{c}_{I} =  \delta \bar{d}_{I} =0;
\end{align}
where $ \eta $ is an anticommuting parameter, and $ \rho_{\pm}  $ is a real parameter. When  $ \rho_{\pm} = 0$, we find the pure (anti)BRST-like symmetry. This is a internal symmetry of the Lee-Yang-like ghost sector itself, leaving the standard LM formalism fields unaffected.

 \chapter{Discussion} \label{section:discussion}

In this work, we studied the first-order formulation of the YM theory and gravity. In particular, the quantum equivalence between these formulations using the path integral quantization procedure. We have shown that we can obtain identities that relate Green's functions of the auxiliary field to Green's functions in the second-order formulation of composite fields. These structural identities, which are directly associated with the quantum equivalence between those  formalisms, are complementary to the standard Ward identities. 

On the other hand, the first-order formalism has additional symmetries due to the auxiliary fields which lead to similar identities. However, we also have shown that one cannot obtain the equivalence between Green's functions of the gauge field from the Ward identities derived from these symmetries.  

We have also investigated the quantum equivalence using other generating functionals, extending the previous works \cite{McKeon:2020lqp, Brandt:2020vre}. We have demonstrated that we can show the quantum equivalence via the effective action $ \Gamma $, which provides a proper definition of self-energy in the first-order formulation. In YM theory, we have shown that the self-energy of the field $ A_{\mu}^{a} $ remains identical in both formulations. The proper self-energy in the first-order formalism is obtained from the naive self-energies defined as the sum of 1PI diagrams with two amputated external legs. This resolves the apparent inconsistency of the self-energies computed in the first-order formalism. 

At finite temperature, the proper self-energy of the field $ A_{\mu}^{a} $, at the high-temperature limit, is identical in both formulations of the YM theory \cite{vasconcelos:2020}. If Eq.~\eqref{eq:effbyQplus} holds, then \begin{equation}\label{eq:effbyQplusatT}
- \mathbf{\Sigma}_{A \mathcal{F}} \mathbf{D}^{-1}\mathbf{Q}_{-} - \mathbf{Q}_{+} \mathbf{D}^{-1}\mathbf{\Sigma}_{\mathcal{F} A}
+  \mathbf{Q}_{+} \mathbf{D}^{-1}\mathbf{\Sigma}_{\mathcal{F} \mathcal{F}} \mathbf{D}^{-1}\mathbf{Q}_{-} = 0
\end{equation}
at the high-temperature limit. Recall that $ \mathbf{\Sigma} $ represents the sum of 1PI diagrams with two amputated legs and $ \mathbf{Q}_{\pm} $ and $ \mathbf{D} $ are components of the Hessian of the FOYM action (see Eq.~\eqref{eq:hessionFOYM}).\footnote{The identity \eqref{eq:effbyQplusatT}  can be written in component-form: $
         \Sigma_{ \mu  }^{ac\, \alpha \beta } Q_{\alpha \beta \, \nu}^{cb} + Q_{\mu }^{ac\, \alpha \beta }   \Sigma_{\alpha \beta\, \nu}^{cb}   + Q^{ac}_{\mu\, \alpha \beta }  \Sigma^{cd \, \alpha \beta \rho \sigma } Q^{db}_{\rho \sigma \, \nu }=0 
$ (for definitions, see Section~\ref{section:SEYM}).} Indeed, the identity~\eqref{eq:effbyQplusatT} is valid, which can be directly verified by using the results in Ref. \cite{vasconcelos:2020}. 

In Ref. \cite{vasconcelos:2020}, it is also shown that the self-energies involving the auxiliary field $ \mathcal{F}_{\mu \nu}^{a} $ and the FP ghost fields $ c^{a} $, $ \bar{c}^{a} $ are sub-leading in the high-temperature limit.  Despite the quantum equivalence, the decomposition of Green's functions as in Eq.~\eqref{eq:effbyQplus} is not arbitrary. The auxiliary field has a physical interpretation richer than initially expected. This is an important subject that may be investigated in the future.

We have provided a systematic procedure to obtain the path integral of the first-order formalism through the path integral of the second-order formalism. This is particularly interesting in gravity since it gives the Senjanovi\'{c} determinant in a manifestly covariant form. With the generating functional of the FOGR theory obtained from this procedure, we have shown that the structural identities can be verified at the integrand level. This can be understood as follow: in the naive path integral, there are tadpole-like contributions arising from the determinant factor $ (\det M (h) )^{-1/2} $, which are exactly canceled by contributions coming from the Senjanovi\'{c} determinant proposed in Eq.~\eqref{eq:measureFOGR}. 

Moreover, this path integral is better suitable at finite temperature, since the contributions from the Senjanovi\'{c} determinant may become relevant (see Appendix~\ref{section:tadpoles}). Nevertheless, it is essential to account for second-class constraints in the first-order formalism to count correctly the degrees of freedom of the theory.

We also have demonstrated that first-order formulations of the YM theory with non-minimal couplings can be systematically derived. These couplings depend on the auxiliary fields, specifically the curvature $F_{\mu \nu}^{a} $. These first-order formulations are equivalent to their corresponding second-order formulations, including at the quantum level. This procedure can be easily generalized to gravity. In this case, it is most relevant since the minimal coupling of fermions to gravity leads to terms that are dependent on the auxiliary field, the connection $ \Gamma^{\lambda} {}_{\mu \nu} $. This would result in a first-order formulation of gravity coupled to fermions that is quantum equivalent to the SOGR theory in any dimension $D>2$.\footnote{This restriction is because of the nature of the HE action. In $D=2$, the HE action is purely topological.} 

In YM theory, we have the Pauli interaction (in the fundamental representation)
\begin{equation}\label{eq:pauli}
    \bar{\psi} T^{c} \tensor{F}{_{c}^{ \alpha \beta }}  \sigma_{\alpha \beta } \psi.
\end{equation}
In gravity, the fermion is minimally coupled by the covariant derivative that leads to a similar term
\begin{equation}\label{eq:devcov}
    \bar{\psi} \gamma^{\mu} \tensor{\omega}{_{\mu}^{ab}} \sigma_{a b} \psi,
\end{equation}
where $ \tensor{\omega}{_{\mu}^{ab}} $ is the spin connection and $ \gamma^{\mu} = \gamma^{a} \tensor{e}{_{a}^{\mu}}$. The following relation holds between them: 
\begin{equation}\label{eq:relation}
    ab \longleftrightarrow \alpha \beta , \quad \mu \longleftrightarrow c, \quad \omega \longleftrightarrow F \quad \text{and} \quad \gamma  \longleftrightarrow T.
\end{equation}
Thus, our investigation of the YM theory can be directly translated to gravity using this analogy. In contrast, the Palatini approach leads to an inequivalent theory. This theory is the well-known {Einstein-Cartan} theory, which describes more general spacetimes with torsion and curvature.

We also have presented an alternative gravity theory that is renormalizable and unitary. This theory is obtained from the standard LM formalism in which LM fields are introduced to constrain the path integral restricting the loop expansion to one-loop order. We have shown that this model is consistent with the quantum equivalence between the first- and second-order formulations of gravity. This model coupled to matter fields was investigated in \cite{Brandt:2021qgh}. 
This could provide another way to address the issue of coupling fermions to the FOGR theory, without breaking its equivalence to the SOGR theory.  We hope to investigate this in the future.

In \cite{Brandt:2022kjo}, we have identified the lack of field redefinition invariance in the path integral of the standard LM formalism. To address this issue, we propose introducing a determinant factor into its measure to restore this invariance. This modification effectively resolves several drawbacks of the standard LM formalism, particularly the doubling of degrees of freedom and the additional one-loop contributions from the LM fields. 

The determinant factor can be exponentiated using ghost fields, which are analogous to the Lee-Yang ghost fields in the worldline formalism \cite{Bastianelli:1991be, Bastianelli:1998jb}. 
These ghosts are responsible for canceling the additional one-loop contributions arising from the LM fields. We have provided a diagrammatic analysis to verify our results. In particular, we demonstrated that the ghost fields do not spoil the restriction of the loop expansion to one-loop order. In addition, they are responsible for removing the unphysical contributions coming from the LM fields. We have shown that the LM fields are linked to Ostrogradsky instabilities, which can lead to an unbounded Hamiltonian and break the unitarity of the theory. 

In the framework of the modified LM formalism, this issue is resolved, since the Lee-Yang-like ghosts cancel out unphysical states associated with the Ostrogradsky ghosts. Indeed, we have shown that the degrees of freedom associated with the Ostrogradsky ghosts are removed effectively by the Lee-Yang-like ghost fields. We also recall that this issue could be resolved using the indefinite metric quantization approach, as shown in Ref.~\cite{Brandt:2021qgh}. 

Finally, we have extended the modified LM formalism to theories with singular actions. We have shown that the Lee-Yang-like ghosts become gauge fields and the FP quantization must be extended to accommodate these additional gauge fields. This implies that we need to  introduce ghosts of ghosts in order to obtain a consistent covariant path integral. We have presented the corresponding BRST symmetry of the resulting effective action. This procedure is applied to the first-order formulation of the YM theory, and the Feynman rules are detailed in Appendix~\ref{section:FeynmanRulesmLMYM}. At the end, we have established that the LM formalism commutes with the FP quantization. This implies that the LM fields can be viewed as purely quantum fields ensuring that the classical equations of motion are preserved. 

One can use the Hilbert-Einstein action within this framework to derive a quantum gravity theory that is both renormalizable and unitary. The classical limit of this theory should yield GR, and the quantum effects are restricted to one-loop order. This theory has two degrees of freedom corresponding to the two polarizations of the graviton, which are identical to the degrees of freedom of the Hilbert-Einstein gravity theory.

\appendix

\chapter{Derivation of Eq.~(\texorpdfstring{\ref{eq:b1:totalderivative})}{2.27)}}\label{section:DerivationTHETA}

In this appendix, we use a different notation than the rest of the work: $ {F} \to G$. Thus, Eq.~\eqref{eq:b1:totalderivative} can be rewritten as 
\begin{equation}\label{eq:b1:totalderivative1}
    G^{a \, \mu \nu} {\star}{G}^{a}_{\mu \nu} = \partial_{\mu} K^{\mu},  
\end{equation}
where 
\begin{equation}\label{eq:b21}
    K^{\mu} = \epsilon^{\mu \nu \rho \sigma} \left( A_{\nu}^{a} {G}^{a}_{\rho \sigma} - \frac{2}{3} g f^{a b c} A_{\nu}^{a} A_{\rho}^{b} {A}_{\sigma}^{c} \right).
\end{equation}
Recall that $ {\star} G^{a \mu \nu} = \epsilon^{\mu \nu \alpha \beta} G^{a}_{\alpha \beta} /2$.
In the Abelian case ($g=0$), the first term in the left-hand side of Eq.~\eqref{eq:b1:totalderivative} yields 
\begin{equation}\label{eq:b3a}
    \frac{1}{2}\epsilon^{\mu \nu \rho \sigma} ( \partial_{\mu} A_{\nu}^{a} - \partial_{\nu} A_{\mu}^{a} ) G^{a}_{\rho \sigma}= G^{a}_{\mu \nu}  {\star}{G}^{a \, \mu \nu} , 
\end{equation}
while the second term vanishes 
\begin{equation}\label{eq:b3b}
    - A_{\nu} \epsilon^{\nu \mu \rho \sigma } \partial_{\mu} G_{\rho \sigma} =0,
\end{equation}
since $G_{\rho \sigma} $ satisfies the Bianchi identity 
\begin{equation}\label{eq:b4:bianchi}
    \epsilon^{\mu \nu \rho \sigma} \partial_{\nu} G_{\rho \sigma} = \epsilon^{\mu \nu \rho \sigma} (\partial_{\nu} G_{\rho \sigma} + \partial_{\rho} G_{\sigma \nu} + \partial_{\sigma} G_{\nu \rho} )=0. 
\end{equation}
To demonstrate it we only needed the definition of $G$ (see Eq.~\eqref{eq:tensor}) and the Bianchi identity \eqref{eq:b4:bianchi}. It is also true (with appropriate modifications) in the general non-Abelian case.

In the general case (non-Abelian), the RHS of Eq.~\eqref{eq:b1:totalderivative} yields
\begin{equation}\label{eq:RHSofTD_1}
    F^{a}_{\mu \nu} {\star}{G}^{a \, \mu \nu} + A_{\nu} \epsilon^{\mu \nu \rho \sigma} \partial_{\mu} G_{\rho \sigma} - \frac{g}{3}  \epsilon^{\mu \nu \rho \sigma} \partial_{\nu} (f^{abc} A_{\nu}^{a} A_{\rho}^{b} A_{\sigma}^{c}).
\end{equation}

We see that the contribution in Eq.~\eqref{eq:b3a} is now equal to $ F^{a}_{\mu \nu} {\star}{G}^{a \, \mu \nu} $, where $ F^{a}_{\mu \nu} = G^{a}_{\mu \nu}  - g f^{abc} A^{b}_{\mu}  A^{c}_{\nu} $ (Abelian part of the field strength). Besides that, the second term in Eq.~\eqref{eq:RHSofTD_1} does not vanish since the Bianchi identity now is given by 
\begin{equation}\label{eq:BNAID}
    A_{ \nu} \epsilon^{\mu \nu \rho \sigma} D_{\mu} G_{\rho \sigma} = 0.
\end{equation}
By Eq.~\eqref{eq:RHSofTD_1}, we have that Eq.~\eqref{eq:b1:totalderivative1} is valid as long as
\begin{equation}\label{eq:RHSofTD_2}
    -g f^{abc} A_{\mu}^{b}  A_{\nu}^{c} {\star}{G}_{\rho \sigma}^{a} + A_{\nu}^{a} \epsilon^{\mu \nu \rho \sigma} \partial_{\nu} G_{\rho \sigma}^{a} - \frac{1}{3} g \epsilon^{\mu \nu \rho \sigma} \partial_{\nu} \left ( f^{abc} A_{\nu}^{a} A_{\rho}^{b} A_{\sigma}^{c}\right ) =0. 
\end{equation}
Using the Bianchi identity \eqref{eq:BNAID}, ignoring the terms of order $ g^{2} $ that vanish 
\begin{equation}\label{eq:TofOg2}
    g^{2} \epsilon^{\mu \nu \rho \sigma} A^{b}_{\mu} A^{c}_{\nu} A^{l}_{\rho} A^{m}_{\sigma} \left ( f^{abc} f^{alm} - f^{abl} f^{acm} + f^{abm} f^{acl}\right ) = 0 
\end{equation}
by using the Jacobi identity, in Eq.~\eqref{eq:RHSofTD_2}; we end up with 
\begin{equation}\label{eq:RHSofTD_3}
    g f^{abc} A_{\mu}^{b} A_{\nu}^{c} {\star}{F}_{\rho \sigma} - \frac{1}{3} g \epsilon^{\mu \nu \rho \sigma} \partial_{\mu} \left ( f^{abc} A_{\nu}^{a} A_{\rho}^{b} A_{\sigma}^{c}\right ),
\end{equation}
which vanishes since
\begin{equation}\label{eq:Derivative}
    \begin{split}
        g \epsilon^{\mu \nu \rho \sigma} \partial_{\mu} \left ( f^{abc} A_{\nu}^{a} A_{\rho}^{b} A_{\sigma}^{c}\right ) &=  g \epsilon^{\rho \sigma \mu \nu} \partial_{\rho} \left ( f^{abc} A_{\sigma}^{a} A^{b}_{\mu} A^{c}_{\nu}\right ) \\
                                                                                                                          &= \frac{g}{2 } \epsilon^{\mu \nu \rho \sigma} f^{abc} \partial_{[ \rho} A_{\sigma ]}^{a} A^{b}_{\mu} A^{c}_{\nu} + \text{ perms.} \\ 
                                                                                                                          &= 3 {\star}{F}_{\rho \sigma}^{a} g f^{abc} A_{\mu}^{b} A_{\nu}^{c}.
    \end{split}
\end{equation}

\chapter{Feynman rules} \label{section:FeynmanRules}

In this appendix, we present the Feynman rules used in this work. For more details, we refer to \cite{Capper:1973pv, McKeon:2020lqp, Brandt:2020vre, martins-filho:2021}.
In our convention, all vertex momenta are directed inward, and their sum vanishes. 

\section{Yang-Mills theory}

The propagators in the SOYM (see Section~\ref{sec:SOYM}) are given by
\begin{subequations}\label{eq:propYMappen}
\begin{align}
    \label{fig:FRpg}
    \vcenter{\hbox{\includegraphics[scale=0.6]{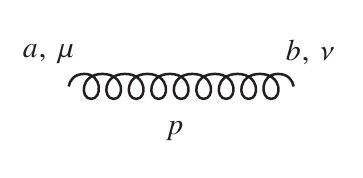}}} 
   &\qquad   -\frac{i}{p^{2}} \left ( \eta_{\mu \nu} -(1 - \alpha) \frac{p_{\mu} p_{\nu}}{p^{2}} \right ) \delta^{ab},
\\\label{fig:FRpgh}
    \vcenter{\hbox{\includegraphics[scale=0.6]{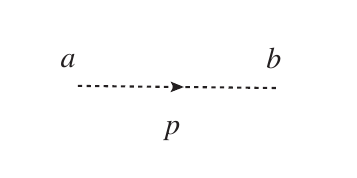}}} 
   &\qquad   \frac{i}{p^{2}}  \delta^{ab}.
\end{align}
\end{subequations}

The interaction vertices are 
\begin{subequations}\label{eq:FR2orYM}
    \allowdisplaybreaks
\begin{align}
    \label{fig:FRAAAc}
    & \vcenter{\hbox{\includegraphics[scale=0.65]{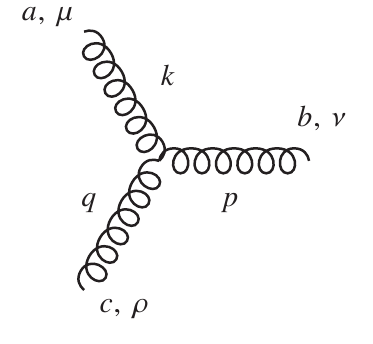}}} 
    \qquad      V^{abc}_{\mu \nu \rho} (k,p,q),
    \\\label{fig:FRAAAAc}
    & \vcenter{\hbox{\includegraphics[scale=0.6]{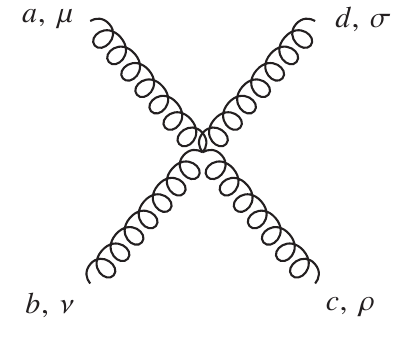}}} 
    \qquad V^{abcd}_{\mu \nu \rho \sigma },
    \\\label{fig:FRAccc}
    & \vcenter{\hbox{\includegraphics[scale=0.6]{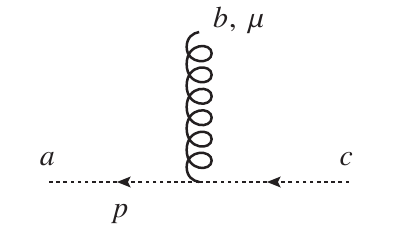}}} 
    \qquad gf^{abc} p_{\mu},
\end{align}
\end{subequations}
where
\begin{equation}\label{eq:FRAAA}
    V^{abc}_{\mu \nu \rho} (k,p,q) = g f^{abc} \left [ \eta_{\mu \nu} (k-p)_{\rho} + \eta_{\nu \rho} (p-q)_{\mu} + \eta_{\rho \mu} (q-k)_{\nu}  \right ]
\end{equation}
and
\begin{equation}\label{eq:FRAAAA}
\begin{split} 
    V^{abcd}_{\mu \nu \rho \sigma} \equiv -i & g^{2}  \big[  f^{abe} f^{cde} (  \eta_{\mu \rho} \eta_{\nu \sigma} - \eta_{\mu \sigma} \eta_{\nu \rho} ) \\ & \quad + f^{ace} f^{bde} ( \eta_{\mu \nu} \eta_{\rho \sigma} - \eta_{\mu \sigma} \eta_{\rho \nu} ) \\ & \quad + f^{ade} f^{bce} ( \eta^{\mu \nu} \eta^{\rho \sigma} - \eta_{\mu \rho} \eta_{\nu \sigma} )\big].
\end{split}
\end{equation}
The spring lines represent the gauge field $A$, and the dotted lines the ghosts $c $ and $ \bar{c} $.

In the FOYM (see Section~\ref{section:FOYM}), we have the following Feynman rules:
\begin{subequations}\label{eq:FR1orderYM1}
\begin{align}
    \label{fig:FRpgA}
    \vcenter{\hbox{\includegraphics[scale=0.60]{fig/FR/propA-eps-converted-to.pdf}}}& \quad \quad 
   -\frac{i}{p^{2}} \left ( \eta_{\mu \nu} -(1 - \alpha ) \frac{p_{\mu} p_{\nu}}{p^{2}} \right ) \delta^{ab},
       \\
    \label{fig:FRpgAF}
    \vcenter{\hbox{\includegraphics[scale=0.6]{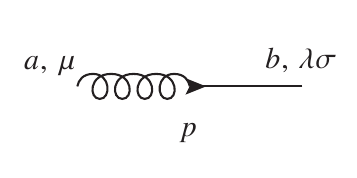}}}& \quad \quad   
- \frac{p_{\lambda } \eta_{\sigma \mu} - p_{\sigma} \eta_{\lambda \mu}}{p^{2}}\delta^{ab} , \\
    \label{fig:FRpgFA}
    \vcenter{\hbox{\includegraphics[scale=0.6]{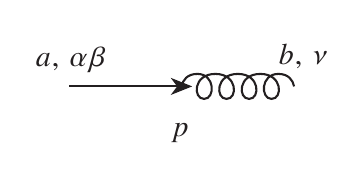}}}& \quad \quad   
+\frac{p_{\alpha } \eta_{\beta \nu} - p_{\beta} \eta_{\alpha \nu}}{p^{2}}\delta^{ab} 
,\\
    \label{fig:FRpgFF}
    \vcenter{\hbox{\includegraphics[scale=0.6]{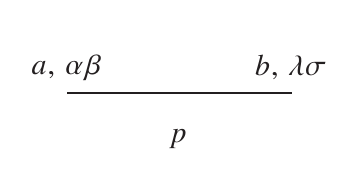}}}& \quad \quad   
    2 i  [ I_{\alpha \beta \lambda \sigma} + L_{\alpha \beta \lambda \sigma}  (p)] \delta^{ab}
    ,\\
    \label{fig:FRpghs}
    \vcenter{\hbox{\includegraphics[scale=0.60]{fig/FR/propghost-eps-converted-to.pdf}}} & \quad \quad  \quad \quad \quad 
\frac{i}{p^{2}}  \delta^{ab},
\end{align}
\end{subequations}
where 
\begin{equation}\label{eq:defLYM}
        L_{\alpha \beta \lambda \sigma} (p) = -\frac{1}{2 p^{2} } \left(p^{\alpha }
   p^{\lambda } \eta^{\beta \sigma
   }-p^{\beta } p^{\lambda }
   \eta^{\alpha \sigma }-p^{\alpha }
   p^{\sigma } \eta^{\beta \lambda
   }+p^{\beta } p^{\sigma }
   \eta^{\alpha \lambda }\right),
    \end{equation}
\begin{equation}\label{eq:defIDa}
    I^{\mu \nu \rho \sigma} \equiv \frac{\eta^{\mu \rho} \eta^{\nu \sigma} - \eta^{\mu \sigma} \eta^{\nu \rho} }{2}.
\end{equation}
The auxiliary field $F^{a \, \mu \nu} $ is represented by solid lines.

The vertices reads
\begin{subequations}\label{eq:vertYM1}
    \allowdisplaybreaks
\begin{align}
    \label{fig:FRFAAdiag}
    & \vcenter{\hbox{\includegraphics[scale=0.65]{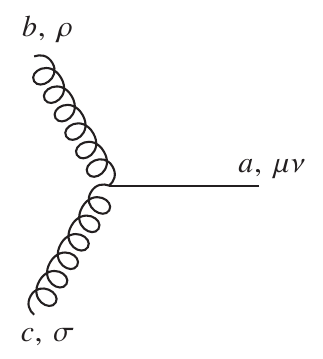}}} \qquad 
    \mathcal{V}^{abc}_{\mu \nu \, \rho \sigma },\\
    \label{fig:FRcAc}
    & \vcenter{\hbox{\includegraphics[scale=0.6]{fig/FR/ghostvertex-eps-converted-to.pdf}}} \quad \quad   
    gf^{abc} p_{\mu};
\end{align}
\end{subequations}
where 
        \begin{equation}\label{eq:FRFAA}
            \mathcal{V}^{abc}_{\mu \nu \, \rho \sigma} = -i g f^{abc} I_{\mu \nu \rho \sigma }. 
        \end{equation}
        Recall that these rules are only valid in the Lorenz gauge $ \partial_{\mu} A^{a \mu} =0$.

\section{Gravity}

The propagator of the ghost in the De Donder gauge is given by 
\begin{equation}
    \label{fig:FRpghEH}
    \vcenter{\hbox{\includegraphics[scale=0.6]{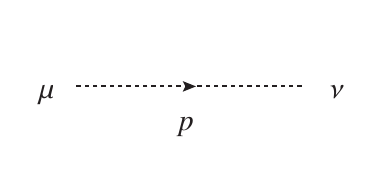}}}
\qquad
    -\frac{i}{p^{2}} \eta^{\mu \nu}
\end{equation}
and the ghost vertex by
\begin{equation}\label{eq:EHghost}
    \vcenter{\hbox{\includegraphics[scale=0.6]{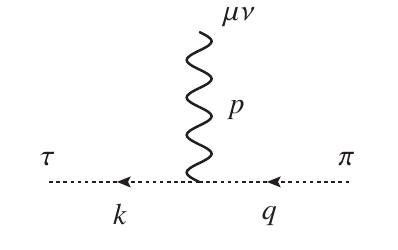}}} 
\qquad
   \frac{ i \kappa }{2}\left[\eta_{\pi \tau} 
    (q^{\mu} k^{\nu } + q^{\nu} k^{\mu} ) - q_{\tau} (p^{\mu} \delta_{\pi}^{ \nu} + p^{\nu} \delta_{\pi}^{\mu} )\right].
\end{equation}
The graviton field $ \phi^{\mu \nu} $ is represented by wavy lines. The auxiliary field $ H_{\mu \nu}^{\lambda} $ of the diagonal first-order formulation of gravity is represented by solid lines. The ghost fields $ d_{\mu} $, $ \bar{d}_{\mu} $ are represented by dotted lines.

The remaining rules of the diagonal FOGR (see Section~\ref{sec:DFOEH}) are 
\begin{subequations}\label{eq:FREH1dprapp}
    \allowdisplaybreaks
    \begin{align}
     \vcenter{\hbox{\includegraphics[scale=0.6]{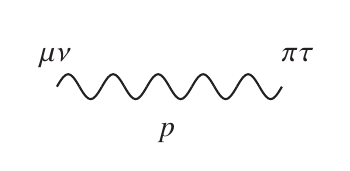}}} 
    &\qquad \mathcal{P}^{\mu \nu \pi \tau} (p),
    \\
     \vcenter{\hbox{\includegraphics[scale=0.6]{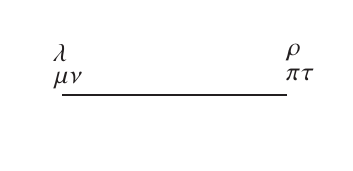}}} 
    & \qquad \tensor*{\mathcal{D}}{*_{\mu \nu}^{\lambda}_{\pi \tau}^{\rho}},
\\
     \vcenter{\hbox{\includegraphics[scale=0.6]{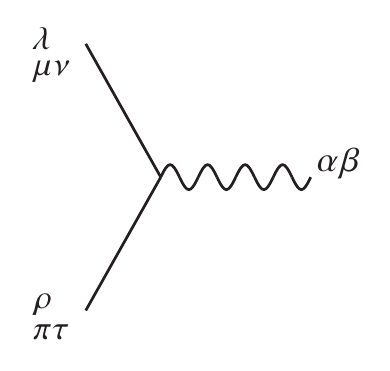}}} 
  & 
\qquad
\tensor*{V}{*_{\pi \tau}^{\rho}^{\lambda}_{\mu \nu}^{\alpha \beta}},
\\
     \vcenter{\hbox{\includegraphics[scale=0.6]{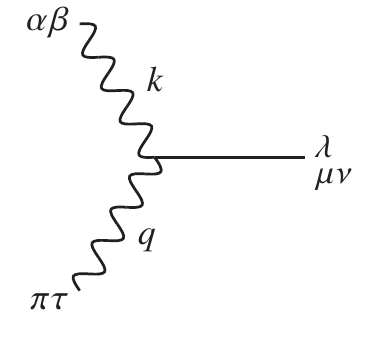}}} 
 & 
\qquad
\tensor*{V}{*^{\lambda}_{\mu \nu}^{\pi \tau \alpha \beta}} (q,k),
\\
     \vcenter{\hbox{\includegraphics[scale=0.6]{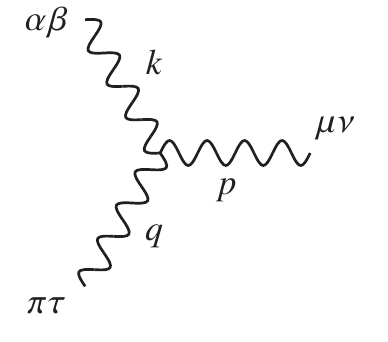}}} 
   & 
   \qquad 
    V^{\mu \nu \pi \tau \alpha \beta } (p,q,k);
\end{align}
\end{subequations}
where the propagators are 
\begin{equation}\label{eq:propPEH}
    \begin{split}
        \mathcal{P}^{\mu \nu \pi \tau} (k) ={}& 
 -i\frac{\eta^{\pi \nu } \eta^{\mu \tau }+\eta^{\pi \mu } \eta^{\nu
 \tau } -(2- \alpha )\eta^{\pi \tau } \eta^{\mu \nu }}{k^2}  
   \\
   & +i(1-\alpha)\frac{ 
       k^{\pi } k^{\mu } \eta^{\nu \tau }+k^{\pi
   } k^{\nu } \eta^{\mu \tau }+\eta^{\pi \nu } k^{\mu } k^{\tau }+\eta^{\pi \mu } k^{\nu }
   k^{\tau }}{k^2}, 
   \\ & -2i(1-\alpha)\frac{ 
\eta^{\pi \tau } k^{\mu } k^{\nu }+k^{\pi } k^{\tau } \eta^{\mu \nu}
}{k^2},
   \end{split}
\end{equation}
\begin{equation}\label{eq:propHEH}
    \begin{split}
        \tensor*{\mathcal{D}}{*_{\mu \nu}^{\lambda}_{\pi \tau}^{\rho}} ={}& 
        \frac{i \kappa^{2}}{4}\eta^{\lambda \rho}
   \left (
     \eta_{\pi \mu} \eta_{\tau \nu} + \eta_{\pi \nu} \eta_{\tau \mu}
   -\frac{2}{(D-2)}  \eta_{\mu \nu} \eta_{\pi \tau} 
 \right ) 
    \\
                                                                                      & - \frac{i \kappa^{2} }{4} \left ( \eta_{\tau \mu} \delta_{\nu}^{\rho} \delta_{\pi}^{\lambda} + \eta_{\pi \mu} \delta_{\nu}^{\rho} \delta_{\tau}^{\lambda} +  \eta_{\tau \nu} \delta_{\mu}^{\rho} \delta_{\pi}^{\lambda} +  \eta_{\pi \nu} \delta_{\mu}^{\rho} \delta_{\tau}^{\lambda}\right );
   \end{split}
\end{equation}
and the vertices by
\begin{equation}\label{eq:FRHHphieq}
    \begin{split}
        V_{\pi \tau}^{\rho} {}{}_{\mu \nu}^{\lambda} {}_{}^{\alpha \beta} ={}&
    \frac{i}{8 \kappa } \left \{ \left [ \left (
                \frac{\delta^{\rho}_{\pi } \delta^{\alpha}_{\nu } \delta^{\beta}_{\tau } \delta^{\lambda}_{ \mu
            }}{D-1}-\delta^{\lambda}_{\pi } \delta^{\alpha}_{\nu } \delta^{\beta }_{\tau } \delta^{\rho }_{\mu }
    + \alpha \leftrightarrow \beta \right )  + \mu \leftrightarrow \nu \right ]  + \pi \leftrightarrow \tau
\right \}  \\ &  + ( \mu \nu, \lambda ) \leftrightarrow (\pi \tau , \rho ),
\end{split}
\end{equation}
\begin{equation}\label{eq:FRHphiphieq}
    \begin{split}
        V^{\lambda}_{\mu \nu}{}^{\pi \tau \alpha \beta} (q,k) ={}&\frac{k^{\theta} }{4} \left \{ \left [ \left (
                \frac{\delta^{\pi}_{\mu } \delta_{\nu}^{ \lambda } 
        }{D-1} \tensor*{\mathcal{D}}{*^{\alpha \beta}_{\theta}^{\tau \rho}_{\rho}}  - \delta^{\pi}_{\mu} \tensor*{\mathcal{D}}{*^{\alpha \beta}_{\theta}^{\tau \lambda}_{\nu}}   + \pi\leftrightarrow \tau\right )  + \alpha  \leftrightarrow \beta  \right ]  + \mu \leftrightarrow \nu
    \right \}  \\ & + ( k, \alpha \beta ) \leftrightarrow (q, \pi \tau  ), 
\end{split} 
\end{equation}
\begin{equation}\label{eq:FRphiphiphieq}
    \begin{split}
        V^{\mu \nu \pi \tau \alpha \beta } (p,q,k) ={}&
    -i \kappa \frac{q^{\kappa} k^{\theta} }{8} \Bigg\{ \left [ \left (
                \frac{ \tensor*{\mathcal{D}}{*^{\alpha \beta}_{\theta}^{\mu \rho}_{\rho}}  \tensor*{\mathcal{D}}{*^{\pi \tau}_{\kappa}^{\nu \gamma}_{\gamma}} 
}{D-1} 
- \tensor*{\mathcal{D}}{*^{\alpha \beta}_{\kappa}^{\mu \rho}_{\gamma}}   
\mathcal{D}^{\pi \tau}_{\theta} {}^{\nu \gamma }_{ \rho }
+ \pi\leftrightarrow \tau\right )  
+ \alpha  \leftrightarrow \beta  \right ]  
\\ & + \mu \leftrightarrow \nu \Bigg\} + (p,\mu \nu ) \leftrightarrow (q, \pi \tau ) + ( p , \mu \nu ) \leftrightarrow ( k, \alpha \beta ).
\end{split}
\end{equation}
The symbol $ \leftrightarrow$ denotes the first term with an interchange of indices (and momenta). 

The FOGR (see Section~\ref{section:FOEH}) Feynman rules are given in Ref. \cite[Eqs. (28--30)]{Brandt:2015nxa}. 

\section{Modified Lagrange multiplier formalism}\label{section:FeynmanRulesmLMYM}

\begin{figure}[h]
    \centering
    \includegraphics[width=0.95\textwidth]{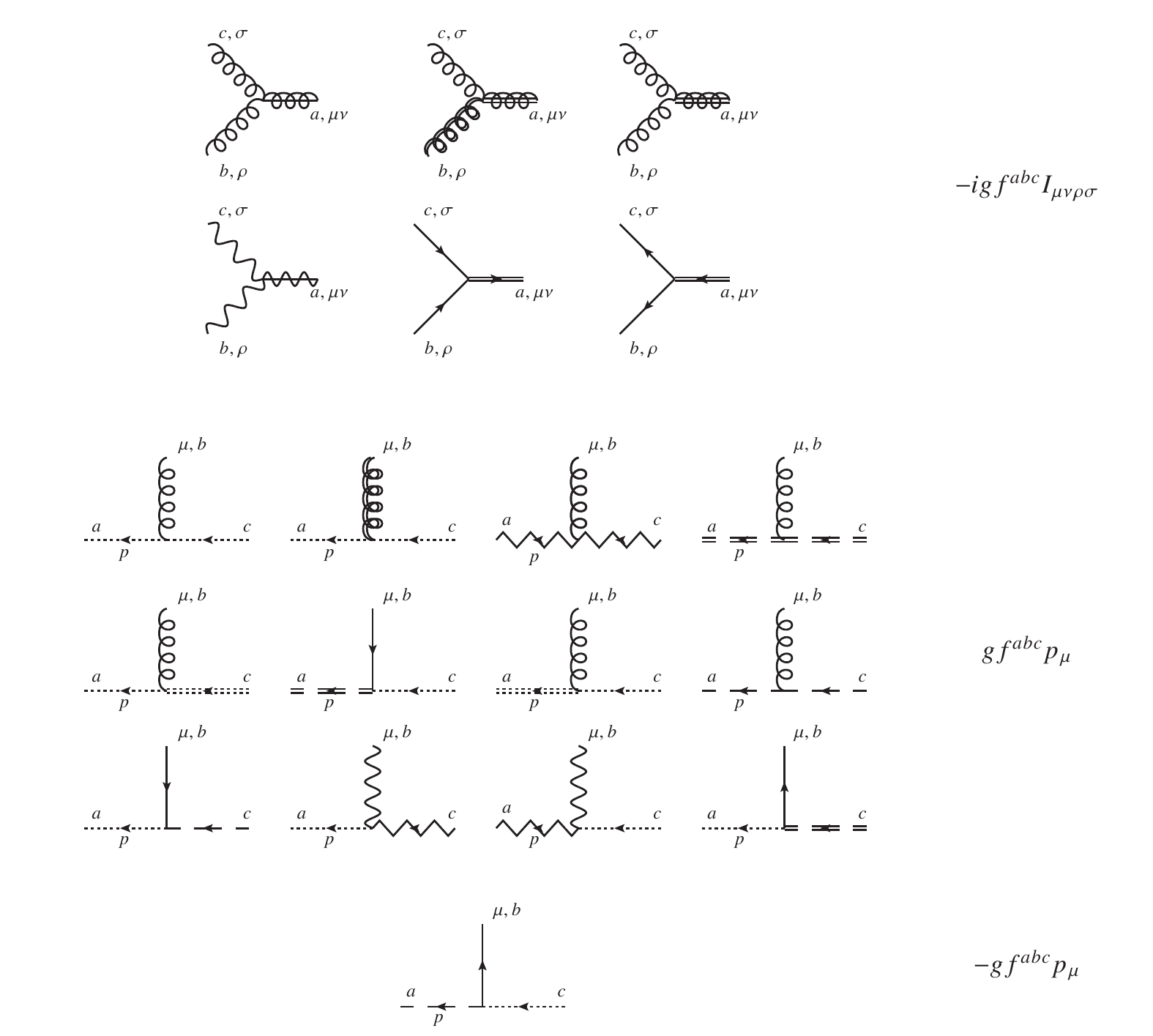}
    \caption{Vertices of the FOYM theory in the modified LM formalism.}
\label{fig:FR2mLMYM}
\end{figure}

\begin{figure}[h]
    \subbottom[Standard LM formalism.]{\includegraphics[width=0.475\textwidth]{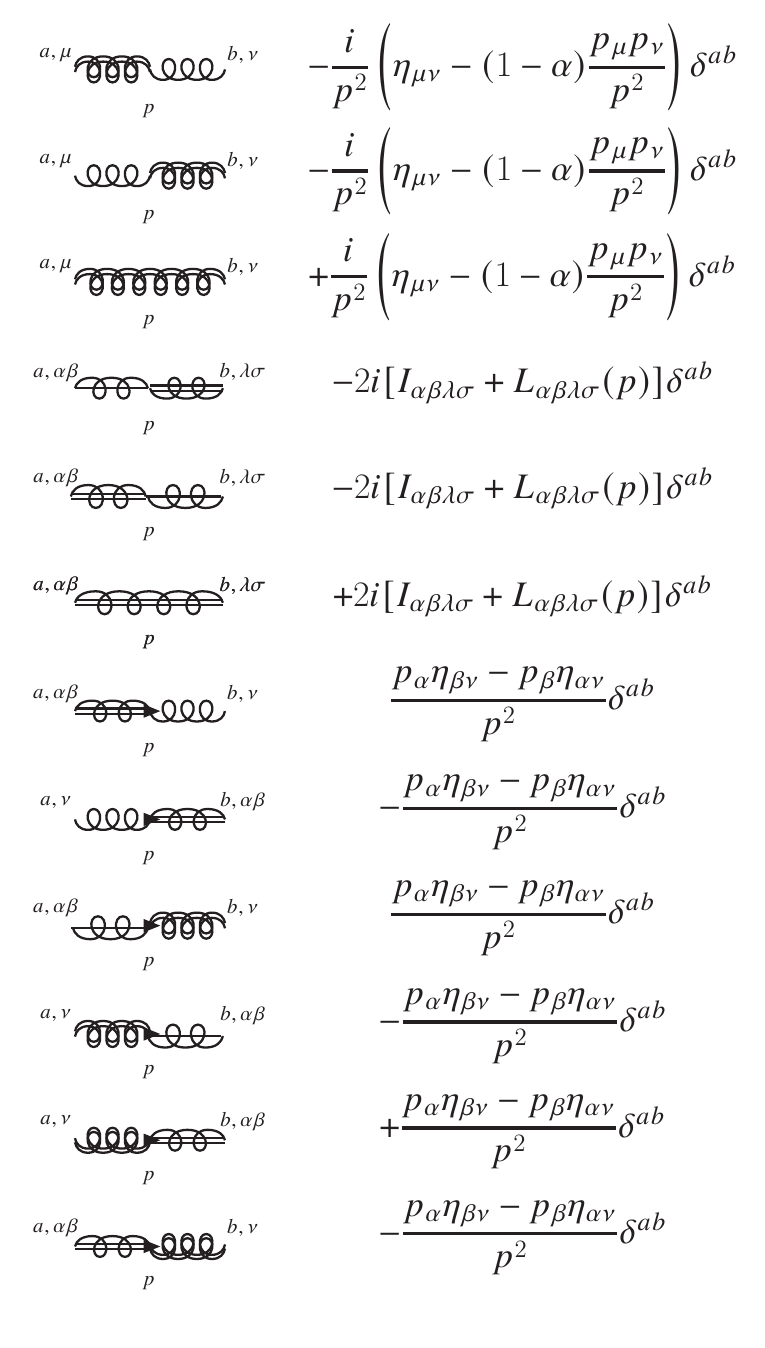}}
    \subbottom[Additional rules in the modified LM formalism.]{\includegraphics[width=0.475\textwidth]{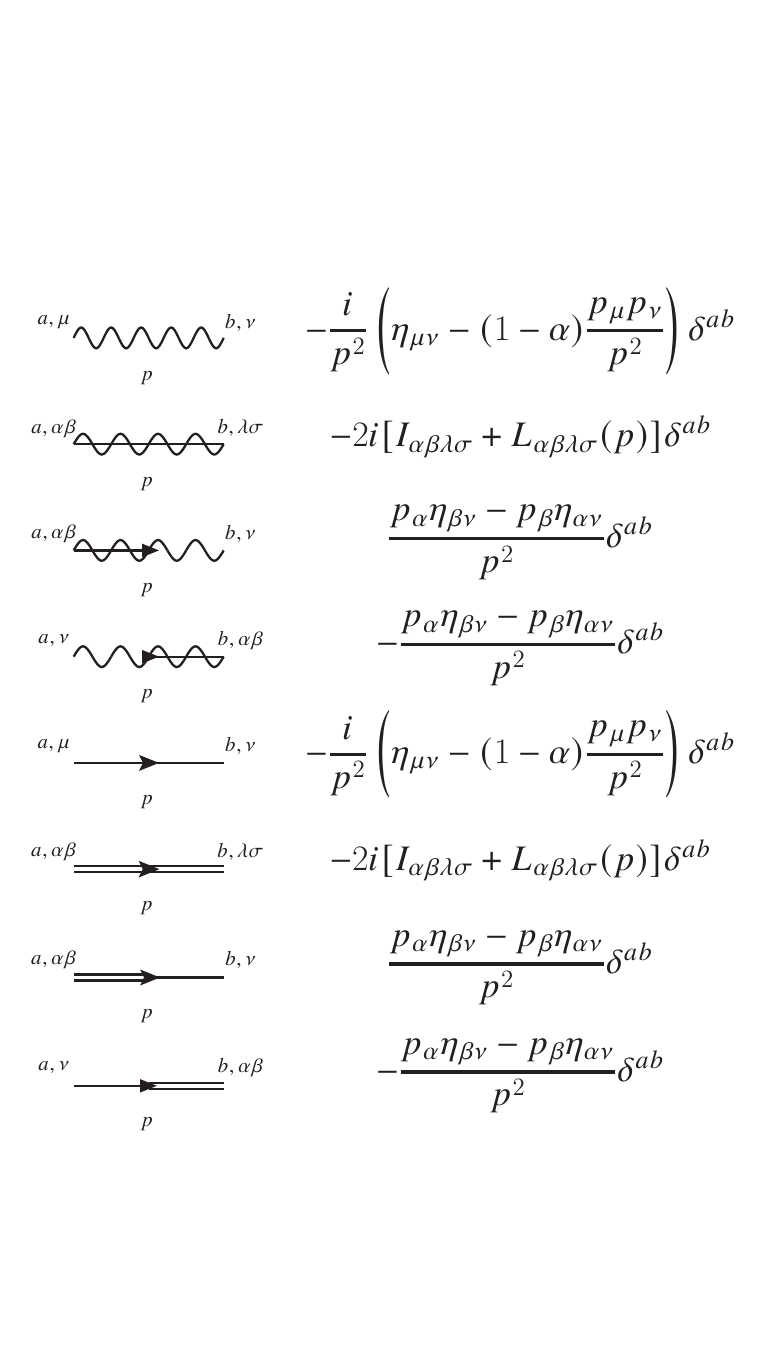}}

    \subbottom[FP ghost sector.]{\includegraphics[width=0.95\textwidth]{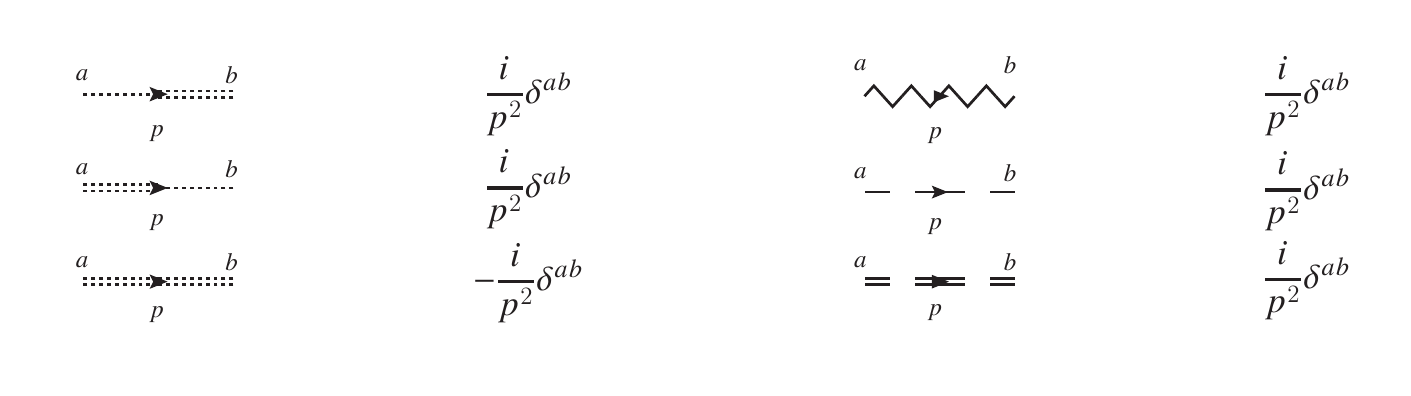}}
\caption{Propagators of the FOYM theory in the framework of the modified LM formalism. We assume that momenta flow to the right.}
\label{fig:FR1mLMYM}
\end{figure}

The Feynman rules of the FOYM in the modified LM formalism (see Section~\ref{section:YMinthemodLM}) are shown in Figs.~\ref{fig:FR2mLMYM} and~\ref{fig:FR1mLMYM}. We represented the fields $ A_{\mu}^{a} $, $ \lambda_{\mu}^{a} $, $ {\theta}_{\mu}^{a} $ ($ \bar{\theta}_{\mu}^{a} $) and $ \chi_{\mu}^{a} $ by spring, double spring, solid, and wavy lines, respectively. The auxiliary fields $ \mathcal{F}_{\mu \nu}^{a} $, $ \Lambda_{\mu \nu}^{a} $, $ \chi_{\mu \nu}^{a} $ and $ \theta_{\mu \nu}^{a} $ ($ \bar{\theta}_{\mu \nu}^{a} $) are represented by spring-solid, spring-double-solid, wavy-solid, double solid lines. The associated FP ghost fields $c^{a} $ ($ \bar{c}^{a}$), $ d^{a} $ ($ \bar{d}^{a}$), $ e^{a} $ ($ \bar{e}^{a}$), $ \gamma^{a} $ ($ \tilde{\gamma}^{a}$), and $ \epsilon^{a} $ ($ \tilde{\epsilon}^{a}$) by dotted, double dotted, zigzag, dashed, and double dashed lines.

\chapter{First-Order formulation of the scalar theory}\label{section:FOspin0}

In this short appendix, we present the first-order formulation of the standard Lagrangian for a spin-0 field $ \phi $ with potential $V( \phi )$.

The second-order Lagrangian is given by
\begin{equation}\label{eq:SOscalarLag}
    \mathcal{L}_{\text{scalar}}^{(2)} = \frac{1}{2} ( \partial_{\mu} \phi )^{2} - V ( \phi ).
\end{equation}
Then, the first-order Lagrangian will be expressed as 
\begin{equation}\label{eq:FOscalarLag}
    \mathcal{L}_{\text{scalar}}^{(1)} = - \frac{1}{2} G_{\mu} G^{\mu} +  G_{\mu} \partial^{\mu} \phi - V (\phi ).
\end{equation}
The field $ G_{\mu} $ is an auxiliary field analog to $ \mathcal{F}_{\mu \nu}^{a} $ of the FOYM theory.

The quantum equivalence can be verified straightforwardly. We need to shift the field $ G_{\mu} $ in the generating functional of the first-order spin-0 theory by its classical value 
\begin{equation}\label{eq:shiftG}
    G_{\mu} \to G_{\mu} + \partial_{\mu} \phi +  J_{\mu} ,
\end{equation}
where $ J_{\mu} $ is the source of the auxiliary field. This shift is analog to Eq.~\eqref{eq:shiftws}. The resulting generating functional will be analog to Eq.~\eqref{eq:fgYM1wsp2}, which demonstrates the quantum equivalence between the two formulations. Besides that, structural identities can also be obtained, for instance,
\begin{equation}\label{eq:SIS0}
    \langle 0|T \phi G_{\mu} | 0 \rangle = \langle 0|T \phi \partial_{\mu} \phi | 0 \rangle \quad \text{and} \quad \langle 0|T G_{\mu} G_{\nu}| 0 \rangle = \langle 0|T \partial_{\mu} \phi \partial_{\nu} \phi| 0 \rangle - i \eta_{\mu \nu}.
\end{equation}

\chapter{Dimensional regularization of massless tadpoles}\label{section:tadpoles}

In this appendix, we compute a massless tadpole-like diagram (a loop diagram independent of external momenta) at finite temperature, and at zero temperature $ T = 0 $, using the imaginary time formalism \cite{Bellac:2011kqa}. To achieve this, we apply dimensional regularization to regularize loop integrals in the $ \phi^{4} $ theory. This comparison illustrates that massless tadpole-like diagrams, which vanish at zero temperature when using dimensional regularization, may not vanish at finite temperature.

\begin{figure}[ht]
    \centering
    \includegraphics[scale=0.8]{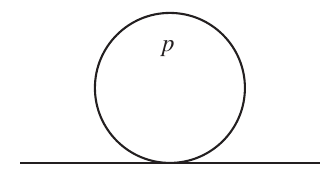}
\caption{One-loop tadpole-like diagram in the $ \phi^4$ theory. Solid lines represent the scalar field $ \phi $.}
    \label{fig:tadpole1}
\end{figure}

\begin{figure}[ht]
    \centering
    \includegraphics[scale=0.8]{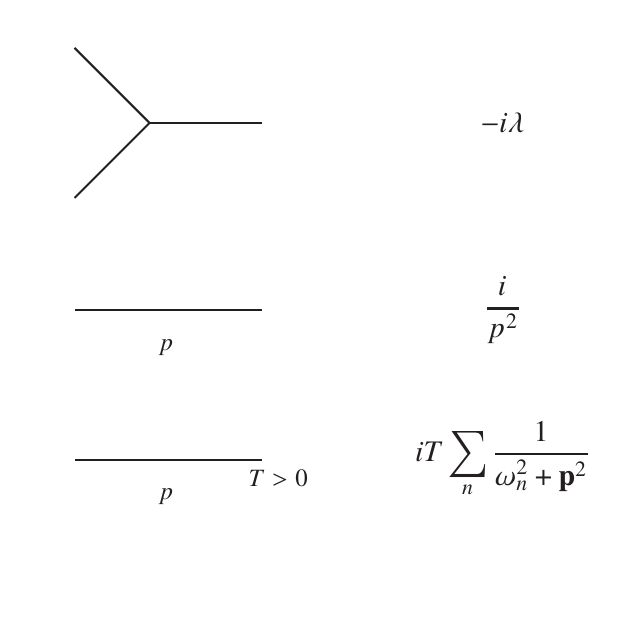}
    \caption{Feynman rules in the massless $ \phi^{4} $ theory at finite and zero temperature.  The Matsubara frequencies are defined as $ \omega_{n} = 2 \pi n T$ ($ k_{B} =1 )$, where $n$ is an integer \cite{Bellac:2011kqa}. Note that, $ p^{2} \equiv p_{\mu} p^{\mu} = p_{0}^{2} - \mathbf{p}^{2} $, $ \mathbf{p} $ is the momentum vector. However, in the imaginary time formalism (after a Wick rotation $ p_{0} \to i p_{0} $) we have that $ -p_{E}^{2} = p_{0}^{2} + \mathbf{p}^{2} $. }
    \label{fig:FRscalar}
\end{figure}

We consider the one-loop tadpole-like diagram in Fig.~\ref{fig:tadpole1}. This is the one-loop contribution to the self-energy of the scalar field $ \phi $ in the standard massless $ \phi^{4} $ theory described by the action 
\begin{equation}\label{eq:lagphi4}
    S_{\phi^{4} }  = \int \mathop{d^{4} x} \left( \frac{1}{2} \partial_{\mu} \phi \partial^{\mu} \phi - \frac{ \lambda }{4!} \phi^{4}\right).
\end{equation}
The Feynman rules required to compute the diagram in Fig.~\ref{fig:tadpole1} are presented in Fig.~\ref{fig:FRscalar}.

\section{Zero temperature}\label{section:ZTemp}

At zero temperature, we can use the Feynman rules given in Fig.~\ref{fig:FRscalar} to compute the diagram in Fig.~\ref{fig:tadpole1} as 
\begin{equation}\label{eq:I[D]}
    I[D] = \frac{\lambda}{2}  \int \frac{\mathop{d^{D} p}}{( 2 \pi )^{D} } \frac{1}{p^2- M^2},
\end{equation}
where $1/2$ is the combinatorial factor and $ M$ is a mass parameter. Note that, we extended the loop integral to $D$ dimensional which is required to proceed with the dimensional regularization. To obtain the correct result, we set $ M = 0$ at the end of the computation of Eq.~\eqref{eq:I[D]}, since the $ \phi^{4} $ theory described by EQ.~\eqref{eq:lagphi4} is massless. 

In order to compute the integral $ I [D] $, we can use a Wick rotation: $ p^2 = - p_{E}^{2} $ and $ d^{d} p = i d^{D} p_{E} $. After some algebra, we arrive at \cite[Eq. (D.10)]{martins-filho:2021} 
\begin{equation}\label{eq:I21}
    I[D] = 
    -i \frac{2 \pi^{D/2}}{ \Gamma (D/2)} \frac{1}{(2 \pi )^{D} } M^{D-2} \frac{\Gamma (1-D/2) }{\Gamma (1)} \to 0,
\end{equation}
where the arrow indicates its value when $ M =0$ (we set $D =4 -2 \epsilon $ in which $ \epsilon $ is a small nonzero value). This shows that the tadpole-like diagram in Fig.~\ref{fig:tadpole1} vanishes using dimensional regularization at zero temperature.

Now, note that the scaleless integral 
\begin{equation}\label{eq:I[D]1}
    I_0 = 
    \int \mathop{d^{D} p} \frac{1}{p^{2n} } = 4 \pi \int_{0}^{\infty} dp p^{D -1 - 2n} 
\end{equation}
should vanish as there is no mass parameter. However, we may have both IR (infrared) and UV (ultraviolet) divergences. Thus, we split up the above integral into two pieces as follows
\begin{equation}\label{eq:splitID}
    \underbrace{\int_{0}^{ \Lambda } dp p^{D -1 -2n}}_{ \text{IR} }
+ 
\underbrace{\int_{\Lambda }^{\infty} dp p^{D -1-2n} }_{ \text{UV} }
     \xrightarrow[D \, = \,  4 +2 \epsilon_{ \text{IR} } ]{D \, = \,  4 - 2 \epsilon_{ \text{UV} } }
     \int_{0}^{ \Lambda } dp p^{3-2n +2 \epsilon_{\text{IR}} } 
+ 
\int_{\Lambda }^{\infty} dp p^{3-2n - 2 \epsilon_{\text{UV}} },
\end{equation}
where $ \Lambda > 0$.
Setting $ \epsilon_{\text{UV}} = \epsilon_{ \text{IR} } = \epsilon>0$ leads to  
\begin{equation}\label{eq:splitID2}
     \int_{0}^{ \Lambda } dp p^{3-2n +2 \epsilon_{\text{IR}} } 
+ 
\int_{\Lambda }^{\infty} dp p^{3-2n - 2 \epsilon_{\text{UV}} }
=
\int_{0+}^{\infty} dp p^{3-2n - 2 \epsilon_{} } \propto \epsilon \to 0
\end{equation}
by dimensional regularization \cite{Leibbrandt:1975dj}.
For instance, $n=2$, we have that 
\begin{equation}\label{eq:splitID2n2}
     \int_{0}^{ \Lambda } dp p^{-1 +2 \epsilon_{\text{IR}} } 
+ 
\int_{\Lambda }^{\infty} dp p^{-1 - 2 \epsilon_{\text{UV}} }
\propto \left ( \frac{1}{\epsilon_{\text{UV}}} -\frac{1}{ \epsilon_{IR}}\right )=0.
\end{equation}
The IR and UV divergences cancel each other.

\section{Finite temperature}\label{section:FTscalar}
In this section, we compute the diagram in Fig.~\ref{fig:tadpole1} at finite temperature using the imaginary time formalism. Using the Feynman rules in Fig.~\ref{fig:FRscalar}, we have that 
\begin{equation}\label{eq:IT1}
    I_{T} [D] = \frac{ \lambda }{2}  T \sum_{n} \int \frac{\mathop{d^{D-1} p}}{(2 \pi )^{D-1}} \frac{1}{\omega_{n}^{2} + \mathbf{p}^{2}}.
\end{equation}

\begin{figure}[ht]
    \centering
    \includegraphics[scale=0.8]{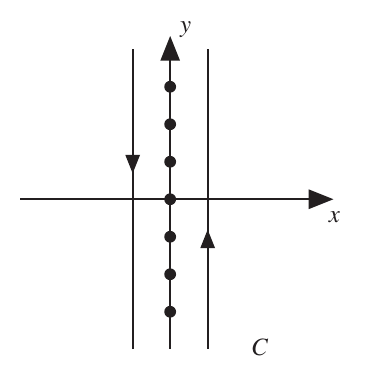}
    \caption{The counterclockwise contour $C$, which closes at infinite.}
    \label{fig:contour}
\end{figure}

We can express the sum as a contour integral \cite{Kapusta:2006pm}
\begin{equation}\label{eq:IT122}
    I_{T} [D]
    = \frac{\lambda}{2} \frac{1}{2 \pi i} \ointctrclockwise_{C} d p_{0} \frac{1}{2} \int \frac{\mathop{d^{D-1} p}}{(2 \pi )^{D-1}} \frac{1}{   \mathbf{p}^{2} - p_{0}^{2} } \coth \left ( \frac{p_{0}}{2 T}\right ),
\end{equation}
where the contour $C$ is shown in Fig.~\ref{fig:contour}.
We have used the Cauchy integral theorem. The poles of $ \coth(p_0/2T)$ are $ 2 \pi i T n$, where $n$ is an integer. Note that, the poles are equal to the Matsubara frequency in the imaginary axis $ i \omega_{n} $. 

Using that 
\begin{equation}\label{eq:IDcoth}
    \frac{1}{2} \coth \left( \frac{x}{2} \right) = 
    \frac{e^{x/2} + e^{-x/2}}{e^{x/2} - e^{-x/2}} = \frac{1}{2} + \frac{1}{e^{x} -1} = -\frac{1}{2} - \frac{1}{e^{-x} -1},
\end{equation}
we obtain 
\begin{equation}\label{eq:IT12}
    \begin{split}
    I_{T} [D]
    = \frac{\lambda}{2} \frac{1}{2 \pi i} 
    & \Bigg[
        \int_{-i \infty + \eta }^{i \infty + \eta } d p_{0}  \int \frac{\mathop{d^{D-1} p}}{(2 \pi )^{D-1}} \frac{1}{   \mathbf{p}^{2} - p_{0}^{2} } \left( \frac{1}{2} + \frac{1}{\exp  ( p_{0} / T  ) 
    -1 }  
\right) 
\\
    & + \int_{i \infty - \eta }^{-i \infty - \eta } d p_{0}  \int \frac{\mathop{d^{D-1} p}}{(2 \pi )^{D-1}} \frac{1}{   \mathbf{p}^{2} - p_{0}^{2} } \left( -\frac{1}{2} -\frac{1}{\exp  ( -p_{0} / T  ) 
    -1 }  
\right)
\Bigg], 
\end{split}
\end{equation}
where $ \eta $ is a small positive real number. We can simplify the above expression by changing the variable of the second integral in Eq.~\eqref{eq:IT12} to $ - p_{0} $ yielding 
\begin{equation} \label{eq:IT13}
\begin{split}
    I_{T} [D]
    = \frac{\lambda}{2} \frac{1}{2 \pi i} 
     \Bigg[
        \int_{-i \infty }^{i \infty } d p_{0}  \int \frac{\mathop{d^{D-1} p}}{(2 \pi )^{D-1}} \frac{1}{   \mathbf{p}^{2} - p_{0}^{2} } 
        + \int_{-i \infty + \eta }^{i \infty + \eta } d p_{0}  \int \frac{\mathop{d^{D-1} p}}{(2 \pi )^{D-1}} \frac{2}{   \mathbf{p}^{2} - p_{0}^{2} } N_{\text{B}} ( p_{0} )
\Bigg], 
\end{split}
\end{equation}
where $ N_{\text{B}} ({p_{0}}) = ( \exp( p_0 /T) -1 )^{-1}$ is the Bose-Einstein distribution function. 
Note that, this split up the integral Eq.~\eqref{eq:IT1} into two pieces. The first piece is temperature independent, which is equivalent to Eq.~\eqref{eq:I[D]}. It vanishes using dimensional regularization. 
The second piece is temperature dependent. 

\begin{figure}[ht]
    \centering 
    \includegraphics[scale=0.8]{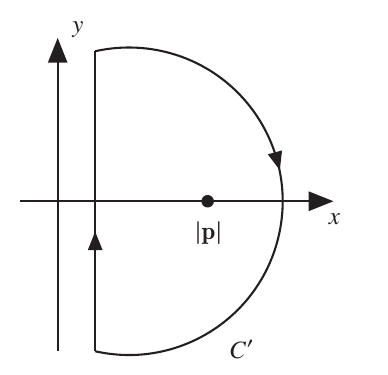}
    \caption{The contour $ C'$ of integral in Eq.~\eqref{eq:CIF}. The radius of the semicircle goes to infinity.}
    \label{fig:contour2}
\end{figure}

To integrate $p_{0} $ in Eq.~\eqref{eq:IT13}, we use the Cauchy integral formula 
\begin{equation}\label{eq:CIF}
    \varointclockwise_{C'} \mathop{d p_{0}} \frac{f(p_{0} )}{\mathbf{p}^{2} - p_{0}^{2}}  =-2 \pi i \frac{f ( |\mathbf{p}| )}{ -2 |\mathbf{p}| } ,
\end{equation}
where the contour $ C'$ is shown in Fig.~\ref{fig:contour2} and $ f(z) = 2 N_{\text{B}}{(z)} $. The sign on the right-hand side of Eq.~\eqref{eq:CIF} comes from the clockwise orientation of the contour used. The only contribution to the contour integral is from the pole $ p_{0}  = | \mathbf{p} | $. 

The temperature-dependent piece is then given by
\begin{equation}\label{eq:IT1contour}
    I_{T} [D] = 
    \frac{\lambda}{2}  \int \frac{\mathop{d^{D-1} p}}{(2 \pi )^{D-1}} \frac{1}{\omega} \frac{1}{ e^{\omega /T} -1},
\end{equation}
where we have used that $ | \mathbf{p} | \equiv \omega   $. This is valid since the mass of $ \phi $ vanishes \cite{Brandt:2021nse}. The expression in Eq.~\eqref{eq:IT1contour} also holds for the massive theory with mass $M$. However, we would have that $ \omega = \sqrt{ \mathbf{p}^{2} + M^{2}}$. This shows that $ \omega $ is associated with the on-shell energy of the particle $ \phi $.

Proceeding, we integrate over $p$ in Eq.~\eqref{eq:IT1contour}, which yields
\begin{equation}\label{eq:IT1contournext}
    I_{T} [D] = 
    \frac{\lambda}{2}  \left(  \frac{ 2 \pi^{(D-1)/2} }{\Gamma[(D-1)/2]} \int_{0}^{\infty} \frac{\mathop{d \omega }}{(2 \pi )^{D-1}} \frac{\omega^{D-3} }{ e^{\omega /T} -1}\right),
\end{equation}
where $ \Gamma ( x) $ is the Gamma function.
Now, we should define $D = 4 - 2 \epsilon $ to regularize the above integral. However, the integral in Eq.~\eqref{eq:IT1contournext} is not divergent. Thus, we can set $ D = 4$ from now on. 

Upon the change of variable $  \omega \to \omega T$, we obtain that
\begin{equation}\label{eq:IT1contour4}
    I_{T} [4] = 
    \frac{\lambda}{2}  \frac{T^{2}}{ 2 \pi^{2}  } \underbrace{\int_{0}^{\infty}  \mathop{d \omega}  \frac{\omega^{D-3} }{ e^{\omega } -1} }_{ \Gamma (D-2) \zeta (D-2)}=
    \frac{\lambda}{2}  \frac{T^{2}}{ 2 \pi^{2} } \Gamma(2) \zeta (2) = \frac{\lambda}{24} T^{2}, 
\end{equation}
where $ \Gamma (2) = 1 $ and $ \zeta (2) = \pi^{2} /6 $. 
Although it vanishes at zero temperature as we have seen in Eq.~\eqref{eq:I21}, the tadpole-like diagram in Fig.~\ref{fig:tadpole1} does not vanish at finite temperature.

\backmatter
\begin{SingleSpace}
\bibliographystyle{apsrev4-2}
\refstepcounter{chapter}
\bibliography{Thesis_arxiv.bib}

\begin{thebibliography}{105}%
\makeatletter
\providecommand \@ifxundefined [1]{%
 \@ifx{#1\undefined}
}%
\providecommand \@ifnum [1]{%
 \ifnum #1\expandafter \@firstoftwo
 \else \expandafter \@secondoftwo
 \fi
}%
\providecommand \@ifx [1]{%
 \ifx #1\expandafter \@firstoftwo
 \else \expandafter \@secondoftwo
 \fi
}%
\providecommand \natexlab [1]{#1}%
\providecommand \enquote  [1]{``#1''}%
\providecommand \bibnamefont  [1]{#1}%
\providecommand \bibfnamefont [1]{#1}%
\providecommand \citenamefont [1]{#1}%
\providecommand \href@noop [0]{\@secondoftwo}%
\providecommand \href [0]{\begingroup \@sanitize@url \@href}%
\providecommand \@href[1]{\@@startlink{#1}\@@href}%
\providecommand \@@href[1]{\endgroup#1\@@endlink}%
\providecommand \@sanitize@url [0]{\catcode `\\12\catcode `\$12\catcode `\&12\catcode `\#12\catcode `\^12\catcode `\_12\catcode `\%12\relax}%
\providecommand \@@startlink[1]{}%
\providecommand \@@endlink[0]{}%
\providecommand \url  [0]{\begingroup\@sanitize@url \@url }%
\providecommand \@url [1]{\endgroup\@href {#1}{\urlprefix }}%
\providecommand \urlprefix  [0]{URL }%
\providecommand \Eprint [0]{\href }%
\providecommand \doibase [0]{https://doi.org/}%
\providecommand \selectlanguage [0]{\@gobble}%
\providecommand \bibinfo  [0]{\@secondoftwo}%
\providecommand \bibfield  [0]{\@secondoftwo}%
\providecommand \translation [1]{[#1]}%
\providecommand \BibitemOpen [0]{}%
\providecommand \bibitemStop [0]{}%
\providecommand \bibitemNoStop [0]{.\EOS\space}%
\providecommand \EOS [0]{\spacefactor3000\relax}%
\providecommand \BibitemShut  [1]{\csname bibitem#1\endcsname}%
\let\auto@bib@innerbib\@empty
\bibitem [{\citenamefont {Gaillard}\ \emph {et~al.}(1999)\citenamefont {Gaillard}, \citenamefont {Grannis},\ and\ \citenamefont {Sciulli}}]{Gaillard:1998ui}%
  \BibitemOpen
  \bibfield  {author} {\bibinfo {author} {\bibfnamefont {M.~K.}\ \bibnamefont {Gaillard}}, \bibinfo {author} {\bibfnamefont {P.~D.}\ \bibnamefont {Grannis}},\ and\ \bibinfo {author} {\bibfnamefont {F.~J.}\ \bibnamefont {Sciulli}},\ }\href {https://doi.org/10.1103/RevModPhys.71.S96} {\bibfield  {journal} {\bibinfo  {journal} {Reviews of Modern Physics}\ }\textbf {\bibinfo {volume} {71}},\ \bibinfo {pages} {S96} (\bibinfo {year} {1999})},\ \Eprint {https://arxiv.org/abs/hep-ph/9812285} {arXiv:hep-ph/9812285} \BibitemShut {NoStop}%
\bibitem [{\citenamefont {Beringer}\ \emph {et~al.}(2012)\citenamefont {Beringer}, \citenamefont {Arguin}, \citenamefont {Barnett}, \citenamefont {Copic}, \citenamefont {Dahl} \emph {et~al.}}]{ParticleDataGroup:2012pjm}%
  \BibitemOpen
  \bibfield  {author} {\bibinfo {author} {\bibfnamefont {J.}~\bibnamefont {Beringer}}, \bibinfo {author} {\bibfnamefont {J.~F.}\ \bibnamefont {Arguin}}, \bibinfo {author} {\bibfnamefont {R.~M.}\ \bibnamefont {Barnett}}, \bibinfo {author} {\bibfnamefont {K.}~\bibnamefont {Copic}}, \bibinfo {author} {\bibfnamefont {O.}~\bibnamefont {Dahl}}, \emph {et~al.},\ }\href {https://doi.org/10.1103/PhysRevD.86.010001} {\bibfield  {journal} {\bibinfo  {journal} {Physical Review D}\ }\textbf {\bibinfo {volume} {86}},\ \bibinfo {pages} {010001} (\bibinfo {year} {2012})}\BibitemShut {NoStop}%
\bibitem [{\citenamefont {Aiko}(2023)}]{aiko:2023}%
  \BibitemOpen
  \bibfield  {author} {\bibinfo {author} {\bibfnamefont {M.}~\bibnamefont {Aiko}},\ }\bibinfo {title} {Review of the {{Standard Model}}},\ in\ \href {https://doi.org/10.1007/978-981-99-1324-4_2} {\emph {\bibinfo {booktitle} {Theoretical {{Studies}} on {{Extended Higgs Sectors Towards Future Precision Measurements}}}}}\ (\bibinfo  {publisher} {Springer Nature Singapore},\ \bibinfo {address} {Singapore},\ \bibinfo {year} {2023})\ pp.\ \bibinfo {pages} {11--34}\BibitemShut {NoStop}%
\bibitem [{\citenamefont {Donoghue}(2012)}]{Donoghue:2012zc}%
  \BibitemOpen
  \bibfield  {author} {\bibinfo {author} {\bibfnamefont {J.~F.}\ \bibnamefont {Donoghue}},\ }in\ \href {https://doi.org/10.1063/1.4756964} {\emph {\bibinfo {booktitle} {The Sixth International School on Field Theory and Gravitation}}}\ (\bibinfo {address} {Petr{\'o}polis - RJ, Brazil},\ \bibinfo {year} {2012})\ pp.\ \bibinfo {pages} {73--94}\BibitemShut {NoStop}%
\bibitem [{\citenamefont {Niederle}\ \emph {et~al.}(1983)\citenamefont {Niederle}, \citenamefont {Popov},\ and\ \citenamefont {Hlavat{\`y}}}]{niederle:1983}%
  \BibitemOpen
  \bibfield  {author} {\bibinfo {author} {\bibfnamefont {J.}~\bibnamefont {Niederle}}, \bibinfo {author} {\bibfnamefont {V.}~\bibnamefont {Popov}},\ and\ \bibinfo {author} {\bibfnamefont {L.}~\bibnamefont {Hlavat{\`y}}},\ }\href@noop {} {\emph {\bibinfo {title} {Functional {{Integrals}} in {{Quantum Field Theory}} and {{Statistical Physics}}}}},\ Mathematical {{Physics}} and {{Applied Mathematics}}\ (\bibinfo  {publisher} {Springer Netherlands},\ \bibinfo {year} {1983})\BibitemShut {NoStop}%
\bibitem [{\citenamefont {McKeon}(2010)}]{McKeon:2010nf}%
  \BibitemOpen
  \bibfield  {author} {\bibinfo {author} {\bibfnamefont {D.~G.~C.}\ \bibnamefont {McKeon}},\ }\href {https://doi.org/10.1142/S0217751X10050093} {\bibfield  {journal} {\bibinfo  {journal} {International Journal of Modern Physics A}\ }\textbf {\bibinfo {volume} {25}},\ \bibinfo {pages} {3453} (\bibinfo {year} {2010})},\ \Eprint {https://arxiv.org/abs/1005.3001} {arXiv:1005.3001 [gr-qc]} \BibitemShut {NoStop}%
\bibitem [{\citenamefont {Lavrov}(2021)}]{Lavrov:2021pqh}%
  \BibitemOpen
  \bibfield  {author} {\bibinfo {author} {\bibfnamefont {P.}~\bibnamefont {Lavrov}},\ }\href {https://doi.org/10.1016/j.physletb.2021.136182} {\bibfield  {journal} {\bibinfo  {journal} {Physics Letters B}\ }\textbf {\bibinfo {volume} {816}},\ \bibinfo {pages} {136182} (\bibinfo {year} {2021})},\ \Eprint {https://arxiv.org/abs/2101.06868} {arXiv:2101.06868 [hep-th]} \BibitemShut {NoStop}%
\bibitem [{\citenamefont {Chishtie}\ and\ \citenamefont {McKeon}(2012{\natexlab{a}})}]{Chishtie:2012sq}%
  \BibitemOpen
  \bibfield  {author} {\bibinfo {author} {\bibfnamefont {F.}~\bibnamefont {Chishtie}}\ and\ \bibinfo {author} {\bibfnamefont {D.~G.~C.}\ \bibnamefont {McKeon}},\ }\href {https://doi.org/10.1088/0264-9381/29/23/235016} {\bibfield  {journal} {\bibinfo  {journal} {Classical and Quantum Gravity}\ }\textbf {\bibinfo {volume} {29}},\ \bibinfo {pages} {235016} (\bibinfo {year} {2012}{\natexlab{a}})},\ \Eprint {https://arxiv.org/abs/1207.2302} {arXiv:1207.2302 [hep-th]} \BibitemShut {NoStop}%
\bibitem [{\citenamefont {Brandt}\ \emph {et~al.}(2020{\natexlab{a}})\citenamefont {Brandt}, \citenamefont {Frenkel}, \citenamefont {{Martins-Filho}},\ and\ \citenamefont {McKeon}}]{McKeon:2020lqp}%
  \BibitemOpen
  \bibfield  {author} {\bibinfo {author} {\bibfnamefont {F.~T.}\ \bibnamefont {Brandt}}, \bibinfo {author} {\bibfnamefont {J.}~\bibnamefont {Frenkel}}, \bibinfo {author} {\bibfnamefont {S.}~\bibnamefont {{Martins-Filho}}},\ and\ \bibinfo {author} {\bibfnamefont {D.~G.~C.}\ \bibnamefont {McKeon}},\ }\href {https://doi.org/10.1103/PhysRevD.101.085013} {\bibfield  {journal} {\bibinfo  {journal} {Physical Review D}\ }\textbf {\bibinfo {volume} {101}},\ \bibinfo {pages} {085013} (\bibinfo {year} {2020}{\natexlab{a}})},\ \Eprint {https://arxiv.org/abs/2003.06819} {arXiv:2003.06819 [hep-th]} \BibitemShut {NoStop}%
\bibitem [{\citenamefont {Brandt}\ \emph {et~al.}(2020{\natexlab{b}})\citenamefont {Brandt}, \citenamefont {Frenkel}, \citenamefont {{Martins-Filho}},\ and\ \citenamefont {McKeon}}]{Brandt:2020vre}%
  \BibitemOpen
  \bibfield  {author} {\bibinfo {author} {\bibfnamefont {F.~T.}\ \bibnamefont {Brandt}}, \bibinfo {author} {\bibfnamefont {J.}~\bibnamefont {Frenkel}}, \bibinfo {author} {\bibfnamefont {S.}~\bibnamefont {{Martins-Filho}}},\ and\ \bibinfo {author} {\bibfnamefont {D.~G.~C.}\ \bibnamefont {McKeon}},\ }\href {https://doi.org/10.1103/PhysRevD.102.045013} {\bibfield  {journal} {\bibinfo  {journal} {Physical Review D}\ }\textbf {\bibinfo {volume} {102}},\ \bibinfo {pages} {045013} (\bibinfo {year} {2020}{\natexlab{b}})},\ \Eprint {https://arxiv.org/abs/2007.04841} {arXiv:2007.04841 [hep-th]} \BibitemShut {NoStop}%
\bibitem [{\citenamefont {{Martins-Filho}}(2021)}]{martins-filho:2021}%
  \BibitemOpen
  \bibfield  {author} {\bibinfo {author} {\bibfnamefont {S.}~\bibnamefont {{Martins-Filho}}},\ }\emph {\bibinfo {title} {{Equival{\^e}ncia qu{\^a}ntica da formula{\c c}{\~a}o de primeira e segunda ordem da teoria de Yang-Mills e da gravita{\c c}{\~a}o}}},\ \href {https://doi.org/10.11606/D.43.2021.tde-16032021-172847} {\bibinfo {type} {{Mestrado em F{\'i}sica}}},\ \bibinfo  {school} {Universidade de S{\~a}o Paulo}, \bibinfo {address} {S{\~a}o Paulo} (\bibinfo {year} {2021})\BibitemShut {NoStop}%
\bibitem [{\citenamefont {McKeon}(1994)}]{McKeon:1994ds}%
  \BibitemOpen
  \bibfield  {author} {\bibinfo {author} {\bibfnamefont {D.~G.~C.}\ \bibnamefont {McKeon}},\ }\href {https://doi.org/10.1139/p94-077} {\bibfield  {journal} {\bibinfo  {journal} {Canadian Journal of Physics}\ }\textbf {\bibinfo {volume} {72}},\ \bibinfo {pages} {601} (\bibinfo {year} {1994})}\BibitemShut {NoStop}%
\bibitem [{\citenamefont {Brandt}\ and\ \citenamefont {McKeon}(2015)}]{Brandt:2015nxa}%
  \BibitemOpen
  \bibfield  {author} {\bibinfo {author} {\bibfnamefont {F.~T.}\ \bibnamefont {Brandt}}\ and\ \bibinfo {author} {\bibfnamefont {D.~G.~C.}\ \bibnamefont {McKeon}},\ }\href {https://doi.org/10.1103/PhysRevD.91.105006} {\bibfield  {journal} {\bibinfo  {journal} {Physical Review D}\ }\textbf {\bibinfo {volume} {91}},\ \bibinfo {pages} {105006} (\bibinfo {year} {2015})},\ \Eprint {https://arxiv.org/abs/1503.02598} {arXiv:1503.02598 [hep-th]} \BibitemShut {NoStop}%
\bibitem [{\citenamefont {Taylor}(1971)}]{Taylor:1971ff}%
  \BibitemOpen
  \bibfield  {author} {\bibinfo {author} {\bibfnamefont {J.~C.}\ \bibnamefont {Taylor}},\ }\href {https://doi.org/10.1016/0550-3213(71)90297-5} {\bibfield  {journal} {\bibinfo  {journal} {Nuclear Physics B}\ }\textbf {\bibinfo {volume} {33}},\ \bibinfo {pages} {436} (\bibinfo {year} {1971})}\BibitemShut {NoStop}%
\bibitem [{\citenamefont {Slavnov}(1972)}]{Slavnov:1972fg}%
  \BibitemOpen
  \bibfield  {author} {\bibinfo {author} {\bibfnamefont {A.~A.}\ \bibnamefont {Slavnov}},\ }\href {https://doi.org/10.1007/BF01090719} {\bibfield  {journal} {\bibinfo  {journal} {Theoretical and Mathematical Physics}\ }\textbf {\bibinfo {volume} {10}},\ \bibinfo {pages} {99} (\bibinfo {year} {1972})}\BibitemShut {NoStop}%
\bibitem [{\citenamefont {Leibbrandt}(1975)}]{Leibbrandt:1975dj}%
  \BibitemOpen
  \bibfield  {author} {\bibinfo {author} {\bibfnamefont {G.}~\bibnamefont {Leibbrandt}},\ }\href {https://doi.org/10.1103/RevModPhys.47.849} {\bibfield  {journal} {\bibinfo  {journal} {Reviews of Modern Physics}\ }\textbf {\bibinfo {volume} {47}},\ \bibinfo {pages} {849} (\bibinfo {year} {1975})}\BibitemShut {NoStop}%
\bibitem [{\citenamefont {Senjanovi{\'c}}(1976)}]{Senjanovic:1976br}%
  \BibitemOpen
  \bibfield  {author} {\bibinfo {author} {\bibfnamefont {P.}~\bibnamefont {Senjanovi{\'c}}},\ }\href {https://doi.org/10.1016/0003-4916(76)90062-2} {\bibfield  {journal} {\bibinfo  {journal} {Annals of Physics}\ }\textbf {\bibinfo {volume} {100}},\ \bibinfo {pages} {227} (\bibinfo {year} {1976})}\BibitemShut {NoStop}%
\bibitem [{\citenamefont {Faddeev}\ and\ \citenamefont {Popov}(1967)}]{Faddeev:1967fc}%
  \BibitemOpen
  \bibfield  {author} {\bibinfo {author} {\bibfnamefont {L.}~\bibnamefont {Faddeev}}\ and\ \bibinfo {author} {\bibfnamefont {V.}~\bibnamefont {Popov}},\ }\href {https://doi.org/10.1016/0370-2693(67)90067-6} {\bibfield  {journal} {\bibinfo  {journal} {Physics Letters B}\ }\textbf {\bibinfo {volume} {25}},\ \bibinfo {pages} {29} (\bibinfo {year} {1967})}\BibitemShut {NoStop}%
\bibitem [{\citenamefont {Lagraa}\ \emph {et~al.}(2010)\citenamefont {Lagraa}, \citenamefont {Mebarki},\ and\ \citenamefont {Mimouni}}]{Lagraa:2010cza}%
  \BibitemOpen
  \bibfield  {author} {\bibinfo {author} {\bibfnamefont {M.~H.}\ \bibnamefont {Lagraa}}, \bibinfo {author} {\bibfnamefont {N.}~\bibnamefont {Mebarki}},\ and\ \bibinfo {author} {\bibfnamefont {J.}~\bibnamefont {Mimouni}},\ }in\ \href {https://doi.org/10.1063/1.3518339} {\emph {\bibinfo {booktitle} {The {{Third Algerian Workshop}} on {{Astronomy}} and {{Astrophysics}}}}}\ (\bibinfo {address} {Constantine, (Algeria)},\ \bibinfo {year} {2010})\ pp.\ \bibinfo {pages} {222--229}\BibitemShut {NoStop}%
\bibitem [{\citenamefont {Casadio}\ \emph {et~al.}(2023)\citenamefont {Casadio}, \citenamefont {Kamenshchik},\ and\ \citenamefont {Kuntz}}]{Casadio:2022ozp}%
  \BibitemOpen
  \bibfield  {author} {\bibinfo {author} {\bibfnamefont {R.}~\bibnamefont {Casadio}}, \bibinfo {author} {\bibfnamefont {A.}~\bibnamefont {Kamenshchik}},\ and\ \bibinfo {author} {\bibfnamefont {I.}~\bibnamefont {Kuntz}},\ }\href {https://doi.org/10.1016/j.aop.2022.169203} {\bibfield  {journal} {\bibinfo  {journal} {Annals of Physics}\ }\textbf {\bibinfo {volume} {449}},\ \bibinfo {pages} {169203} (\bibinfo {year} {2023})}\BibitemShut {NoStop}%
\bibitem [{\citenamefont {J{\"a}rv}\ \emph {et~al.}(2020)\citenamefont {J{\"a}rv}, \citenamefont {Karam}, \citenamefont {Kozak}, \citenamefont {Lykkas}, \citenamefont {Racioppi},\ and\ \citenamefont {Saal}}]{Jarv:2020qqm}%
  \BibitemOpen
  \bibfield  {author} {\bibinfo {author} {\bibfnamefont {L.}~\bibnamefont {J{\"a}rv}}, \bibinfo {author} {\bibfnamefont {A.}~\bibnamefont {Karam}}, \bibinfo {author} {\bibfnamefont {A.}~\bibnamefont {Kozak}}, \bibinfo {author} {\bibfnamefont {A.}~\bibnamefont {Lykkas}}, \bibinfo {author} {\bibfnamefont {A.}~\bibnamefont {Racioppi}},\ and\ \bibinfo {author} {\bibfnamefont {M.}~\bibnamefont {Saal}},\ }\href {https://doi.org/10.1103/PhysRevD.102.044029} {\bibfield  {journal} {\bibinfo  {journal} {Physical Review D}\ }\textbf {\bibinfo {volume} {102}},\ \bibinfo {pages} {044029} (\bibinfo {year} {2020})}\BibitemShut {NoStop}%
\bibitem [{\citenamefont {Verner}(2021)}]{Verner:2020gfa}%
  \BibitemOpen
  \bibfield  {author} {\bibinfo {author} {\bibfnamefont {S.}~\bibnamefont {Verner}},\ }\href {https://doi.org/10.1088/1475-7516/2021/04/001} {\bibfield  {journal} {\bibinfo  {journal} {Journal of Cosmology and Astroparticle Physics}\ }\textbf {\bibinfo {volume} {2021}}\bibinfo  {number} { (04)},\ \bibinfo {pages} {001}}\BibitemShut {NoStop}%
\bibitem [{\citenamefont {Karam}\ \emph {et~al.}(2021)\citenamefont {Karam}, \citenamefont {Raidal},\ and\ \citenamefont {Tomberg}}]{Karam:2020rpa}%
  \BibitemOpen
\bibfield  {number} {  }\bibfield  {author} {\bibinfo {author} {\bibfnamefont {A.}~\bibnamefont {Karam}}, \bibinfo {author} {\bibfnamefont {M.}~\bibnamefont {Raidal}},\ and\ \bibinfo {author} {\bibfnamefont {E.}~\bibnamefont {Tomberg}},\ }\href {https://doi.org/10.1088/1475-7516/2021/03/064} {\bibfield  {journal} {\bibinfo  {journal} {Journal of Cosmology and Astroparticle Physics}\ }\textbf {\bibinfo {volume} {2021}}\bibinfo  {number} { (03)},\ \bibinfo {pages} {064}}\BibitemShut {NoStop}%
\bibitem [{\citenamefont {Antoniadis}\ \emph {et~al.}(2018)\citenamefont {Antoniadis}, \citenamefont {Karam}, \citenamefont {Lykkas},\ and\ \citenamefont {Tamvakis}}]{Antoniadis:2018ywb}%
  \BibitemOpen
\bibfield  {number} {  }\bibfield  {author} {\bibinfo {author} {\bibfnamefont {I.}~\bibnamefont {Antoniadis}}, \bibinfo {author} {\bibfnamefont {A.}~\bibnamefont {Karam}}, \bibinfo {author} {\bibfnamefont {A.}~\bibnamefont {Lykkas}},\ and\ \bibinfo {author} {\bibfnamefont {K.}~\bibnamefont {Tamvakis}},\ }\href {https://doi.org/10.1088/1475-7516/2018/11/028} {\bibfield  {journal} {\bibinfo  {journal} {Journal of Cosmology and Astroparticle Physics}\ }\textbf {\bibinfo {volume} {2018}}\bibinfo  {number} { (11)},\ \bibinfo {pages} {028}}\BibitemShut {NoStop}%
\bibitem [{\citenamefont {Pinto}\ \emph {et~al.}(2018)\citenamefont {Pinto}, \citenamefont {Vecchio}, \citenamefont {Fatibene},\ and\ \citenamefont {Ferraris}}]{Pinto:2018rfg}%
  \BibitemOpen
\bibfield  {number} {  }\bibfield  {author} {\bibinfo {author} {\bibfnamefont {P.}~\bibnamefont {Pinto}}, \bibinfo {author} {\bibfnamefont {L.~D.}\ \bibnamefont {Vecchio}}, \bibinfo {author} {\bibfnamefont {L.}~\bibnamefont {Fatibene}},\ and\ \bibinfo {author} {\bibfnamefont {M.}~\bibnamefont {Ferraris}},\ }\href {https://doi.org/10.1088/1475-7516/2018/11/044} {\bibfield  {journal} {\bibinfo  {journal} {Journal of Cosmology and Astroparticle Physics}\ }\textbf {\bibinfo {volume} {2018}}\bibinfo  {number} { (11)},\ \bibinfo {pages} {044}}\BibitemShut {NoStop}%
\bibitem [{\citenamefont {Dioguardi}\ \emph {et~al.}(2022)\citenamefont {Dioguardi}, \citenamefont {Racioppi},\ and\ \citenamefont {Tomberg}}]{Dioguardi:2021fmr}%
  \BibitemOpen
\bibfield  {number} {  }\bibfield  {author} {\bibinfo {author} {\bibfnamefont {C.}~\bibnamefont {Dioguardi}}, \bibinfo {author} {\bibfnamefont {A.}~\bibnamefont {Racioppi}},\ and\ \bibinfo {author} {\bibfnamefont {E.}~\bibnamefont {Tomberg}},\ }\href {https://doi.org/10.1007/JHEP06(2022)106} {\bibfield  {journal} {\bibinfo  {journal} {Journal of High Energy Physics}\ }\textbf {\bibinfo {volume} {2022}},\ \bibinfo {pages} {106} (\bibinfo {year} {2022})}\BibitemShut {NoStop}%
\bibitem [{\citenamefont {McKeon}\ and\ \citenamefont {Sherry}(1992)}]{McKeon:1992rq}%
  \BibitemOpen
  \bibfield  {author} {\bibinfo {author} {\bibfnamefont {D.~G.~C.}\ \bibnamefont {McKeon}}\ and\ \bibinfo {author} {\bibfnamefont {T.~N.}\ \bibnamefont {Sherry}},\ }\href {https://doi.org/10.1139/p92-074} {\bibfield  {journal} {\bibinfo  {journal} {Canadian Journal of Physics}\ }\textbf {\bibinfo {volume} {70}},\ \bibinfo {pages} {441} (\bibinfo {year} {1992})}\BibitemShut {NoStop}%
\bibitem [{\citenamefont {Brandt}\ \emph {et~al.}(2020{\natexlab{c}})\citenamefont {Brandt}, \citenamefont {Frenkel},\ and\ \citenamefont {McKeon}}]{Brandt:2018lbe}%
  \BibitemOpen
  \bibfield  {author} {\bibinfo {author} {\bibfnamefont {F.~T.}\ \bibnamefont {Brandt}}, \bibinfo {author} {\bibfnamefont {J.}~\bibnamefont {Frenkel}},\ and\ \bibinfo {author} {\bibfnamefont {D.~G.~C.}\ \bibnamefont {McKeon}},\ }\href {https://doi.org/10.1139/cjp-2019-0037} {\bibfield  {journal} {\bibinfo  {journal} {Canadian Journal of Physics}\ }\textbf {\bibinfo {volume} {98}},\ \bibinfo {pages} {344} (\bibinfo {year} {2020}{\natexlab{c}})}\BibitemShut {NoStop}%
\bibitem [{\citenamefont {McKeon}\ \emph {et~al.}(2019)\citenamefont {McKeon}, \citenamefont {Brandt}, \citenamefont {Frenkel},\ and\ \citenamefont {Sakoda}}]{Brandt:2019ymg}%
  \BibitemOpen
  \bibfield  {author} {\bibinfo {author} {\bibfnamefont {D.~G.~C.}\ \bibnamefont {McKeon}}, \bibinfo {author} {\bibfnamefont {F.~T.}\ \bibnamefont {Brandt}}, \bibinfo {author} {\bibfnamefont {J.}~\bibnamefont {Frenkel}},\ and\ \bibinfo {author} {\bibfnamefont {G.~S.~S.}\ \bibnamefont {Sakoda}},\ }\href {https://doi.org/10.1103/PhysRevD.100.125014} {\bibfield  {journal} {\bibinfo  {journal} {Physical Review D}\ }\textbf {\bibinfo {volume} {100}},\ \bibinfo {pages} {125014} (\bibinfo {year} {2019})}\BibitemShut {NoStop}%
\bibitem [{\citenamefont {Brandt}\ \emph {et~al.}(2021{\natexlab{a}})\citenamefont {Brandt}, \citenamefont {Frenkel}, \citenamefont {{Martins-Filho}},\ and\ \citenamefont {McKeon}}]{Brandt:2020gms}%
  \BibitemOpen
  \bibfield  {author} {\bibinfo {author} {\bibfnamefont {F.~T.}\ \bibnamefont {Brandt}}, \bibinfo {author} {\bibfnamefont {J.}~\bibnamefont {Frenkel}}, \bibinfo {author} {\bibfnamefont {S.}~\bibnamefont {{Martins-Filho}}},\ and\ \bibinfo {author} {\bibfnamefont {D.~G.~C.}\ \bibnamefont {McKeon}},\ }\href {https://doi.org/10.1016/j.aop.2021.168426} {\bibfield  {journal} {\bibinfo  {journal} {Annals of Physics}\ }\textbf {\bibinfo {volume} {427}},\ \bibinfo {pages} {168426} (\bibinfo {year} {2021}{\natexlab{a}})},\ \Eprint {https://arxiv.org/abs/2009.09553} {arXiv:2009.09553 [hep-th]} \BibitemShut {NoStop}%
\bibitem [{\citenamefont {McKeon}\ \emph {et~al.}(2021)\citenamefont {McKeon}, \citenamefont {Brandt}, \citenamefont {Frenkel},\ and\ \citenamefont {{Martins-Filho}}}]{Brandt:2021qgh}%
  \BibitemOpen
  \bibfield  {author} {\bibinfo {author} {\bibfnamefont {D.~G.~C.}\ \bibnamefont {McKeon}}, \bibinfo {author} {\bibfnamefont {F.~T.}\ \bibnamefont {Brandt}}, \bibinfo {author} {\bibfnamefont {J.}~\bibnamefont {Frenkel}},\ and\ \bibinfo {author} {\bibfnamefont {S.}~\bibnamefont {{Martins-Filho}}},\ }\href {https://doi.org/10.1016/j.aop.2021.168659} {\bibfield  {journal} {\bibinfo  {journal} {Annals of Physics}\ }\textbf {\bibinfo {volume} {434}},\ \bibinfo {pages} {168659} (\bibinfo {year} {2021})},\ \Eprint {https://arxiv.org/abs/2102.02854} {arXiv:2102.02854 [hep-th]} \BibitemShut {NoStop}%
\bibitem [{\citenamefont {Brandt}\ \emph {et~al.}(2021{\natexlab{b}})\citenamefont {Brandt}, \citenamefont {Frenkel}, \citenamefont {{Martins-Filho}}, \citenamefont {McKeon},\ and\ \citenamefont {Sakoda}}]{Brandt:2021nev}%
  \BibitemOpen
  \bibfield  {author} {\bibinfo {author} {\bibfnamefont {F.~T.}\ \bibnamefont {Brandt}}, \bibinfo {author} {\bibfnamefont {J.}~\bibnamefont {Frenkel}}, \bibinfo {author} {\bibfnamefont {S.}~\bibnamefont {{Martins-Filho}}}, \bibinfo {author} {\bibfnamefont {D.~G.~C.}\ \bibnamefont {McKeon}},\ and\ \bibinfo {author} {\bibfnamefont {G.~S.~S.}\ \bibnamefont {Sakoda}},\ }\href {https://doi.org/10.1139/cjp-2021-0248} {\bibfield  {journal} {\bibinfo  {journal} {Canadian Journal of Physics}\ }\textbf {\bibinfo {volume} {100}},\ \bibinfo {pages} {139} (\bibinfo {year} {2021}{\natexlab{b}})},\ \Eprint {https://arxiv.org/abs/2105.00318} {arXiv:2105.00318 [hep-th]} \BibitemShut {NoStop}%
\bibitem [{\citenamefont {Ostrogradsky}(1850)}]{Ostrogradsky:1850fid}%
  \BibitemOpen
  \bibfield  {author} {\bibinfo {author} {\bibfnamefont {M.}~\bibnamefont {Ostrogradsky}},\ }\href@noop {} {\bibfield  {journal} {\bibinfo  {journal} {Mem. Acad. St. Petersbourg}\ }\textbf {\bibinfo {volume} {6}},\ \bibinfo {pages} {385} (\bibinfo {year} {1850})}\BibitemShut {NoStop}%
\bibitem [{\citenamefont {Aoki}\ and\ \citenamefont {Motohashi}(2020)}]{Aoki:2020gfv}%
  \BibitemOpen
  \bibfield  {author} {\bibinfo {author} {\bibfnamefont {K.}~\bibnamefont {Aoki}}\ and\ \bibinfo {author} {\bibfnamefont {H.}~\bibnamefont {Motohashi}},\ }\href {https://doi.org/10.1088/1475-7516/2020/08/026} {\bibfield  {journal} {\bibinfo  {journal} {Journal of Cosmology and Astroparticle Physics}\ }\textbf {\bibinfo {volume} {2020}}\bibinfo  {number} { (08)},\ \bibinfo {pages} {026}}\BibitemShut {NoStop}%
\bibitem [{\citenamefont {Pauli}(1943)}]{pauli:1943}%
  \BibitemOpen
\bibfield  {number} {  }\bibfield  {author} {\bibinfo {author} {\bibfnamefont {W.}~\bibnamefont {Pauli}},\ }\href {https://doi.org/10.1103/RevModPhys.15.175} {\bibfield  {journal} {\bibinfo  {journal} {Reviews of Modern Physics}\ }\textbf {\bibinfo {volume} {15}},\ \bibinfo {pages} {175} (\bibinfo {year} {1943})}\BibitemShut {NoStop}%
\bibitem [{\citenamefont {Sudarshan}(1961)}]{Sudarshan:1961vs}%
  \BibitemOpen
  \bibfield  {author} {\bibinfo {author} {\bibfnamefont {E.~C.~G.}\ \bibnamefont {Sudarshan}},\ }\href {https://doi.org/10.1103/PhysRev.123.2183} {\bibfield  {journal} {\bibinfo  {journal} {Physical Review}\ }\textbf {\bibinfo {volume} {123}},\ \bibinfo {pages} {2183} (\bibinfo {year} {1961})}\BibitemShut {NoStop}%
\bibitem [{\citenamefont {Brandt}\ and\ \citenamefont {{Martins-Filho}}(2023)}]{Brandt:2022kjo}%
  \BibitemOpen
  \bibfield  {author} {\bibinfo {author} {\bibfnamefont {F.~T.}\ \bibnamefont {Brandt}}\ and\ \bibinfo {author} {\bibfnamefont {S.}~\bibnamefont {{Martins-Filho}}},\ }\href {https://doi.org/10.1016/j.aop.2023.169323} {\bibfield  {journal} {\bibinfo  {journal} {Annals of Physics}\ }\textbf {\bibinfo {volume} {453}},\ \bibinfo {pages} {169323} (\bibinfo {year} {2023})},\ \Eprint {https://arxiv.org/abs/2212.05629} {arXiv:2212.05629 [hep-th]} \BibitemShut {NoStop}%
\bibitem [{\citenamefont {Becchi}\ \emph {et~al.}(1974)\citenamefont {Becchi}, \citenamefont {Rouet},\ and\ \citenamefont {Stora}}]{Becchi:1974xu}%
  \BibitemOpen
  \bibfield  {author} {\bibinfo {author} {\bibfnamefont {C.}~\bibnamefont {Becchi}}, \bibinfo {author} {\bibfnamefont {A.}~\bibnamefont {Rouet}},\ and\ \bibinfo {author} {\bibfnamefont {R.}~\bibnamefont {Stora}},\ }\href {https://doi.org/10.1016/0370-2693(74)90058-6} {\bibfield  {journal} {\bibinfo  {journal} {Physics Letters B}\ }\textbf {\bibinfo {volume} {52}},\ \bibinfo {pages} {344} (\bibinfo {year} {1974})}\BibitemShut {NoStop}%
\bibitem [{\citenamefont {Tyutin}(1975)}]{Tyutin:1975qk}%
  \BibitemOpen
  \bibfield  {author} {\bibinfo {author} {\bibfnamefont {I.~V.}\ \bibnamefont {Tyutin}},\ }\bibfield  {journal} {\bibinfo  {journal} {P. N. Lebedev Physical Institute}\ }\textbf {\bibinfo {volume} {39}},\ \href {https://doi.org/10.48550/arXiv.0812.0580} {10.48550/arXiv.0812.0580} (\bibinfo {year} {1975})\BibitemShut {NoStop}%
\bibitem [{\citenamefont {Rubakov}\ and\ \citenamefont {Wilson}(2002)}]{rubakov:2002}%
  \BibitemOpen
  \bibfield  {author} {\bibinfo {author} {\bibfnamefont {V.}~\bibnamefont {Rubakov}}\ and\ \bibinfo {author} {\bibfnamefont {S.}~\bibnamefont {Wilson}},\ }\href@noop {} {\emph {\bibinfo {title} {Classical Theory of Gauge Fields}}}\ (\bibinfo  {publisher} {Princeton University Press},\ \bibinfo {address} {New Jersey},\ \bibinfo {year} {2002})\BibitemShut {NoStop}%
\bibitem [{\citenamefont {Yang}\ and\ \citenamefont {Mills}(1954)}]{Yang:1954ek}%
  \BibitemOpen
  \bibfield  {author} {\bibinfo {author} {\bibfnamefont {C.-N.}\ \bibnamefont {Yang}}\ and\ \bibinfo {author} {\bibfnamefont {R.~L.}\ \bibnamefont {Mills}},\ }\href {https://doi.org/10.1103/PhysRev.96.191} {\bibfield  {journal} {\bibinfo  {journal} {Physics Review}\ }\textbf {\bibinfo {volume} {96}},\ \bibinfo {pages} {191} (\bibinfo {year} {1954})}\BibitemShut {NoStop}%
\bibitem [{\citenamefont {Tong}(2017)}]{Tong:2017oea}%
  \BibitemOpen
  \bibfield  {author} {\bibinfo {author} {\bibfnamefont {D.}~\bibnamefont {Tong}},\ }\href {https://doi.org/10.1007/JHEP07(2017)104} {\bibfield  {journal} {\bibinfo  {journal} {Journal of High Energy Physics}\ }\textbf {\bibinfo {volume} {2017}},\ \bibinfo {pages} {104} (\bibinfo {year} {2017})}\BibitemShut {NoStop}%
\bibitem [{\citenamefont {Wald}(1984)}]{wald:1984}%
  \BibitemOpen
  \bibfield  {author} {\bibinfo {author} {\bibfnamefont {R.~M.}\ \bibnamefont {Wald}},\ }\href@noop {} {\emph {\bibinfo {title} {General Relativity}}},\ \bibinfo {edition} {1st}\ ed.\ (\bibinfo  {publisher} {University of Chicago Press},\ \bibinfo {address} {London},\ \bibinfo {year} {1984})\BibitemShut {NoStop}%
\bibitem [{\citenamefont {Georgi}\ and\ \citenamefont {Glashow}(1974)}]{Georgi:1974sy}%
  \BibitemOpen
  \bibfield  {author} {\bibinfo {author} {\bibfnamefont {H.}~\bibnamefont {Georgi}}\ and\ \bibinfo {author} {\bibfnamefont {S.~L.}\ \bibnamefont {Glashow}},\ }\href {https://doi.org/10.1103/PhysRevLett.32.438} {\bibfield  {journal} {\bibinfo  {journal} {Physical Review Letters}\ }\textbf {\bibinfo {volume} {32}},\ \bibinfo {pages} {438} (\bibinfo {year} {1974})}\BibitemShut {NoStop}%
\bibitem [{\citenamefont {Forkel}(2000)}]{Forkel:2000sq}%
  \BibitemOpen
  \bibfield  {author} {\bibinfo {author} {\bibfnamefont {H.}~\bibnamefont {Forkel}},\ }\href {https://doi.org/10.48550/arxiv.hep-ph/0009136} {\bibinfo {title} {A {{Primer}} on {{Instantons}} in {{QCD}}}} (\bibinfo {year} {2000}),\ \Eprint {https://arxiv.org/abs/hep-ph/0009136} {arXiv:hep-ph/0009136} \BibitemShut {NoStop}%
\bibitem [{\citenamefont {Kim}\ and\ \citenamefont {Carosi}(2010)}]{Kim:2008hd}%
  \BibitemOpen
  \bibfield  {author} {\bibinfo {author} {\bibfnamefont {J.~E.}\ \bibnamefont {Kim}}\ and\ \bibinfo {author} {\bibfnamefont {G.}~\bibnamefont {Carosi}},\ }\href {https://doi.org/10.1103/RevModPhys.82.557} {\bibfield  {journal} {\bibinfo  {journal} {Reviews of Modern Physics}\ }\textbf {\bibinfo {volume} {82}},\ \bibinfo {pages} {557} (\bibinfo {year} {2010})}\BibitemShut {NoStop}%
\bibitem [{\citenamefont {Fried}\ and\ \citenamefont {M{\"u}ller}(1993)}]{fried:1993}%
  \BibitemOpen
  \bibfield  {author} {\bibinfo {author} {\bibfnamefont {H.~M.}\ \bibnamefont {Fried}}\ and\ \bibinfo {author} {\bibfnamefont {B.}~\bibnamefont {M{\"u}ller}},\ }in\ \href {https://doi.org/10.1142/1923} {\emph {\bibinfo {booktitle} {{{QCD Vacuum Structure}}}}}\ (\bibinfo  {publisher} {World Scientific},\ \bibinfo {address} {American University Paris},\ \bibinfo {year} {1993})\BibitemShut {NoStop}%
\bibitem [{\citenamefont {Das}(2006)}]{das:2006}%
  \BibitemOpen
  \bibfield  {author} {\bibinfo {author} {\bibfnamefont {A.}~\bibnamefont {Das}},\ }\href {https://doi.org/10.1142/6145} {\emph {\bibinfo {title} {Field {{Theory}}: {{A Path Integral Approach}}}}},\ \bibinfo {edition} {2nd}\ ed.,\ \bibinfo {series} {World {{Scientific Lecture Notes}} in {{Physics}}}, Vol.~\bibinfo {volume} {75}\ (\bibinfo  {publisher} {World Scientific},\ \bibinfo {year} {2006})\BibitemShut {NoStop}%
\bibitem [{\citenamefont {Peskin}(2018)}]{peskin:2018}%
  \BibitemOpen
  \bibfield  {author} {\bibinfo {author} {\bibfnamefont {M.~E.}\ \bibnamefont {Peskin}},\ }\href {https://doi.org/10.1201/9780429503559} {\emph {\bibinfo {title} {An {{Introduction To Quantum Field Theory}}}}},\ \bibinfo {edition} {0th}\ ed.\ (\bibinfo  {publisher} {CRC Press},\ \bibinfo {year} {2018})\BibitemShut {NoStop}%
\bibitem [{\citenamefont {Ryder}(1996)}]{Ryder:1985wq}%
  \BibitemOpen
  \bibfield  {author} {\bibinfo {author} {\bibfnamefont {L.~H.}\ \bibnamefont {Ryder}},\ }\href {https://doi.org/10.1017/CBO9780511813900} {\emph {\bibinfo {title} {Quantum {{Field Theory}}}}},\ \bibinfo {edition} {2nd}\ ed.\ (\bibinfo  {publisher} {Cambridge University Press},\ \bibinfo {year} {1996})\BibitemShut {NoStop}%
\bibitem [{\citenamefont {Das}(2020)}]{dasLecturesQuantumField2020}%
  \BibitemOpen
  \bibfield  {author} {\bibinfo {author} {\bibfnamefont {A.}~\bibnamefont {Das}},\ }\href {https://doi.org/10.1142/11845} {\emph {\bibinfo {title} {Lectures on {{Quantum Field Theory}}}}},\ \bibinfo {edition} {2nd}\ ed.\ (\bibinfo  {publisher} {World Scientific},\ \bibinfo {year} {2020})\BibitemShut {NoStop}%
\bibitem [{\citenamefont {{Zinn-Justin}}(1975)}]{Zinn-Justin:1974ggz}%
  \BibitemOpen
  \bibfield  {author} {\bibinfo {author} {\bibfnamefont {J.}~\bibnamefont {{Zinn-Justin}}},\ }in\ \href {https://doi.org/10.1007/3-540-07160-1_1} {\emph {\bibinfo {booktitle} {Trends in {{Elementary Particle Theory}}}}},\ Vol.~\bibinfo {volume} {37},\ \bibinfo {editor} {edited by\ \bibinfo {editor} {\bibfnamefont {H.}~\bibnamefont {Rollnik}}\ and\ \bibinfo {editor} {\bibfnamefont {K.}~\bibnamefont {Dietz}}}\ (\bibinfo  {publisher} {Springer Berlin Heidelberg},\ \bibinfo {address} {Berlin, Heidelberg},\ \bibinfo {year} {1975})\ pp.\ \bibinfo {pages} {1--39}\BibitemShut {NoStop}%
\bibitem [{\citenamefont {Taylor}(1976)}]{taylor:1976}%
  \BibitemOpen
  \bibfield  {author} {\bibinfo {author} {\bibfnamefont {J.~C.}\ \bibnamefont {Taylor}},\ }\href@noop {} {\emph {\bibinfo {title} {Gauge {{Theories}} of {{Weak Interactions}}}}}\ (\bibinfo  {publisher} {Cambridge University Press},\ \bibinfo {address} {Cambridge, UK},\ \bibinfo {year} {1976})\BibitemShut {NoStop}%
\bibitem [{\citenamefont {Kiriushcheva}\ and\ \citenamefont {Kuzmin}(2006)}]{Kiriushcheva:2005yu}%
  \BibitemOpen
  \bibfield  {author} {\bibinfo {author} {\bibfnamefont {N.}~\bibnamefont {Kiriushcheva}}\ and\ \bibinfo {author} {\bibfnamefont {S.}~\bibnamefont {Kuzmin}},\ }\href {https://doi.org/10.1016/j.aop.2005.09.009} {\bibfield  {journal} {\bibinfo  {journal} {Annals of Physics}\ }\textbf {\bibinfo {volume} {321}},\ \bibinfo {pages} {958} (\bibinfo {year} {2006})}\BibitemShut {NoStop}%
\bibitem [{\citenamefont {Henneaux}\ and\ \citenamefont {Teitelboim}(1992)}]{henneaux1992quantization}%
  \BibitemOpen
  \bibfield  {author} {\bibinfo {author} {\bibfnamefont {M.}~\bibnamefont {Henneaux}}\ and\ \bibinfo {author} {\bibfnamefont {C.}~\bibnamefont {Teitelboim}},\ }\href@noop {} {\emph {\bibinfo {title} {Quantization of Gauge Systems}}},\ Princeton Paperbacks\ (\bibinfo  {publisher} {Princeton University Press},\ \bibinfo {address} {Princeton, NJ},\ \bibinfo {year} {1992})\BibitemShut {NoStop}%
\bibitem [{\citenamefont {{K. Sundermeyer}}(1982)}]{k.sundermeyer:1982}%
  \BibitemOpen
  \bibfield  {author} {\bibinfo {author} {\bibnamefont {{K. Sundermeyer}}},\ }\href {https://doi.org/10.1007/BFb0036225} {\emph {\bibinfo {title} {Constrained {{Dynamics}}}}},\ \bibinfo {series} {Lecture {{Notes}} in {{Physics}}}, Vol.\ \bibinfo {volume} {169}\ (\bibinfo  {publisher} {Springer-Verlag},\ \bibinfo {address} {Berlin/Heidelberg},\ \bibinfo {year} {1982})\BibitemShut {NoStop}%
\bibitem [{\citenamefont {Cattaneo}\ \emph {et~al.}(1995)\citenamefont {Cattaneo}, \citenamefont {{Cotta-Ramusino}}, \citenamefont {Gamba},\ and\ \citenamefont {Martellini}}]{Cattaneo:1995xa}%
  \BibitemOpen
  \bibfield  {author} {\bibinfo {author} {\bibfnamefont {A.~S.}\ \bibnamefont {Cattaneo}}, \bibinfo {author} {\bibfnamefont {P.}~\bibnamefont {{Cotta-Ramusino}}}, \bibinfo {author} {\bibfnamefont {A.}~\bibnamefont {Gamba}},\ and\ \bibinfo {author} {\bibfnamefont {M.}~\bibnamefont {Martellini}},\ }\href {https://doi.org/10.1016/0370-2693(95)00718-Z} {\bibfield  {journal} {\bibinfo  {journal} {Physics Letters B}\ }\textbf {\bibinfo {volume} {355}},\ \bibinfo {pages} {245} (\bibinfo {year} {1995})}\BibitemShut {NoStop}%
\bibitem [{\citenamefont {Brandt}\ \emph {et~al.}(2018)\citenamefont {Brandt}, \citenamefont {Frenkel},\ and\ \citenamefont {McKeon}}]{Brandt:2018avq}%
  \BibitemOpen
  \bibfield  {author} {\bibinfo {author} {\bibfnamefont {F.~T.}\ \bibnamefont {Brandt}}, \bibinfo {author} {\bibfnamefont {J.}~\bibnamefont {Frenkel}},\ and\ \bibinfo {author} {\bibfnamefont {D.~G.~C.}\ \bibnamefont {McKeon}},\ }\href {https://doi.org/10.1103/PhysRevD.98.025024} {\bibfield  {journal} {\bibinfo  {journal} {Physical Review D}\ }\textbf {\bibinfo {volume} {98}},\ \bibinfo {pages} {025024} (\bibinfo {year} {2018})},\ \Eprint {https://arxiv.org/abs/1807.09823} {arXiv:1807.09823 [hep-th]} \BibitemShut {NoStop}%
\bibitem [{\citenamefont {Frenkel}\ and\ \citenamefont {Taylor}(2017)}]{Frenkel:2017xvm}%
  \BibitemOpen
  \bibfield  {author} {\bibinfo {author} {\bibfnamefont {J.}~\bibnamefont {Frenkel}}\ and\ \bibinfo {author} {\bibfnamefont {J.~C.}\ \bibnamefont {Taylor}},\ }\href {https://doi.org/10.1016/j.aop.2017.10.002} {\bibfield  {journal} {\bibinfo  {journal} {Annals of Physics}\ }\textbf {\bibinfo {volume} {387}},\ \bibinfo {pages} {1} (\bibinfo {year} {2017})},\ \Eprint {https://arxiv.org/abs/1703.10394} {arXiv:1703.10394 [hep-th]} \BibitemShut {NoStop}%
\bibitem [{\citenamefont {Shapiro}(2002)}]{Shapiro:2001rz}%
  \BibitemOpen
  \bibfield  {author} {\bibinfo {author} {\bibfnamefont {I.}~\bibnamefont {Shapiro}},\ }\href {https://doi.org/10.1016/S0370-1573(01)00030-8} {\bibfield  {journal} {\bibinfo  {journal} {Physics Reports}\ }\textbf {\bibinfo {volume} {357}},\ \bibinfo {pages} {113} (\bibinfo {year} {2002})},\ \Eprint {https://arxiv.org/abs/hep-th/0103093} {arXiv:hep-th/0103093} \BibitemShut {NoStop}%
\bibitem [{\citenamefont {Hehl}\ \emph {et~al.}(1976)\citenamefont {Hehl}, \citenamefont {Von Der~Heyde}, \citenamefont {Kerlick},\ and\ \citenamefont {Nester}}]{Hehl:1976kj}%
  \BibitemOpen
  \bibfield  {author} {\bibinfo {author} {\bibfnamefont {F.~W.}\ \bibnamefont {Hehl}}, \bibinfo {author} {\bibfnamefont {P.}~\bibnamefont {Von Der~Heyde}}, \bibinfo {author} {\bibfnamefont {G.~D.}\ \bibnamefont {Kerlick}},\ and\ \bibinfo {author} {\bibfnamefont {J.~M.}\ \bibnamefont {Nester}},\ }\href {https://doi.org/10.1103/RevModPhys.48.393} {\bibfield  {journal} {\bibinfo  {journal} {Reviews of Modern Physics}\ }\textbf {\bibinfo {volume} {48}},\ \bibinfo {pages} {393} (\bibinfo {year} {1976})}\BibitemShut {NoStop}%
\bibitem [{\citenamefont {Hehl}\ \emph {et~al.}(1995)\citenamefont {Hehl}, \citenamefont {McCrea}, \citenamefont {Mielke},\ and\ \citenamefont {Ne'eman}}]{Hehl:1994ue}%
  \BibitemOpen
  \bibfield  {author} {\bibinfo {author} {\bibfnamefont {F.~W.}\ \bibnamefont {Hehl}}, \bibinfo {author} {\bibfnamefont {J.}~\bibnamefont {McCrea}}, \bibinfo {author} {\bibfnamefont {E.~W.}\ \bibnamefont {Mielke}},\ and\ \bibinfo {author} {\bibfnamefont {Y.}~\bibnamefont {Ne'eman}},\ }\href {https://doi.org/10.1016/0370-1573(94)00111-F} {\bibfield  {journal} {\bibinfo  {journal} {Physics Reports}\ }\textbf {\bibinfo {volume} {258}},\ \bibinfo {pages} {1} (\bibinfo {year} {1995})}\BibitemShut {NoStop}%
\bibitem [{\citenamefont {Pereira}(2014)}]{Pereira:2013qza}%
  \BibitemOpen
  \bibfield  {author} {\bibinfo {author} {\bibfnamefont {J.~G.}\ \bibnamefont {Pereira}},\ }in\ \href {https://doi.org/10.1007/978-3-642-41992-8_11} {\emph {\bibinfo {booktitle} {Springer {{Handbook}} of {{Spacetime}}}}},\ \bibinfo {editor} {edited by\ \bibinfo {editor} {\bibfnamefont {A.}~\bibnamefont {Ashtekar}}\ and\ \bibinfo {editor} {\bibfnamefont {V.}~\bibnamefont {Petkov}}}\ (\bibinfo  {publisher} {Springer Berlin Heidelberg},\ \bibinfo {address} {Berlin, Heidelberg},\ \bibinfo {year} {2014})\ pp.\ \bibinfo {pages} {197--212}\BibitemShut {NoStop}%
\bibitem [{\citenamefont {Capper}\ \emph {et~al.}(1973)\citenamefont {Capper}, \citenamefont {Leibbrandt},\ and\ \citenamefont {Medrano}}]{Capper:1973pv}%
  \BibitemOpen
  \bibfield  {author} {\bibinfo {author} {\bibfnamefont {D.~M.}\ \bibnamefont {Capper}}, \bibinfo {author} {\bibfnamefont {G.}~\bibnamefont {Leibbrandt}},\ and\ \bibinfo {author} {\bibfnamefont {M.~R.}\ \bibnamefont {Medrano}},\ }\href {https://doi.org/10.1103/PhysRevD.8.4320} {\bibfield  {journal} {\bibinfo  {journal} {Physical Review D}\ }\textbf {\bibinfo {volume} {8}},\ \bibinfo {pages} {4320} (\bibinfo {year} {1973})}\BibitemShut {NoStop}%
\bibitem [{\citenamefont {Lautrup}(1967)}]{lautrup:1967}%
  \BibitemOpen
  \bibfield  {author} {\bibinfo {author} {\bibfnamefont {B.}~\bibnamefont {Lautrup}},\ }\href@noop {} {\bibfield  {journal} {\bibinfo  {journal} {Kong. Dan. Vid. Sel. Mat. Fys. Med.}\ }\textbf {\bibinfo {volume} {35}} (\bibinfo {year} {1967})}\BibitemShut {NoStop}%
\bibitem [{\citenamefont {Nakanishi}(1966)}]{Nakanishi:1966zz}%
  \BibitemOpen
  \bibfield  {author} {\bibinfo {author} {\bibfnamefont {N.}~\bibnamefont {Nakanishi}},\ }\href {https://doi.org/10.1143/PTP.35.1111} {\bibfield  {journal} {\bibinfo  {journal} {Progress of Theoretical Physics}\ }\textbf {\bibinfo {volume} {35}},\ \bibinfo {pages} {1111} (\bibinfo {year} {1966})}\BibitemShut {NoStop}%
\bibitem [{\citenamefont {Palatini}(1919)}]{palatini:1919}%
  \BibitemOpen
  \bibfield  {author} {\bibinfo {author} {\bibfnamefont {A.}~\bibnamefont {Palatini}},\ }\href {https://doi.org/10.1007/BF03014670} {\bibfield  {journal} {\bibinfo  {journal} {Rendiconti del Circolo Matematico di Palermo}\ }\textbf {\bibinfo {volume} {43}},\ \bibinfo {pages} {203} (\bibinfo {year} {1919})}\BibitemShut {NoStop}%
\bibitem [{\citenamefont {Chishtie}\ and\ \citenamefont {McKeon}(2012{\natexlab{b}})}]{Chishtie:2011wd}%
  \BibitemOpen
  \bibfield  {author} {\bibinfo {author} {\bibfnamefont {F.}~\bibnamefont {Chishtie}}\ and\ \bibinfo {author} {\bibfnamefont {D.~G.~C.}\ \bibnamefont {McKeon}},\ }\href {https://doi.org/10.1142/S0217751X12500777} {\bibfield  {journal} {\bibinfo  {journal} {International Journal of Modern Physics A}\ }\textbf {\bibinfo {volume} {27}},\ \bibinfo {pages} {1250077} (\bibinfo {year} {2012}{\natexlab{b}})},\ \Eprint {https://arxiv.org/abs/1110.1425} {arXiv:1110.1425} \BibitemShut {NoStop}%
\bibitem [{\citenamefont {Le~Bellac}(2011)}]{Bellac:2011kqa}%
  \BibitemOpen
  \bibfield  {author} {\bibinfo {author} {\bibfnamefont {M.}~\bibnamefont {Le~Bellac}},\ }\href {https://doi.org/10.1017/CBO9780511721700} {\emph {\bibinfo {title} {Thermal {{Field Theory}}}}},\ \bibinfo {edition} {1st}\ ed.\ (\bibinfo  {publisher} {Cambridge University Press},\ \bibinfo {address} {Cambridge, UK},\ \bibinfo {year} {2011})\BibitemShut {NoStop}%
\bibitem [{\citenamefont {Chishtie}\ and\ \citenamefont {McKeon}(2013)}]{Chishtie:2013fna}%
  \BibitemOpen
  \bibfield  {author} {\bibinfo {author} {\bibfnamefont {F.}~\bibnamefont {Chishtie}}\ and\ \bibinfo {author} {\bibfnamefont {D.~G.~C.}\ \bibnamefont {McKeon}},\ }\href {https://doi.org/10.1088/0264-9381/30/15/155002} {\bibfield  {journal} {\bibinfo  {journal} {Classical and Quantum Gravity}\ }\textbf {\bibinfo {volume} {30}},\ \bibinfo {pages} {155002} (\bibinfo {year} {2013})}\BibitemShut {NoStop}%
\bibitem [{\citenamefont {Romero}\ \emph {et~al.}(2022)\citenamefont {Romero}, \citenamefont {Montesinos},\ and\ \citenamefont {Escobedo}}]{Romero:2022kjh}%
  \BibitemOpen
  \bibfield  {author} {\bibinfo {author} {\bibfnamefont {J.}~\bibnamefont {Romero}}, \bibinfo {author} {\bibfnamefont {M.}~\bibnamefont {Montesinos}},\ and\ \bibinfo {author} {\bibfnamefont {R.}~\bibnamefont {Escobedo}},\ }\href {https://doi.org/10.1103/PhysRevD.106.084009} {\bibfield  {journal} {\bibinfo  {journal} {Physical Review D}\ }\textbf {\bibinfo {volume} {106}},\ \bibinfo {pages} {084009} (\bibinfo {year} {2022})},\ \Eprint {https://arxiv.org/abs/2210.07378} {arXiv:2210.07378 [gr-qc]} \BibitemShut {NoStop}%
\bibitem [{\citenamefont {Vasconcelos}(2020)}]{vasconcelos:2020}%
  \BibitemOpen
  \bibfield  {author} {\bibinfo {author} {\bibfnamefont {X.~M.}\ \bibnamefont {Vasconcelos}},\ }\emph {\bibinfo {title} {{Teorias de gauge no formalismo de primeira ordem com efeitos de temperatura finita}}},\ \href {https://doi.org/10.11606/D.43.2020.tde-27042020-170559} {\bibinfo {type} {{Mestrado em F{\'i}sica}}},\ \bibinfo  {school} {Universidade de S{\~a}o Paulo}, \bibinfo {address} {S{\~a}o Paulo} (\bibinfo {year} {2020})\BibitemShut {NoStop}%
\bibitem [{\citenamefont {Takahashi}(1957)}]{Takahashi:1957xn}%
  \BibitemOpen
  \bibfield  {author} {\bibinfo {author} {\bibfnamefont {Y.}~\bibnamefont {Takahashi}},\ }\href {https://doi.org/10.1007/BF02832514} {\bibfield  {journal} {\bibinfo  {journal} {Nuovo Cim.}\ }\textbf {\bibinfo {volume} {6}},\ \bibinfo {pages} {371} (\bibinfo {year} {1957})}\BibitemShut {NoStop}%
\bibitem [{\citenamefont {Dyson}(1949)}]{Dyson:1949ha}%
  \BibitemOpen
  \bibfield  {author} {\bibinfo {author} {\bibfnamefont {F.~J.}\ \bibnamefont {Dyson}},\ }\href {https://doi.org/10.1103/PhysRev.75.1736} {\bibfield  {journal} {\bibinfo  {journal} {Phys. Rev.}\ }\textbf {\bibinfo {volume} {75}},\ \bibinfo {pages} {1736} (\bibinfo {year} {1949})}\BibitemShut {NoStop}%
\bibitem [{\citenamefont {Schwinger}(1951)}]{Schwinger:1951ex}%
  \BibitemOpen
  \bibfield  {author} {\bibinfo {author} {\bibfnamefont {J.~S.}\ \bibnamefont {Schwinger}},\ }\href {https://doi.org/10.1073/pnas.37.7.452} {\bibfield  {journal} {\bibinfo  {journal} {Proc. Nat. Acad. Sci.}\ }\textbf {\bibinfo {volume} {37}},\ \bibinfo {pages} {452} (\bibinfo {year} {1951})}\BibitemShut {NoStop}%
\bibitem [{\citenamefont {Cornwall}\ \emph {et~al.}(1974)\citenamefont {Cornwall}, \citenamefont {Jackiw},\ and\ \citenamefont {Tomboulis}}]{Cornwall:1974vz}%
  \BibitemOpen
  \bibfield  {author} {\bibinfo {author} {\bibfnamefont {J.~M.}\ \bibnamefont {Cornwall}}, \bibinfo {author} {\bibfnamefont {R.}~\bibnamefont {Jackiw}},\ and\ \bibinfo {author} {\bibfnamefont {E.}~\bibnamefont {Tomboulis}},\ }\href {https://doi.org/10.1103/PhysRevD.10.2428} {\bibfield  {journal} {\bibinfo  {journal} {Physical Review D}\ }\textbf {\bibinfo {volume} {10}},\ \bibinfo {pages} {2428} (\bibinfo {year} {1974})}\BibitemShut {NoStop}%
\bibitem [{\citenamefont {Frank}\ \emph {et~al.}(2007)\citenamefont {Frank}, \citenamefont {Hahn}, \citenamefont {Heinemeyer}, \citenamefont {Hollik}, \citenamefont {Rzehak},\ and\ \citenamefont {Weiglein}}]{Frank:2006yh}%
  \BibitemOpen
  \bibfield  {author} {\bibinfo {author} {\bibfnamefont {M.}~\bibnamefont {Frank}}, \bibinfo {author} {\bibfnamefont {T.}~\bibnamefont {Hahn}}, \bibinfo {author} {\bibfnamefont {S.}~\bibnamefont {Heinemeyer}}, \bibinfo {author} {\bibfnamefont {W.}~\bibnamefont {Hollik}}, \bibinfo {author} {\bibfnamefont {H.}~\bibnamefont {Rzehak}},\ and\ \bibinfo {author} {\bibfnamefont {G.}~\bibnamefont {Weiglein}},\ }\href {https://doi.org/10.1088/1126-6708/2007/02/047} {\bibfield  {journal} {\bibinfo  {journal} {Journal of High Energy Physics}\ }\textbf {\bibinfo {volume} {2007}},\ \bibinfo {pages} {047} (\bibinfo {year} {2007})}\BibitemShut {NoStop}%
\bibitem [{\citenamefont {Aoyama}\ \emph {et~al.}(2018)\citenamefont {Aoyama}, \citenamefont {Kinoshita},\ and\ \citenamefont {Nio}}]{Aoyama:2017uqe}%
  \BibitemOpen
  \bibfield  {author} {\bibinfo {author} {\bibfnamefont {T.}~\bibnamefont {Aoyama}}, \bibinfo {author} {\bibfnamefont {T.}~\bibnamefont {Kinoshita}},\ and\ \bibinfo {author} {\bibfnamefont {M.}~\bibnamefont {Nio}},\ }\href {https://doi.org/10.1103/PhysRevD.97.036001} {\bibfield  {journal} {\bibinfo  {journal} {Physical Review D}\ }\textbf {\bibinfo {volume} {97}},\ \bibinfo {pages} {036001} (\bibinfo {year} {2018})}\BibitemShut {NoStop}%
\bibitem [{\citenamefont {Eides}\ and\ \citenamefont {Shelyuto}(2022)}]{Eides:2021wuv}%
  \BibitemOpen
  \bibfield  {author} {\bibinfo {author} {\bibfnamefont {M.~I.}\ \bibnamefont {Eides}}\ and\ \bibinfo {author} {\bibfnamefont {V.~A.}\ \bibnamefont {Shelyuto}},\ }\href {https://doi.org/10.1103/PhysRevA.105.012803} {\bibfield  {journal} {\bibinfo  {journal} {Physical Review A}\ }\textbf {\bibinfo {volume} {105}},\ \bibinfo {pages} {012803} (\bibinfo {year} {2022})}\BibitemShut {NoStop}%
\bibitem [{\citenamefont {Hanneke}\ \emph {et~al.}(2008)\citenamefont {Hanneke}, \citenamefont {Fogwell},\ and\ \citenamefont {Gabrielse}}]{Hanneke:2008tm}%
  \BibitemOpen
  \bibfield  {author} {\bibinfo {author} {\bibfnamefont {D.}~\bibnamefont {Hanneke}}, \bibinfo {author} {\bibfnamefont {S.}~\bibnamefont {Fogwell}},\ and\ \bibinfo {author} {\bibfnamefont {G.}~\bibnamefont {Gabrielse}},\ }\href {https://doi.org/10.1103/PhysRevLett.100.120801} {\bibfield  {journal} {\bibinfo  {journal} {Physical Review Letters}\ }\textbf {\bibinfo {volume} {100}},\ \bibinfo {pages} {120801} (\bibinfo {year} {2008})}\BibitemShut {NoStop}%
\bibitem [{\citenamefont {Crivelli}(2018)}]{Crivelli:2018vfe}%
  \BibitemOpen
  \bibfield  {author} {\bibinfo {author} {\bibfnamefont {P.}~\bibnamefont {Crivelli}},\ }\href {https://doi.org/10.1007/s10751-018-1525-z} {\bibfield  {journal} {\bibinfo  {journal} {Hyperfine Interactions}\ }\textbf {\bibinfo {volume} {239}},\ \bibinfo {pages} {49} (\bibinfo {year} {2018})}\BibitemShut {NoStop}%
\bibitem [{\citenamefont {Witten}(1988)}]{Witten:1988hc}%
  \BibitemOpen
  \bibfield  {author} {\bibinfo {author} {\bibfnamefont {E.}~\bibnamefont {Witten}},\ }\href {https://doi.org/10.1016/0550-3213(88)90143-5} {\bibfield  {journal} {\bibinfo  {journal} {Nuclear Physics B}\ }\textbf {\bibinfo {volume} {311}},\ \bibinfo {pages} {46} (\bibinfo {year} {1988})}\BibitemShut {NoStop}%
\bibitem [{\citenamefont {Green}\ \emph {et~al.}(2011)\citenamefont {Green}, \citenamefont {Kiriushcheva},\ and\ \citenamefont {Kuzmin}}]{Green:2008qa}%
  \BibitemOpen
  \bibfield  {author} {\bibinfo {author} {\bibfnamefont {K.~R.}\ \bibnamefont {Green}}, \bibinfo {author} {\bibfnamefont {N.}~\bibnamefont {Kiriushcheva}},\ and\ \bibinfo {author} {\bibfnamefont {S.~V.}\ \bibnamefont {Kuzmin}},\ }\href {https://doi.org/10.1140/epjc/s10052-011-1678-2} {\bibfield  {journal} {\bibinfo  {journal} {The European Physical Journal C}\ }\textbf {\bibinfo {volume} {71}},\ \bibinfo {pages} {1678} (\bibinfo {year} {2011})}\BibitemShut {NoStop}%
\bibitem [{\citenamefont {Chamseddine}\ and\ \citenamefont {Wyler}(1989)}]{Chamseddine:1989yz}%
  \BibitemOpen
  \bibfield  {author} {\bibinfo {author} {\bibfnamefont {A.}~\bibnamefont {Chamseddine}}\ and\ \bibinfo {author} {\bibfnamefont {D.}~\bibnamefont {Wyler}},\ }\href {https://doi.org/10.1016/0370-2693(89)90528-5} {\bibfield  {journal} {\bibinfo  {journal} {Physics Letters B}\ }\textbf {\bibinfo {volume} {228}},\ \bibinfo {pages} {75} (\bibinfo {year} {1989})}\BibitemShut {NoStop}%
\bibitem [{\citenamefont {Borisov}\ \emph {et~al.}(2016)\citenamefont {Borisov}, \citenamefont {Mamaev},\ and\ \citenamefont {Bizyaev}}]{borisov:2016}%
  \BibitemOpen
  \bibfield  {author} {\bibinfo {author} {\bibfnamefont {A.~V.}\ \bibnamefont {Borisov}}, \bibinfo {author} {\bibfnamefont {I.~S.}\ \bibnamefont {Mamaev}},\ and\ \bibinfo {author} {\bibfnamefont {I.~A.}\ \bibnamefont {Bizyaev}},\ }\href {https://doi.org/10.1134/S1560354716040055} {\bibfield  {journal} {\bibinfo  {journal} {Regular and Chaotic Dynamics}\ }\textbf {\bibinfo {volume} {21}},\ \bibinfo {pages} {455} (\bibinfo {year} {2016})}\BibitemShut {NoStop}%
\bibitem [{\citenamefont {{de Le{\'o}n}}(2012)}]{deleon:2012}%
  \BibitemOpen
  \bibfield  {author} {\bibinfo {author} {\bibfnamefont {M.}~\bibnamefont {{de Le{\'o}n}}},\ }\href {https://doi.org/10.1007/s13398-011-0046-2} {\bibfield  {journal} {\bibinfo  {journal} {Revista de la Real Academia de Ciencias Exactas, Fisicas y Naturales. Serie A. Matematicas}\ }\textbf {\bibinfo {volume} {106}},\ \bibinfo {pages} {191} (\bibinfo {year} {2012})}\BibitemShut {NoStop}%
\bibitem [{\citenamefont {Arnold}\ \emph {et~al.}(2010)\citenamefont {Arnold}, \citenamefont {Valery},\ and\ \citenamefont {Kozlov}}]{kozlov:2010}%
  \BibitemOpen
  \bibfield  {author} {\bibinfo {author} {\bibfnamefont {V.~I.}\ \bibnamefont {Arnold}}, \bibinfo {author} {\bibfnamefont {A.~I.~N.}\ \bibnamefont {Valery}},\ and\ \bibinfo {author} {\bibfnamefont {E.~K.}\ \bibnamefont {Kozlov}},\ }\href {https://doi.org/10.1007/978-3-540-48926-9} {\emph {\bibinfo {title} {Mathematical Aspects of Classical and Celestial Mechanics}}},\ \bibinfo {edition} {3rd}\ ed.,\ Encyclopaedia of Mathematical Sciences\ (\bibinfo  {publisher} {Springer},\ \bibinfo {address} {Berlin, Heidelberg},\ \bibinfo {year} {2010})\BibitemShut {NoStop}%
\bibitem [{\citenamefont {Pagani}\ \emph {et~al.}(1987)\citenamefont {Pagani}, \citenamefont {Tecchiolli},\ and\ \citenamefont {Zerbini}}]{Pagani:1987ue}%
  \BibitemOpen
  \bibfield  {author} {\bibinfo {author} {\bibfnamefont {E.}~\bibnamefont {Pagani}}, \bibinfo {author} {\bibfnamefont {G.}~\bibnamefont {Tecchiolli}},\ and\ \bibinfo {author} {\bibfnamefont {S.}~\bibnamefont {Zerbini}},\ }\href {https://doi.org/10.1007/BF00402140} {\bibfield  {journal} {\bibinfo  {journal} {Letters in Mathematical Physics}\ }\textbf {\bibinfo {volume} {14}},\ \bibinfo {pages} {311} (\bibinfo {year} {1987})}\BibitemShut {NoStop}%
\bibitem [{\citenamefont {Koopman}(1931)}]{koopman:1931}%
  \BibitemOpen
  \bibfield  {author} {\bibinfo {author} {\bibfnamefont {B.~O.}\ \bibnamefont {Koopman}},\ }\href {https://doi.org/10.1073/pnas.17.5.315} {\bibfield  {journal} {\bibinfo  {journal} {Proceedings of the National Academy of Sciences of the United States of America}\ }\textbf {\bibinfo {volume} {17}},\ \bibinfo {pages} {315} (\bibinfo {year} {1931})}\BibitemShut {NoStop}%
\bibitem [{\citenamefont {V.~Neumann}(1932)}]{v.neumann:1932}%
  \BibitemOpen
  \bibfield  {author} {\bibinfo {author} {\bibfnamefont {J.}~\bibnamefont {V.~Neumann}},\ }\href {https://doi.org/10.2307/1968537} {\bibfield  {journal} {\bibinfo  {journal} {The Annals of Mathematics}\ }\textbf {\bibinfo {volume} {33}},\ \bibinfo {pages} {587} (\bibinfo {year} {1932})}\BibitemShut {NoStop}%
\bibitem [{\citenamefont {Ray}(1966{\natexlab{a}})}]{ray:1966}%
  \BibitemOpen
  \bibfield  {author} {\bibinfo {author} {\bibfnamefont {J.~R.}\ \bibnamefont {Ray}},\ }\href {https://doi.org/10.1119/1.1972703} {\bibfield  {journal} {\bibinfo  {journal} {American Journal of Physics}\ }\textbf {\bibinfo {volume} {34}},\ \bibinfo {pages} {1202} (\bibinfo {year} {1966}{\natexlab{a}})}\BibitemShut {NoStop}%
\bibitem [{\citenamefont {Ray}(1966{\natexlab{b}})}]{ray:1966a}%
  \BibitemOpen
  \bibfield  {author} {\bibinfo {author} {\bibfnamefont {J.~R.}\ \bibnamefont {Ray}},\ }\href {https://doi.org/10.1119/1.1973007} {\bibfield  {journal} {\bibinfo  {journal} {American Journal of Physics}\ }\textbf {\bibinfo {volume} {34}},\ \bibinfo {pages} {406} (\bibinfo {year} {1966}{\natexlab{b}})}\BibitemShut {NoStop}%
\bibitem [{\citenamefont {Flannery}(2005)}]{flannery:2005}%
  \BibitemOpen
  \bibfield  {author} {\bibinfo {author} {\bibfnamefont {M.~R.}\ \bibnamefont {Flannery}},\ }\href {https://doi.org/10.1119/1.1830501} {\bibfield  {journal} {\bibinfo  {journal} {American Journal of Physics}\ }\textbf {\bibinfo {volume} {73}},\ \bibinfo {pages} {265} (\bibinfo {year} {2005})}\BibitemShut {NoStop}%
\bibitem [{\citenamefont {Lemos}(2022)}]{lemos:2022}%
  \BibitemOpen
  \bibfield  {author} {\bibinfo {author} {\bibfnamefont {N.~A.}\ \bibnamefont {Lemos}},\ }\href {https://doi.org/10.1007/s00707-021-03106-1} {\bibfield  {journal} {\bibinfo  {journal} {Acta Mechanica}\ }\textbf {\bibinfo {volume} {233}},\ \bibinfo {pages} {47} (\bibinfo {year} {2022})}\BibitemShut {NoStop}%
\bibitem [{\citenamefont {Lewis}\ and\ \citenamefont {Murray}(1995)}]{lewis:1995}%
  \BibitemOpen
  \bibfield  {author} {\bibinfo {author} {\bibfnamefont {A.~D.}\ \bibnamefont {Lewis}}\ and\ \bibinfo {author} {\bibfnamefont {R.~M.}\ \bibnamefont {Murray}},\ }\href {https://doi.org/10.1016/0020-7462(95)00024-0} {\bibfield  {journal} {\bibinfo  {journal} {International Journal of Non-Linear Mechanics}\ }\textbf {\bibinfo {volume} {30}},\ \bibinfo {pages} {793} (\bibinfo {year} {1995})}\BibitemShut {NoStop}%
\bibitem [{\citenamefont {Bloch}\ and\ \citenamefont {Rojo}(2008)}]{bloch:2008}%
  \BibitemOpen
  \bibfield  {author} {\bibinfo {author} {\bibfnamefont {A.~M.}\ \bibnamefont {Bloch}}\ and\ \bibinfo {author} {\bibfnamefont {A.~G.}\ \bibnamefont {Rojo}},\ }\href {https://doi.org/10.1103/PhysRevLett.101.030402} {\bibfield  {journal} {\bibinfo  {journal} {Physical Review Letters}\ }\textbf {\bibinfo {volume} {101}},\ \bibinfo {pages} {030402} (\bibinfo {year} {2008})}\BibitemShut {NoStop}%
\bibitem [{\citenamefont {Abud~Filho}\ \emph {et~al.}(1983)\citenamefont {Abud~Filho}, \citenamefont {Gomes}, \citenamefont {Simao},\ and\ \citenamefont {Coutinho}}]{abudfilho:1983}%
  \BibitemOpen
  \bibfield  {author} {\bibinfo {author} {\bibfnamefont {M.}~\bibnamefont {Abud~Filho}}, \bibinfo {author} {\bibfnamefont {L.}~\bibnamefont {Gomes}}, \bibinfo {author} {\bibfnamefont {F.}~\bibnamefont {Simao}},\ and\ \bibinfo {author} {\bibfnamefont {F.}~\bibnamefont {Coutinho}},\ }\href {https://doi.org/http://inis.iaea.org/search/search.aspx?orig_q=RN:16058866} {\bibfield  {journal} {\bibinfo  {journal} {Revista Brasileira de F{\'i}sica}\ }\textbf {\bibinfo {volume} {13}},\ \bibinfo {pages} {384} (\bibinfo {year} {1983})}\BibitemShut {NoStop}%
\bibitem [{\citenamefont {Kamefuchi}\ \emph {et~al.}(1961)\citenamefont {Kamefuchi}, \citenamefont {O'Raifeartaigh},\ and\ \citenamefont {Salam}}]{Kamefuchi:1961sb}%
  \BibitemOpen
  \bibfield  {author} {\bibinfo {author} {\bibfnamefont {S.}~\bibnamefont {Kamefuchi}}, \bibinfo {author} {\bibfnamefont {L.}~\bibnamefont {O'Raifeartaigh}},\ and\ \bibinfo {author} {\bibfnamefont {A.}~\bibnamefont {Salam}},\ }\href {https://doi.org/10.1016/0029-5582(61)90056-6} {\bibfield  {journal} {\bibinfo  {journal} {Nuclear Physics}\ }\textbf {\bibinfo {volume} {28}},\ \bibinfo {pages} {529} (\bibinfo {year} {1961})}\BibitemShut {NoStop}%
\bibitem [{\citenamefont {Bastianelli}(1992)}]{Bastianelli:1991be}%
  \BibitemOpen
  \bibfield  {author} {\bibinfo {author} {\bibfnamefont {F.}~\bibnamefont {Bastianelli}},\ }\href {https://doi.org/10.1016/0550-3213(92)90070-R} {\bibfield  {journal} {\bibinfo  {journal} {Nuclear Physics B}\ }\textbf {\bibinfo {volume} {376}},\ \bibinfo {pages} {113} (\bibinfo {year} {1992})}\BibitemShut {NoStop}%
\bibitem [{\citenamefont {Bastianelli}\ and\ \citenamefont {Corradini}(1999)}]{Bastianelli:1998jb}%
  \BibitemOpen
  \bibfield  {author} {\bibinfo {author} {\bibfnamefont {F.}~\bibnamefont {Bastianelli}}\ and\ \bibinfo {author} {\bibfnamefont {O.}~\bibnamefont {Corradini}},\ }\href {https://doi.org/10.1103/PhysRevD.60.044014} {\bibfield  {journal} {\bibinfo  {journal} {Physical Review D}\ }\textbf {\bibinfo {volume} {60}},\ \bibinfo {pages} {044014} (\bibinfo {year} {1999})}\BibitemShut {NoStop}%
\bibitem [{\citenamefont {McKeon}\ \emph {et~al.}(2024)\citenamefont {McKeon}, \citenamefont {Brandt},\ and\ \citenamefont {{Martins-Filho}}}]{McKeon:2024psy}%
  \BibitemOpen
  \bibfield  {author} {\bibinfo {author} {\bibfnamefont {D.~G.~C.}\ \bibnamefont {McKeon}}, \bibinfo {author} {\bibfnamefont {F.~T.}\ \bibnamefont {Brandt}},\ and\ \bibinfo {author} {\bibfnamefont {S.}~\bibnamefont {{Martins-Filho}}},\ }\href {https://doi.org/10.1140/epjc/s10052-024-12764-z} {\bibfield  {journal} {\bibinfo  {journal} {The European Physical Journal C}\ }\textbf {\bibinfo {volume} {84}},\ \bibinfo {pages} {399} (\bibinfo {year} {2024})},\ \Eprint {https://arxiv.org/abs/2401.11133} {arXiv:2401.11133 [hep-th]} \BibitemShut {NoStop}%
\bibitem [{\citenamefont {Brandt}\ \emph {et~al.}(2024)\citenamefont {Brandt}, \citenamefont {Frenkel}, \citenamefont {{Martins-Filho}},\ and\ \citenamefont {McKeon}}]{Brandt:2024rsy}%
  \BibitemOpen
  \bibfield  {author} {\bibinfo {author} {\bibfnamefont {F.~T.}\ \bibnamefont {Brandt}}, \bibinfo {author} {\bibfnamefont {J.}~\bibnamefont {Frenkel}}, \bibinfo {author} {\bibfnamefont {S.}~\bibnamefont {{Martins-Filho}}},\ and\ \bibinfo {author} {\bibfnamefont {D.~G.~C.}\ \bibnamefont {McKeon}},\ }\href {https://doi.org/10.1016/j.aop.2024.169607} {\bibfield  {journal} {\bibinfo  {journal} {Annals of Physics}\ }\textbf {\bibinfo {volume} {462}},\ \bibinfo {pages} {169607} (\bibinfo {year} {2024})},\ \Eprint {https://arxiv.org/abs/2401.16343} {arXiv:2401.16343 [hep-th]} \BibitemShut {NoStop}%
\bibitem [{\citenamefont {Berezin}(1987)}]{berezin:1987}%
  \BibitemOpen
  \bibfield  {author} {\bibinfo {author} {\bibfnamefont {F.~A.}\ \bibnamefont {Berezin}},\ }\href {https://doi.org/10.1007/978-94-017-1963-6} {\emph {\bibinfo {title} {Introduction to {{Superanalysis}}}}},\ edited by\ \bibinfo {editor} {\bibfnamefont {A.~A.}\ \bibnamefont {Kirillov}}\ (\bibinfo  {publisher} {Springer Netherlands},\ \bibinfo {address} {Dordrecht},\ \bibinfo {year} {1987})\BibitemShut {NoStop}%
\bibitem [{\citenamefont {Kapusta}\ and\ \citenamefont {Gale}(2006)}]{Kapusta:2006pm}%
  \BibitemOpen
  \bibfield  {author} {\bibinfo {author} {\bibfnamefont {J.~I.}\ \bibnamefont {Kapusta}}\ and\ \bibinfo {author} {\bibfnamefont {C.}~\bibnamefont {Gale}},\ }\href {https://doi.org/10.1017/CBO9780511535130} {\emph {\bibinfo {title} {Finite-{{Temperature Field Theory}}: {{Principles}} and {{Applications}}}}},\ \bibinfo {edition} {2nd}\ ed.\ (\bibinfo  {publisher} {Cambridge University Press},\ \bibinfo {year} {2006})\BibitemShut {NoStop}%
\bibitem [{\citenamefont {Brandt}\ \emph {et~al.}(2021{\natexlab{c}})\citenamefont {Brandt}, \citenamefont {Frenkel}, \citenamefont {{Martins-Filho}}, \citenamefont {McKeon},\ and\ \citenamefont {Sakoda}}]{Brandt:2021nse}%
  \BibitemOpen
  \bibfield  {author} {\bibinfo {author} {\bibfnamefont {F.~T.}\ \bibnamefont {Brandt}}, \bibinfo {author} {\bibfnamefont {J.}~\bibnamefont {Frenkel}}, \bibinfo {author} {\bibfnamefont {S.}~\bibnamefont {{Martins-Filho}}}, \bibinfo {author} {\bibfnamefont {D.~G.~C.}\ \bibnamefont {McKeon}},\ and\ \bibinfo {author} {\bibfnamefont {G.~S.~S.}\ \bibnamefont {Sakoda}},\ }\href {https://doi.org/10.1103/PhysRevD.104.105007} {\bibfield  {journal} {\bibinfo  {journal} {Physical Review D}\ }\textbf {\bibinfo {volume} {104}},\ \bibinfo {pages} {105007} (\bibinfo {year} {2021}{\natexlab{c}})},\ \Eprint {https://arxiv.org/abs/2110.07694} {arXiv:2110.07694 [hep-th]} \BibitemShut {NoStop}%
\end{thebibliography}%
\clearemptydoublepage
\end{SingleSpace} 
\end{document}